\documentclass[11pt,a4paper]{article}
\usepackage[T1]{fontenc}
\usepackage[utf8]{inputenc}
\usepackage{lmodern,microtype}
\usepackage[margin=25mm]{geometry}
\usepackage{amsmath,amssymb,amsthm,mathtools,mathrsfs}
\usepackage{booktabs,longtable,array,enumitem,tabularx}
\usepackage[dvipsnames]{xcolor}
\usepackage[numbers,sort&compress]{natbib}
\usepackage{xurl}
\usepackage[hidelinks,bookmarksnumbered]{hyperref}
\usepackage[nameinlink,noabbrev]{cleveref}
\usepackage{graphicx,capt-of,needspace,float}
\usepackage[font=small,labelfont=bf]{caption}
\setlist{itemsep=0.2em,topsep=0.5em,parsep=0em}
\newtheorem{theorem}{Theorem}[section]
\newtheorem{proposition}[theorem]{Proposition}
\newtheorem{lemma}[theorem]{Lemma}
\newtheorem{corollary}[theorem]{Corollary}
\theoremstyle{definition}

\newtheorem{assumption}[theorem]{Assumption}
\theoremstyle{remark}
\newtheorem{remark}[theorem]{Remark}
\newcommand{\R}{\mathbb R}
\newcommand{\C}{\mathbb C}
\newcommand{\Z}{\mathbb Z}
\newcommand{\N}{\mathbb N}
\newcommand{\Q}{\mathbb Q}
\newcommand{\eps}{\varepsilon}
\newcommand{\nnz}{\operatorname{nnz}}
\newcommand{\polylog}{\operatorname{polylog}}
\newcommand{\Ot}{\widetilde O}
\newcommand{\norm}[1]{\left\lVert #1\right\rVert}
\newcommand{\abs}[1]{\left\lvert #1\right\rvert}
\newcommand{\ip}[2]{\left\langle #1,#2\right\rangle}
\newcommand{\SI}{Supplementary Information}
\usepackage{titletoc}

\newcommand{\spec}{\operatorname{spec}}

\newcommand{\dist}{\operatorname{dist}}
\newcommand{\curl}{\operatorname{curl}}

\newcommand{\diag}{\operatorname{diag}}

\newcommand{\tO}{\widetilde O}
\newcommand{\ee}{\eta_\Gamma}
\newcommand{\Cs}{C_\star}
\newcommand{\cs}{c_\star}
\newcommand{\Ls}{L_\star}
\newcommand{\ds}{d_\star}
\newcommand{\Cc}{\mathcal C}

\newcommand{\MM}{\mathcal M}
\newcommand{\one}{\mathbf 1}
\newcommand{\dd}{\,\mathrm d}
\newcommand{\sipart}[2]{%
  \par\bigskip\phantomsection
  \addcontentsline{toc}{part}{Part #1\texorpdfstring{\quad}{: }#2}%
  {\centering\Large\bfseries Part #1\par\smallskip\large #2\par}%
  \nopagebreak\medskip}

\hypersetup{pdftitle={Maxwell's equations are universal for sparse linear algebra},pdfauthor={Yannick Augenstein}}
\title{\textbf{Maxwell's equations are universal for sparse linear algebra}}
\author{Yannick Augenstein}
\date{}
\begin{document}
\maketitle
\begin{abstract}
Optical and wave-based analog computing has re-emerged as a route to
accelerating linear algebra, with metamaterials and photonic circuits shown to
invert matrices and solve integral equations. However, it has remained unclear
which problems such passive structures can represent, and at what cost, once
encoding, precision, conditioning and readout are accounted for. Here, we show
that every polynomially conditioned sparse linear system can be compiled into a
three-dimensional dielectric -- a cavity filled with lossless, isotropic
material of bounded permittivity -- whose time-harmonic response encodes its
solution. For $m$ nonzero coefficients, the volume of the structure and the
cost of writing it down and reading out the answer scale as $\Ot(m^{3/2})$,
coefficients require only $O(\log m)$ bits, and every field that approximately
satisfies Maxwell's equations decodes to an approximate solution. We prove the
construction for the continuum equations and give a computer-assisted version
for the Yee discretization underlying finite-difference solvers. Read in reverse, it
implies that electromagnetic simulation is, up to this overhead, as hard as
general sparse linear algebra: a sufficiently fast general-purpose Maxwell
solver would yield a faster sparse linear solver.

\end{abstract}
\section{Introduction}

Solving large sparse systems of linear equations is the central task of
scientific computing, and electromagnetics is no exception: discretizing
Maxwell's equations with finite differences or finite elements turns the
analysis and design of a photonic device into a sparse linear system with
millions to billions of unknowns~\citep{yee1966,Oskooi2010,ShinFan2012}.
Optical analog computing runs this correspondence in the opposite direction.
Instead of computing the response of a given structure, one designs a structure
whose response computes something else -- e.g., a derivative, a convolution or
the solution of an equation~\citep{Silva2014,ZangenehNejad2021,McMahon2023}.
Metamaterials and inverse-designed metastructures have been shown to solve
integral equations~\citep{estakhri2019,Camacho2021,cordaro2023}, reconfigurable
networks invert matrices~\citep{tzarouchis2022,chen2024}, and integrated
photonic circuits solve differential equations~\citep{yang2014,ye2024}. These
build on a broader effort to perform linear algebra with light, from universal
interferometer meshes~\citep{reck1994,Miller2013SelfConfiguring,Carolan2015,Clements2016,Bogaerts2020}
to optical neural networks and photonic tensor
cores~\citep{Shen2017,Lin2018,Feldmann2021,Wetzstein2020,Shastri2021,Wright2022}.

Mathematical constructions also establish the expressive power of
electromagnetic networks. \citet{MiltonSeppecher2010} characterize
terminal-response matrices at a fixed frequency, including arbitrary real
symmetric matrices in the lossless case. Their continuum construction uses
dielectric and magnetic elements with parameters tending to zero or infinity,
and leaves the accuracy of approximations with finite material parameters
open. Reconfigurable wave networks support general matrix inversion,
including generalized inverses, with bounds on convergence
time~\citep{tzarouchis2022}.

Whether such devices can ever offer a computational advantage is decided by
details that are easy to overlook in a demonstration: the cost of programming
the device, the precision with which it must be built and read out, the
sensitivity of the answer to imperfections, and the size of the structure as
the problem grows~\citep{Aaronson2015FinePrint,McMahon2023}. The same questions
were asked of analog computers long before the current interest in photonics,
and they have a precise formulation in the theory of
computation~\citep{Shannon1941,Vergis1986,BSS1989,SiegelmannSontag1994,BournezPouly2021}.
Physical systems as simple as waves, rays and fluids can embed universal
computation~\citep{PourElRichards1981,Moore1990,ReifTygarYoshida1994,Cubitt2015,Tao2016,Cardona2021},
but these constructions establish what is computable in principle, not what a
physical system computes efficiently. For optical equation solvers, the
relevant question is quantitative: can general sparse systems be represented
by a passive dielectric with fixed material bounds while controlling
construction size, conditioning and precision?

Electromagnetic simulation faces the same question from the other side. The
linear systems produced by Maxwell solvers are highly structured -- local,
three-dimensional and determined by a scalar material distribution -- and fast
solvers exploit this structure wherever they
can~\citep{George1973,LiptonRoseTarjan1979,EngquistYing2011CPAM,Martinsson2019}.
However, time-harmonic wave problems are notoriously hard to solve
iteratively~\citep{ErnstGander2012,GanderZhang2019,MoiolaSpence2014}, and it
is not known whether their structure makes them fundamentally easier than
general sparse systems. For other structured classes this question has sharp
answers. Graph Laplacians can be solved in nearly linear
time~\citep{SpielmanTeng2014}, whereas the stiffness matrices of trusses and
multi-commodity flow problems are as hard as general sparse
systems~\citep{kyngzhang2017,DingKyngProbstGutenbergZhang2022,DingKyngZhang2022}.
To the best of our knowledge, no reduction from general sparse systems to
continuum Maxwell equations with fixed permeability and bounded scalar
permittivity has established explicit bounds on size, conditioning and
precision.

Both questions ask how expressive a dielectric is, i.e., which equations can be
written into a material distribution, and at what cost. Here, we answer this
question with an explicit compiler. Given a sparse system $Mx=b$ with $m$
nonzero coefficients, polynomial condition number and coefficients of
$O(\log m)$ bits, the compiler outputs a three-dimensional dielectric inside a
perfectly conducting box, a source and a finite list of field measurements,
such that every field approximately satisfying Maxwell's equations yields an
approximate solution of $Mx=b$. The matrix may be nonsymmetric, its entries may
have either sign and its sparsity pattern is arbitrary, while the dielectric is
lossless, isotropic and has a permittivity between 1 and 25. The volume of the
structure and the computational cost of encoding and decoding all scale as
$\Ot(m^{3/2})$, where the tilde hides logarithmic factors. We prove this for
the continuum Maxwell equations and give a computer-assisted construction with
the same scaling for the Yee discretization~\citep{yee1966} used by
finite-difference solvers.

Physically, the construction is the familiar picture of coupled defect modes
in a photonic crystal~\citep{Yariv1999,Busch2003,Joannopoulos2008}, made
quantitative. Each unknown is carried by the amplitude of a localized mode,
couplings between unknowns arise from the overlap of neighboring modes, and
auxiliary modes mediate the signed and nonsymmetric couplings that a
reciprocal dielectric can not provide directly. The proof controls everything
this picture neglects -- the remaining field components, the unwanted
couplings summed over the whole device, the finite boundary and near-resonances
of the cavity -- tightly enough that the encoded equations survive. Read in
reverse, the compiler turns any electromagnetic solver into a sparse linear
solver. A solver that handles the resulting family of structures in time
$N^\theta$, with $N$ the size of the structure, therefore solves sparse linear
systems in time $\Ot(m^{3/2}+m^{3\theta/2})$.

We first describe the compiler and its guarantees, then the physical mechanism
that makes it work, the recovery of answers from approximate fields and two
small numerical realizations. We then
discuss what the result implies for optical computing and for electromagnetic
simulation. Methods summarizes the formal statements, and the \SI\ contains the
complete proofs.

\section{Results}

\subsection{From sparse equations to a dielectric}

The compiler takes a nonsingular rational system $Mx=b$ with $m$ nonzero
matrix entries and a requested accuracy $\eta$, measured as the relative
residual $\norm{Mx-b}_2\le\eta\norm b_2$. Two numerical conditions define the
class of inputs: the condition number $\kappa_2(M)$ grows at most as a fixed
power of $m$, and $\eta$ decreases at most as a fixed inverse power of $m$.
Each coefficient has $O(\log m)$ bits. These conditions fix the accuracy
regime, while the pattern of couplings between unknowns is unrestricted.

\begin{figure}[tbp]
\centering
\includegraphics[width=\linewidth]{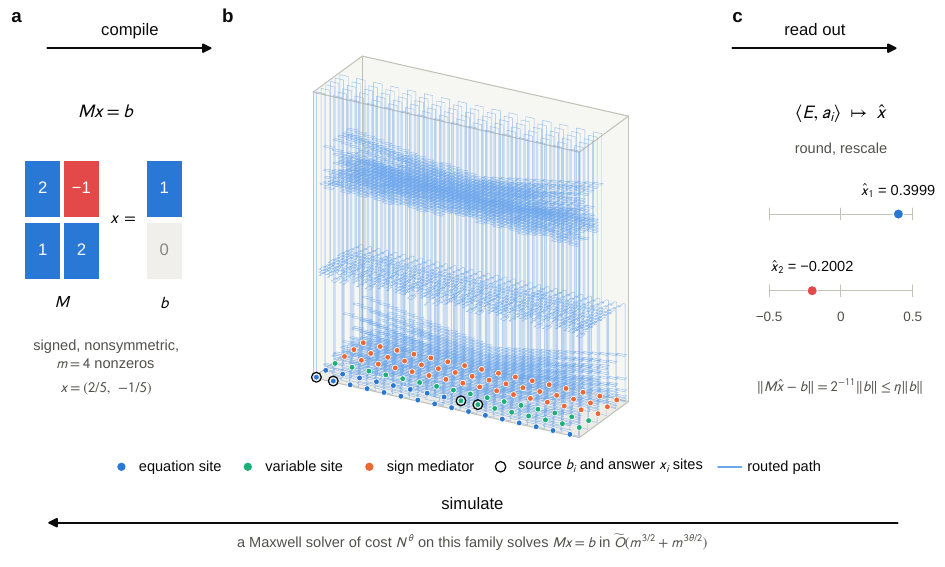}
\caption{\textbf{A sparse linear system and a dielectric cavity encode each other.}
\textbf{a}, The input is a sparse system $Mx=b$ with $m$ signed rational coefficients; shown is the
$2\times2$ example $M=\bigl(\begin{smallmatrix}2&-1\\1&2\end{smallmatrix}\bigr)$, $b=(1,0)$, whose
solution is $x=(2/5,\,-1/5)$. Blue and red cells are positive and negative entries.
\textbf{b}, The compiler expresses the equations through a local network of 96 sites (equation sites,
variable sites and sign mediators, each connected to at most three others) and routes its 96
connections as paths through a $505\times169\times3105$ integer grid. The drawing is the network the
compiler produced for the example in a, with the routing axis compressed sixfold for display; the paths
are long because the compiler equalizes their lengths. Each site becomes a broad, weak defect in the
periodic dielectric that fills the box, each path becomes a chain of such defects, and the walls are
perfect conductors. Ringed sites carry the source $b$ and the answer $x$.
\textbf{c}, A finite list of field moments $\langle E,a_i\rangle$ at the answer sites, rounded and
rescaled, returns $\hat x$. For the example, exact arithmetic on the compiled network gives
$\hat x=(819/2048,\,-205/1024)$ with relative residual $2^{-11}$, within the requested tolerance
$\eta=1/4$. Read from left to right (compile, read out), the cavity is an optical computer for $Mx=b$;
read from right to left (simulate), a Maxwell solver that handles this family of cavities in time
$N^{\theta}$, where $N$ counts grid cells, is a sparse linear solver of cost
$\widetilde O(m^{3/2}+m^{3\theta/2})$.}
\label{fig:concept}
\end{figure}

The translation proceeds in three stages (\cref{fig:concept}). First, the
system is rewritten as a \emph{logical network}, i.e., a sparse symmetric
system in which every site couples to at most three neighbors with a coupling
from a small fixed set of values. Auxiliary sites copy variables, add
contributions and mediate signs, so that a row or column of $M$ with many
nonzero entries is spread over many sites of low degree. The network
determines the original solution through a fixed set of its sites, and a small
residual in the network implies a small residual in $Mx=b$. Second, the network
is routed through a three-dimensional cubic grid, with every edge replaced by a
path of nearest-neighbor steps and with enough space between paths that they
do not interact. Third, every site of the routed network becomes a localized
electromagnetic mode in a dielectric background, and the physical couplings between
neighboring modes reproduce the couplings of the network.

The volume is set by the routing stage. The logical network has
$O(m\log m)$ sites, and it can be routed in a cubic box whose sides grow as
$O(\sqrt{m\log m})$, which gives a volume of $\Ot(m^{3/2})$. This is the same exponent that Kolmogorov and
Barzdin found for embedding general networks with wires of finite thickness in
three dimensions, where it is optimal for networks with good expansion
properties~\citep{KolmogorovBarzdin1967,GromovGuth2012}. The separation between
modes and the buffer to the boundary add only logarithmic factors.

The output of the compiler is a complete electromagnetic problem. For the
continuum construction, it consists of the permittivity $\varepsilon(r)$ in a
rectangular box $\Omega$ with perfectly electrically conducting (PEC) walls, a
square-integrable current source $f$ and a list of weighting functions
$a_i$, such that the time-harmonic electric field $E$ solves
\begin{equation}
 \nabla\times\nabla\times E-\varepsilon E=f
 \quad\text{in }\Omega,\qquad
 \hat n\times E=0\quad\text{on }\partial\Omega,
 \label{eq:maxwell}
\end{equation}
in units in which the frequency and the permeability are one. The answer is
read from the moments $\ip{E}{a_i}$, i.e., from weighted volume integrals of
the field. The permittivity is real, scalar and bounded, $1\le\varepsilon\le25$;
the continuum construction in fact uses only values between 6 and 10. The
discrete construction outputs the corresponding Yee system with one
permittivity value per grid cell, at a fixed resolution of ten cells per
shortest wavelength in the material, and reads the answer from prescribed
field coordinates. Both constructions use the same logical network and routing
and differ in how the network is realized physically. The guarantees they
share are summarized in \cref{tab:guarantees}.

\begin{table}[tbp]
\centering
\caption{\textbf{Guarantees of the compiler.} The left column lists the
quantities that determine whether the translation between a sparse system
$Mx=b$ with $m$ nonzero coefficients and an electromagnetic problem is
efficient; the right column gives the bound proved for each. Size, encoding
and decoding bounds hold for every run of the randomized encoder, and count
the complete box including empty background, routing space and the buffer to
the walls. Conditioning and recovery hold together with probability at least
$1-3/(64m)$ for $m\ge2$: on this event, every field whose relative Maxwell
residual is at most $\delta$ decodes to an approximate solution with relative
residual at most $\eta$. The bounds apply to both the continuum and the Yee
construction. Constants and powers of logarithms depend only on the fixed
input class and are not optimized.}
\label{tab:guarantees}
\small
\begin{tabularx}{\linewidth}{@{}p{0.32\linewidth}X@{}}
\toprule
Quantity & Guarantee \\
\midrule
Size of the structure & $\Ot(m^{3/2})$, counting the whole box: background, routing space and buffer \\
Encoding and decoding & $\Ot(m^{3/2})$ bit operations \\
Coefficients and readout & $O(\log m)$ bits each \\
Permittivity & Real, scalar, $1\le\varepsilon\le25$; PEC only at the outer walls \\
Conditioning & Polynomial in $m$ for the full electromagnetic operator \\
Field accuracy required & Relative residual $\delta\ge m^{-c_\delta}$ for a fixed $c_\delta$ \\
Recovery & Every field meeting this residual decodes to an answer with relative residual $\le\eta$ \\
\bottomrule
\end{tabularx}
\end{table}

\subsection{Defect modes as unknowns}

The central difficulty is the third stage. A network of mode amplitudes is
useful only if the actual Maxwell field -- including all components that are
not part of the network -- reproduces its equations to the required accuracy.
The required couplings shrink as the problem grows, since the encoded errors
must fall below an inverse polynomial of $m$, and the unwanted couplings must
remain smaller still after summation over all $\Ot(m^{3/2})$ sites.

\begin{figure}[tbp]
\centering
\includegraphics[width=\linewidth]{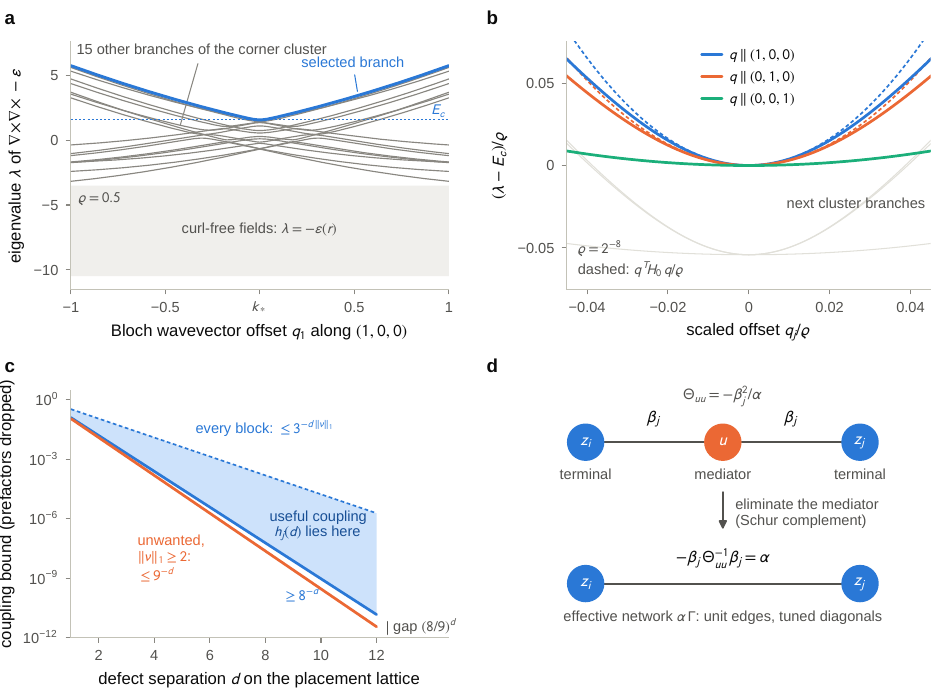}
\caption{\textbf{Broad defects in a weakly modulated dielectric interact like a network.}
\textbf{a}, Bloch spectrum of $\nabla\times\nabla\times-\varepsilon(r)$ for the periodic background
$\varepsilon=7+\varrho\,[\cos\pi r_1+2\cos\pi r_2+4\cos\pi r_3]$ along the line $k=k_*+q_1(1,0,0)$
through $k_*=(\pi/2,\pi/2,\pi/2)$, computed with a vector plane-wave expansion (1029 plane-wave
components; eigenvalues converged to $10^{-7}$). The modulation is exaggerated to $\varrho=0.5$ for
visibility; the construction uses $\varrho=2^{-100}$. At $k_*$ eight plane waves of the unmodulated
medium are degenerate, forming a cluster of sixteen transverse branches that the modulation splits.
The selected branch (blue) is the highest of them and has an isolated local minimum $E_c$ at $k_*$.
Curl-free fields contribute the band $\lambda=-\varepsilon(r)$ (gray), and the other cluster branches
cross $E_c$ away from $k_*$: there is no complete gap, and the proof bounds leakage into these states
instead of assuming its absence.
\textbf{b}, The selected branch near $k_*$ along the three axes, computed at $\varrho=2^{-8}$ and drawn
in scaled variables, in which the curves no longer change as $\varrho\to0$. Dashed: the quadratic form
$q^{T}\!H_0\,q/\varrho$ with the exact leading-order Hessian of the proof,
$H_0=\pi^2\,\mathrm{diag}(4.12,\,3.26,\,0.44)$; the next branches of the cluster (light gray) stay
$0.054\,\varrho$ below the minimum.
\textbf{c}, Proven decay rates of the interaction between two defects $d$ placement-lattice steps
apart, with constant prefactors dropped: the useful nearest-neighbor coupling $h_j(d)$ is at least
$8^{-d}$ and, like every interaction block, at most $3^{-d}$, whereas every block between defects
that are not nearest neighbors is at most $9^{-d}$. The gap between the two rates is $(8/9)^{d}$:
because $9>8$, the summed unwanted couplings fall below the useful ones once $d=O(\log m)$.
\textbf{d}, A sign mediator $u$ between two terminals $z_i$, $z_j$ with couplings $\beta_j$ and tuned
self-term $\Theta_{uu}=-\beta_j^2/\alpha$; eliminating it leaves the effective coupling $\alpha$, so the
defect amplitudes obey $\alpha\Gamma z=\alpha g$ for the compiled network matrix $\Gamma$, which has
unit edges and tuned diagonals.}
\label{fig:mechanism}
\end{figure}

The background is a weakly modulated periodic dielectric,
\begin{equation}
 \varepsilon_{\rm per}(r)=7+\varrho
 \bigl[\cos(\pi r_1)+2\cos(\pi r_2)+4\cos(\pi r_3)\bigr],
 \label{eq:background}
\end{equation}
with $\varrho=2^{-100}$ and period two in each direction. In a
homogeneous medium, the eight plane waves with wavevectors
$(\pm\pi/2,\pm\pi/2,\pm\pi/2)$ are degenerate at the corner
$k_*=(\pi/2,\pi/2,\pi/2)$ of the Brillouin zone, which gives a cluster of
sixteen states once both polarizations are counted. The modulation, with its
three distinct amplitudes, splits this cluster, and we prove that one of the
resulting branches of the operator $\nabla\times\nabla\times-\varepsilon_{\rm per}$
has an isolated, nondegenerate local minimum at $k_*$, i.e., near $k_*$ the
branch rises quadratically in every direction (\cref{fig:mechanism}a,b). The
splitting of the cluster is verified in exact rational arithmetic, and
analytic estimates over all other Fourier components extend the statement to
the continuum.

A band minimum acts locally like a band edge. Slightly raising the
permittivity over a broad region binds a mode just below the minimum,
in the same way as a defect in a photonic crystal binds a mode inside a band
gap~\citep{figotinklein1997,figotinklein1998,Joannopoulos2008}. The mode is
the envelope of the Bloch wave at $k_*$ and decays exponentially away from
the defect. Two such modes at distance $d$ couple through their overlapping
tails, which gives a tight-binding description of the defect lattice in the
spirit of coupled-resonator waveguides and Wannier-function models of
photonic-crystal circuits~\citep{Yariv1999,Busch2003}. We prove matching
bounds on this coupling. Along the three routing directions, the useful
nearest-neighbor coupling is at least of order $8^{-d}$, while the coupling
between defects that are $\nu$ lattice steps apart is at most of order
$3^{-d\norm{\nu}_1}$. The routing places the sites such that
only sites connected in the network are lattice neighbors, so every unwanted
coupling spans at least two lattice steps. Since $(1/3)^2<1/8$, each of these
couplings is smaller than the useful one by a factor that grows exponentially
in $d$ (\cref{fig:mechanism}c). A separation $d$ that grows
only logarithmically in $m$ therefore suppresses the sum of all unwanted
couplings below any required inverse polynomial.

Signs and asymmetry are produced by elimination. A dielectric is reciprocal,
and the coupling between two identical modes has a fixed sign, whereas $M$ is
an arbitrary signed, nonsymmetric matrix. The logical network solves this at
the algebraic level: the original system is embedded in a larger symmetric
system, and auxiliary \emph{mediator} sites, detuned by small changes of their
local permittivity, produce effective couplings of either sign once their
amplitudes are eliminated, i.e., through a Schur complement (\cref{fig:mechanism}d). The physical
construction only has to realize nearest-neighbor couplings of one sign and
tunable on-site energies, which is exactly what the defect lattice provides.

A local band minimum is not a complete band gap. Other Bloch waves propagate
at the operating frequency elsewhere in momentum space, and a defect mode can
leak into them. The defects are therefore made broad and smooth, with radius
$L$ and permittivity change of order $L^{-2}$, so that their momentum content
is concentrated near $k_*$. The leakage then decays as $e^{-cL^{1/3}}$, and a
defect radius growing as a power of $\log m$ suffices. The momentum window
used to separate the useful channel from the remaining waves is a tool of the
analysis, not a feature of the structure, and all leakage enters the residual
of the actual Maxwell equation, together with the errors from the finite
boundary and the finite description of the material.

The last step controls the response of the cavity as a whole. A lossless
cavity has resonances, and near a resonance a small residual can produce a
large field error. A small uniform shift of the permittivity, drawn at random
after all other parameters are fixed, avoids every resonance by a margin that
is polynomial in $m$ with high probability. On this event, the inverse of the
full Maxwell operator is bounded polynomially in $m$, which is what allows the
leakage to be tolerated instead of eliminated.

\subsection{Approximate fields give approximate answers}

The recovery guarantee is stated in terms of quantities that a solver can
check. Writing $A$ for the Maxwell operator of \cref{eq:maxwell}, the compiler
outputs a field tolerance $\delta$ such that
\begin{equation}
 \norm{AE-f}\le\delta\norm f
 \quad\Longrightarrow\quad
 \norm{Mx-b}_2\le\eta\norm b_2,
 \label{eq:recovery}
\end{equation}
where $x$ is decoded from the moments of $E$. The implication holds for every
field $E$ that meets the residual bound, irrespective of how it was obtained,
including complex fields of any polarization. The tolerance $\delta$ decreases
only polynomially with $m$, and the decoder needs each moment only to
$O(\log m)$ bits of precision. Since the bounded inverse makes the field
error at most polynomially larger than its residual, every admissible field
lies close to a reference field constructed in the proof, and the moments of
the reference field encode a solution of the logical network to within the
budget of \cref{eq:recovery}.

The material, source and weighting functions are themselves outputs of finite
size. The permittivity consists of the fixed trigonometric background,
dyadic shifts and polynomial profiles on balls whose radii and spacing grow
with $m$; there are no shrinking features, thin gaps or interior conductors.
Computing these descriptions does not require solving the assembled Maxwell
problem. Instead, the encoder calibrates the defect parameters once, from
finite algebra at $k_*$ and controlled Fourier tails, and every approximation
error is propagated to the final answer.

\subsection{Numerical illustrations}

\begin{figure}[tbp]
\centering
\includegraphics[width=\linewidth]{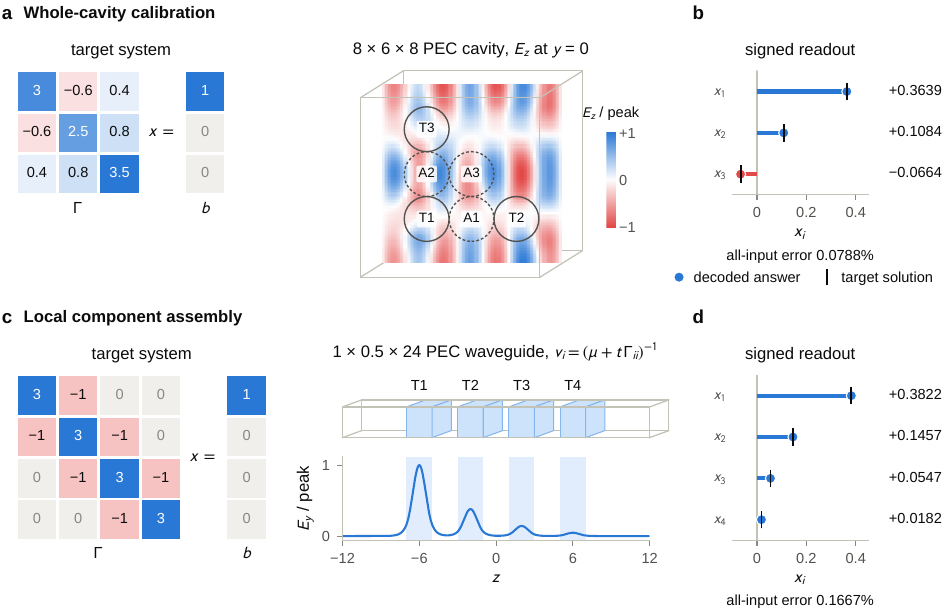}
\caption{\textbf{Calibrated cavities and reusable components give two numerical realizations.}
\textbf{a}, A prescribed signed matrix is encoded by six smooth spherical
dielectric defects in an $8\times6\times8$ PEC cavity. The left block shows
the target system $\Gamma x=b$; blue, red and gray cells mark positive, negative
and zero entries. The plotted section is
the signed electric-field phasor $E_z$ at $y=0$ for $b=(1,0,0)$, normalized by
its peak absolute value in the section. Solid circles mark the support of the
three terminal defects T1--T3; dashed circles mark the auxiliary defects A1--A3.
The full three-dimensional material strengths were calibrated against the
cavity response and held fixed during refinement.
\textbf{b}, Signed volume moments give the decoded answer (circles), compared
with the target solution (black ticks). The reported error $\norm{\Gamma X-I}_2$
covers all inputs; the displayed field was computed at $h=1/7$.
\textbf{c}, Four translated copies of one dielectric well realize a chain
using the local rule $v_i=(\mu+t\Gamma_{ii})^{-1}$; the target system is
shown as in \textbf{a}. Blue regions mark the
supports of smooth wells spanning the waveguide cross-section; the transverse
dimensions are enlarged. The curve is the computed $E_y$ along the centerline
for $b=(1,0,0,0)$ at $h=1/22$, normalized by its peak absolute value.
\textbf{d}, The same local source and readout construction recovers the
four-component answer, with all-input error $0.1667\%$.}
\label{fig:numerical}
\end{figure}

Two small numerical realizations illustrate complementary parts of the encoding
(\cref{fig:numerical}). First, we calibrate six smooth dielectric defects in a
three-dimensional periodic cavity to a prescribed signed $3\times3$ matrix
$\Gamma$. Three defects carry the sources and readout functions, while three
auxiliary defects adjust the cavity response. The material remains scalar and
positive, with $4.77<\varepsilon<21.24$. Once calibrated, the structure,
source profiles and decoder are fixed, and changing the source amplitudes
changes the right-hand side. A full-vector Maxwell solve on a grid finer than
the calibration grid gives $\norm{\Gamma X-I}_2=7.88\times10^{-4}$, where $X$ is the decoded
response to all basis inputs; this bounds the relative equation error for
every right-hand side.

Second, we assemble several systems from one locally calibrated component in
a rectangular PEC waveguide. Its transverse symmetry gives a single invariant
channel, with confinement supplied by waveguide cutoff. An isolated well
determines a resonance parameter $\mu$, and its evanescent tail determines
the coupling $t$ between neighboring wells. For a chain with off-diagonal
entries $-1$, the material strengths follow directly from
$v_i=(\mu+t\Gamma_{ii})^{-1}$. This rule transfers from three to four sites
without changing the common strength, and a detuned auxiliary well produces
a positive effective coupling by elimination. Five prescribed targets,
including an indefinite chain, have equation errors below $0.55\%$ on the two
finer of three full-vector grids, with the assembled responses used only for
verification. The component calibration, refinement tests and material-precision
study are given in Section~\ref{n:sec:numerical} of the \SI.

\subsection{Consequences for electromagnetic solvers}

\begin{figure}[tbp]
\centering
\includegraphics[width=\linewidth]{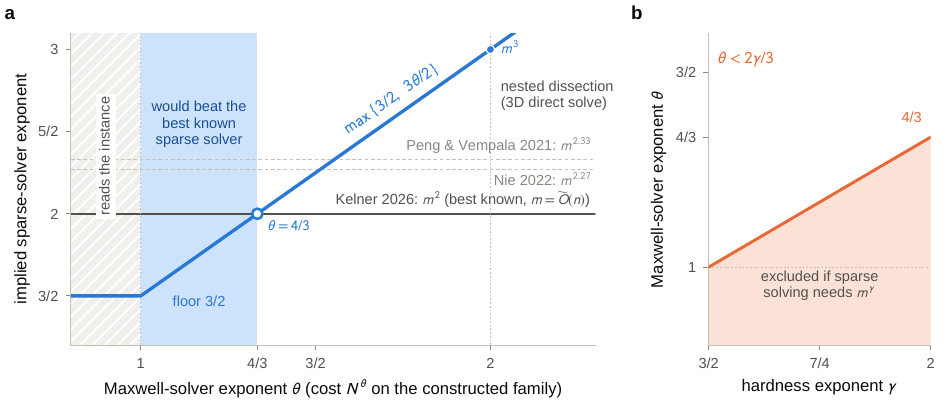}
\caption{\textbf{What a fast Maxwell solver would mean for sparse linear algebra.}
\textbf{a}, If a solver handles every cavity of the constructed family in time $N^{\theta}$ up to
logarithmic factors, where $N$ counts grid cells and description length, then encoding, solving and
decoding solve every polynomially conditioned sparse system with $m$ nonzeros in time
$m^{\max\{3/2,\,3\theta/2\}}$ up to logarithmic factors (blue; the floor $3/2$ is the cost of encoding
and decoding). Horizontal lines are known bit-complexity upper bounds for the same problem class
($m=\widetilde O(n)$ nonzeros, polynomial condition number): the
$\widetilde O(mn\log(\kappa/\eta))$ algorithm of Kelner~\cite{kelner2026} gives $m^{2}$ and
supersedes the $m^{2.33}$ of Peng and Vempala~\cite{pengvempala2021,PengVempala2025} and the
$m^{2.27}$ of Nie~\cite{nie2022} (both stated with the bound $2.3729$ on the matrix-multiplication exponent
used in those papers). Vertical
lines mark $\theta=1$, below which a solver can not read its input, and $\theta=2$, the cost of a
nested-dissection direct solve on a three-dimensional grid~\cite{George1973,LiptonRoseTarjan1979}. A
uniform solver with $1\le\theta<4/3$ (shaded) would improve the best known sparse-solver exponent.
The statement concerns worst-case solvers on this family of cavities.
\textbf{b}, Conversely, if solving sparse systems requires time $m^{\gamma}$ for some
$\gamma\in(3/2,2]$, the range compatible with the $m^{2}$ upper bounds (conjugate gradients in exact
arithmetic in the real-arithmetic model of the hypothesis; Kelner's algorithm in the bit model), no
uniform Maxwell solver for the family can have $\theta<2\gamma/3$ (shaded).}
\label{fig:consequence}
\end{figure}

The compiler is an algorithm in its own right: encode $Mx=b$, call an
electromagnetic solver, and decode its output. Suppose a solver handles the
constructed family uniformly, in time $O(N^\theta)$ for a structure of size $N$,
with only logarithmic dependence on the condition number and on the requested
accuracy, and returns the required moments. Substituting $N=\Ot(m^{3/2})$ gives
\begin{equation}
 T_{\rm sparse}(m)=\Ot\!\left(m^{3/2}+m^{3\theta/2}\right),
 \label{eq:transfer}
\end{equation}
where the first term accounts for encoding and decoding, and the second for
the electromagnetic solve (\cref{fig:consequence}). This implication is
unconditional. For polynomially conditioned systems with $O(n)$ nonzero
entries, the best known algorithm runs in $\Ot(n^2)$ bit
operations~\citep{kelner2026}, improving on earlier algorithms that beat matrix
multiplication~\citep{pengvempala2021,nie2022}. A uniform Maxwell solver with
$\theta<4/3$ would therefore solve sparse linear systems faster than any known
method. For comparison, nested dissection solves three-dimensional grid
problems directly in time $O(N^2)$, i.e., $\theta=2$~\citep{George1973,LiptonRoseTarjan1979}.

A lower bound on Maxwell solvers requires, in addition, a hypothesis about the
difficulty of sparse linear algebra. Assuming that no randomized algorithm
solves the input class in time $\Ot(m^{\gamma'})$ for any $\gamma'<\gamma$, for some
$\gamma>3/2$ -- a randomized version of the hypothesis used for
structured linear systems and packing linear
programs~\citep{kyngzhang2017,kyng2020} -- \cref{eq:transfer} excludes uniform
Maxwell solvers with $\theta<2\gamma/3$. Known algorithms restrict the
hypothesis to $\gamma\le\min\{2,1+c_\kappa\}$, where $m^{c_\kappa}$ bounds the condition
number, so the strongest consequence of this kind excludes $\theta<4/3$ --
the same threshold below which a Maxwell solver would improve on the best known
sparse algorithms. The
statement concerns solvers that work on the whole constructed family with the
stated dependence on conditioning and accuracy; restricted geometries,
well-conditioned problems or less accurate outputs can admit faster
algorithms.

\section{Discussion}

The compiler shows that a lossless, isotropic dielectric of bounded contrast is
expressive enough to represent every polynomially conditioned sparse linear
system, at the price of a polynomial increase in size. Locality, reciprocity
and a scalar, positive permittivity restrict which couplings a structure can
realize directly, but they do not restrict which equations it can encode once
auxiliary modes and space are available. In this sense, the material
distribution becomes a program: the compiler specifies the structure together
with its source, its readout and the accuracy to which its response must be
known.

For optical computing, the result complements electromagnetic circuit theory
and reconfigurable equation solvers (\cref{tab:prior}).
\citet{MiltonSeppecher2010} establish which terminal responses ideal circuits
can realize, while \citet{tzarouchis2022} give general matrix-inversion
architectures and convergence bounds. Here, the compiler realizes sparse
systems in a continuum dielectric with fixed permeability and bounded scalar
permittivity, and controls errors from the electromagnetic problem through to
the recovered solution. It accounts for the costs that decide whether an
analog solver can be useful as the problem
grows~\citep{Aaronson2015FinePrint,McMahon2023}: programming the structure,
its size, the precision of every coefficient and the readout. The construction
retains the full sparse input class while requiring only logarithmic precision
per coefficient. The same accounting has
been decisive for quantum linear-system
algorithms~\citep{HHL2009,ChildsKothariSomma2017,Aaronson2015FinePrint}, whose
application to electromagnetic scattering and finite-element problems yields at
most polynomial speedups once input and output are
included~\citep{clader2013,MontanaroPallister2016}. Unlike optical Ising
machines, which target NP-hard
optimization~\citep{Inagaki2016,Pierangeli2019,Mohseni2022}, the problem here
is solvable in polynomial time, and the relevant question is the exponent.

\begin{table}[tbp]
\centering
\caption{\textbf{Physical linear-equation solvers and electromagnetic response
synthesis.} The problem column describes the scope of each construction or
architecture; the example column gives the largest reported size $n$ in
experiment (expt.) or simulation (sim.), where available. The table also lists
the physical mechanism -- an iterative feedback loop or a stationary
response -- together with hardware scaling where stated, and the treatment of
accuracy and input and output costs. Theoretical constructions are marked as
such. This work gives a sparse-system reduction with bounds on size, encoding
and decoding cost and accuracy; its simulated examples
(\cref{fig:numerical}) are a signed $3\times3$ system and a four-site chain.}
\label{tab:prior}
\small
\setlength{\tabcolsep}{3pt}
\renewcommand{\arraystretch}{1.15}
\begin{tabularx}{\linewidth}{@{}>{\raggedright\arraybackslash}p{25mm}>{\raggedright\arraybackslash}X>{\raggedright\arraybackslash}p{19mm}>{\raggedright\arraybackslash}p{32mm}>{\raggedright\arraybackslash}p{32mm}@{}}
\toprule
Work & Problem & Largest example & Mechanism & Accuracy and I/O \\
\midrule
\citet{MiltonSeppecher2010} & Prescribed symmetric terminal response & -- (theory) & Stationary, dielectric and magnetic elements & Ideal material limits; finite-parameter approximation open \\
\citet{estakhri2019} & Integral equation, kernel fixed in the structure & 5 (expt.) & Feedback loop, Neumann series & Measured accuracy \\
\citet{Camacho2021} & Several fixed kernels, nonsymmetric & 5 (expt.) & Feedback loop & Measured accuracy \\
\citet{cordaro2023} & Integral equation, fixed symmetric kernel & 3 (expt.) & Feedback in free space & Fit to simulation \\
\citet{tzarouchis2022} & Programmable complex matrix, generalized inverse & 11 (sim.) & Iteration, $2n^2$ active elements & Convergence time in $\kappa$, $\epsilon$ \\
\citet{chen2024} & Programmable matrix, Richardson iteration & 4 (expt.), 16 by blocking & Iteration, $n^2$ interferometers & Accuracy metric; I/O traffic counted \\
\citet{ye2024} & Fixed diffusion operator on a mesh & $5\times5$ (expt.) & Stationary, one node per grid point & Comparison with simulation \\
\citet{Sun2019PNAS} & Programmable matrix, signs by differences & 4 (expt.), 100 (sim.) & Stationary electronic feedback, $\sim n^2$ cells & Empirical, grows with $\kappa$ \\
\midrule
This work & Any sparse nonsingular system, signed and nonsymmetric & 4 (sim.) & Stationary response, volume $\Ot(m^{3/2})$ & Residual guarantee; encoding and decoding $\Ot(m^{3/2})$ \\
\bottomrule
\end{tabularx}
\end{table}

For electromagnetic simulation, the result places a limit on what general
solvers can achieve. Fast Maxwell solvers exploit structure beyond locality and
a bounded permittivity, e.g., smooth or piecewise-constant geometry, absorbing
boundaries, moderate quality factors or a small number of distinct
materials~\citep{EngquistYing2011CPAM,Martinsson2019,ShinFan2013}. The
constructed structures have none of these properties. They are closed,
lossless cavities with polynomially many weakly coupled resonant modes, and a
solver that is fast on all of them is, by \cref{eq:transfer}, a fast sparse
linear solver. Inverse design operates in exactly this space of permittivity
distributions~\citep{JensenSigmund2011,Molesky2018}, so an optimizer can, in
principle, produce instances that are as hard to simulate as general sparse
systems. In practice, the difficulty of a given device will depend on how far
it is from this worst case, which the present construction does not quantify.

The exponent $3/2$ coincides with the volume needed to embed a general network
with wires of finite thickness in three
dimensions~\citep{KolmogorovBarzdin1967,GromovGuth2012}. This suggests that the
exponent is natural for any construction that routes couplings through local
paths, but it is not a proof of optimality, since wave-mediated couplings
need not behave like wires. A compiler with nearly linear size would strengthen
the consequence for simulation to $\theta<\gamma$. Conversely, a lower bound on
the volume needed to encode general systems would identify a property of
Maxwell's equations that makes them easier than general sparse linear algebra.

The construction is a mathematical existence proof with deliberately
conservative constants. The modulation of the background is
$2^{-100}$, the proven bounds contain large powers of logarithms, and the fixed
constants are astronomically large -- e.g., the additive constant in the
$O(\log m)$ coefficient length has on the order of $2^{2^{50}}$ bits. As such,
the bounds describe the asymptotic scaling and not the size of any realistic
instance, and the permittivity must be specified to a precision that grows with
the problem size.
The model is an ideal lossless cavity at a single frequency. The theorem
relates the stationary response to the algebraic answer but does not address
the time or energy needed to reach and measure that response, nor fabrication
errors, material dispersion, absorption or detector noise. The result
therefore establishes what a dielectric can represent and what a solver must
handle, not a practical speed or energy advantage for optical computing.
The numerical examples use practical parameter choices: the periodic cavity
illustrates signed encoding after whole-device calibration, while the waveguide
illustrates reusable local assembly in a chain with exact transverse symmetry.
Their scope is the encoding mechanism at small sizes; the general
three-dimensional construction and its asymptotic resource bounds follow from
the analysis. Reusable junctions that preserve
their calibrated response in branching networks would connect the two numerical
settings more closely. Other wave platforms -- acoustic, elastic or quantum --
may admit similar compilers with their own interaction and stability analyses.

In summary, we have shown that sparse linear systems and electromagnetic
fields can be translated into each other at a cost of $\Ot(m^{3/2})$, with
every source of error accounted for. The same construction specifies how
general equations can be written into a passive dielectric and which
algebraic difficulty a general-purpose Maxwell solver inherits, providing a
common, quantitative reference point for computing with light and computing
light.

\section{Methods}

\subsection{Input class}

Fix constants $c_\kappa,c_\eta,B_{\rm in}>0$. An input consists of a nonsingular
$M\in\Q^{n\times n}$ with $n\ge1$, a right-hand side $b\in\Q^n$ and a tolerance $\eta$, with
\begin{equation}
 m=\nnz(M),\qquad \kappa_2(M)\le m^{c_\kappa},\qquad m^{-c_\eta}\le\eta\le1.
 \label{eq:methods-input}
\end{equation}
Every numerator and denominator has at most $B_{\rm in}\log_2(2+m)$ bits, and
indices have $O(\log m)$ bits. Nonsingularity implies $n\le m$. The output is a
rational vector $x$ with $\norm{Mx-b}_2\le\eta\norm b_2$. The encoder may
replace $\eta$ by $\eta_0=m^{-\lceil c_\eta\rceil}$, so that the tolerance never has to be
read to more than logarithmic precision. The empty system is solved directly.
Constants in all asymptotic bounds depend only on $c_\kappa$, $c_\eta$ and $B_{\rm in}$.

\subsection{Continuum realization}

The domain $\Omega$ is a rectangular box with integer side lengths, and the
field satisfies \cref{eq:maxwell} with $1\le\varepsilon(r)\le25$. Let
$\mathcal C$ denote the self-adjoint curl-curl operator defined by the closed
form $\langle\nabla\times E,\nabla\times F\rangle$ on
$H_0(\operatorname{curl};\Omega)$, let $X$ be its domain with the graph norm
$\norm E_X^2=\norm E_{L^2}^2+\norm{\mathcal CE}_{L^2}^2$, and set
$A=\mathcal C-\varepsilon$. For fixed $c_\kappa,c_\eta,B_{\rm in}$, there are a randomized
encoder, a decoder and explicit constants $C_{\rm res},c_A,c_\delta$ such that, with
$\ell=1+\lceil\log_2(2+m)\rceil$, every run of the encoder satisfies
\begin{align}
 |\Omega|&\le C_{\rm res}m^{3/2}\ell^{24},&
 N_{\rm desc}&\le C_{\rm res}m^{3/2}\ell^{64},\notag\\
 T_{\rm enc}+T_{\rm dec}&\le C_{\rm res}m^{3/2}\ell^{4096},&
 B_{\rm out}&\le C_{\rm res}\ell,\qquad \delta\ge m^{-c_\delta}.
 \label{eq:methods-continuum-cost}
\end{align}
Here $N_{\rm desc}$ counts the unit cells of the box together with all
coefficients and indices of the description, and $B_{\rm out}$ bounds the bit
length of every output coefficient. For $m\ge2$, with probability at least
$1-3/(64m)$, the operator $A:X\to L^2$ is invertible with
\begin{equation}
 \norm{A^{-1}}_{L^2\to L^2}\le L_b,\qquad
 \norm A_{X\to L^2}\norm{A^{-1}}_{L^2\to X}\le702L_b\le m^{c_A},
 \label{eq:methods-inverse}
\end{equation}
and every $E\in X$ with $\norm{AE-f}_{L^2}\le\delta\norm f_{L^2}$ decodes to a
valid answer for every allowed error in the moment messages, whose range and
precision both have $O(\ell)$ bits. The condition bound refers to the graph
norm, since $A$ is unbounded on $L^2$. The complete statement is Theorem~\ref{c:thm:main}
of the \SI, proved in Part~II.

\subsection{Yee realization}

The discrete target uses the Yee curl incidence matrix $C_{\rm Y}$ with unit
grid spacing, frequency $\omega=1/8$ and PEC termination; $A$ now denotes the
resulting Yee matrix. Every edge mass is
the average of the permittivities of its four adjacent cells, and
\begin{equation}
 A=C_{\rm Y}^{\mathsf T}C_{\rm Y}-\omega^2D_\varepsilon,
 \qquad 1\le\varepsilon_c\le25.
 \label{eq:methods-discrete}
\end{equation}
The grid is uniform, without coordinate stretching or absorbing layers, and
resolves the shortest wavelength in the material with ten cells. The
realization has $\Ot(m^{3/2})$ unknowns, the same encoding and decoding cost,
$O(\log m)$-bit coefficients and, with probability at least $1-3/(64m)$, a
full matrix with $\kappa_2(A)\le m^{c_A}$ for which every vector meeting
the residual tolerance decodes correctly. The decoder reads prescribed field
coordinates to $O(\log m)$ bits. The spectral analysis of the discrete model is
independent of the continuum one and does not rely on the resolution to
approximate the continuum equations (Theorem~\ref{d:thm:main} and Part~III
of the \SI). Its spectral input is a computer-assisted certificate for a
periodic Yee background. The finite inequalities of this certificate have been
checked in full; the final step, which binds the checked data to the exact Yee
operator near the band minimum, is being completed as part of an ongoing formal
verification, and the Yee result is conditional on it
(Section~\ref{si:sec:yee-certificate}).

\subsection{Logical network and routing}

The compiler normalizes and rounds the coefficients within the error budget,
expresses their arithmetic through a fixed alphabet of local relations, and
adds variables to bound the degree of every site by three. The original system
is embedded in a symmetric system, and sign mediators make all couplings of the
realized network positive, with the required signs appearing after
elimination of the mediators. Balanced routing then replaces every edge by a
nearest-neighbor path in a cubic grid of side $O(\sqrt{m\log m})$. Every step
preserves a polynomial inverse bound and a linear map from the residual of the
network to the residual of $Mx=b$ (\SI, Part~I).

\subsection{Band minimum, interactions and leakage}

The background of \cref{eq:background} has a simple Bloch branch with
$\lambda(k_*+q)=E_c+q^{\mathsf T}Hq+O(|q|^4)$ and $H$ positive definite.
Exact rational arithmetic identifies the branch in the finite corner cluster,
and estimates over all other Fourier components establish the expansion for
the continuum operator. An integer placement matrix adapts the routing grid to
the anisotropy of $H$.

Defects are smooth bumps of radius $L$ and strength of order $L^{-2}$, analyzed
through an auxiliary resolvent localized in momentum near $k_*$. Along the
three routing directions, the coupling between neighboring defects at
separation $d$ satisfies $h_j(d)\ge2^{-1024}L^2 8^{-d}$, while the full coupling
block between defects displaced by $\nu\ne0$ lattice steps satisfies
$\norm{G_h(\nu)}\le C_*L^5(3^{-d\norm \nu_1}+e^{-c_*L})$. The margin
$(1/3)^2<1/8$ lets the useful couplings dominate the summed unwanted ones after
diagonal tuning. The momentum cutoff leaves a residual in the actual Maxwell
equation bounded by $e^{-c_*L^{1/3}}$ times explicit polynomial factors.
Choosing $L=O(\ell^6)$, $d=O(\ell)$ and a boundary buffer of width $O(\ell^7)$
makes all errors smaller than the required inverse-polynomial budgets. The
uniform permittivity shift is sampled after all profiles are fixed, and a
count of the eigenvalues of the finite-box operator gives
\cref{eq:methods-inverse} (\SI, Part~II).

\subsection{Finite description and recovery}

The permittivity consists of the trigonometric background, dyadic shifts and
polynomial profiles on balls of radius growing with $m$. Calibration uses
finite algebra at $k_*$ and controlled Fourier tails and does not solve the
assembled problem. Sources and weighting functions use tensor polynomials of
coordinate degree $O(\ell^8)$, with $O(\ell)$ bits per coefficient and error
bounds for the complete functions. If $E_{\rm ref}$ is the reference field
constructed in the proof, with $\norm{AE_{\rm ref}-f}_{L^2}\le r_{\rm ref}$, then
every field meeting the residual tolerance satisfies
\begin{equation}
 \norm{E-E_{\rm ref}}_{L^2}\le L_b\bigl(\delta\norm f_{L^2}+r_{\rm ref}\bigr).
 \label{eq:methods-comparison}
\end{equation}
The decoder reads the moments $z_i=v_i\langle E,\widehat a_i\rangle$, whose
real test fields $\widehat a_i$, weights $v_i$ and tolerances are part of the
output. \Cref{eq:methods-comparison}, the approximation bounds for the test
fields and the residual map of the logical network give \cref{eq:recovery}.

\subsection{Algorithm transfer and conditional limits}

For the continuum transfer, $N$ counts the volume and the description, and $B_{\rm out}$
the coefficient and message lengths. The solver is assumed to satisfy
\begin{equation}
 T_{\rm Max}(N,B_{\rm out},\kappa_X,\delta)
 =O\!\left(N^\theta B_{\rm out}\,\polylog(N,\kappa_X,\delta^{-1})\right),
 \label{eq:methods-solver}
\end{equation}
where $\kappa_X$ is the condition bound of \cref{eq:methods-inverse}, and to
return the prescribed moments of a residual field with a fixed positive
probability, the cost of computing the moments included. A polynomial
dependence on $\kappa_X$, as in unpreconditioned iterative methods, does not
meet this assumption. Solver calls are capped, candidate answers are checked
against $Mx=b$, and repetition amplifies the success probability, which gives
\cref{eq:transfer} in the bit model.

The conditional statement uses a real-number formulation: a full-column-rank
$F_0$ with $m$ nonzeros and $\kappa_2(F_0)\le m^{c_\kappa}$, and the target
\begin{equation}
 \norm{F_0x-\Pi_{F_0}b}_2^2\le n^{-10c_\kappa}\norm{\Pi_{F_0}b}_2^2,
 \label{eq:methods-realram}
\end{equation}
where $\Pi_{F_0}$ projects onto the column space. A reduction through a sparse
saddle-point system preserves this criterion even when
$\norm{\Pi_{F_0}b}\ll\norm b$. The hypothesis states that no randomized
algorithm solves every such instance with probability at least $2/3$ in time
$\Ot(m^{\gamma'})$ for any $\gamma'<\gamma$; it is a randomized
exponent-gap version of the hypothesis used for structured linear systems and
packing linear programs~\citep{kyngzhang2017,kyng2020}. For $\gamma>3/2$ it
excludes $\theta<2\gamma/3$ in \cref{eq:methods-solver}. Conjugate gradients in
exact arithmetic require $O(mn)\le O(m^2)$ operations, and their convergence
under the condition-number bound limits the hypothesis further, so that
\begin{equation}
 \tfrac32<\gamma\le\min\{2,1+c_\kappa\}.
 \label{eq:methods-gamma}
\end{equation}
Faster algorithms for the same model would restrict this range further
(Theorems~\ref{d:thm:algorithm-transfer} and~\ref{c:thm:transfer} and
Part~IV of the \SI).

\subsection{Computer-assisted steps}

The finite part of the continuum band analysis -- the decomposition of the
corner cluster, orthogonality, ordering of eigenvalues, velocity and Hessian
identities and placement inequalities -- is checked in exact rational
arithmetic. The discrete spectral analysis uses exact matrix inequalities on a
covering of the Brillouin zone. In both cases, the analytic arguments in the
\SI\ show how these finite inequalities imply the required spectral
properties (\SI, Part~V). The formal verification of the continuum
construction in Lean is substantially complete, with proofs for the continuum
operator, band and defect estimates, leakage, finite representations and
residual-to-solution recovery. A theorem map and verification records in the
companion repository specify the checked statements and their dependencies.

\subsection{Use of AI tools}

The constructions, proofs and computer-assisted certificates in this work were
developed with AI systems under the direction of the author: OpenAI Astra~6 in
ChatGPT (Pro reasoning mode), Astra~6 in the Codex app (max and ultra
settings), and Anthropic Claude Opus~5.5 in Claude Code, used in 2026. The
author defined the problem and its requirements, steered the approach and
reviewed the results. The same tools were used to prepare the manuscript text,
the figures and the literature search. The proofs in the \SI\ were checked by
independent AI-based audits, the finite certificates are verified by computer
programs in exact or outward-rounded arithmetic, and a formal verification in
Lean is in progress. Every reference was checked against its publisher or arXiv
record. The author takes full responsibility for the content of this paper.

\section*{Data and code availability}

The programs that check the computer-assisted steps and their input data, the
Lean development, the code and data of the numerical illustrations and the
scripts that generate the figures are available in the
\href{https://github.com/yaugenst/maxwell-universality}{\texttt{yaugenst/maxwell-universality}}
GitHub repository.

\begingroup\renewcommand{\bibfont}{\footnotesize}\setlength{\bibsep}{1pt}
\bibliographystyle{unsrtnat}
\bibliography{references}
\endgroup
\clearpage
\setcounter{section}{0}
\setcounter{equation}{0}
\setcounter{figure}{0}
\setcounter{table}{0}
\setcounter{theorem}{0}
\renewcommand{\thesection}{S\arabic{section}}
\renewcommand{\theHsection}{si.\arabic{section}}
\numberwithin{equation}{section}
\renewcommand{\theHequation}{si.\theequation}
\renewcommand{\thefigure}{S\arabic{figure}}
\renewcommand{\theHfigure}{si.\arabic{figure}}
\renewcommand{\thetable}{S\arabic{table}}
\renewcommand{\theHtable}{si.\arabic{table}}
\begin{center}
{\LARGE\bfseries Supplementary Information}
\end{center}
\bigskip
\contentsmargin{2em}
\titlecontents{section}[3.2em]{\addvspace{0.2em}}{\contentslabel{3.2em}}{}{\titlerule*[0.6pc]{.}\contentspage}
\titlecontents{part}[0em]{\addvspace{0.8em}\bfseries}{}{}{\hfill\contentspage}
\startcontents[si]
\begingroup\small
\printcontents[si]{}{-1}{\setcounter{tocdepth}{1}}
\endgroup
\clearpage
\section{Setting and main statements}\label{si:sec:setting}

This section fixes the sparse input model and states the two realization
theorems, for the continuum Maxwell operator
(Theorem~\ref{c:thm:main}) and for the scalar Yee discretization
(Theorem~\ref{d:thm:main}), together with the resulting algorithm
transfers. Both realizations share the finite algebraic compiler and the
three-dimensional routing of Part~I
(Sections~\ref{d:sec:logical}--\ref{d:app:logical-residual}). Parts~II
and~III give their separate spectral, interaction, boundary and recovery
analyses. Part~IV proves the algorithm transfers and their conditional
consequences, and Part~V describes the finite computer-assisted
certificates and the formal proofs. Part~VI details the numerical
illustrations.

\subsection{Conventions and notation}\label{si:sec:notation}

All vector spaces are complex unless real scalars are specified. Inner
products are linear in the first argument. Operator norms on $L^2$ and
Euclidean norms on finite vectors are denoted by the same symbol. The
Fourier transform $\mathcal F$ on $\R^3$ is unitary, with factor $(2\pi)^{-3/2}$.
The notation $\tO$ hides only fixed powers of $\log(2+m)$.
Constants may depend on the fixed input constants $c_\kappa,c_\eta,B_{\rm in}$ of
Section~\ref{si:sec:input} and on the fixed
electromagnetic platform, but never on $m$.

Every symbol below has one meaning throughout the main text and this
supplement; a hat ($\widehat F$, $\widehat a$, $\widehat x$) always marks a finite
(rounded or truncated) approximation of the unhatted object. The numerical
illustrations in the main text and in Part~VI define their own symbols, e.g.,
the grid spacing $h$ and the decoded response matrix $X$.
\begin{itemize}
\item \emph{Input.} $M\in\Q^{n\times n}$, $b$, $x$, $m=\nnz(M)$; tolerance
$\eta$, its rounded value $\eta_0$, the logical tolerances $\eta_*$ and
$\eta_\Gamma=\eta_*/16$, the routed-network tolerance $\eta_{\rm res}$ and the
saddle tolerance $\eta_s$; exponents $c_\kappa$ (condition), $c_\eta$
(tolerance) and the bit constants $B_{\rm in}$, $B_{\rm out}$.
\item \emph{Output constants.} $C_{\rm res}$ (resource constant), $c_A$
(condition exponent of the Maxwell operator), $c_\delta$ (tolerance
exponent); $\ell=1+\lceil\log_2(2+m)\rceil$; $N$ (size of the structure, for
the Yee model its dimension), $N_{\rm desc}$, $T_{\rm enc},T_{\rm dec},T_{\rm Max},T_{\rm sparse}$;
solver exponent $\theta$, hardness exponent $\gamma$.
\item \emph{Electromagnetics.} $r=(r_1,r_2,r_3)\in\R^3$ the spatial coordinate, $r_i$ the well centers; $\varepsilon$ permittivity ($\varepsilon_{\rm per}$ the
periodic background, $\varepsilon_c$ a Yee cell value, $\varrho=2^{-100}$ the
continuum modulation amplitude), $\omega$ frequency, $\Omega$ the box with
outward unit normal $\hat n$, $\Cc$ the curl--curl operator, $C_{\rm Y}$ the Yee
curl, $D_\varepsilon$ the Yee mass matrix; $A$ the Maxwell operator of the
respective back end ($A=\Cc-\varepsilon$ or $A=C_{\rm Y}^{\mathsf T}C_{\rm Y}-\omega^2D_\varepsilon$),
$A_{\rm per}$ its periodic-background version; $E$ the electric field (for the
Yee model the vector of edge unknowns), $f$ the source, $f_*$ the ideal source,
$f_{\rm a}$ the auxiliary source, $E_{\rm a}$ the auxiliary field, $E_{\rm ref}$ the
reference field; $\delta$ the field residual tolerance; $V=H_0(\curl;\Omega)$,
$X=D(\Cc)$, $\kappa$ condition numbers ($\kappa_2$, $\kappa_X$); $L_b$ the
inverse bound, $\sigma$ the random detuning, $\tau$ its range, $J_{\rm grid}$
the number of grid points, $\mathcal G$ the good event.
\item \emph{Band structure.} $k$ wavevector, $k_*$ the band minimum, $q$ the
Bloch offset, $\nu\in\Z^3$ lattice offsets and Fourier aliases, $p$ physical
momentum, $\lambda$ eigenvalues and band functions, $E_c$ the band minimum
energy, $H=\tfrac12\lambda''(k_*)$ the Hessian ($H_0$ its leading-order value),
$\Phi_q$ Bloch cell functions, $\chi$ the momentum cutoff, $R_\sigma$ the
auxiliary resolvent, $q_0$ the chart radius, $k_\parallel,k_\perp$ ($q_\parallel,q_\perp$ for Yee) the
longitudinal and transverse momentum components in the pole analysis.
\item \emph{Defects and network.} $L$ defect radius, $h=L^{-1}$, $d$ placement
separation, $B$ the integer placement matrix, $w$ the bump, $u_h$ ($u$ for Yee)
the defect mode, $\mu_h$ ($\mu$) its eigenvalue, $a_h$ the source mode, $a_i$
the test fields, $v_i$ the tunings, $z_i$ the moments, $G_h(\nu)$ ($G(\nu)$)
the interaction blocks, $h_j(d)$ the useful couplings, $\beta_j$ their
calibrated values, $\alpha$ and $\zeta$ the useful and unwanted coupling
scales, $\Gamma$ the logical network matrix with source $g$,
$\Theta$ the target matrix on terminals and mediators, $g_\Theta=(g,0)$,
$S_b=\Theta+\Delta$ the bright block, $\MM$ the Birman--Schwinger matrix, $Q$ the
restriction to the defect union, $U$ the insertion of the mode, $\overline K$
the inverse bound of $\Gamma$, $J$ the number of sites, $\ell_{\rm buf}$ the
boundary buffer, $h_y$ the readout accuracy.
\item \emph{Part I.} $F=M/(ma_M)$ the normalized matrix, $g=b/\nu_M$, $P=m^{\lceil c_\kappa\rceil+1}$,
$N_{\rm r}$ rounding bits, $\Gamma_{\rm c}$ the circuit matrix of size $S$,
$\Gamma_0$, $\Gamma_+$ the symmetrized and mediated matrices, $S_+$ vertices,
$\ell_{\rm p}$ the port array side, $\ell_w$ the routed length, $Q_{\rm c},Q_+$
inverse bounds, $T_{\rm res}$ the residual amplification.
\item \emph{Constants.} $\Cs,\cs,c_u,\Ls,\ds$ and $Z$ are the fixed
continuum constants of Part~II; $C_{\rm d},c_{\rm d},r_{\rm d},a_{\rm d}$ the fixed
Yee platform constants of Part~III; $\rho$ with a subscript is a residual
vector. Inside proofs, $C$ and $c$ with primes denote fixed positive constants
local to the estimate, $K$ is a derivative or truncation order, and $i,j,a,s,t,y$
are bound indices, parameters and integration variables local to a displayed
formula; none of them coincides with a global symbol used in the same section.
High-symmetry points of the Brillouin zone are typeset upright,
$\mathrm{L}_{++-}$ and $\mathrm{\Gamma}$, and are distinct from the defect radius
$L$ and the network matrix $\Gamma$.
\end{itemize}

\subsection{Sparse input and accuracy}\label{si:sec:input}
Fix constants $c_\kappa,c_\eta>0$ and an input bit-length constant $B_{\rm in}$. An
input consists of a nonsingular matrix $M\in\Q^{n\times n}$, $n\ge1$, and
$b\in\Q^n$, with
\begin{equation}
 m=\nnz(M),\qquad \kappa_2(M)\le m^{c_\kappa},
 \qquad m^{-c_\eta}\le\eta\le1.                         \label{c:eq:input}
\end{equation}
Each numerator and denominator of every entry of $M$ and $b$ has at most
$B_{\rm in}\log_2(2+m)$ bits, and sparse indices have $O(\log(2+m))$
bits. Nonsingularity implies $n\le m$. The target accuracy is the
relative residual
\begin{equation}
 \norm{Mx-b}_2\le\eta\norm b_2.                    \label{c:eq:target}
\end{equation}
The encoder may replace $\eta$ by $\eta_0=m^{-\lceil c_\eta\rceil}$, so it
need not read an arbitrarily long representation of the requested
tolerance. The empty system is answered directly before this construction;
for $n\ge1$, the case $b=0$ is answered by $x=0$.

\subsection{The continuum model and its realization}\label{c:sec:statement}

For a rectangular box $\Omega$, let
\begin{equation}
 V=H_0(\curl;\Omega),\qquad
 c(E,F)=\ip{\curl E}{\curl F}_{L^2}.
                                                               \label{c:eq:form}
\end{equation}
Let $\Cc$ be the operator represented by this closed nonnegative form.
Explicitly,
\begin{equation}
 \begin{split}
 D(\Cc)=\{E\in V:\text{ there is }g\in L^2\text{ such that }
 c(E,F)=\ip{g}{F}\text{ for every }F\in V\},\\
 \Cc E=g,
 \qquad X=D(\Cc),\qquad
 \norm E_X^2=\norm E_2^2+\norm{\Cc E}_2^2.
 \end{split}                                                   \label{c:eq:domain}
\end{equation}
Thus PEC means zero tangential trace. For bounded real $\varepsilon$,
$A=\Cc-\varepsilon$ is self-adjoint on $D(\Cc)$. A weak solution with
an $L^2$ source belongs to this domain, since
$\Cc E=f+\varepsilon E\in L^2$.

\begin{theorem}[Continuum realization]
\label{c:thm:main}
For every fixed choice of the constants $c_\kappa,c_\eta,B_{\rm in}$ above there is a uniform
randomized encoder and a finite-bit decoder with the following properties.
For every input satisfying~\eqref{c:eq:input}, they produce an integer-sided
rectangular box $\Omega\subset\R^3$, a finitely represented real scalar
$1\le\varepsilon\le25$, a real $f\in L^2(\Omega;\R^3)$, a positive
rational tolerance $\delta$, and finitely represented real $L^2$ test fields.
The physical equation is exactly
\begin{equation}
 \curl\curl E-\varepsilon E=f\quad\text{in }\Omega,
 \qquad \hat n\times E=0\quad\text{on }\partial\Omega.       \label{c:eq:physical}
\end{equation}
There are no internal conductors, anisotropic coefficients, coordinate
stretches, absorbing layers, or distributional sources.

Put $\ell=1+\lceil\log_2(2+m)\rceil$. Effectively specified constants
$C_{\rm res},c_A,c_\delta$, depending only on $c_\kappa,c_\eta,B_{\rm in}$, satisfy, on every random draw,
\begin{align}
 |\Omega|&\le C_{\rm res}m^{3/2}\ell^{24},
 &N_{\rm desc}&\le C_{\rm res}m^{3/2}\ell^{64},       \label{c:eq:volume-description}\\
 T_{\rm enc}+T_{\rm dec}&\le C_{\rm res}m^{3/2}\ell^{4096},
 &B_{\rm out}&\le C_{\rm res}\ell,
 &\delta&\ge m^{-c_\delta}.                              \label{c:eq:resources}
\end{align}
Here $N_{\rm desc}$ counts the whole ambient box in unit-volume cells and
all coefficient and index slots in the finite source, material, and readout
format. Times are bit-operation bounds. All calibration, empty space,
routing, mediators, buffering, and finite output are included.

For $m\ge2$, one event $\mathcal G$ has probability at least
$1-3/(64m)$. On this event,
\begin{equation}
 \begin{gathered}
 A:X\longrightarrow L^2\text{ is an isomorphism},\qquad
 \norm{A^{-1}}_{L^2\to L^2}\le L_b,\\
 \norm A_{X\to L^2}\norm{A^{-1}}_{L^2\to X}
 \le702L_b\le m^{c_A}.
 \end{gathered}                                    \label{c:eq:maininverse}
\end{equation}
where $L_b$ is the explicit rational number in~\eqref{c:eq:detuning}.
For \emph{every} $E\in X$ satisfying
\begin{equation}
 \norm{AE-f}_2\le\delta\norm f_2,                    \label{c:eq:physicalres}
\end{equation}
and every collection of moment messages within the prescribed absolute
accuracies, the decoder returns a rational $x$ satisfying~\eqref{c:eq:target}.
Both integer range and fractional precision of each message are
$O(\log(2+m))$ bits. In particular, the guarantee covers every polarization and every field,
independently of the solver that produced it.
\end{theorem}

The material format is a fixed period-two trigonometric background plus
dyadic polynomial profiles on disjoint explicitly specified balls. The
balls have radius $L$, their centers and the box sides are integers, and
$L=O(\ell^6)$. No ball, gap, or boundary buffer shrinks with $m$.
The material profile degree is $O(\ell^2)$; source and readout degrees
are at most a fixed multiple of $\ell^8$. These degrees and every
coefficient are counted in~\eqref{c:eq:volume-description}
and~\eqref{c:eq:resources}. Section~\ref{c:sec:representation}
gives the exact finite format, its evaluation algorithms, and its
approximation bounds. The shortest background period is two; the vacuum
wavelength is $2\pi$. Therefore~\eqref{c:eq:volume-description} is also the physical
volume bound in wavelength units, up to a fixed factor.

Section~\ref{c:sec:cost} supplies one entirely explicit choice of the
theorem-wide constants; it is far from optimized and does not describe a
practical device size. These constants are tower exponentials: the
analytic majorant is $\Cs=2^{2^{40}}$ in~\eqref{c:eq:masterconstants},
and $C_{\rm res}=2^{Z^{16}}$ with $Z\ge\Cs^{128}$ in~\eqref{c:eq:theoremconstants}.
In particular, the coupling scale $3^{-d}$ with $d\ge\Cs^8=2^{2^{43}}$
already needs more than $2^{2^{43}}$ bits, so the $O(\log(2+m))$ bit
lengths in the theorem carry a fixed constant of astronomical size, and
the theorem is a purely asymptotic statement.

\subsection{The Yee model and its realization}\label{d:sec:model}

Let a rectangular cubical cell complex have side lengths $N_x,N_y,N_z$ and unit grid spacing. We use the Yee placement of electric unknowns on edges and curls on faces~\citep{yee1966}. Electric edges lying tangentially in the boundary are set to zero. Thus retained $x$-oriented edges have indices
\begin{equation}\label{d:eq:pec-index}
 0\le i_x<N_x,\qquad 1\le i_y<N_y,\qquad 1\le i_z<N_z,
\end{equation}
with cyclic versions for the other orientations. Let $C_{\rm Y}$ denote the resulting face-edge curl incidence matrix. We keep all face rows, allowing identically zero boundary-face rows.

Assign one scalar $\eps_c$ to each physical cell. The electric mass matrix is diagonal, with
\begin{equation}\label{d:eq:cell-average}
 (D_\eps)_{ee}=\frac14\sum_{c\sim e}\eps_c.
\end{equation}
Every retained edge has exactly four incident cells inside the domain. Fix
\begin{equation}\label{d:eq:maxwell-model}
 \omega=\frac18,\qquad
 A=C_{\rm Y}^{\mathsf T}C_{\rm Y}-\omega^2D_\eps,
 \qquad 1\le\eps_c\le25.
\end{equation}
All coordinate stretches equal one. The construction uses the physical box itself with no absorbing layer. Its dimension is
\begin{align}\label{d:eq:yee-dimension}
 N={}&N_x(N_y-1)(N_z-1)+(N_x-1)N_y(N_z-1)\notag\\
 &+(N_x-1)(N_y-1)N_z.
\end{align}
Every cell, including background cells between defects and the physical buffer, is included in the size accounting.

The maximal material wavenumber obeys the fixed sampling condition
\begin{equation}\label{d:eq:resolution}
 \omega\sqrt{25}\le\frac{2\pi}{10},
 \qquad \frac{25}{64}<\frac{4\pi^2}{100}.
\end{equation}
This is an admissibility condition on the discrete family. It does not itself supply a continuum error estimate.

\begin{theorem}[Sparse systems in scalar discrete Maxwell operators]\label{d:thm:main}
For fixed $c_\kappa,c_\eta,B_{\rm in}$, there are a uniform randomized encoder, a finite-precision decoder, and effectively computable constants $c_A,c_\delta>0$ with the following property. Given an input satisfying~\eqref{c:eq:input}, the encoder produces a finite instance $AE=f$ of~\eqref{d:eq:maxwell-model}, a tolerance $\delta$, and decoding data. On every random draw,
\begin{align}\label{d:eq:main-resources}
 N&=\Ot_{c_\kappa,c_\eta,B_{\rm in}}(m^{3/2}), &
 T_{\rm enc}+T_{\rm dec}&=\Ot_{c_\kappa,c_\eta,B_{\rm in}}(m^{3/2}),\notag\\
 B_{\rm out}&=O_{c_\kappa,c_\eta,B_{\rm in}}(\log(2+m)), &
 \delta&\ge m^{-c_\delta},
\end{align}
where $B_{\rm out}$ bounds the output material and source coefficient lengths and the time bound is in bit operations. For $m\ge2$, there is a common event $\mathcal G$ of probability at least $1-3/(64m)$ on which
\begin{equation}\label{d:eq:main-good}
 \kappa_2(A)\le m^{c_A}
\end{equation}
and, for every $E\in\C^N$ satisfying
\begin{equation}\label{d:eq:maxwell-residual}
 \norm{AE-f}_2\le\delta\norm{f}_2,
\end{equation}
the decoder returns $x\in\mathbb Q^n$ satisfying~\eqref{c:eq:target}. The decoder uses the real parts of prescribed coordinates of $E$ with an explicit absolute precision of $O(\log(2+m))$ bits. The bounds include all auxiliary variables, empty cells, buffering, calibration, source construction, and readout. Finitely many base cases are handled by a counted fixed-size procedure.
\end{theorem}

\begin{remark}[Status of the Yee certificate]
The spectral premise of Theorem~\ref{d:thm:main} is supplied by a
computer-assisted certificate whose finite inequalities have been checked in
full. Its final binding step to the exact operator is being completed
(Section~\ref{si:sec:yee-certificate}); Theorem~\ref{d:thm:main} is
conditional on this step.
\end{remark}

Once the encoder's draw belongs to $\mathcal G$, the recovery statement holds for every vector meeting~\eqref{d:eq:maxwell-residual}, irrespective of how that vector is found. The theorem does not require an electromagnetic solver to return a particular approximation. The output size and encoder/decoder runtime bounds also hold on unsuccessful draws.

The finite-precision interface is part of the statement. The encoder fixes
the width of each signed fixed-point coordinate message, and the decoder
consumes only these prescribed bits on every draw. On $\mathcal G$, the
inverse bound gives a known polynomial bound on $\norm E$, so every
field satisfying~\eqref{d:eq:maxwell-residual} fits the reserved range and can be supplied at the stated
absolute accuracy. Both the integer range and fractional precision are
logarithmic. Certified intervals of that accuracy give the same interface.
An inaccessible arbitrary-real output is not counted as a finite-bit solution.

\begin{table}[tb]
\centering
\caption{\textbf{Scope of the Yee realization, Theorem~\ref{d:thm:main}.} The
left column lists the input class, the form of the electromagnetic instance
and the resources and numerical properties of the reduction; the right column
gives the guarantee proved for each. Size and cost bounds hold on every random
draw; conditioning and recovery hold on the common good event. Constants
depend only on the fixed input constants and on the selected fixed platform.}
\label{d:tab:resources}\small
\begin{tabular}{p{0.30\linewidth}p{0.58\linewidth}}
\toprule
Resource or assumption & Guarantee \\
\midrule
Sparse input & Nonsingular rational square system with $m$ nonzeros, short coefficients, and $\kappa_2(M)\le m^{c_\kappa}$ \\
Electromagnetic instance & Resolved unstretched three-dimensional Yee box with $1\le\eps_c\le25$ and PEC boundaries \\
Ambient size & $\Ot(m^{3/2})$, including all background and auxiliary unknowns \\
Preparation and recovery & $\Ot(m^{3/2})$ bit operations on every draw \\
Numerical guarantees & Inverse-polynomial residual tolerance on every draw; polynomial full-matrix conditioning on the common good event \\
Recovery quantifier & Every field satisfying the output residual tolerance on that same event \\
Fixed preprocessing & A proved terminating effective search, independent of $m$, included in the uniform bound \\
\bottomrule
\end{tabular}
\end{table}

Uniformity allows an effectively computable fixed startup cost. The spectral arguments select one background, defect, operating shift, and set of coupling bounds by terminating searches. Once selected, their support sizes and spectral dimensions do not depend on the sparse input. The theorem includes their fixed costs in the asymptotic constants. The spectral certificates determine the background, while the strict defect and interaction estimates guarantee that the fixed search terminates.

\subsection{Algorithm transfer}\label{si:sec:transfer-statements}

The realization theorems transfer algorithms in the direction opposite
to the construction. An algorithm for every Maxwell instance in the
constructed class can be applied to the encoding of a sparse linear system, after which
the decoder recovers the algebraic solution. Quantifying this implication
requires both the cost of the Maxwell solve and the cost of constructing
its input. It also requires agreement between the arithmetic model,
accuracy criterion, and randomness in the reduction and in the hardness
hypothesis.

We state the hypothesis explicitly. The underlying approximation problem and
conditioning scale follow the sparse linear-equation formulation of Kyng,
Wang and Zhang~\cite{kyng2020}. Their Definition~3.2 excludes a running time
$\Ot(m^\gamma)$ and does not specify a randomized success probability.
We use a randomized exponent-gap variant: success at least $2/3$ is required,
and every exponent $\gamma'<\gamma$ is excluded.
These two conventions are part of our hypothesis.

\begin{assumption}[Matching randomized sparse linear-system hardness]
\label{d:ass:slth}
For fixed constants $\gamma>0$ and $c_\kappa>0$, consider real full-column-rank
matrices $F_0\in\R^{n_0\times n}$ with $m\ge2$ nonzeros and
$\kappa_2(F_0)\le m^{c_\kappa}$. The input is a sparse list for $F_0$ and a
random-access vector $b\in\R^{n_0}$. Let $\Pi_{F_0}$ denote orthogonal
projection onto the column space of $F_0$. For every $\gamma'<\gamma$,
there is no randomized Real-RAM algorithm with worst-case running time
$\Ot(m^{\gamma'})$ that, on every such input, returns with
probability at least $2/3$ a vector $x$ satisfying
\begin{equation}
 \norm{F_0x-\Pi_{F_0}b}_2^2
 \le n^{-10c_\kappa}\norm{\Pi_{F_0}b}_2^2.
 \label{d:eq:slth-target}
\end{equation}
\end{assumption}

The randomized success requirement in this assumption is part of the
hypothesis made here. A deterministic hardness statement alone would not
imply it. The relative error is measured against the projected right-hand
side, which can be much smaller than $\norm{b}_2$. That distinction is
essential for the transfer.

\begin{theorem}[Algorithm transfer for the Yee family]
\label{d:thm:algorithm-transfer}
Suppose a uniform Maxwell solver, on all instances produced by
Theorem~\ref{d:thm:main}, with their condition and accuracy bounds,
returns a residual solution with a fixed positive success probability
and has the known running-time bound
\begin{equation}
 T_{\rm Max}(N,B_{\rm out},\kappa,\delta)
 \le O\!\left(N^\theta B_{\rm out}\,
        \operatorname{polylog}(N,\kappa,\delta^{-1})\right).
 \label{d:eq:maxwell-solver-model}
\end{equation}
Here $B_{\rm out}$ bounds the coefficient and source bit lengths, and the output
uses the finite-precision interface of Theorem~\ref{d:thm:main}. Then the
problem in~\eqref{d:eq:slth-target}, for every fixed $c_\kappa$, has a randomized
Real-RAM algorithm with running time
\begin{equation}
 T_{\rm sparse}(m)
 =\Ot\!\left(m^{3/2}+m^{3\theta/2}\right),
 \label{d:eq:sparse-composition-time}
\end{equation}
and failure probability smaller than any prescribed inverse polynomial.
\end{theorem}

The continuum realization gives the analogous transfer. Its condition
number is measured between the graph-norm space $X$ and $L^2$.

\begin{theorem}[Costed continuum-to-sparse transfer]
\label{c:thm:transfer}
Suppose a uniform continuum solver, for every instance produced by
Theorem~\ref{c:thm:main}, has a known worst-case bound
\begin{equation}
 T_{\rm Max}(N,B_{\rm out},\kappa_X,\delta)
 \le O\!\left(N^\theta B_{\rm out}\,
       \operatorname{polylog}(N,\kappa_X,\delta^{-1})\right),
 \quad \kappa_X=\|A\|_{X\to L^2}\|A^{-1}\|_{L^2\to X}.
                                                               \label{c:eq:solverbound}
\end{equation}
Here $N$ includes the full ambient volume and finite description slots;
$B_{\rm out}$ includes coefficient lengths and prescribed message widths. With a
fixed positive probability the solver supplies the prescribed accurate
moments of some field meeting~\eqref{c:eq:physicalres}. Producing those
moments is part of~\eqref{c:eq:solverbound}.
Then the rational problem~\eqref{c:eq:input}--\eqref{c:eq:target} has a
randomized bit algorithm of cost
\begin{equation}
 \boxed{\ T_{\rm sparse}(m)=
          \tO(m^{3/2}+m^{3\theta/2})\ }.             \label{c:eq:transfer}
\end{equation}
The same bound holds in the Real-RAM model of Assumption~\ref{d:ass:slth} for a
full-column-rank real $F_0$ with $m$ nonzeros, $\kappa_2(F_0)\le m^{c_\kappa}$,
random-access real source $b$, and target
\begin{equation}
 \|F_0x-\Pi_{F_0}b\|^2
      \le n^{-10c_\kappa}\|\Pi_{F_0}b\|^2.                 \label{c:eq:projectiontarget}
\end{equation}
Failure can be reduced below any prescribed inverse polynomial.
\end{theorem}

In both transfer theorems the solver cost depends polylogarithmically on
the condition number and on $\delta^{-1}$. On the constructed families
these quantities are polynomial in $m$, in the continuum case with a fixed
degree of the astronomical size just described, so methods whose cost
grows polynomially in $\kappa$, such as Krylov iterations, are not covered by
\eqref{d:eq:maxwell-solver-model} or~\eqref{c:eq:solverbound}.

\begin{remark}[Weak residuals]
\label{si:rem:weak-residual}
Theorems~\ref{c:thm:main} and~\ref{c:thm:transfer} also hold for fields
$E\in V$ that satisfy the weak residual bound
$\norm{AE-f}_{V^*}\le(\delta/140)\norm f_2$ in place
of~\eqref{c:eq:physicalres}, with the same decoder. Here
$\norm E_V^2=\norm E_2^2+\norm{\curl E}_2^2$, $V^*$ is the
antidual of $V$, and $AE\in V^*$ is
$F\mapsto c(E,F)-\ip{\varepsilon E}{F}$. This is the residual a
conforming $H_0(\curl)$ finite-element solution satisfies directly.
On the event $\mathcal G$ of Theorem~\ref{c:thm:main}, $A$ is an
isomorphism $V\to V^*$ with
\begin{equation}
 \norm{A^{-1}}_{V^*\to V}\le140L_b.          \label{si:eq:weakinverse}
\end{equation}
Indeed, the form $c(E,E)-\ip{\varepsilon E}{E}+26\norm E_2^2$ is
coercive on $V$, so $V$ is the form domain of the self-adjoint operator
$A+26$, and the spectral theorem extends $A^{-1}$ to a bounded map
$V^*\to V$ because $\spec A\subset[-25,\infty)$ keeps distance at
least $L_b^{-1}$ from zero. For the bound, let $g\in V^*$, $E=A^{-1}g$
and $F=A^{-1}E\in X$. Then $\norm E_2^2=g(F)\le\norm g_{V^*}\norm F_V$.
Lemma~\ref{c:lem:detuning} gives $\norm F_2\le L_b\norm E_2$ and
$\norm{\Cc F}_2\le(1+25L_b)\norm E_2$, so
$\norm{\curl F}_2^2=\ip{\Cc F}{F}\le26L_b^2\norm E_2^2$ and
$\norm E_2\le\sqrt{27}L_b\norm g_{V^*}\le27L_b\norm g_{V^*}$.
Next, $\norm{\curl E}_2^2=g(E)+\ip{\varepsilon E}{E}
\le\norm g_{V^*}\norm E_V+25\norm E_2^2$, hence
$\norm E_V^2\le26\cdot27^2L_b^2\norm g_{V^*}^2+\norm g_{V^*}\norm E_V$,
which gives~\eqref{si:eq:weakinverse} since $L_b\ge1$.
Section~\ref{c:sec:recovery} uses the residual only through
$\norm{E-E_{\rm ref}}_2$ and $\norm E_2$. Writing
$E-E_{\rm ref}=A^{-1}(AE-f)+A^{-1}(f-AE_{\rm ref})$ with
$f-AE_{\rm ref}\in L^2$ gives
$\norm{E-E_{\rm ref}}_2\le140L_b(\delta/140)\norm f_2
+L_b\norm{AE_{\rm ref}-f}_2$, and in the same way
$\norm E_2\le L_b(1+\delta)\norm f_2$. These are
\eqref{c:eq:fieldcomparison} and~\eqref{c:eq:fieldrange}; the rest of the
recovery argument is unchanged.
\end{remark}

\subsection{Organization of the proofs}

Three interfaces organize both proofs. First, the physical platform supplies a simple scalar defect with a separated mode and quantitative interaction bounds on a fixed translation lattice. Second, a logical compiler turns $Mx=b$ into a bounded-degree symmetric matrix $\Gamma$ embedded in a balanced cubic-grid box, with a polynomial inverse bound and an explicit residual map. Third, finite assembly realizes $\Gamma$ through scalar defect tunings, controls the whole Maxwell inverse, and implements a local source and decoder.

In the continuum back end (Part~II) the platform is an explicit weak
periodic scalar modulation with a simple local Bloch-band minimum, and
broad weak defects interact through a localized auxiliary resolvent; the
proof of Theorem~\ref{c:thm:main} concludes Section~\ref{c:sec:cost}.
In the Yee back end (Part~III) the platform is a certified band edge of
a period-20 scalar dielectric with finite scalar defects; the proof of
Theorem~\ref{d:thm:main} concludes Section~\ref{d:sec:composition}.
The two back ends are proved independently. The resolved-grid
condition~\eqref{d:eq:resolution} fixes a wavelength sampling ratio, but
it does not transfer the Yee construction to the continuum at the
accuracy required by the reduction, because the useful interactions and
the permitted errors shrink with the sparse input size.

\sipart{I}{Shared front end: algebraic compiler and routing}
\section{Compiling sparse algebra into local networks}
\label{d:sec:logical}

The physical constructions of Parts~II and~III provide controlled
families of local interactions. To use it for an arbitrary sparse system, we
must first express that system in a form compatible with those interactions.
An input matrix can have large vertex degree, signed coefficients, and
connections between variables that are far apart in any chosen geometric
layout. Its coefficients also vary with the input. We remove these freedoms
through an explicit algebraic compilation and spatial routing procedure.
The resulting matrix has positive unit off-diagonal entries, degree at most
three, and a fixed finite diagonal alphabet. Its dependence on the input is
carried by the network geometry, choices from the fixed diagonal alphabet,
and the source.

The compilation must preserve numerical information as well as exact
solutions. In particular, perturbations of auxiliary equations are part of
the residual presented by a downstream solver. We therefore bound the
inverse of every enlarged system and track arbitrary residuals through
each elimination. This supplies a stable interface between sparse algebra
and the electromagnetic construction.

\begin{proposition}[A local algebraic normal form]
\label{d:thm:logical}
Under the input assumptions~\eqref{c:eq:input}, a nonzero-source system
$Mx=b$ can be transformed deterministically into
\begin{equation}
    \Gamma z=g_\Gamma
    \label{d:eq:logical-normal-form}
\end{equation}
with the following properties. The matrix $\Gamma$ is real symmetric and
supported on a subgraph of the three-dimensional unit cubic lattice. Its
graph degree is at most three, every nonzero off-diagonal entry equals
$+1$, and its diagonal belongs to a fixed finite subset of $[-3,3]$.
Moreover,
\begin{equation}
    \norm{\Gamma}_2\le 7,
    \qquad
    \norm{\Gamma^{-1}}_2\le \overline K,
    \label{d:eq:logical-norms}
\end{equation}
where $\overline K$ is an explicit polynomial bound in $m$.
The bounding rectangular box has $O((m\log m)^{3/2})$ lattice sites.
The source coordinates have $O(\log m)$ bits, and a prescribed
inverse-polynomial relative residual for~\eqref{d:eq:logical-normal-form}
suffices for coordinate selection, rounding, and scalar rescaling to
recover an answer to $Mx=b$ at the requested relative residual.
Construction and decoding, including writing the full occupancy of the
bounding box, require $\Ot(m^{3/2})$ bit operations.
\end{proposition}

The constants in this proposition depend only on the fixed input constants $c_\kappa,c_\eta,B_{\rm in}$.
The construction does not call a solver for the original system. We give
the main transformations here and prove their quantitative claims in
Sections~\ref{d:app:logical}--\ref{d:app:logical-residual}.

\subsection{Short coefficients through a stable arithmetic circuit}

Let $P=m^{\lceil c_\kappa\rceil+1}$ and
$\eta_0=m^{-\lceil c_\eta\rceil}$. For $b\ne0$, define
\begin{equation}
 a_M=\max_{i,j}|M_{ij}|,\qquad \nu_M=\norm{b}_\infty,
 \qquad F=M/(ma_M),\qquad g=b/\nu_M.
 \label{d:eq:logical-normalization}
\end{equation}
Then $\norm{F}_2\le1$, $\sigma_{\min}(F)\ge P^{-1}$, and
$1\le\norm{g}_2\le\sqrt m$. Solving $Fu=g$ gives
$x=(\nu_M/(ma_M))u$. Round the existing entries of $F$ to a dyadic grid
of spacing $2^{-N_{\rm r}}\le\eta_0/(64mP)$, without introducing entries outside
the original sparsity pattern. The rounded matrix $\widehat F$ satisfies
\begin{equation}
 \norm{\widehat F-F}_2\le \frac{\eta_0}{64P},
 \qquad \norm{\widehat F^{-1}}_2\le2P,
 \qquad N_{\rm r}=O(\log m).
 \label{d:eq:logical-rounding}
\end{equation}

We express multiplication by each dyadic coefficient through binary
Horner evaluation. Copy trees give a separate occurrence of an input to
each nonzero coefficient bit, and binary addition trees sum the resulting
terms within each row. A sign gate implements each coefficient's sign.
The equations defining these gates, together with the output equations,
form a square system
\begin{equation}
 \Gamma_{\rm c}\begin{pmatrix}u\\w\end{pmatrix}=g_{\rm c}
 \label{d:eq:logical-circuit}
\end{equation}
of size $S\le6m(N_{\rm r}+1)$. Every coefficient of $\Gamma_{\rm c}$ belongs to
$\{0,\pm1,\pm\tfrac12\}$, and each row and column has at most three
nonzero entries. In particular, $\norm{\Gamma_{\rm c}}_2\le3$.

The circuit is arranged so that an auxiliary perturbation cannot be
amplified exponentially by repeated branching and recombination. If its
gate equations are $T_{\rm g}w-B_{\rm g}u=\rho_{\rm aux}$, then $T_{\rm g}$ is unit lower triangular
and each entry of $T_{\rm g}^{-1}$ is a sum over directed paths. Branching occurs
only in the initial copy trees; after leaving a copy tree, a path follows
a Horner chain, a sign gate, and a row-sum tree without further branching.
There are at most $S$ paths between any two gates. Consequently,
\begin{equation}
 \norm{T_{\rm g}^{-1}}_2\le S^2,
 \qquad \norm{\Gamma_{\rm c}^{-1}}_2\le16PS^4=:Q_{\rm c}.
 \label{d:eq:logical-circuit-stability}
\end{equation}
More specifically, for any approximate solution of the circuit equations,
\begin{equation}
 \norm{\widehat F u-g}_2
 \le \sqrt n\bigl((S+1)^2+1\bigr)
      \norm{\Gamma_{\rm c}(u,w)-g_{\rm c}}_2.
 \label{d:eq:logical-circuit-residual}
\end{equation}
This estimate treats all gate equations as potentially inexact. It is the
property needed when the circuit is later recovered from an approximate
electromagnetic field.

\subsection{Signed algebra with positive unit edges}

Symmetrization gives
\begin{equation}
 \Gamma_0=\begin{pmatrix}0&\Gamma_{\rm c}\\\Gamma_{\rm c}^\top&0\end{pmatrix},
 \qquad\text{with source }\begin{pmatrix}g_{\rm c}\\0\end{pmatrix}.
 \label{d:eq:logical-symmetrization}
\end{equation}
The singular values of $\Gamma_{\rm c}$ become the absolute eigenvalues of $\Gamma_0$.
Thus this step preserves the relevant norm and inverse-norm bounds,
while a block of its residual is exactly the circuit residual.

To replace a signed edge of weight $w_e\in\{\pm1,\pm\tfrac12\}$,
introduce a mediator connected to its two endpoints by edges of weight
$+1$. Give that mediator diagonal $-1/w_e$, and give each original vertex
the negative sum of its incident old weights. If $\mathcal I$ is the unsigned
incidence matrix, $(D_{\rm med})_{ee}=-1/w_e$, and $D_{\rm end}$ is this endpoint diagonal,
the enlarged matrix is
\begin{equation}
 \Gamma_+=\begin{pmatrix}D_{\rm end}&\mathcal I\\\mathcal I^\top&D_{\rm med}\end{pmatrix},
 \qquad D_{\rm end}-\mathcal ID_{\rm med}^{-1}\mathcal I^\top=\Gamma_0.
 \label{d:eq:logical-sign-mediators}
\end{equation}
The identity in~\eqref{d:eq:logical-sign-mediators} explains how signs are
represented using positive unit edges: they appear in the effective
interaction after mediator elimination. The mediator diagonals belong to
$\{-2,-1,1,2\}$, while endpoint diagonals are half-integers in $[-3,3]$.
The degree remains at most three.

The graph of $\Gamma_+$ is bipartite, with original vertices on one side and
mediators on the other. It has at most $S_+=5S$ vertices. Since
$\norm{\mathcal I}_2\le\sqrt6$ and $\norm{D_{\rm med}^{-1}}_2\le1$, block factorization
gives
\begin{equation}
 \norm{\Gamma_+^{-1}}_2\le16Q_{\rm c}=:Q_+.
 \label{d:eq:logical-positive-inverse}
\end{equation}
Eliminating a mediator residual amplifies its norm by less than four.
Thus neither the sign conversion nor symmetrization loses polynomial
control.

\subsection{Routing the network in three dimensions}

We next embed a subdivision of this bounded-degree bipartite graph in a
cubic lattice. Each incidence receives its own port in a square array.
The two ports of each edge are routed to a designated adjacent pair at
the top of a three-dimensional box. A row--column--row permutation
schedule realizes all these assignments. The schedule is obtained by
recursively splitting even-regular bipartite multigraphs into perfect
matchings and using adjacent odd--even transposition sorting within rows
and columns. Each swap occupies a fixed-size lattice box, with two
vertex-disjoint paths inside it.

For a power of two $\ell_{\rm p}$ with $2\sqrt{S_+}\le\ell_{\rm p}<4\sqrt{S_+}$, the construction
uses a box with all three side lengths $O(\ell_{\rm p})$. Each routed edge has length
at most $\ell_w=1024\ell_{\rm p}$. The box therefore contains $O(S_+^{3/2})$ sites,
including the sites not occupied by a path. This accounting is the source
of the exponent $3/2$ in the construction. The volume $O(S_+^{3/2})$
matches the classical bound for realizing bounded-degree networks in
three dimensions~\citep{KolmogorovBarzdin1967,LeightonRosenberg1986}.

Path length also controls the sign of the effective endpoint interaction.
The bipartition permits endpoints of opposite lattice parity. We then add
a local two-edge detour whenever needed so that every path length is
$1$ modulo $4$. Put zero diagonal on the internal vertices and $+1$ on
every path edge. If a path has $4j$ internal vertices, the inverse of its
internal adjacency matrix has vanishing endpoint diagonal entries and
far-corner entry $-1$. Its Schur complement is therefore exactly a
positive unit endpoint edge, with no diagonal correction. The routed
matrix $\Gamma$ has $\Gamma_+$ as its exact Schur complement.

The same calculation yields the explicit bound
\begin{equation}
 \overline K=9\ell_w^2(Q_++\ell_w)
 \quad\text{with}\quad
 \norm{\Gamma^{-1}}_2\le\overline K.
 \label{d:eq:logical-final-inverse}
\end{equation}
Since $S=O(m\log m)$ and $\ell_w=O(\sqrt S)$, this is polynomial in $m$.
The geometric layout uses $\Ot(m^{3/2})$ volume and bit operations.
We do not know whether this routing exponent is optimal.

\subsection{Residuals and decoding}

Let
\begin{equation}
 T_{\rm res}=4n\bigl((S+1)^2+1\bigr)(1+2\ell_w),
 \qquad \eta_{\rm res}=\frac{\eta_0}{16T_{\rm res}}.
 \label{d:eq:logical-tolerance}
\end{equation}
Eliminating internal path vertices, eliminating sign mediators, selecting
the circuit block, and eliminating gate variables transfers a relative
$\eta_{\rm res}$ residual for $\Gamma$ into an error at most
$\eta_0\norm{g}_2/16$ for $\widehat F u=g$. Equation
\eqref{d:eq:logical-rounding} then controls the difference between
$\widehat F$ and $F$. Rounding the selected input coordinates on a grid
of spacing $\eta_0/(16m)$ leaves the normalized residual below
$\eta_0\norm{g}_2/4$. Scalar rescaling by $\nu_M/(ma_M)$ produces the
required answer to the original system.

The decoder uses coordinate restriction and short rational arithmetic.
The selected coordinates have polynomial magnitude, so their required
integer and fractional precision is $O(\log m)$ bits. Every auxiliary
variable, path vertex, empty site in the bounding box, and decoding
operation has been included in the resource estimate. The output of this
section is thus a quantitative normal form suitable for the finite
Maxwell assembly.

\begin{figure}[tbp]
\centering
\includegraphics[width=\linewidth]{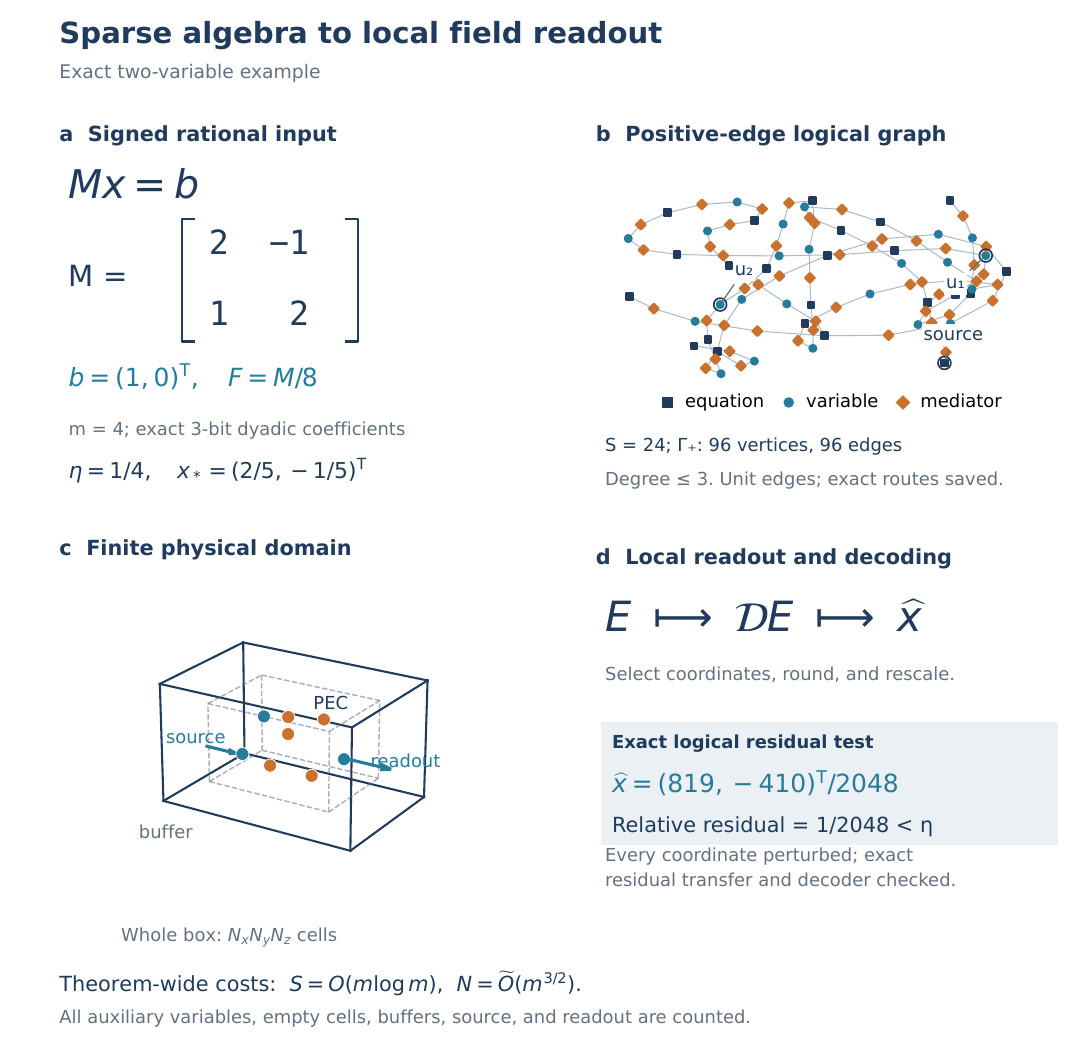}
\caption{An exact sparse-algebra example and the schematic physical
translation. (a) The signed nonsymmetric matrix is normalized to
$F=M/8$ with no coefficient rounding error. (b) The actual compiled
circuit has $S=24$ variables; symmetrization and sign mediators give the
shown $\Gamma_+$ graph with 96 vertices and 96 positive unit edges, degree
at most three, and fixed half-integer diagonals. Its verified cubic
routing, supplied as exact figure data, has 760\,320 occupied lattice
sites in a $505\times169\times3105$ bounding box containing
264\,996\,225 sites in total. (c) Scalar dielectric defects, the complete
PEC box, buffer, source, and local readout are schematic. The physical construction counts all
$N_xN_yN_z$ cells. (d) An exact perturbation of every logical coordinate,
in place of a computed Maxwell field, passes the prescribed logical
residual tolerance; coordinate selection,
rounding, and rescaling return $\widehat x=(819,-410)^\top/2048$ with
relative source-system residual $1/2048<\eta=1/4$.}
\label{d:fig:overview}
\end{figure}

\subsection{The interface used by the continuum back end}
The continuum assembly in Part~II uses the following form of
Proposition~\ref{d:thm:logical}, with a logical tolerance
$\eta_\Gamma$ fixed in advance. Sections~\ref{d:app:logical}--\ref{d:app:logical-residual}
prove it; Remark~\ref{si:rem:continuum-compiler} records the
parameter-specific details.

\begin{lemma}[Logical interface]
\label{c:lem:logical}
In $\tO(m^{3/2})$ bit operations one constructs a real symmetric matrix
$\Gamma$ on a subgraph of the unit cubic lattice, of degree at most three,
with positive unit edges and half-integer diagonals in $[-3,3]$.
It has $\norm\Gamma\le7$, an explicit polynomial bound
$\norm{\Gamma^{-1}}\le \overline K$, and a source $g$ with
$\norm g_\infty=1$ and at most $m$ nonzero entries. All three side lengths
of its bounding box are $O(\sqrt{m\log(2+m)})$.
An explicit $\ee>0$, with $\ee^{-1}$ polynomial in $m$, has the property
that every $z$ satisfying
$\norm{\Gamma z-g}\le\ee\norm g$ decodes by coordinate selection,
rounding, and rational rescaling to~\eqref{c:eq:target}.

The complete list of parameters is
\begin{align}
 P&=m^{\lceil c_\kappa\rceil+1},\qquad \eta_0=m^{-\lceil c_\eta\rceil},
 \qquad 2^{-N_{\rm r}}\le\eta_0/(64mP),\\
 S&\le6m(N_{\rm r}+1),\quad S_+\le5S,\quad
 2\sqrt{S_+}\le\ell_{\rm p}<4\sqrt{S_+},\quad \ell_{\rm p}\text{ a power of two},\\
 \ell_w&=1024\ell_{\rm p},\quad Q_{\rm c}=16PS^4,\quad Q_+=16Q_{\rm c},\\
 \overline K&=9\ell_w^2(Q_++\ell_w),\qquad
 T_{\rm res}=4n\bigl((S+1)^2+1\bigr)(1+2\ell_w),\notag\\
 \eta_*&=\min\left\{\frac12,
 \frac{\eta_0}{\max\{1,T_{\rm res}\}\max\{1,m^2\}}\right\},
 \qquad \ee=\eta_*/16.                         \label{c:eq:logicalparameters}
\end{align}
Thus $T_{\rm res}\ee\le\eta_0/16$. The factor $1/16$ reserves the
logical residual margin. The factor $\max\{1,m^2\}$ is a spare margin:
the continuum chain budgets its absolute readout errors separately,
through $h_y$ in~\eqref{c:eq:momentaccuracy} and $e_0$
in~\eqref{c:eq:e0}. All choices use rational arithmetic.
A bounding box has at most $2^{17}\ell_{\rm p}^3$ sites. These include its empty
sites, not only its graph vertices.
\end{lemma}

\section{A stable compiler for sparse linear systems}
\label{d:app:logical}

We prove Proposition~\ref{d:thm:logical} with explicit parameters. All
norms in this section are Euclidean operator or vector norms unless a
subscript states otherwise. The case $b=0$ is answered by $x=0$; bounded
input sizes can be handled by a fixed finite computation. We therefore
assume $m\ge2$ and $b\ne0$.

\subsection{Normalization and perturbation bounds}

Set $P=m^{\lceil c_\kappa\rceil+1}$,
$\eta_0=m^{-\lceil c_\eta\rceil}$, and use the normalization in
\eqref{d:eq:logical-normalization}. Since
\begin{equation}
 a_M\le\norm{M}\le\norm{M}_{\mathrm F}\le\sqrt m\,a_M,
 \qquad \sigma_{\min}(M)\ge a_M m^{-c_\kappa},
\end{equation}
we have $\norm{F}\le1$ and $\norm{F^{-1}}\le P$.
The vector $g$ has infinity norm one and at most $m$ coordinates.
Choose an integer $N_{\rm r}\ge1$ such that
$2^{-N_{\rm r}}\le\eta_0/(64mP)$ and round each existing nonzero of $F$
to its nearest multiple of $2^{-N_{\rm r}}$, with a fixed tie rule. Delete entries
that become zero. Then
\begin{equation}
 \norm{\widehat F-F}
 \le\norm{\widehat F-F}_{\mathrm F}
 \le\sqrt m\,2^{-N_{\rm r}}/2\le\eta_0/(64P).
 \label{d:eq:app-dyadic-perturbation}
\end{equation}
In particular,
$\norm{F^{-1}(\widehat F-F)}\le1/64$.
The Neumann series for
$[I+F^{-1}(\widehat F-F)]^{-1}$ gives
$\norm{\widehat F^{-1}}\le2P$.
Each coefficient has $N_{\rm r}=O_{c_\kappa,c_\eta}(\log m)$ fractional bits. Normalization
and rounding can be performed independently on each short rational
input entry, in $O(m\operatorname{polylog}m)$ bit operations; a common
denominator across all entries is unnecessary.

\subsection{An acyclic circuit with a polynomial inverse}

Write each nonzero absolute dyadic coefficient as
\begin{equation}
 |\widehat F_{ij}|=\sum_{a=1}^{N_{\rm r}} t_{ij,a}2^{-a},
 \qquad t_{ij,a}\in\{0,1\}.
\end{equation}
For each original input variable $u_j$, build a binary copy tree with
one leaf for each nonzero bit occurrence referring to $u_j$. An input
or copy output has at most two later uses. If there is only one use,
the original input can supply it directly. Nonsingularity of
$\widehat F$ ensures that every input has a use.

For each coefficient, use the Horner equations
\begin{equation}
 v_{N_{\rm r}+1}=0,\qquad
 v_a=\tfrac12 v_{a+1}
       +\tfrac12 t_{ij,a}u_{j,a},\qquad a=N_{\rm r},N_{\rm r}-1,\ldots,1.
 \label{d:eq:app-horner}
\end{equation}
The value $v_{N_{\rm r}+1}=0$ is a constant and not an additional variable.
Each nonzero bit uses its designated copy-tree leaf. A sign gate
multiplies $v_1$ by $\operatorname{sign}(\widehat F_{ij})$, and a binary
addition tree sums the terms of each row. Let its output be $o_i$.
There is at least one term per row. The defining gate equations and
the $n$ equations $o_i=g_i$ give the square system
$\Gamma_{\rm c}(u,w)=g_{\rm c}$.

There are at most $mN_{\rm r}$ bit occurrences. The copy trees use at most
$2mN_{\rm r}$ gates, the Horner chains at most $mN_{\rm r}$, sign gates at most $m$,
and row additions at most $m$. Including original inputs gives
\begin{equation}
 S=n+\dim w\le6m(N_{\rm r}+1).
\end{equation}
Every equation has at most three coefficients of absolute value at
most one. Every variable occurs in its defining equation, if present,
and at most two later uses. Thus the row and column absolute sums of
$\Gamma_{\rm c}$ are at most three and $\norm{\Gamma_{\rm c}}\le3$.

In topological order, the auxiliary equations take the form
$T_{\rm g}w-B_{\rm g}u=\rho_{\rm aux}$, where $T_{\rm g}$ is unit lower triangular. Each row of
$B_{\rm g}$ has at most one nonzero and each column at most two, so
$\norm{B_{\rm g}}\le2$. If $R_{\rm g}$ selects the distinct row outputs, then
$\norm{R_{\rm g}}\le1$ and
\begin{equation}
 \widehat F=R_{\rm g}T_{\rm g}^{-1}B_{\rm g}.
 \label{d:eq:app-circuit-factorization}
\end{equation}
The inverse $T_{\rm g}^{-1}$ is a finite weighted-path expansion. Edge weights
have absolute value at most one. Within a copy tree there is a unique
path between any compatible pair of gates. A path that leaves a copy
tree is determined by its exit leaf and then has at most one
continuation: Horner chains, sign gates, and addition trees all have
outdegree at most one. Hence at most $S$ paths connect any pair of
auxiliary gates. Every inverse entry has absolute value at most $S$,
which gives $\norm{T_{\rm g}^{-1}}\le S^2$.

For an arbitrary right-hand side $\rho=(\rho_{\rm out},\rho_{\rm aux})$,
eliminating $w$ in the equations defining $\Gamma_{\rm c}^{-1}\rho$ yields
\begin{equation}
 \widehat F u=\rho_{\rm out}-R_{\rm g}T_{\rm g}^{-1}\rho_{\rm aux}.
\end{equation}
Therefore
\begin{align}
 \norm{u}&\le2P(1+S^2)\norm{\rho}\le4PS^2\norm{\rho},\\
 \norm{w}&\le S^2(2\norm{u}+\norm{\rho})
             \le9PS^4\norm{\rho}.
\end{align}
Combining these estimates proves
$\norm{\Gamma_{\rm c}^{-1}}\le16PS^4=Q_{\rm c}$.
For residual transport we use an explicit coordinate bound. Put
$e_{\rm c}=\norm{\Gamma_{\rm c}(u,w)-g_{\rm c}}$. Every gate residual has magnitude at most $e_{\rm c}$.
A copy leaf differs from its input by at most $Se_{\rm c}$. In each row expression,
weights have magnitude at most one and every occurrence has its own leaf.
Induction through its addition and scaling gates bounds the output error
by the number of gates times $e_{\rm c}$, plus the sum of its leaf errors. There
are at most $S$ gates and $S+1$ leaves, so this is at most $(S+1)^2e_{\rm c}$.
The output equation contributes one further $e_{\rm c}$ per row. Taking the
Euclidean norm over the $n$ rows proves
\eqref{d:eq:logical-circuit-residual}. The bound includes residuals in
every auxiliary equation.

\subsection{Symmetrization and sign mediators}

Form $\Gamma_0$ and its source as in~\eqref{d:eq:logical-symmetrization}.
Its graph is bipartite and has $2S$ vertices and at most $3S$ edges.
Direct multiplication gives
$\Gamma_0^2=\operatorname{diag}(\Gamma_{\rm c}\Gamma_{\rm c}^\top,\Gamma_{\rm c}^\top\Gamma_{\rm c})$, so
$\norm{\Gamma_0}=\norm{\Gamma_{\rm c}}$ and
$\norm{\Gamma_0^{-1}}=\norm{\Gamma_{\rm c}^{-1}}$.
For a residual solution $(v,(u,w))$, the first block of the residual
is $\Gamma_{\rm c}(u,w)-g_{\rm c}$ and needs no amplification to be recovered.

For each edge $e=\{i,j\}$ with weight $w_e$, introduce a mediator with
diagonal $(D_{\rm med})_{ee}=-1/w_e$. Let $\mathcal I$ be the unsigned incidence matrix of
the old graph and set $(D_{\rm end})_{ii}=-\sum_{e\ni i}w_e$. Since
$(D_{\rm med})_{ee}^{-1}=-w_e$, the diagonal entries of $\mathcal ID_{\rm med}^{-1}\mathcal I^\top$ equal
$-\sum_{e\ni i}w_e$, and its off-diagonal $ij$ entry is $-w_e$.
It follows that $D_{\rm end}-\mathcal ID_{\rm med}^{-1}\mathcal I^\top=\Gamma_0$ exactly.

The enlarged graph has $S_+\le5S$ vertices, degree at most three, and
positive unit edges. Its endpoint diagonals lie in
$\{j/2:-6\le j\le6\}$; its mediator diagonals lie in
$\{-2,-1,1,2\}$. The incidence bounds give
$\norm{\mathcal I}\le\sqrt{\norm{\mathcal I}_1\norm{\mathcal I}_\infty}\le\sqrt6$ and
$\norm{D_{\rm med}^{-1}}\le1$.
The factorization
\begin{equation}
 \Gamma_+=
 \begin{pmatrix}I&\mathcal ID_{\rm med}^{-1}\\0&I\end{pmatrix}
 \begin{pmatrix}\Gamma_0&0\\0&D_{\rm med}\end{pmatrix}
 \begin{pmatrix}I&0\\D_{\rm med}^{-1}\mathcal I^\top&I\end{pmatrix}
\end{equation}
implies
\begin{equation}
 \norm{\Gamma_+^{-1}}
 \le(1+\sqrt6)^2\max(1,Q_{\rm c})\le16Q_{\rm c}=Q_+.
\end{equation}
The source is $(g_{\rm c},0,0)$. Subtracting $\mathcal ID_{\rm med}^{-1}$ times the mediator
residual from the endpoint residual shows that the recovered $\Gamma_0$
residual has norm at most
$(1+\sqrt6)\norm{\rho_{\Gamma_+}}<4\norm{\rho_{\Gamma_+}}$.

\section{Explicit three-dimensional routing}
\label{d:app:routing}

\begin{lemma}[A counted cubic embedding]
\label{d:lem:cubic-routing}
A bipartite graph with $S_+$ vertices and degree at most three admits an
explicit subdivision in a rectangular unit cubic lattice box containing
$O(S_+^{3/2})$ sites. Distinct edge paths meet only at common graph
endpoints. For a power of two $\ell_{\rm p}$ with
$2\sqrt{S_+}\le\ell_{\rm p}<4\sqrt{S_+}$, every path has length at most $1024\ell_{\rm p}$
and length congruent to $1$ modulo $4$. The paths and the full bounding
box can be written in $O(S_+^{3/2}\log S_+)$ bit operations.
\end{lemma}

\begin{proof}
Put $\ell_{\rm p}=2^{1+\lceil(\log_2S_+)/2\rceil}$, twice the least power of two whose square is at least $S_+$.
Place the graph vertices in an $(\ell_{\rm p}/2)$ by $(\ell_{\rm p}/2)$ array and allocate a $2$ by
$2$ tile of an $\ell_{\rm p}$ by $\ell_{\rm p}$ port array to each vertex. Three positions in
the tile serve its incidences; unused positions are dummy ports. Assign
the two incidence tokens of each edge to one designated horizontal
pair of adjacent target ports. There are $\ell_{\rm p}^2/2\ge2S_+$ pairs, whereas
the graph has at most $3S_+/2$ edges. Assign dummy tokens to the remaining
targets to complete a permutation of all $\ell_{\rm p}^2$ ports.

Construct a bipartite multigraph whose two vertex sets are the $\ell_{\rm p}$
source rows and the $\ell_{\rm p}$ target rows, with one edge per token. This graph
is $\ell_{\rm p}$-regular. At an even-degree stage, pair consecutive incidences at
each source vertex and at each target vertex. The pairings define
fixed-point-free involutions $\iota_1,\iota_2$ on the tokens. Their alternating
components are even cycles, including the two-token case. For $\varpi=\iota_1\iota_2$,
the $\varpi$-orbits of $i$ and $\iota_1(i)$ are the two disjoint halves of one
component. Color $i$ according to which half has the smaller token label.
Both pairings reverse this color, so each color receives half the
incidences at every vertex.

The orbit minima are computed by stored arrays, not repeated traversal.
With $N_t$ tokens, initialize $j_0(i)=\varpi(i)$ and $\mu_0(i)=i$ and perform
\[
 j_{t+1}(i)=j_t(j_t(i)),\qquad
 \mu_{t+1}(i)=\min\{\mu_t(i),\mu_t(j_t(i))\}.
\]
Each pass reads the previous arrays and writes fresh arrays. After $t$
passes, $j_t(i)=\varpi^{2^t}(i)$ and $\mu_t(i)$ is the least label in the first
$2^t$ orbit positions. Thus $\lceil\log_2 N_t\rceil$ passes suffice.
The final comparison is $\mu_t(i)<\mu_t(\iota_1(i))$. Split the incidence lists
by this color and recurse until degree one; this gives $\ell_{\rm p}$ perfect
matchings. Each recursion level has $\ell_{\rm p}^2$ tokens in total and needs
$O(\ell_{\rm p}^2\log\ell_{\rm p})$ array operations. Across all levels the cost is
$O(\ell_{\rm p}^2(1+\log\ell_{\rm p})^2)$ array operations, including the pairings and
materialized output lists. Token labels have $O(\log\ell_{\rm p})$ bits.
Consequently the coloring bit cost is $O(\ell_{\rm p}^2(1+\log\ell_{\rm p})^3)$, which is
$O(\ell_{\rm p}^3\log\ell_{\rm p})$ after adjusting a fixed constant for small $\ell_{\rm p}$.

Give each token the color $j\in\{0,\ldots,\ell_{\rm p}-1\}$ of its matching.
First permute tokens within each source row so that color $j$ lies in
column $j$. Second, permute tokens within each column by their target
row. Third, permute tokens within each target row by their target
column. Each step is a collection of line permutations: a matching
contains each source and target row once, and the target ports were
assigned bijectively.

Each line permutation can be implemented by odd--even transposition
sort, i.e., parallel layers of adjacent comparisons that alternate
between odd and even positions. This network sorts any line of length
$\ell_{\rm p}$ in $\ell_{\rm p}$ layers~\citep[Section~5.3.4]{knuth1998}. We use $2\ell_{\rm p}$
layers in each of the three phases, giving $6\ell_{\rm p}$ layers in total.

To realize these layers geometrically, put port $(i,j)$ at
$(16i,16j,0)$ and allocate height $16$ per layer. A stationary token
travels vertically. For a swap between ports $(0,0,0)$ and $(16,0,0)$,
use the following two polygonal paths, filling each segment with unit
lattice edges:
\begin{align*}
 \mathcal P_A:\ &(0,0,0),(0,0,4),(0,4,4),(16,4,4),\\
       &(16,4,12),(16,0,12),(16,0,16),\\
 \mathcal P_B:\ &(16,0,0),(16,0,8),(16,-4,8),(0,-4,8),\\
       &(0,-4,12),(0,0,12),(0,0,16).
\end{align*}
Both paths have length $40$. Their vertices are disjoint, their
interiors lie strictly between the two slab faces, and their only
face points are the four ports. Translate this gadget for row swaps
and interchange the first two coordinates for column swaps. In a
single layer, swapped pairs are disjoint; row or column spacing is
$16$, while the sideways excursion has magnitude at most $4$.
Thus different swap boxes and stationary token paths are disjoint.
Successive layers meet only at their ports, occupied by a permutation
of the tokens.

After the final layer, extend each real incidence vertically by $16$
and join each designated adjacent pair by a horizontal segment at that
height. These segments are disjoint. Discard all dummy-token paths.

It remains to connect the incidences belonging to each graph vertex.
For tile $(i,j)$, its three port positions
are $(32i,32j,0)$, $(32i+16,32j,0)$, and $(32i,32j+16,0)$.
Place the graph vertex at $(32i,32j,-16)$ for one
bipartition and at $(32i,32j,-15)$ for the other. Connect to the first port
vertically. For either other port, first travel horizontally at the
height of the vertex, then vertically. These three arms meet only at their graph
vertex; different bottom stars lie in disjoint tiles. An original
edge now follows its first bottom arm and token trajectory, its top
pairing segment, and the reversed second trajectory and bottom arm.
It is simple, and two such paths intersect only at a shared endpoint.

The two bipartitions have opposite lattice parity, so each complete
edge path has odd length. For a path of length $3$ modulo $4$, replace
the unit segment from height $-8$ to $-7$ in one of its incidence arms
by a three-edge detour displaced one unit in the positive first
coordinate. These vertical arms are distinct and at least $16$ apart.
The detour lies above the bottom spokes and below the sorting layers,
so it is disjoint from all other paths. It adds two to the length,
giving $1$ modulo $4$.

An incidence trajectory has length at most $40\cdot6\ell_{\rm p}=240\ell_{\rm p}$. The two bottom
arms total at most $64$, the two top extensions contribute $32$, the
top pairing contributes $16$, and the parity detour contributes at
most two. Hence every edge length is at most
$480\ell_{\rm p}+114\le1024\ell_{\rm p}$. All coordinates fit in a box with side lengths
at most $16\ell_{\rm p}+9$, $16\ell_{\rm p}+9$, and $96\ell_{\rm p}+33$, whose number of lattice sites
is $O(\ell_{\rm p}^3)=O(S_+^{3/2})$. Token sorting uses $O(\ell_{\rm p}^3)$ comparisons on
$O(\log S_+)$-bit keys, and explicit path output uses $O(S_+\ell_{\rm p})$ steps.
Writing the entire occupancy array also costs $O(\ell_{\rm p}^3)$ word operations.
This proves all claims.
\end{proof}

\section{Residual transfer through the routed system}
\label{d:app:logical-residual}

Apply Lemma~\ref{d:lem:cubic-routing} to $\Gamma_+$. On a path of length
$4j+1$, keep all edge weights equal to $+1$ and give its $4j$ internal
vertices zero diagonal. Denote their adjacency matrix by $\Gamma_{4j}$.
For $j\ge1$, the recurrence
\begin{equation}
 v_{a-1}+v_{a+1}=\rho_a,\qquad v_0=v_{4j+1}=0,
\end{equation}
expresses the even coordinates as alternating sums of odd right-hand
side entries from the left, and the odd coordinates as alternating
sums of even entries from the right. Every inverse entry is $0$ or
$\pm1$, and
\begin{equation}
 (\Gamma_{4j}^{-1})_{1,1}=(\Gamma_{4j}^{-1})_{4j,4j}=0,
 \quad (\Gamma_{4j}^{-1})_{1,4j}=-1,
 \quad \norm{\Gamma_{4j}^{-1}}\le4j.
 \label{d:eq:app-wire-inverse}
\end{equation}
The endpoint Schur complement therefore has zero diagonal correction
and off-diagonal contribution $+1$. For $j=0$, the direct edge is
retained. Eliminating all internal path vertices recovers $\Gamma_+$
exactly.

Let $D_{\rm wire}$ be the block diagonal matrix of all nonempty
internal paths and $\mathcal I_{\rm wire}$ the endpoint-to-internal incidence
block. With $\ell_w=1024\ell_{\rm p}$,
\begin{equation}
 \norm{D_{\rm wire}^{-1}}\le\ell_w,
 \qquad \norm{\mathcal I_{\rm wire}}\le\sqrt3<2.
\end{equation}
The second estimate follows because an endpoint has at most three
incidences, while an internal vertex next to an endpoint belongs to
only that endpoint. A nonempty internal path has at least four
vertices. Block factorization, as in the mediator step, gives
\begin{equation}
 \norm{\Gamma^{-1}}
 \le (1+2\ell_w)^2\max(Q_+,\ell_w)
 \le9\ell_w^2(Q_++\ell_w)=\overline K.
\end{equation}
All retained diagonals have magnitude at most three and every row has
at most three unit edges, so $\norm{\Gamma}\le6\le7$.

The source $g_\Gamma$ is the circuit source with zeros appended in all
other coordinates, and $\norm{g_\Gamma}=\norm{g}$.
Eliminating an internal path residual gives a $\Gamma_+$ residual bounded
by $(1+2\ell_w)\norm{\rho_\Gamma}$. The mediator, symmetry, and circuit
estimates then yield
\begin{equation}
 \norm{\widehat F u-g}
 \le4n\bigl((S+1)^2+1\bigr)(1+2\ell_w)\norm{\rho_\Gamma}
 =T_{\rm res}\norm{\rho_\Gamma}.
 \label{d:eq:app-full-logical-residual}
\end{equation}
Taking the relative tolerance $\eta_{\rm res}=\eta_0/(16T_{\rm res})$ gives
$\norm{\widehat F u-g}\le\eta_0\norm{g}/16$ and hence
$\norm{u}\le3P\norm{g}$. By
\eqref{d:eq:app-dyadic-perturbation},
\begin{equation}
 \norm{Fu-g}
 \le \frac{\eta_0}{16}\norm{g}
       +\frac{\eta_0}{64P}\,3P\norm{g}
 =\frac{7\eta_0}{64}\norm{g}.
\end{equation}
Read each selected coordinate of $u$ to absolute error at most
$\eta_0/(32m)$, then round the approximation to the nearest multiple of
$\eta_0/(16m)$ and call the result $u_{\rm round}$. The total error per
coordinate is at most $\eta_0/(16m)$. Since $n\le m$,
its Euclidean perturbation is at most $\eta_0/(16\sqrt m)$ under a
conservative rounding bound. Using $\norm{F}\le1$ and
$\norm{g}\ge1$ gives
\begin{equation}
 \norm{Fu_{\rm round}-g}<\tfrac14\eta_0\norm{g}.
\end{equation}
Finally, $x=(\nu_M/(ma_M))u_{\rm round}$ satisfies
\begin{equation}
 \frac{\norm{Mx-b}}{\norm{b}}
 =\frac{\norm{Fu_{\rm round}-g}}{\norm{g}}
 <\eta_0/4\le\eta.
\end{equation}
The coordinates being rounded have magnitude at most $3P\sqrt m$.
Their required integer and fractional bit lengths, and those of the
short rational rescaling, are $O(\log m)$. The decoder performs no
linear solve. Together with $S=O(m\log m)$ and the counted routing
cost, this completes Proposition~\ref{d:thm:logical}.

\begin{remark}[Parameters of the continuum interface]
\label{si:rem:continuum-compiler}
Lemma~\ref{c:lem:logical} uses the construction of
Sections~\ref{d:app:logical}--\ref{d:app:logical-residual} with the same
parameters. Coefficient rounding uses a fixed downward tie rule, and
coordinate rounding uses the same fixed tie rule. The routed box has
side lengths at most $16\ell_{\rm p}+9$, $16\ell_{\rm p}+9$ and $96\ell_{\rm p}+33$, and hence at most
$2^{17}\ell_{\rm p}^3$ lattice sites, including its empty sites. Sorting, explicit
path output, and writing the entire occupancy array take
$O(\ell_{\rm p}^3\log S_+)=\tO(m^{3/2})$ bit operations. The tolerance in
\eqref{c:eq:logicalparameters} satisfies
$T_{\rm res}\eta_\Gamma\le\eta_0/16$, i.e.\
$\eta_\Gamma\le\eta_{\rm res}$ with $\eta_{\rm res}$ from~\eqref{d:eq:logical-tolerance}.
Therefore every $z$ with $\norm{\Gamma z-g}\le\eta_\Gamma\norm g$
satisfies $\norm{\widehat Fu-g}\le\eta_0\norm g/16$ by
\eqref{d:eq:app-full-logical-residual}, and the rounding and rescaling
of Section~\ref{d:app:logical-residual} return a solution
of~\eqref{c:eq:target}, with the margin $\eta_0/4$.
\end{remark}

\sipart{II}{Continuum back end}
\subsection*{Proof structure}
The finite compiler and balanced cubic routing of Part~I turn the sparse
system into a local network with controlled residual transport;
Lemma~\ref{c:lem:logical} states the interface used here. The physical construction supplies a
local Bloch-band minimum, scalar defects and interaction estimates that
realize this network. A finite-box inverse bound, including the
infinite-dimensional gradient space, controls the remaining physical modes.
Finite material, source and readout approximations complete the reduction.

\subsection*{Why a local band minimum is sufficient}
The periodic medium constructed below has a simple minimum on a small
Bloch chart. The selected band is the top of the sixteen-fold corner
cluster of Section~\ref{c:sec:edge}, and its minimum is local: the
quadratic regime extends only over a chart of radius $q_0=\varrho2^{-40}$
in Bloch momentum. A defect of radius $L$ must be localized in momentum
inside this chart, which forces the defect radius to satisfy
$L\ge\Ls$ in~\eqref{c:eq:masterconstants}. No assertion is made that its energy lies in a complete
three-dimensional gap. A cutoff resolvent on that chart is an
\emph{auxiliary proof operator}, not an implemented material or a physical
filter. The auxiliary response differs from a genuine Maxwell response
by a quantified $L^2$ residual. The full finite-box inverse bound controls
that difference. This distinction allows all three spatial directions to
be used without assuming an invariant physical sector.

\section{A full-operator inverse bound on any finite PEC box}
\label{c:sec:detuning}

This section is independent of the particular dielectric platform.

\begin{lemma}[The finite spectral count]
\label{c:lem:count}
Let $\Omega=\prod_{j=1}^3(0,N_j)$ with integer $N_j\ge1$, and let
$1\le\varepsilon_0\le25$ be any bounded measurable real scalar function.
For $A_0=\Cc-\varepsilon_0$, there are at most
\begin{equation}
 N_\Omega=2\prod_{j=1}^3(1+2N_j)\le54|\Omega|             \label{c:eq:count}
\end{equation}
spectral values in $[-1/4,1/4]$, counting multiplicity. These are isolated
finite-multiplicity eigenvalues; there is no essential spectrum in that
interval.
\end{lemma}
\begin{proof}
Use the component families
\[
 \begin{array}{ll}
 E_1:&\cos(i_1\pi r_1/N_1)\sin(i_2\pi r_2/N_2)\sin(i_3\pi r_3/N_3),\\
 E_2:&\sin(i_1\pi r_1/N_1)\cos(i_2\pi r_2/N_2)\sin(i_3\pi r_3/N_3),\\
 E_3:&\sin(i_1\pi r_1/N_1)\sin(i_2\pi r_2/N_2)\cos(i_3\pi r_3/N_3).
 \end{array}
\]
The cosine index may be zero; the two sine indices are positive. After
normalization these are a complete orthonormal component basis of $L^2$.
Grouping by $(i_1,i_2,i_3)$ diagonalizes the curl form into the matrices
$|p|^2I-pp^T$, with fixed harmless component signs. With three
positive indices there is one zero direction and two eigenvalues
$|p|^2$; with one zero index there is one positive direction.

Here is a domain justification. Integration by parts against a compactly
supported smooth vector field and Parseval show convergence of its block
expansion in the curl-form norm. Every finite expansion has zero
tangential trace. Its zero extension has no surface term in its curl;
inward dilation and mollification approximate it in the form norm by
compactly supported smooth fields, i.e., compactly supported smooth fields
are dense in the fields with zero tangential trace~\citep[Chapter~3]{monk2003}. Taking closures therefore gives
exactly $H_0(\curl;\Omega)$, not a smaller smooth subspace.

Let $P_{\rm g}$ and $P_{\rm t}$ be the orthogonal projections onto $\ker\Cc$ and its
orthogonal complement. The restriction $\Cc_{\rm t}$ to the complement has compact resolvent, with positive eigenvalues
\[
 \pi^2\sum_{j=1}^3(\nu_j/N_j)^2,
\]
with at least two positive indices and multiplicity at most two per
triple. Write the bounded material blocks as
$\varepsilon_{\rm gg}=P_{\rm g}\varepsilon_0P_{\rm g}\ge I$, $\varepsilon_{\rm tt}=P_{\rm t}\varepsilon_0P_{\rm t}\le25I$, and
$\varepsilon_{\rm tg}=P_{\rm t}\varepsilon_0P_{\rm g}$. The compression of $A_0$ to $\ker\Cc$ is
$-\varepsilon_{\rm gg}$, with spectrum in $[-25,-1]$, so the window $[-1/4,1/4]$ keeps
distance $3/4$ from it. For $|\lambda|\le1/4$, the gradient block
$-(\varepsilon_{\rm gg}+\lambda)$ is invertible. Its Schur operator is
\begin{equation}
 S_{\rm t}(\lambda)=\Cc_{\rm t}-\varepsilon_{\rm tt}-\lambda+\varepsilon_{\rm tg}(\varepsilon_{\rm gg}+\lambda)^{-1}\varepsilon_{\rm tg}^*.
                                                               \label{c:eq:transverseschur}
\end{equation}
It has domain $D(\Cc_{\rm t})$, compact resolvent, and
$S_{\rm t}(\lambda)\ge\Cc_{\rm t}-26I$. Moreover, for $\lambda_2>\lambda_1$,
\[
 S_{\rm t}(\lambda_2)\le S_{\rm t}(\lambda_1)-(\lambda_2-\lambda_1)I.
\]
Its ordered eigenvalues are continuous by the bounded-perturbation
min--max inequality and strictly decrease in this sense. Each crosses
zero at most once. Only as many of them as there are eigenvalues of $\Cc_{\rm t}$ at most $26$ can do so, because the others
are positive throughout the interval. Schur factorization shows that
these crossings are exactly the spectrum of $A_0$ in the interval, with
the same nullity; at every other point it gives a bounded inverse into
$D(\Cc)$. This also excludes essential spectrum there.

Finally $\sqrt{26}/\pi<2$, so each index of a free eigenvalue at most
26 is less than $2N_j$. Counting triples with multiplicity two gives
\eqref{c:eq:count}. The infinite gradient space has not been counted as a
finite collection of modes; it was eliminated as a uniformly invertible
operator block.
\end{proof}

\begin{lemma}[Uniform scalar detuning]
\label{c:lem:detuning}
For $0<\tau\le1/64$ dyadic, let $J_{\rm grid}$ be the least power of two at least
$64mN_\Omega$. Choose $j$ uniformly in $\{0,\ldots,J_{\rm grid}-1\}$ and put
\begin{equation}
 \sigma=\frac{(j+1/2)\tau}{J_{\rm grid}},\qquad
 L_b=\frac{64mN_\Omega}{\tau}.                         \label{c:eq:detuning}
\end{equation}
Provided the material interval is preserved, $A=A_0-\sigma I$ satisfies
\eqref{c:eq:maininverse}, apart from its final comparison
$702L_b\le m^{c_A}$, except on an event of probability at most
$3/(64m)$. That comparison is proved in Section~\ref{c:sec:cost}.
\end{lemma}
\begin{proof}
An interval of radius $L_b^{-1}$ around one real eigenvalue contains at
most $2J_{\rm grid}/(L_b\tau)+1$ grid points. Spectrum outside $[-1/4,1/4]$ is
farther than $L_b^{-1}$ from the detuning interval. Lemma~\ref{c:lem:count}
and a union bound give
\[
 \Pr\{\dist(\sigma,\spec A_0)<L_b^{-1}\}
 \le\frac{2N_\Omega}{L_b\tau}+\frac{N_\Omega}{J_{\rm grid}}
 \le\frac3{64m}.
\]
Singular draws are included; equality at the endpoint is harmless.
Self-adjointness gives $\norm{A^{-1}}_{2\to2}\le L_b$ on the
complementary event. If $E=A^{-1}f$, then
$\norm{\Cc E}\le\norm f+25\norm E$, whence
$\norm{A^{-1}}_{2\to X}\le27L_b$ and
$\norm A_{X\to2}<26$. Their product is at most $702L_b$.
\end{proof}

The form representation in~\eqref{c:eq:domain} uses the first representation
theorem~\citep[Chapter~VI, Section~2, Theorem~2.1]{kato1976}. Its hypotheses
hold because $V$ is dense in $L^2$, complete in the form norm, and the
form is nonnegative and symmetric. Bounded real multiplication preserves
self-adjointness on its operator domain. The displayed Fourier argument
also proves the required constant-coefficient decomposition directly.

\section{An explicit scalar periodic medium with a simple local minimum}
\label{c:sec:edge}

Set
\begin{equation}
 \varrho=2^{-100},\qquad
 V_0(r)=\cos(\pi r_1)+2\cos(\pi r_2)+4\cos(\pi r_3),
 \qquad \varepsilon_{\rm per}=7+\varrho V_0.           \label{c:eq:background}
\end{equation}
The period is two in each coordinate. At the antiperiodic corner
$k_*=(\pi/2,\pi/2,\pi/2)$, use local Bloch momentum $q$ and Fourier
aliases $p_\nu(q)=k_*+q+\pi\nu$, $\nu\in\Z^3$. The free fiber is the direct
sum of
\begin{equation}
 |p_\nu|^2I-p_\nu p_\nu^T-7I.                                \label{c:eq:freefiber}
\end{equation}
For complex momentum the squares and transpose in this formula are
bilinear, not Hermitian. Real momenta give the self-adjoint fibers.
The potential $V_0$ connects only neighboring aliases, with weights
$1/2,1,2$.

For complex $q$, the free fiber is the maximal multiplication operator
on $\ell^2(\Z^3;\C^3)$ with domain
\[
 \left\{u:\sum_\nu\big\|[p_\nu(q)^2I-p_\nu(q)p_\nu(q)^T-7I]u_\nu\big\|^2
                    <\infty\right\}.
\]
It is closed, since convergence of a sequence and of its image gives
the defining identity in each finite alias block. The material
convolution is bounded, so its addition uses this same domain.
For real $q$ this is the self-adjoint Bloch fiber obtained by the
unitary Fourier decomposition of the physical curl form. The Bloch
transform groups ordinary Fourier frequencies modulo $\pi\Z^3$;
Parseval proves this direct-integral decomposition, initially on
smooth compactly supported fields and then by graph closure.

We do not assume that these unbounded operator domains are independent
of momentum. All complex inverses used below are constructed from the
explicit free inverse
\begin{equation}
 (p^2I-pp^T-\mathcal EI)^{-1}
   =\frac{I-pp^T/\mathcal E}{p^2-\mathcal E},                           \label{c:eq:freeinverse}
\end{equation}
and a bounded Neumann series. Multiplying the block formulas verifies
both inverse identities; the image belongs to the maximal domain
because the product with the free operator is the identity plus a
bounded operator. Uniform bounds on the alias blocks and on the
Neumann series prove holomorphy in the bounded-operator norm.
This constructs the complex inverse and finite Schur operators
without a common-domain perturbation assumption.

\subsection{The corner cluster}
At $q=0$, the eight wavevectors $p_\varsigma=\tfrac\pi2\varsigma$, $\varsigma\in\{\pm1\}^3$,
have a sixteen-dimensional transverse eigenspace at
$E_0=3\pi^2/4-7$. Let $P_\varsigma=I-\varsigma\varsigma^T/3$ and let $P_{\rm cl}$ be the projection onto this cluster. The first-order perturbation on
this space is $K_{\rm cl}=-P_{\rm cl}V_0P_{\rm cl}$. Reflection parity $\varpi\in\{\pm1\}^3$ reduces
it to the plane $\one^\perp\subset\R^3$. With
$b_\varpi=(\varpi_1,2\varpi_2,4\varpi_3)$ and $P_1=I-\one\one^T/3$, the block is
\begin{equation}
 K_{\rm cl}(\varpi)=P_1\diag(b_\varpi)P_1-\frac{\varpi_1+2\varpi_2+4\varpi_3}{2}P_1.
                                                               \label{c:eq:parityblock}
\end{equation}
To verify this, an amplitude $\phi\in\one^\perp$ at $\varsigma=(1,1,1)$ has
amplitude $\varpi^{\iota(\varsigma)}\diag(\varsigma)\phi/\sqrt8$ at $\varsigma$, where
$\iota_j(\varsigma)=(1-\varsigma_j)/2$. Toggling $\varsigma_j$ multiplies this amplitude by the
polar reflection $I-2e_je_j^T$ and parity $\varpi_j$. Applying the coupling
$-\varpi_j2^{j-2}$ gives~\eqref{c:eq:parityblock} after projection. Distinct parity embeddings are orthogonal, because
the sum of a nontrivial sign character on $\{\pm1\}^3$ is zero.
Each projected parity space has dimension two, so the eight spaces
exhaust the sixteen-dimensional transverse cluster.

The two roots of this block are
\begin{equation}
 -\frac{\varpi_1+2\varpi_2+4\varpi_3}{6}
 \ \,\pm\,\frac{\sqrt{21-2\varpi_1\varpi_2-4\varpi_1\varpi_3-8\varpi_2\varpi_3}}3.
                                                               \label{c:eq:clusterroots}
\end{equation}
The unique largest root is
\begin{equation}
 \lambda_*=(5+2\sqrt{19})/6,
 \quad \varpi=(1,-1,-1),\qquad
 \lambda_*-\lambda_{\rm next}>1/20.                  \label{c:eq:clustergap}
\end{equation}
The next root is $1/2+\sqrt3$. Substitution in the eight displayed
quadratics proves the ordering; the accompanying exact program checks
all sixteen intervals and also rebuilds the $24\times24$ corner
projection and its parity intertwiners.

Let $u_0$ be the normalized top cluster vector. Its eight wavevector
blocks have equal norm by reflection symmetry. Projecting the free
velocity onto the transverse cluster gives
\[
 \Upsilon(t)=\pi\diag_{\varsigma}(\varsigma\cdot t)I_{\varsigma^\perp}.
\]
Consequently
\begin{equation}
 \ip{u_0}{\Upsilon(t)u_0}=0,\qquad
 \norm{\Upsilon(t)u_0}^2=\pi^2|t|^2.                    \label{c:eq:velocity}
\end{equation}
If $\lambda_K(t)$ is the top eigenvalue of $K_{\rm cl}+\Upsilon(t)$ near zero, ordinary
finite-dimensional Schur elimination gives
\begin{equation}
 t^TH_0t=\ip{\Upsilon(t)u_0}{(\lambda_*-K_{\rm cl})^{-1}_{u_0^\perp}\Upsilon(t)u_0},
 \qquad H_0=\tfrac12\lambda_K''(0).
                                                               \label{c:eq:A0}
\end{equation}
Since $\norm{K_{\rm cl}}\le\norm{V_0}_\infty\le7$, the spectrum of $K_{\rm cl}$ lies in
$[-7,7]$, so $(\lambda_*-K_{\rm cl})^{-1}\ge1/14$ on $u_0^\perp$. The gap
\eqref{c:eq:clustergap} gives $(\lambda_*-K_{\rm cl})^{-1}\le20$ there.
With~\eqref{c:eq:velocity} this gives
\begin{equation}
 \frac{\pi^2}{14}I\le H_0\le20\pi^2I.           \label{c:eq:A0bounds}
\end{equation}
The exact cluster roots lie in $[-2.29,2.29]$, so the lower constant is
far from sharp. Reflections make $H_0$
diagonal. Its exact entries are
\begin{equation}
 H_0=\pi^2\diag
 \left(\frac{47}{31}+\frac{352\sqrt{19}}{589},\ 
 2+\frac{11\sqrt{19}}{38},\ 
 \frac18+\frac{11\sqrt{19}}{152}\right).              \label{c:eq:explicitA0}
\end{equation}
For a direct verification, take
$\phi^{(0)}_i=(\lambda^{\rm d}_i-\lambda_*)^{-1}$ with $\lambda^{\rm d}=(7/2,1/2,-3/2)$, normalize $\phi^{(0)}$,
and evaluate~\eqref{c:eq:A0} in the parity block with the corresponding
sign flipped. This uses only arithmetic in $\Q(\sqrt{19})$.

\subsection{The continuum remainder, including every Fourier alias}
\begin{lemma}[Local continuum band]
\label{c:lem:edge}
The actual continuum operator $A_{\rm per}=\curl\curl-\varepsilon_{\rm per}$
has a simple analytic Bloch eigenvalue $\lambda(q)$ for
$|q_j|<R_q:=\varrho2^{-12}$. It is real for real $q$ and even in each
coordinate. With $E_c=\lambda(0)$ and $H=\tfrac12\lambda''(0)$,
\begin{align}
 |E_c-E_0-\varrho\lambda_*|&\le2^{16}\varrho^2,\\
 \norm{\varrho H-H_0}&\le2^{40}\varrho,\\
 (2\varrho)^{-1}I&\le H\le(256/\varrho)I.              \label{c:eq:edgebounds}
\end{align}
For $|q|\le R_q/2$,
\begin{equation}
 |\lambda(q)-E_c-q^THq|\le M_4|q|^4,
 \qquad M_4=4\varrho R_q^{-4}.                       \label{c:eq:quartic}
\end{equation}
On the smaller chart $|q_j|\le2q_0$, $q_0=\varrho2^{-40}$, the selected
band is separated from every other fiber spectral value by at least
$\varrho/100$. The separated inverse and eigenprojection are bounded
holomorphic operators. The selected cell eigenfunction has Fourier
$\ell^1$ norm at most 32 in a complex neighborhood of this chart.
\end{lemma}
\begin{proof}
Put $q=\varrho t$ and $\mathcal E=E_0+\varrho e$. Retain all three components
of the eight central aliases. Every other free alias has transverse
energy at least $11\pi^2/4-7$, while its longitudinal denominator at
energy $\mathcal E$ is $-3\pi^2/4-\varrho e$. Formula~\eqref{c:eq:freeinverse}
shows that its inverse norm is less than one for
$|t_j|\le2^{-12}$ and $|e|\le8$. Indeed the transverse denominator is
at least $2\pi^2-O(\varrho|p|)-8\varrho$; its ratio to the
quadratic numerator is uniformly bounded, also as $|p|\to\infty$.
Using $3<\pi<4$ gives the stated conservative bound. The eight central
longitudinal directions have the same bound, since their diagonal
free correction is $q^2-(s\cdot q)^2/3$.

The potential has norm at most seven. Its complementary inverse is the
Neumann series with ratio at most $7\varrho<1/4$, and has norm at most
two. The cluster--complement coupling, divided by $\varrho$, has norm
at most eight. Eliminating the complement therefore gives the exact
sixteen-dimensional analytic Schur matrix
\begin{equation}
 K_{\rm cl}+\Upsilon(t)-eI+\varrho\Sigma_2(t,e),\qquad \norm{\Sigma_2(t,e)}\le256.
                                                               \label{c:eq:feshbach}
\end{equation}
The bound is $2\cdot8^2$ for the self-energy plus less than one for
the direct quadratic term. This derivation includes the infinitely many
longitudinal aliases; none was discarded by a transverse approximation.

On $|e-\lambda_*|=1/100$,
$\norm{\Upsilon(t)}\le3\pi2^{-12}<1/400$, so the resolvent of $K_{\rm cl}+\Upsilon(t)$ is
less than 140. Its contour projection $\Pi(t)$ has rank one and norm at
most two. Hence
$(K_{\rm cl}+\Upsilon(t)-e)^{-1}=\Pi(t)/(\lambda_K(t)-e)+\widetilde R(e)$, where the reduced
resolvent $\widetilde R(e)=(K_{\rm cl}+\Upsilon(t)-e)^{-1}(I-\Pi(t))$ is analytic inside
the circle. On the circle its norm is at most $140\cdot3$, so by the
maximum principle it is at most $420$ inside. This bounds the resolvent
near its root by $2/|e-\lambda_K(t)|+2048$. On the circle
$|e-\lambda_K(t)|=2^{16}\varrho$, multiplication by
$\varrho\Sigma_2$ has norm less than $1/16$. Homotopy of the finite
Schur determinant, or its contour projection, thus leaves exactly one
root and gives
$|e_\varrho(t)-\lambda_K(t)|\le2^{16}\varrho$.
The root is analytic because its contour projection has rank one.
The complementary free inverse proves that this root is an actual
continuum fiber eigenvalue. It is simple. The homotopy counts zeros of
the Schur determinant with multiplicity, so the root is a simple zero,
and the inverse Schur matrix (adjugate over determinant) has a simple
pole with rank-one residue. The Feshbach formula writes the fiber
resolvent as this inverse between two analytic lifting operators plus an
analytic term; the Riesz projection of the full fiber therefore has rank
one.

Cauchy's estimate on the $t$-polydisc gives the Hessian error
$2^{40}\varrho$. The finite inequalities in~\eqref{c:eq:A0bounds}
then give~\eqref{c:eq:edgebounds}. Coordinate reflections of the scalar
material, with the polar-vector action on fields, send $q_j$ to $-q_j$.
Simplicity gives the asserted evenness. The analytic bound
$|\lambda(q)-E_0|\le3\varrho$ and the one-variable Cauchy estimate
on each complex line give~\eqref{c:eq:quartic}: after removing the
constant and quadratic terms, the even tail is at most
$3\varrho(|q|/R_q)^4/[1-(|q|/R_q)^2]$.

The cluster gap persists on the smaller chart by the same Schur
estimates. The other free spectrum stays a fixed distance away.
For the cell eigenvector, the complementary Neumann series starts in
finitely many neighboring aliases. Each potential application changes
an alias by one, and its $\ell^1$ norm is at most seven. The free
inverse has norm at most one on each noncentral $3\times3$ block.
The geometric series, together with the eight central blocks, bounds
the eigenvector's Fourier $\ell^1$ norm by 32. Cauchy bounds its
derivatives. This proves all assertions without a finite-element
spectral convergence assumption.
\end{proof}

For later normalization choose the analytic cell eigenfunction
$\Phi_q$ with mean squared norm one for real $q$, real $\Phi_0$,
and dual pairing $\langle\Phi_{-q},\Phi_q\rangle_{\rm bil}=1$.
One construction is to apply the analytic rank-one projection to
$\Phi_0$ and divide by the analytic square root of its nonzero pairing
with $\Phi_0$. The transpose identity $A_{\rm per}(q)^T=A_{\rm per}(-q)$ in the
antiperiodic cell gauge gives $\Pi(-q)=\Pi(q)^T$ for the analytic
eigenprojection, and hence this normalization. The pairing stays away
from zero: on $|t_j|\le2^{-39}$ the projection moves by $O(2^{-33})$
from its value at $q=0$, where the pairing is one. At purely imaginary
$q$, both the operator and the normalized eigenfunction are real.
This fact will fix the sign of the useful coupling numerator.

A real zeroth-order field, used only as a starting vector for calibration,
is
\begin{equation}
 \Phi^{(0)}(r)=\sqrt8\begin{pmatrix}
 \phi^{(0)}_1\sin(\pi r_1/2)\sin(\pi r_2/2)\sin(\pi r_3/2)\\
 -\phi^{(0)}_2\cos(\pi r_1/2)\cos(\pi r_2/2)\sin(\pi r_3/2)\\
 -\phi^{(0)}_3\cos(\pi r_1/2)\sin(\pi r_2/2)\cos(\pi r_3/2)
 \end{pmatrix},                                      \label{c:eq:initialphi}
\end{equation}
where $\phi^{(0)}$ is the normalized vector following~\eqref{c:eq:explicitA0}.
It is divergence-free, has mean squared norm one, and the selected
actual field can be signed so that
$\norm{\Phi_0-\Phi^{(0)}}_{\ell^1({\rm Fourier})}\le2^{16}\varrho$.
This follows by the same isolated-cluster Schur equation and geometric
complement series.

\subsection{A fixed integer placement matrix}
Use
\begin{equation}
 B=\diag(7180532256056349,\ 6388184103869743,\ 2347449745487106).
                                                               \label{c:eq:B}
\end{equation}
The exact identities~\eqref{c:eq:explicitA0}, rational enclosures for $\pi$
and $\sqrt{19}$, and~\eqref{c:eq:edgebounds} imply
\begin{equation}
 .99\,|y|\le|H^{-1/2}By|\le1.01\,|y|.               \label{c:eq:Bmetric}
\end{equation}
The verifier checks the stronger squared inequalities, allowing the full
$2^{-60}=2^{40}\varrho$ Hessian error. This is an integer placement
lattice, not a change to the Euclidean metric or to material tensors.

\section{Smooth cutoffs and the auxiliary resolvent}
\label{c:sec:cutoff}

Define the fixed bump and its square root by
\begin{equation}
 w(y)=\begin{cases}\exp[-1/(1-|y|^2)],&|y|<1,\\0,&|y|\ge1,\end{cases}
 \qquad q_w(y)=\sqrt{w(y)}.                           \label{c:eq:well}
\end{equation}
For a one-dimensional cutoff, normalize
$\psi_0(s)=\exp[-1/(1-4s^2)]\one_{|s|<1/2}$ to have integral one and put
$\psi=\one_{[-3/2,3/2]}*\psi_0$. Its normalization integral is between
$1/8$ and one. Thus $0\le\psi\le1$, $\psi=1$ on $[-1,1]$, and
$\psi=0$ outside $[-2,2]$. In the corner Bloch coordinates set
\begin{equation}
 \chi(q)=\prod_{j=1}^3\psi(q_j/q_0).                \label{c:eq:chi}
\end{equation}
It is a scalar multiplier on the Bloch torus, extended by zero outside
this chart. In physical Fourier variables it is periodic modulo the
reciprocal lattice. In particular, it commutes with every translation
and every constant-coefficient derivative.

\begin{lemma}[Quantitative Gevrey estimates]
\label{c:lem:gevrey}
The zero extensions in~\eqref{c:eq:well} are smooth. For every directional
derivative of order $K$,
\begin{equation}
 \norm{\partial^K w}_\infty+\norm{\partial^K q_w}_\infty
 \le2\cdot1024^K(K!)^2.                              \label{c:eq:gevreyder}
\end{equation}
They satisfy the Fourier tail bound
\begin{equation}
 \int_{|p|\ge R}(|(\mathcal Fw)(p)|+|(\mathcal Fq_w)(p)|)\dd p
 \le2^{100}\exp(-\sqrt R/512).                       \label{c:eq:gevreytail}
\end{equation}
The cutoff $\chi$ has derivative bounds
$2^{45}(2^{16}/q_0)^K(K!)^2$. All its values, and the values of $w,q_w$,
can be enclosed to $N_{\rm b}$ bits in time polynomial in $N_{\rm b}$.
\end{lemma}
\begin{proof}
For a real point inside the unit ball, let $s_y=1-|y|^2$.
In the complex polydisc of radius $s_y/100$, the real part of
$(1-z^2)^{-1}$ is at least $1/(2s_y)$; for the square root use
$1/(4s_y)$ in the exponential. Cauchy's estimate gives
$K!(100/s_y)^K e^{-1/(4s_y)}$. Maximizing in $s_y$ and using
$K^K e^{-K}\le K!$ bounds coordinate derivatives by
$400^K(K!)^2$. The directional expansion gives~\eqref{c:eq:gevreyder}
with the displayed larger constant. These estimates also show that every
derivative tends to zero at the boundary, justifying the smooth zero
extension.

Integrate by parts $K$ times in the direction of $p$ and use the
support volume less than eight. Choosing an integer
$K$ within one of $\sqrt{|p|}/64$ gives
$|(\mathcal Fw)(p)|+|(\mathcal Fq_w)(p)|\le
2^{20}e^{-\sqrt{|p|}/256}$. For small $|p|$ use the $L^1$
bound. Radial integration and one further weakening of the exponent
prove~\eqref{c:eq:gevreytail}. Rescaling the one-dimensional bump,
dividing by its normalization at least $1/8$, and taking the three
products gives the claimed cutoff bounds.

For evaluation, if the positive denominator in the bump is less than
$1/[4(N_{\rm b}+10)]$, the value is below $2^{-N_{\rm b}-2}$ and can be enclosed directly.
Otherwise the exponential argument is $O(N_{\rm b})$. Its Taylor series with
range reduction and a geometric factorial-tail bound uses polynomially
many rational operations and bits. A periodic trapezoidal rule applied
to a smooth compactly supported bump has error bounded by the sum of its
Fourier coefficients at the aliased indices;~\eqref{c:eq:gevreytail}
therefore gives $O(N_{\rm b}^2)$ nodes per dimension for fixed intervals.
This encloses the normalization and the convolution defining $\psi$.
Equivalently, one may integrate a polynomial approximation with the same
Gevrey tail. Outward rational rounding gives certified intervals.
\end{proof}

For clarity, the following fixed constants are used throughout:
\begin{equation}
 \Cs=2^{2^{40}},\qquad \cs=2^{-1024},\qquad
 c_u=2^{-1024},\qquad
 \Ls=8^{\lceil64\cdot2^{40}/3\rceil},\qquad
 \ds=\Cs^8.                                         \label{c:eq:masterconstants}
\end{equation}
Their size is only a convenience for a single majorant ledger. The proofs
below give bounds far smaller than these numbers. Section~\ref{c:app:ledger}
records the derivative and arithmetic majorants that justify their use;
none is a supremum over inaccessible data.

Let $L\ge\Ls$ be a power of eight and $h=L^{-1}$. Compute a dyadic $z_h$
with
\begin{equation}
 |z_h-(E_c-h^2)|\le h^4.                              \label{c:eq:z}
\end{equation}
Section~\ref{c:sec:calibration} gives a certified terminating algorithm.
For $0\le\sigma\le h^4/64$, define on the whole space
\begin{equation}
 R_\sigma=\chi(q)\big(A_{\rm per}(q)-z_h-\sigma\big)^{-1}
 \quad\text{on the chart, and zero elsewhere}.         \label{c:eq:R}
\end{equation}
This inverse exists on the chart by Lemma~\ref{c:lem:edge}. It is a real
self-adjoint bounded operator, with
\begin{equation}
 \norm{R_\sigma}\le4L^2,\quad
 \norm{R_\sigma-R_0}\le16\sigma L^4,\quad
 (A_{\rm per}-z_h-\sigma)R_\sigma=\chi.               \label{c:eq:Ridentities}
\end{equation}
The last equality holds as an operator into $L^2$, and
$R_\sigma L^2\subset D(\curl\curl)$ on $\R^3$.
The norm estimates follow from the selected denominator, at least
$h^2/2$, and the fixed separation of all other bands. The denominator
bound uses $\lambda(q)\ge E_c$ on the chart: there
$M_4|q|^2\le M_4(4q_0)^2<10^{-6}/\varrho$, far below the smallest
eigenvalue $(2\varrho)^{-1}$ of $H$, so~\eqref{c:eq:quartic} gives
$\lambda(q)-E_c\ge q^THq/2$. The final factor
$\chi$ in~\eqref{c:eq:Ridentities} can not be replaced by $I$; this is
the main analytic issue of the proof.

\section{A broad scalar defect with an isolated auxiliary mode}
\label{c:sec:defect}

Let $\mathbb B_L=\{r:|r|<L\}$, let $Q_h$ multiply by $q_w(hx)$ and restrict
to $\mathbb B_L$, and define on $L^2(\mathbb B_L;\C^3)$
\begin{equation}
 S_h=Q_hR_0Q_h^*.                                    \label{c:eq:Sh}
\end{equation}
The adjoint includes the same multiplier and extension by zero. This is
an infinite-dimensional self-adjoint operator; it is not declared compact.

\begin{lemma}[An effective scalar operator with a quantitative gap]
\label{c:lem:perron}
Let $H$ satisfy~\eqref{c:eq:edgebounds}. The scalar compact operator
\begin{equation}
 T_H=\sqrt w\,(-\nabla\cdot H\nabla+1)^{-1}\sqrt w
 \quad\text{on }L^2(\mathbb B_1)                              \label{c:eq:TA}
\end{equation}
has a simple largest eigenvalue $\mu_H$ and a real positive normalized
eigenvector $u_H$. It satisfies
\begin{equation}
 2^{-127}\le\mu_H\le2^{-92},\qquad
 \dist(\mu_H,\spec T_H\setminus\{\mu_H\})\ge2^{-190}.
                                                               \label{c:eq:perrongap}
\end{equation}
In addition,
\begin{equation}
 u_H(y)\ge2^{-29}\sqrt{w(y)},\qquad
 \int\sqrt w\,u_H\ge2^{-32}.                         \label{c:eq:perronoverlap}
\end{equation}
\end{lemma}
\begin{proof}
Fourier transformation gives the Yukawa kernel
\[
 Y_H(y)=\frac{e^{-|H^{-1/2}y|}}
 {4\pi\sqrt{\det H}\,|H^{-1/2}y|}.
\]
It can also be obtained by changing variables in the radial fundamental
solution of $-\Delta+1$, whose normalization follows by integrating its
normal derivative over a small sphere. On $\mathbb B_1\times\mathbb B_1$ the bounds
on $H$ imply
\begin{equation}
 Y_H(r-y)\ge Y_-:=\varrho2^{-24},\quad
 Y_H(r-y)\le\frac{4\varrho}{|r-y|},\quad
 \sup_r\norm{Y_H(r-\cdot)}_{L^2(\mathbb B_1)}\le Y_+:=32\varrho.
                                                               \label{c:eq:yukawabounds}
\end{equation}
For example, $|H^{-1/2}(r-y)|<1$, so the exponential is at least
$1/4$; inserting the extreme eigenvalues of $H$ proves the first
inequality. Squaring the second and integrating over a radius-two ball
gives the third. The Schur integral bound gives $\norm{T_H}\le256\varrho$.
Also $\int w\ge1/8$: on the half-radius ball, $w\ge e^{-4/3}>1/4$,
and its volume is $\pi/6$.

The kernel is square integrable on $\mathbb B_1\times\mathbb B_1$, so $T_H$ is
compact. It is nonnegative as a quadratic form and has a strictly
positive kernel in the interior. Its largest Rayleigh quotient is
attained; replacing an eigenvector by its absolute value does not
reduce it, and the integral equation makes the resulting eigenvector
strictly positive. Testing on $\sqrt w$ gives
$\mu_H\ge Y_-\int w\ge\varrho2^{-27}=2^{-127}$.
The integral equation and Cauchy--Schwarz give
\[
 u_H(r)\le(Y_+/\mu_H)\sqrt{w(r)},\qquad
 \int\sqrt w\,u_H\ge\mu_H/Y_+,
 \qquad u_H(r)\ge(Y_-/Y_+)\sqrt{w(r)}.
\]
These imply~\eqref{c:eq:perronoverlap}.

For the spectral gap define the reversible Markov kernel
\[
 \mathcal K(r,y)=\frac{\sqrt{w(r)}Y_H(r-y)\sqrt{w(y)}u_H(y)}
 {\mu_Hu_H(r)}.
\]
It has invariant measure $u_H^2\dd r$. The preceding upper bound on
$u_H$ implies
$\mathcal K(r,y)\ge\varsigma_0u_H(y)^2$, where
$\varsigma_0=Y_-\mu_H/Y_+^2\ge2^{-61}$.
Write $\mathcal K=\varsigma_0\Pi_{\rm c}+(1-\varsigma_0)\mathcal K_1$, where $\Pi_{\rm c}$ is the orthogonal projection onto constants and $\mathcal K_1$ is another
Markov operator preserving that measure. Jensen's inequality makes
$\mathcal K_1$ a contraction on $L^2(u_H^2\dd r)$. On the mean-zero subspace,
$\norm{\mathcal K}\le1-\varsigma_0$. The unitary map $v\mapsto v/u_H$
identifies $T_H/\mu_H$ with $\mathcal K$. This proves simplicity and a gap at
least $\mu_H\varsigma_0\ge\varrho2^{-88}>2^{-190}$.
No uncomputed scalar eigenvalue is a premise of this argument.
\end{proof}

\begin{lemma}[Quantitative effective-mass comparison]
\label{c:lem:effectivemass}
For every $L\ge\Ls$, the largest eigenvalue $\mu_h$ of $S_h$ is simple,
real, and isolated. It and a real normalized eigenvector $u_h$ satisfy
\begin{align}
 2^{-128}L^2&\le\mu_h\le2^{-90}L^2,
 &\dist(\mu_h,\spec S_h\setminus\{\mu_h\})&\ge2^{-192}L^2,
                                                               \label{c:eq:mugap}\\
 \norm{u_h-h^{3/2}\Phi_0(r)u_H(hx)}_{L^2(\mathbb B_L)}&\le2^{921} h\le\Cs h.
                                                               \label{c:eq:modeapprox}
\end{align}
The sign can be chosen so that the normalized starting vector
$h^{3/2}q_w(hx)\Phi^{(0)}(r)$ has overlap at least $2^{-48}$ with $u_h$.
Moreover,
\begin{equation}
 \norm{u_h}_{H^1(\R^3)}\le2^{512}L^2\le\Cs L^2,                   \label{c:eq:modeH1}
\end{equation}
where zero extension is understood.
\end{lemma}
\begin{proof}
Split $R_0$ into its selected-band part and the remaining part. The latter
has norm at most $2^{112}$ by Lemma~\ref{c:lem:edge}; after multiplication
by $h^2$ it is a bounded perturbation of norm at most $2^{112}h^2$.

Write the selected normalized Bloch eigenvector in alias coordinates as
$\phi(q)$. Define an isometry from the scaled momentum chart by
\[
 (\mathcal U_h v)(r)=h^{3/2}(2\pi)^{-3/2}
 \int e^{ihp\cdot r}\Phi_{hp}(r)v(p)\dd p,
\]
and put
\[
 s_h(p)=\frac{h\sqrt{\chi(hp)}}{\sqrt{\lambda(hp)-z_h}},
 \qquad s_H(p)=(1+p^THp)^{-1/2}.
\]
Extend $s_h$ by zero off the chart. The selected part of $h^2S_h$ is
$B_hB_h^*$, where $B_h=Q_h\mathcal U_hs_h$. Its nonzero spectrum is
that of $B_h^*B_h$.

The latter operator has a concrete Fourier kernel. Up to the unitary
Fourier factor $(2\pi)^{-3/2}$, its middle factor is
\begin{equation}
 \sum_{\nu,\nu'}\phi_\nu(hp)^*\phi_{\nu'}(hp')
 (\mathcal Fw)\big(p-p'+\pi(\nu-\nu')/h\big).                \label{c:eq:umklapp}
\end{equation}
For $\nu=\nu'$ this is the kernel of multiplication by $w$ conjugated by
Fourier transform, with the vector multipliers $\phi(hp)$ and $\phi(hp')$.
For $\nu\ne\nu'$, the argument of $\mathcal Fw$ is at least
$1/h$ in norm: both $hp,hp'$ lie in the very small fixed chart.
The two Fourier $\ell^1$ bounds $32$ on $\phi$, the pointwise bound
$|(\mathcal Fw)(p)|\le2^{20}e^{-\sqrt{|p|}/256}$ from the proof of
Lemma~\ref{c:lem:gevrey}, and the volume, less than $h^{-3}$, of the
scaled chart in the row and column Schur integrals make the norm of
this off-alias part at most
\begin{equation}
 2^{100}h^{-3}e^{-1/(512\sqrt h)}\le2^{512}h.           \label{c:eq:umklapperror}
\end{equation}
The last inequality follows by maximizing $s^8e^{-s/512}$ for $s\ge0$;
the stated constant is deliberately larger than that maximum.

The dispersion estimate gives
\begin{equation}
 \sup_p|s_h(p)-s_H(p)|\le2^{128}h,
 \qquad
 \sup_p s_H(p)\norm{\phi(hp)-\phi(0)}\le2^{128}h.            \label{c:eq:multipliererror}
\end{equation}
Here is an explicit check of the first assertion. On the chart the
normalized denominator differs from $1+p^THp$ by at most
$h^2+M_4h^2|p|^4$ and is at least one half of it. Applying the derivative
of $s^{-1/2}$ bounds the difference by
$C h^2(1+M_4|p|^4)/(1+2^{99}|p|^2)^{3/2}$.
For $|p|\le4q_0/h$ its supremum is less than $2^{100}h$.
Where the cutoff is not one, $s_H\le h/(q_0\sqrt{2^{99}})$,
which fits the displayed bound. The second assertion follows from the
Cauchy derivative bound on $\phi$, using
$|p|s_H(p)\le2^{-99/2}$. These calculations use only the explicit
powers in Lemma~\ref{c:lem:edge}.

Multiplication by $w$ has norm at most one. The diagonal-alias
comparison has norm at most $2^{133}h$, so adding the off-alias
bound in~\eqref{c:eq:umklapperror} gives
\begin{equation}
 \norm{B_h^*B_h-
 s_H\mathcal Fw\mathcal F^{-1}s_H}\le2^{513}h\le2^{520}h.
                                                               \label{c:eq:effectivecompare}
\end{equation}
The nonzero spectrum of the comparison operator is exactly that of
$T_H$. The selected part is compact: its momentum chart is bounded,
its multiplier is bounded for fixed $h$, and the spatial support of
$Q_h$ is bounded, so its integral kernel is Hilbert--Schmidt.
The separated part need not be compact, but its norm after scaling is
at most $2^{112}h^2$. The min--max inequalities above that bounded
essential spectrum, or the elementary isolated-eigenvalue Schur
argument, now transfer~\eqref{c:eq:perrongap} to~\eqref{c:eq:mugap}.
The choice $L\ge\Ls$ makes both errors smaller than $2^{-200}$, so the
top eigenvalue of $h^2S_h$ lies in $[\mu_H-2^{-199},\mu_H+2^{-199}]$
and the rest of its spectrum stays at distance at least
$2^{-190}-2^{-198}$ from it. Multiplying by $L^2=h^{-2}$
gives~\eqref{c:eq:mugap}.

For completeness, eigenvectors can be compared without identifying two
different Hilbert spaces. Let $y_H$ be the normalized top eigenvector
of $s_H\mathcal Fw\mathcal F^{-1}s_H$. Projection of the perturbed
eigenvector equation onto $y_H^\perp$ bounds its component there by
twice the error in~\eqref{c:eq:effectivecompare} divided by the gap.
Reconstruct the spatial eigenvector as $B_hy_h/\sqrt{\mu_hh^2}$.
The operator $B_h$ differs by $O(h)$, with the explicit bounds just
used, from
\[
 v\longmapsto h^{3/2}\Phi_0(r)\sqrt{w(hx)}
       [\mathcal F^{-1}(s_Hv)](hx).
\]
To check this operator comparison, use the disjoint Bloch aliases on
$|p|\le q_0/h$,~\eqref{c:eq:multipliererror}, and the bound
$\sup_{|p|>q_0/h}s_H(p)\le2^{100}h$ on the remaining input.
The comparison operator applied to $y_H/\sqrt{\mu_H}$ is exactly
$h^{3/2}\Phi_0(r)u_H(hx)$. Finally the bounded separated-band
perturbation is treated by the same gap projection. The scalar-well
argument above gives the stronger gap $2^{-188}$, so the conservative
Gram error $2^{520}h$ gives an eigenvector error at most $2^{710}h$.
On the eigenvalue interval bounded below by $2^{-128}$, the inverse
square root and its Lipschitz constant are both bounded by $2^{200}$.
Combining these estimates with the reconstruction error gives
$2^{913}h$ for the selected spatial mode. Adding the separated-band
perturbation gives $2^{920}h$; taking its real part and normalizing
increases the error by at most a factor of two. This proves
\eqref{c:eq:modeapprox}, with all constants below the majorant of
Section~\ref{c:app:ledger}.

The starting-vector overlap is positive quantitatively. By
\eqref{c:eq:perronoverlap}, its leading integral is at least
$2^{-29}\int w(y)|\Phi_0(y/h)|^2\dd y$, apart from the
$\Phi^{(0)}-\Phi_0$ error. The mean of $|\Phi_0|^2$ is one.
Its nonconstant Fourier coefficients are absolutely summable, and
\eqref{c:eq:gevreytail} shows that the integral is at least $1/16$.
The field error $2^{16}\varrho$, the mode error $\Cs h$, and
normalization leave the conservative overlap $2^{-48}$.

For~\eqref{c:eq:modeH1}, use translations rather than an assumed elliptic
regularity theorem for all polarizations. The cutoff $\chi$ commutes
with translations. The local inverse without its factor $\chi$ has
norm at most $4L^2$. Its resolvent identity under translation of
$\varepsilon_{\rm per}$ gives
\[
 \norm{\mathcal T_aR_0\mathcal T_a^{-1}-R_0}\le2^{20}L^4|a|.
\]
Together with the derivative bound on $q_w(hx)$ this gives
$\norm{\mathcal T_aS_h\mathcal T_a^{-1}-S_h}\le2^{32}L^4|a|$. Projecting $\mathcal T_au_h$ onto
$u_h^\perp$, using the gap $2^{-192}L^2$, and choosing the real sign
by positive overlap yields
$\norm{\mathcal T_au_h-u_h}\le2^{256}L^2|a|$ for small $a$.
The Fourier difference-quotient characterization then gives
$u_h\in H^1$ with~\eqref{c:eq:modeH1}. Explicitly, Fatou applied to
$|e^{iap_j}-1|^2/a^2$ bounds $\int p_j^2|(\mathcal Fu_h)(p)|^2\dd p$.
This also justifies the zero extension across the ball boundary.
\end{proof}

Define the compactly supported source mode
\begin{equation}
 a_h=Q_h^*u_h=q_w(hx)u_h(r).                          \label{c:eq:source-mode}
\end{equation}
It satisfies the exact identity
\begin{equation}
 a_h=\mu_h^{-1}w(hx)R_0a_h.                           \label{c:eq:source-mode-equation}
\end{equation}
This identity, not a claim of an exact isolated eigenstate of the full
infinite Maxwell operator, will control leakage.

\begin{lemma}[Nonzero analytic source numerator]
\label{c:lem:numerator}
For real $\beta$ with $|\beta|\le h$, define the two analytic band
source factors by
\begin{align*}
 F_h^+(q)&=(2\pi)^{-3/2}\int e^{-iq\cdot r}
             \Phi_{-q}(r)\cdot a_h(r)\dd r,\\
 F_h^-(q)&=(2\pi)^{-3/2}\int e^{iq\cdot r}
             \Phi_q(r)\cdot a_h(r)\dd r.
\end{align*}
At $q=i\beta$ both are real and
\begin{equation}
 h^{3/2}F_h^\pm(i\beta)\ge2^{-48}.                   \label{c:eq:numeratorlower}
\end{equation}
The scaled product
$\mathcal N_h(k)=h^3F_h^-(hk)F_h^+(hk)$ has uniformly bounded
analytic derivatives of every fixed order on the neighborhoods used
below. Up to order 32 the explicit bounds in
Section~\ref{c:app:ledger} apply, independently of $h$.
\end{lemma}
\begin{proof}
At imaginary momentum the cell functions are real by the normalization
following Lemma~\ref{c:lem:edge}. Insert~\eqref{c:eq:modeapprox}, put
$y=hx$, and use $e^{\pm\beta\cdot r}\ge e^{-1}$ on the support.
The leading term is bounded below by
\[
 (2\pi)^{-3/2}e^{-1}2^{-29}
 \int w(y)|\Phi_0(y/h)|^2\dd y.
\]
The last integral is at least $1/16$, as in the preceding proof.
The change from $\Phi_0$ to $\Phi_{\pm i\beta}$ is at most a fixed
explicit constant times $h$, by its Fourier $\ell^1$ Cauchy bound.
The $L^2$ mode error contributes at most $\Cs h$ times a fixed
support-volume and cell-field bound. Since $L\ge\Ls$, these errors
are smaller than half the displayed positive lower bound, which proves
\eqref{c:eq:numeratorlower}.

Differentiation in $k$ introduces factors $hx$ of norm at most one,
and derivatives of $\Phi_{hk}$ introduce $h$ times its fixed Cauchy
constants. Cauchy--Schwarz bounds the rescaled $L^1$ norm of $a_h$ by a
fixed number. Thus all fixed-order derivatives have the stated uniform
bounds, even for real $k$ as large as the scaled chart. No differentiability
of $u_h$ beyond its compact support is needed for this assertion.
\end{proof}

The scalar tuning interval is already explicit. At $v_* =1/\mu_h$,
$v_*^{-1}I-S_h$ has a simple zero mode and complementary separation
$2^{-192}L^2$. Every target diagonal in
$[-2^{-196}L^2,2^{-196}L^2]$ is realized by
$v=(\mu_h+\Theta_{ii})^{-1}$. These positive values are at most
$2^{130}L^{-2}<1$. This is the auxiliary defect resonance used by the
construction; its error as a physical Maxwell mode is established in
Section~\ref{c:sec:leakage}.

\section{Three-dimensional interactions with a strict rate margin}
\label{c:sec:green}

Translations below are ordinary Euclidean translations. For
$R_\nu=2LdB\nu$, $\nu\in\Z^3\setminus\{0\}$, define
\[
 G_h(\nu)=Q_hR_0\mathcal T_{R_\nu}Q_h^*,\qquad
 h_j(d)=\ip{u_h}{G_h(e_j)u_h}.
\]
Since $L$ is even, the corner phase $e^{ik_*\cdot R_\nu}$ equals one.

\begin{lemma}[Full blocks and useful scalar couplings]
\label{c:lem:interactions}
For $L\ge\Ls$ and every integer $d\ge1$,
\begin{equation}
 \norm{G_h(\nu)}\le\Cs L^5
       \big(3^{-d|\nu|_1}+e^{-\cs L}\big).              \label{c:eq:blockdecay}
\end{equation}
If $d\ge\ds$, $d^{15}\ge\Cs^4L$, and
\begin{equation}
 \Cs L^5e^{-\cs L}\le(c_u/4)L^2 8^{-d},              \label{c:eq:couplingtail}
\end{equation}
then, in all three coordinate directions,
\begin{equation}
 h_j(d)\ge c_uL^2 8^{-d}.                            \label{c:eq:couplinglower}
\end{equation}
The same upper bounds hold with $R_\sigma$ in place of $R_0$, after
adding the norm error $16\sigma L^4$ for the whole compressed union.
In particular the useful and unwanted exponential bases obey
$(1/3)^2<1/8$.
\end{lemma}
This margin separates couplings with $|\nu|_1\ge2$ from the useful ones.
Two occupied sites at $|\nu|_1=1$ would couple at the useful scale; the
parity layout of Section~\ref{c:sec:network} places only intended
terminal--mediator incidences at that distance. Condition
\eqref{c:eq:couplingtail} is guaranteed by the parameter tests
\eqref{c:eq:Ltailtest} and~\eqref{c:eq:tail-test}.

\subsection{Contour movement for a smooth, nonanalytic cutoff}
A compactly supported smooth cutoff cannot be treated as an analytic
function. We give the error explicitly. If $\beta$ is real with
$|\beta|\le h$, set $D_\beta=\beta\cdot\nabla_q$ and
\begin{equation}
 \chi_K(q+is\beta)=\sum_{j=0}^K\frac{(is)^j}{j!}D_\beta^j\chi(q),
 \qquad
 (\partial_s-iD_\beta)\chi_K
 =-\frac{i(is)^K}{K!}D_\beta^{K+1}\chi.               \label{c:eq:almostanalytic}
\end{equation}
Use $K=\lfloor[4(2^{16}/q_0)|\beta|]^{-1}\rfloor$; $\beta=0$ needs no
shift. Since $4\cdot2^{16}/q_0=2^{158}$ and $|\beta|\le h$, we have
$K\ge\lfloor L/2^{158}\rfloor\ge1$, because $\Ls\ge2^{158}$.
Lemma~\ref{c:lem:gevrey} and $K!\le K^K$ bound the last expression by
$\Cs e^{-\cs L}$. Its support is in the cutoff transition, where the
fiber inverse is uniformly bounded independently of $h$, including
through the limiting contour used below. The sum $\chi_K$ and each
of its first 32 derivatives are bounded by the fixed derivative
majorants in Section~\ref{c:app:ledger}.

Throughout, Bloch integrals carry the phase $e^{iq\cdot R}$, and the
contour is shifted by $+is\beta$ with $\beta\cdot R\ge0$.
For an analytic operator-valued integrand $Y$, differentiate
\[
 \mathcal J(s)=\int e^{i(q+is\beta)\cdot R}
       Y(q+is\beta)\chi_K(q+is\beta)\dd q.
\]
Holomorphy turns the derivative of the exponential and $Y$ into a
$q$-derivative. Integration by parts leaves exactly the last expression
in~\eqref{c:eq:almostanalytic}. There is no boundary term because all
cutoff derivatives vanish at the chart boundary. If $\beta\cdot R\ge0$,
the exponential has modulus at most one along the movement. Hence the
original integral differs from the shifted one by at most
$\Cs e^{-\cs L}$ times the explicitly bounded source or cell factors.
This is the only error introduced by moving the nonanalytic cutoff.

\subsection{The upper bound}
For real $q$ in the chart and
$|H^{1/2}\beta|\le .99h$, the even Taylor expansion gives
\begin{equation}
 \operatorname{Re}(\lambda(q+i\beta)-z_h-\sigma)
 \ge 2^{-8}(h^2+q^THq).                              \label{c:eq:tube}
\end{equation}
To verify it, the quadratic part is
$h^2+q^THq-\beta^TH\beta$ up to $2h^4$.
The quartic remainder and its mixed terms are absorbed using
$M_4(4q_0)^2<10^{-6}/\varrho$ and $h\le\Ls^{-1}$.
The other fiber directions remain separated by Lemma~\ref{c:lem:edge}.
The inverse on this tube is therefore bounded by a fixed constant times
$L^2$.

For a displacement $R$ choose
$\beta=.99h H^{-1}R/|H^{-1/2}R|$. The preceding contour calculation,
applied between two period cells, gives an upper bound
$\Cs L^2e^{-.99h|H^{-1/2}R|}+\Cs e^{-\cs L}$.
Each radius-$L$ support meets at most a fixed multiple of $L^3$ cells.
The row and column Schur bounds for the matrix of cell-block norms,
and the bounded change of phase across a support, give
\[
 \norm{Q_hR_0\mathcal T_RQ_h^*}
 \le\Cs L^5\big(e^{-.99h|H^{-1/2}R|}+e^{-\cs L}\big).
\]
For $R=2LdB\nu$,~\eqref{c:eq:Bmetric} and
$|\nu|_2\ge|\nu|_1/\sqrt3$ show that the exponent per $d|\nu|_1$ is at least
$2(.99)^2/\sqrt3>\log3$. This proves~\eqref{c:eq:blockdecay}.
The elementary comparison follows, for example, from
$\sqrt3<7/4$ and $\log3<11/10$, both certified by rational series.

\subsection{The lower bound: pole and transverse Gaussian}
For direction $e_j$, let $\beta_{j,h}>0$ be the unique small solution of
\begin{equation}
 E_c-\lambda(i\beta_{j,h}e_j)=E_c-z_h.
                                                               \label{c:eq:betaroot}
\end{equation}
The even expansion, positivity of $H$, and the derivative of its
quadratic term prove existence and uniqueness by monotonicity. They also
give
\begin{equation}
 .99/\sqrt{H_{jj}}\le\beta_{j,h}/h\le1.01/\sqrt{H_{jj}},
 \quad I_{j,h}:=2LB_{jj}\beta_{j,h}
 \le2(1.01)^2<\log8-1/50.                            \label{c:eq:action}
\end{equation}
Coordinate reflection symmetry makes the transverse gradient zero at
this point. For $0\le s<1$ the contour shifted by
$is\beta_{j,h}e_j$ is nonsingular. At $s=1$, its only selected-band
singularity is $q=0$. The real part of the denominator is at least a
fixed positive multiple of $|q|^2$ there, so the inverse is bounded by
$C(1+|q|^{-2})$, integrable in three dimensions. Dominated convergence
justifies the limiting contour. All errors from
\eqref{c:eq:almostanalytic} remain supported away from this singularity.

Put $q=hk$, $b_{j,h}=\beta_{j,h}/h$, and
\[
 D_h(k)=h^{-2}(\lambda(hk)-z_h).
\]
After the shift the selected contribution has the form
\begin{equation}
 L^2e^{-I_{j,h}d}
 \int e^{2iB_{jj}dk_j}
 \frac{\mathcal N_h(k+ib_{j,h}e_j)}
 {D_h(k+ib_{j,h}e_j)}\,
 \chi_K(hk+i\beta_{j,h}e_j)\dd k.                   \label{c:eq:scaledgreen}
\end{equation}
The normalization of $\mathcal N_h$ already includes $(2\pi)^{-3}$.
The other bands have the same exponential prefactor and a smooth
integrand. The cutoff movement adds at most $\Cs L^5e^{-\cs L}$.

Write $k=k_\parallel e_j+k_\perp$ with $k_\perp\perp e_j$. Near zero the denominator in
\eqref{c:eq:scaledgreen} has
\[
 D(0,0)=0,\quad D_\parallel(0,0)=ic_{j,h},\quad D_\perp(0,0)=0,
 \qquad c_{j,h}>0,
\]
with real positive transverse Hessian $Q_{j,h}$. The explicit bounds
above give
\begin{equation}
 1<c_{j,h}<2^{112},\qquad
 2^{98}I<Q_{j,h}<2^{110}I.                            \label{c:eq:polebounds}
\end{equation}
On a complex radius-one neighborhood after scaling, the analytic
functions and their derivatives have bounds independent of $h$;
$h\le\Ls^{-1}$ keeps this neighborhood far inside the Bloch chart.
The elementary analytic implicit-function construction, by contraction
in the $k_\parallel$ variable, gives the unique pole
\begin{equation}
 k_\parallel=p_h(k_\perp)=\frac{i}{2c_{j,h}}k_\perp^TQ_{j,h}k_\perp+O(|k_\perp|^3),
 \qquad \operatorname{Im}p_h(k_\perp)\ge2^{-256}|k_\perp|^2
                                                               \label{c:eq:pole}
\end{equation}
on $|k_\perp|\le2^{-4096}$. The explicit contraction and derivative bounds
are recorded in Section~\ref{c:app:ledger}. In particular this is a
uniform neighborhood, not one chosen separately for each input.

Factor $D(k_\parallel,k_\perp)=(k_\parallel-p_h(k_\perp))d_1(k_\parallel,k_\perp)$ and divide the analytic numerator:
\[
 \frac{\mathcal N_h(k_\parallel e_j+k_\perp+ib_{j,h}e_j)}{D(k_\parallel,k_\perp)}
   =\frac{\mathcal A_h(k_\perp)}{k_\parallel-p_h(k_\perp)}+\mathcal W_h(k_\parallel,k_\perp),
 \quad \mathcal A_h(0)=\frac{\mathcal N_h(ib_{j,h}e_j)}{ic_{j,h}}.
\]
Choose smooth cutoffs equal to one on a smaller fixed neighborhood,
with all poles over the transverse support inside the core of the
longitudinal cutoff. For $T=2B_{jj}d>0$ and $k_\perp\ne0$,
\[
 \int_{\R}\frac{e^{iTu}}{k_\parallel-p_h(k_\perp)}\dd k_\parallel=2\pi i e^{iTp_h(k_\perp)}.
\]
This identity follows by closing the contour in the upper half-plane;
alternatively integrate on a finite interval first. The oscillatory
tails are bounded by $C/(TR)$ by integration by parts and then tend
to zero. Replacing the full line by the longitudinal cutoff has error
$O(T^{-16})$, since the removed region stays a fixed distance from the
poles and its positive-order derivatives are integrable.

Now put $k_\perp=T^{-1/2}y$. Taylor's formula and~\eqref{c:eq:pole} give
\[
 \int \mathcal A_h(k_\perp)e^{iTp_h(k_\perp)}\dd k_\perp
 =\mathcal A_h(0)\frac{2\pi c_{j,h}}{T\sqrt{\det Q_{j,h}}}
    +O(T^{-3/2}).
\]
Indeed the amplitude error is bounded by $C|k_\perp|$ and the exponent error
by $CT|k_\perp|^3$, both against $e^{-2^{-257}T|k_\perp|^2}$. Their radial
integrals are bounded by fixed constants times $T^{-3/2}$.
The omitted Gaussian tails have the same bound after increasing the
fixed constant. Thus the leading coefficient in~\eqref{c:eq:scaledgreen}
is
\begin{equation}
 b^0_{j,h}=\frac{4\pi^2\mathcal N_h(ib_{j,h}e_j)}
 {2B_{jj}\sqrt{\det Q_{j,h}}},\qquad
 2^{-512}\le b^0_{j,h}\le\Cs.                         \label{c:eq:leadingcoefficient}
\end{equation}
The lower bound follows from~\eqref{c:eq:numeratorlower},
$B_{jj}<2^{53}$, and~\eqref{c:eq:polebounds}. In particular the leading
term cannot cancel between polarizations.

We also need a uniform bound on the remainder outside the pole chart,
whose scaled domain grows like $1/h$. For $|k|$ outside a fixed
neighborhood, the selected denominator and its derivatives satisfy
\[
 |D_h(k+ib_{j,h}e_j)|\ge c(1+|k|^2),\qquad
 |\partial^K D_h(k+ib_{j,h}e_j)|
 \le C_K(1+|k|)^{2-K}.
\]
For orders $K\ge3$ use $h^{K-2}$ from differentiation and
$h\le4q_0/|k|$ on the scaled chart. The inverse-derivative recurrence
then gives the corresponding order $-2$ symbol bounds. Each fixed
numerator derivative is bounded by Lemma~\ref{c:lem:numerator}. The
localized regular integrand consists of the regular quotient $\mathcal W_h$ times
the pole-chart cutoffs, the selected quotient outside the pole chart,
including the transition shell of $\chi_K$, and the other bands. Hence the
$L^1$ norm of the sixteenth longitudinal derivative of the localized
regular integrand is at most $\Cs/h$. The integral of
$(1+|k|)^{-2}$ over a radius-$4q_0/h$ ball is $O(1/h)$.
For the other bands the scaled inverse includes a factor $h^2$ and
its integral over that ball is also $O(1/h)$. Derivatives of the
shifted cutoff are bounded uniformly, as shown after
\eqref{c:eq:almostanalytic}. Sixteen integrations by parts therefore
bound this entire regular contribution by $\Cs Ld^{-16}$ inside the
brackets of the following estimate:
\begin{equation}
 \left|h_j(d)-L^2e^{-I_{j,h}d}\frac{b^0_{j,h}}d\right|
 \le\Cs L^2e^{-I_{j,h}d}
       (d^{-3/2}+Ld^{-16})+\Cs L^5e^{-\cs L}.          \label{c:eq:greenremainder}
\end{equation}
This calculation includes the full regular spectral complement, not
only the selected-band numerator. Here, as throughout, $\Cs$ is the
fixed number in~\eqref{c:eq:masterconstants}.

For $d\ge\Cs^8$ and $d^{15}\ge\Cs^4L$, the first two errors in
\eqref{c:eq:greenremainder} are less than one quarter of the main term.
Equation~\eqref{c:eq:action} gives a strict exponential margin; for these
$d$, $e^{d/50}/d$ is larger than $2^{1024}$. The last error is absorbed
by~\eqref{c:eq:couplingtail}. This proves~\eqref{c:eq:couplinglower} and
completes Lemma~\ref{c:lem:interactions}.

\begin{lemma}[Off-support and periodization estimates]
\label{c:lem:cellbounds}
Let $\Box_a=2a+[0,2)^3$ be period cells. For $0\le\sigma\le h^4/64$,
let $\mathscr R_\sigma$ denote the local fiber inverse on the support
of $\chi$, before multiplication by $\chi$, and define
\begin{equation}
 R_\sigma^{(j)}=
 \chi\mathscr R_\sigma(\partial_j\varepsilon_{\rm per})
       \mathscr R_\sigma.
                                                               \label{c:eq:Rderivative}
\end{equation}
For $\mathcal R=R_\sigma$ or $\mathcal R=R_\sigma^{(j)}$,
\begin{align}
 \norm{\one_{\Box_a}\mathcal R\one_{\Box_b}}
 &\le\Cs L^4\big(e^{-\cs\dist(\Box_a,\Box_b)/L}+e^{-\cs L}\big),
                                                               \label{c:eq:cellspatial}\\
 \norm{\one_{\Box_a}\mathcal R\one_{\Box_b}}
 &\le\Cs L^4 e^{-\cs\sqrt{|a-b|/L}}.                 \label{c:eq:cellgevrey}
\end{align}
For $f_{\rm a}\in H^1$,
\begin{equation}
 \partial_j(R_\sigma f_{\rm a})=R_\sigma\partial_jf_{\rm a}+
                         R_\sigma^{(j)}f_{\rm a}.            \label{c:eq:derivativeidentity}
\end{equation}
The auxiliary source $f_{\rm a}$ of Section~\ref{c:sec:assembly} belongs to $H^1$
by Lemma~\ref{c:lem:auxrecovery}.
\end{lemma}
\begin{proof}
The first estimate repeats the nonsingular tube argument. A product of
two local inverses has norm at most a fixed constant times $L^4$ on
a slightly smaller tube. Equation~\eqref{c:eq:Bmetric} is not needed:
the smallest Euclidean decay rate is at least a fixed multiple of
$h/\sqrt{\norm A}$, larger than $\cs h$.

For the second estimate, the cell-fiber integrand is analytic in a
complex strip of width at least $2^{-256}h$ over the real support of
the cutoff. Cauchy's estimate, the Gevrey bounds on $\chi$, and the
bounded cell gauge factors give derivatives of order $K$ at most
\[
 \Cs L^4(2^{512}L)^K(K!)^2.
\]
Integrating by parts in the cell Fourier coefficient and optimizing
$K$ as in Lemma~\ref{c:lem:gevrey} proves~\eqref{c:eq:cellgevrey}, with
the smaller constant $\cs$ in the exponent. This summable estimate,
rather than the nondecaying error term in~\eqref{c:eq:cellspatial}, is
used for periodization.

For~\eqref{c:eq:derivativeidentity}, translate the periodic material and
use the local resolvent identity, then differentiate at zero. The
cutoff commutes with translations. The formula contains
one factor $\chi$, not two, since $\chi$ is not a projection. The local
inverse without $\chi$ in~\eqref{c:eq:Rderivative} is essential for
this identity. The norm bounds justify the difference-quotient limit.
No $H^1$ estimate on individual cell restrictions or artificial cell
traces is used.
\end{proof}

\section{Leakage from the auxiliary channel}
\label{c:sec:leakage}

The estimates in this section are the replacement for a complete bulk
band gap. They control the actual error produced by the cutoff in
\eqref{c:eq:Ridentities}.

\begin{lemma}[A nested momentum-cutoff estimate]
\label{c:lem:leakage}
Let $r_i\in2\Z^3$, $1\le i\le J$, and suppose the balls
$\mathbb B_L+r_i$ are disjoint. Put
\[
 W(r)=\sum_i v_iw(h(r-r_i)),\qquad
 0\le v_i\le2^{130}h^2,
 \qquad a_i=\mathcal T_{r_i}a_h.
\]
If $f_* =\sum_i c_i a_i$ and
\begin{equation}
 f_{\rm a}=f_*+WR_\sigma f_{\rm a},                                  \label{c:eq:tequation}
\end{equation}
then
\begin{equation}
 \norm{(1-\chi)f_{\rm a}}\le
 \Cs(1+J)e^{-\cs L^{1/3}}
       \left(\norm{f_{\rm a}}+\sum_i|c_i|\right).             \label{c:eq:leakage}
\end{equation}
No support or polarization condition is imposed on $f_{\rm a}$ in this lemma.
\end{lemma}
\begin{proof}
Let $P_{\ge s}$ be the Bloch-momentum projection onto torus distance at least
$s$ from the corner. On the support of $\chi$ with $|q|\ge q_0/4$,
\eqref{c:eq:edgebounds} and~\eqref{c:eq:quartic} give
\begin{equation}
 \norm{P_{\ge s}R_\sigma}\le2^{192}\quad(s\ge q_0/4).
                                                               \label{c:eq:highqbound}
\end{equation}
For example the quadratic denominator at $q_0/4$ is at least
$2^{-187}$, and the other bands are separated. Disjointness gives
$\norm W\le2^{130}h^2$, so their product is less than $1/4$.

If the two Bloch-momentum sets are separated by $\Delta_q$, multiplication
by $W$ between them has norm at most
\begin{equation}
 2^{100}J\max_i v_i\,e^{-\sqrt{\Delta_q/h}/512}.         \label{c:eq:fouriercross}
\end{equation}
Indeed its physical Fourier convolution kernel is the sum of translated
copies of $v_ih^{-3}(\mathcal Fw)(p/h)$. Frequencies in the two sets
differ in physical momentum by at least their reciprocal-torus
distance $\Delta_q$. Young's inequality and~\eqref{c:eq:gevreytail}
prove~\eqref{c:eq:fouriercross}. Possible changes of reciprocal alias
do not invalidate this lower bound on physical momentum difference.

Apply $P_{\ge s}$ to~\eqref{c:eq:tequation}, split the input at $s-\Delta_q$,
and use that $R_\sigma$ commutes with these projections. The high-input
part is bounded by $\tfrac14\norm{P_{\ge s-\Delta_q}f_{\rm a}}$. The low-input
part is bounded by~\eqref{c:eq:fouriercross} times $4L^2\norm{f_{\rm a}}$.
The factor $h^2$ in $v_i$ cancels this $L^2$; the resulting constant
is at most $2^{232}J$.

First apply this argument to~\eqref{c:eq:source-mode-equation}. It holds
also for every translated $a_i$, because the centers are period
translations. Use $K=\lceil L^{1/3}\rceil$ nested cutoffs from
$q_0/4$ to $q_0/2$, with step $\Delta_q=q_0/(4K)$ and coefficient
$\mu_h^{-1}\le2^{128}h^2$. Iteration gives
\[
 \norm{P_{\ge q_0/2}a_i}\le
 4^{-K}\norm{a_i}+
 2^{233}e^{-\sqrt{q_0L/(4K)}/512}\norm{a_i}
 \le\Cs e^{-\cs L^{1/3}},
\]
since $\norm{a_i}\le1$. The last comparison uses
$K\le2L^{1/3}$ and $q_0=2^{-140}$; even the much larger exponent
constant $2^{-84}$ would suffice.

Now iterate the same inequality for $f_{\rm a}$ from $q_0/2$ to $q_0$.
Every source term is at most
$\Cs e^{-\cs L^{1/3}}\sum_i|c_i|$ by the preceding estimate.
The geometric sum of the contraction factors is less than two.
The initial remainder is $4^{-K}\norm{f_{\rm a}}$, and all low-input leakage
terms have the same exponential bound. Finally $1-\chi$ vanishes on
$|q|<q_0$, so its norm is bounded by that of $P_{\ge q_0}f_{\rm a}$.
This proves~\eqref{c:eq:leakage} for every component. It also covers
momentum outside the chart, where $R_\sigma$ is zero.
\end{proof}

For $E_{\rm a}=R_\sigma f_{\rm a}$ satisfying~\eqref{c:eq:tequation}, the exact
physical residual in the infinite medium is therefore
\begin{equation}
 (A_{\rm per}-z_h-\sigma-W)E_{\rm a}-f_*
       =-(1-\chi)f_{\rm a}.                                 \label{c:eq:physicalleakidentity}
\end{equation}
The construction below makes the right side smaller than the required
inverse-polynomial budget. The cutoff is not being inserted into the
physical equation.

\section{Finite algebra and assembly of the auxiliary response}
\label{c:sec:assembly}

\subsection{The counted logical network}\label{c:sec:network}
We use the logical interface of Lemma~\ref{c:lem:logical}.

Thus $1\le\norm g\le\sqrt m$.
Double the logical lattice. Put a terminal at each doubled vertex and
one mediator at the midpoint of each actual edge. If $J$ is the total
number of these sites, then $J\le2^{19}\ell_{\rm p}^3$. Multiply their coordinates
by $2LdB$. All centers are in $2\Z^3$. At placement-lattice distance
one the only occupied pairs are the intended terminal--mediator
incidences: terminals have all coordinates even, and a midpoint has
exactly one odd coordinate. Every other occupied pair is at distance
at least two. This includes the two vertices of a parity detour in
Section~\ref{d:app:routing}, which are lattice-adjacent but not joined
by a path edge: after doubling they lie at distance two with no mediator
between them. The physical balls are disjoint.

\subsection{Explicit separation and diagonal tuning}
For the moment suppose $L,d$ satisfy the tests below; their terminating
selection is proved in Section~\ref{c:sec:parameters}. Put
\begin{equation}
 \zeta=3^{-d},\qquad
 \alpha=(c_u/4)L^2 8^{-d}=2^{-1026}L^2 8^{-d}.         \label{c:eq:alpha}
\end{equation}
Require
\begin{align}
 d&\ge\ds, &d^{15}&\ge\Cs^4L, &\alpha&\le1,\\
 \Cs^4L^{12}\zeta&\le\ee/2^{24},
 &\Cs^4L^{12}\zeta^2&\le\alpha\ee/(2^{24}\overline K),          \label{c:eq:dtests}\\
 J\Cs L^5e^{-\cs L}&\le\zeta^2/4.                   \label{c:eq:Ltailtest}
\end{align}
The last two inequalities imply~\eqref{c:eq:couplingtail}.
Calibrate the three real numbers $-h_j(d)$ by dyadic $\beta_j$ with
absolute error at most $\zeta^2/16$. Lemma~\ref{c:lem:interactions} gives
$|\beta_j|\ge3\alpha$.

Order terminals first and mediators second. Define the target matrix
$\Theta$ by incidences $\beta_j$ for a mediator in direction $j$ and
\begin{equation}
 \Theta_{ii}=\alpha(\Gamma_{ii}-\deg i),\qquad
 \Theta_{ee}=-\beta_j^2/\alpha.                            \label{c:eq:T}
\end{equation}
Its mediator Schur complement is exactly $\alpha\Gamma$. If $\Theta_{\rm tm},\Theta_{\rm mm}$
are its incidence and mediator blocks, then
\begin{equation}
 \norm{\Theta_{\rm mm}^{-1}}\le\alpha^{-1},\quad
 \norm{\Theta_{\rm tm}\Theta_{\rm mm}^{-1}}\le\sqrt6,\quad
 \norm{\Theta^{-1}}\le16\overline K/\alpha.                         \label{c:eq:Tinverse}
\end{equation}
Indeed each mediator contributes
$-\beta_j(-\beta_j^2/\alpha)^{-1}\beta_j=\alpha$ to both endpoint
diagonals and to their mutual edge. The endpoint diagonal correction
cancels the degree term in~\eqref{c:eq:T}. Each entry of
$\Theta_{\rm tm}\Theta_{\rm mm}^{-1}$ has magnitude at most $1/3$; its row and column degrees
are at most three and two. Triangular block factorization proves
\eqref{c:eq:Tinverse}. For any residual vector,
\begin{equation}
 \norm{\Gamma z_{\rm t}-g}\le(4/\alpha)
             \norm{\Theta z-\alpha g_\Theta}.                 \label{c:eq:mediatorres}
\end{equation}

Choose positive dyadic $v_i$ so that
\begin{equation}
 |v_i^{-1}-(\mu_h+\Theta_{ii})|\le\zeta^2.                \label{c:eq:vtuning}
\end{equation}
For terminals, $|\Theta_{ii}|\le6\alpha\le6$ and $L\ge\Ls$ give
$|\Theta_{ii}|\le2^{-196}L^2$. For mediators, $|\beta_j|\le2\Cs L^5\zeta$
and the second smallness test give the same bound.
Thus the tuning interval following Lemma~\ref{c:lem:numerator} applies,
and $0<v_i\le2^{130}L^{-2}<1$.
To implement~\eqref{c:eq:vtuning}, enclose $\mu_h$ to $\zeta^2/16$,
form the rational target, and round its reciprocal on a dyadic grid
of spacing at most $\zeta^2/(16L^4)$. The reciprocal derivative is
at most $4L^4$ on this interval. This is a finite precision rule,
not an exact-real tuning instruction.

Let $\tau$ be the largest power of two not exceeding
\begin{equation}
 \frac{\zeta^2}{2^{24}L^4}.                           \label{c:eq:tau}
\end{equation}
All $0\le\sigma\le\tau$ satisfy~\eqref{c:eq:Ridentities}; in particular
$\norm{R_\sigma-R_0}\le\zeta^2/2^{20}$.
Let $Q$ collect the translated $Q_h$ blocks, let $V_{\rm d}=\diag(v_iI)$,
and set
\begin{equation}
 W=Q^*V_{\rm d}Q,\qquad
 \MM=V_{\rm d}^{-1}-QR_\sigma Q^*.                          \label{c:eq:Mdef}
\end{equation}
Here the defect direct sum is infinite dimensional, and $\norm Q\le1$.

\begin{lemma}[Stable auxiliary assembly]
\label{c:lem:assembly}
For every $0\le\sigma\le\tau$, $\MM$ is invertible and
$\norm{\MM^{-1}}\le256\overline K/\alpha$. If $U$ inserts $u_h$ at every site,
then its bright Schur block has the form
\begin{equation}
 S_b=\Theta+\Delta,\qquad
 \norm\Delta\le\Cs^4L^{12}\zeta^2,
 \qquad \norm{S_b^{-1}}\le32\overline K/\alpha.                \label{c:eq:bright}
\end{equation}
All complementary modes and all unwanted interactions are included.
\end{lemma}
The parity layout of Section~\ref{c:sec:network} is essential here:
it places every unwanted occupied pair at $|\nu|_1\ge2$, so that all
unwanted couplings enter the non-neighbor sum of order $\zeta^2$ below.
Lemma~\ref{c:lem:interactions} alone does not exclude unwanted pairs at
$|\nu|_1=1$, whose coupling would be of the useful size.
\begin{proof}
For $0<\zeta\le1/2$, the elementary identity
$\sum_{\nu\in\Z^3}\zeta^{|\nu|_1}=((1+\zeta)/(1-\zeta))^3$ gives
\[
 \sum_{\nu\ne0}\zeta^{|\nu|_1}\le64\zeta,
 \qquad\sum_{|\nu|_1\ge2}\zeta^{|\nu|_1}\le256\zeta^2.
\]
The matrix of block norms bounds the complete operator by its row and
column sums. The geometric part of~\eqref{c:eq:blockdecay} therefore
has norm at most $64\Cs L^5\zeta$, and its non-neighbor part at most
$256\Cs L^5\zeta^2$. The nondecaying contour error is summed only
over the finite collection:~\eqref{c:eq:Ltailtest} bounds it by
$\zeta^2/4$. It has not been summed over an infinite lattice.
The uniform detuning changes the whole compressed union by at most
$\zeta^2/2^{20}$, independently of $J$.

Write the bright/complement decomposition of $\MM$ as
$\left(\begin{smallmatrix}\MM_{bb}&\MM_{bc}\\\MM_{bc}^*&\MM_{cc}\end{smallmatrix}\right)$.
At infinite separation its complementary block is the direct sum of
$\mu_hI-S_h$ on $u_h^\perp$, with gap $2^{-192}L^2$.
On-site scalar shifts have no mixed block. The displayed row sums and
\eqref{c:eq:dtests} give
\[
 \norm{\MM_{cc}^{-1}}\le2^{194}L^{-2},\qquad
 \norm{\MM_{bc}}\le128\Cs L^5\zeta,\qquad
 1+\norm{\MM_{bc}}\norm{\MM_{cc}^{-1}}\le2.
\]
This is an inverse-perturbation argument, valid even though the
complement may be indefinite. Direct bright errors consist of diagonal
rounding, incidence calibration, detuning, and the non-neighbor sum.
Adding $\MM_{bc}\MM_{cc}^{-1}\MM_{bc}^*$ bounds their sum by
$\Cs^4L^{12}\zeta^2$. Equations~\eqref{c:eq:Tinverse} and
\eqref{c:eq:dtests} make $\norm{\Theta^{-1}\Delta}<1/2$, proving the inverse
bound for $S_b$. A second triangular factorization, with factors of
norm at most two, gives $\norm{\MM^{-1}}\le256\overline K/\alpha$.
\end{proof}

\subsection{\texorpdfstring{A local $L^2$ source and its auxiliary logical response}{A local L2 source and its auxiliary logical response}}
Put $g_\Theta=(g,0)$ and define the ideal source
\begin{equation}
 f_*=(\alpha/\mu_h)Q^*Ug_\Theta.                          \label{c:eq:idealsource}
\end{equation}
It is a compactly supported $H^1$ function, hence a genuine $L^2$ source.
No unbounded operator is applied in its definition. Define, only for
analysis,
\begin{align}
 w_d&=\MM^{-1}QR_\sigma f_*,\\
 E_{\rm a}&=R_\sigma(f_*+Q^*w_d),\qquad
 f_{\rm a}=f_*+Q^*w_d.                                      \label{c:eq:auxsolution}
\end{align}
Then $w_d=V_{\rm d}QE_{\rm a}$ and $f_{\rm a}=f_*+WE_{\rm a}$, so
\eqref{c:eq:tequation} holds. The encoder never computes this inverse or
this field.

\begin{lemma}[Auxiliary recovery and magnitude bounds]
\label{c:lem:auxrecovery}
Let $z=U^*V_{\rm d}QE_{\rm a}$, i.e., $z_i=v_i\ip{E_{\rm a}}{a_i}$ are the
volume moments~\eqref{c:eq:idealreadout} of $E_{\rm a}$. Then
\begin{equation}
 \norm{\Gamma z_{\rm t}-g}\le\ee\norm g/1024.                  \label{c:eq:auxlogical}
\end{equation}
Also
\begin{align}
 2^{-256}\alpha L^{-2}\norm g&\le\norm{f_*}\le\norm g,\\
 \norm{f_{\rm a}}&\le\Cs \overline K\norm g,
 &\norm{E_{\rm a}}&\le\Cs \overline KL^2\norm g,\\
 \norm{f_{\rm a}}_{H^1}&\le\Cs^4\overline K^2\alpha^{-1}L^4\norm g,
 &\norm{E_{\rm a}}_{H^1}&\le\Cs^5\overline K^2\alpha^{-1}L^6\norm g.
                                                               \label{c:eq:auxnorms}
\end{align}
\end{lemma}
\begin{proof}
On each diagonal block $Q_hR_0Q_h^*u_h=\mu_hu_h$.
The off-site row sums and the detuning bound therefore give
\[
 QR_\sigma f_*=\alpha Ug_\Theta+\rho_s,
 \qquad\norm{\rho_s}\le\alpha\ee\norm g/2^{24}.
\]
Here the factor $\mu_h^{-1}\le2^{128}L^{-2}$ is absorbed by the first
smallness test in~\eqref{c:eq:dtests}. Eliminating the complement
multiplies this error by at most two. Thus
$\norm{S_bz-\alpha g_\Theta}\le\alpha\ee\norm g/2^{23}$ and
$\norm z\le33\overline K\norm g$. Adding $\Delta z$, using the second test, and
applying~\eqref{c:eq:mediatorres} proves~\eqref{c:eq:auxlogical} with
room to spare.

Since $\mu_h=\ip{a_h}{R_0a_h}\le4L^2\norm{a_h}^2$,
$\norm{a_h}\ge2^{-65}$. Disjoint supports,~\eqref{c:eq:mugap}, and
\eqref{c:eq:alpha} give the source bounds. From
\eqref{c:eq:Ridentities} and Lemma~\ref{c:lem:assembly},
\[
 \norm{QR_\sigma f_*}\le2^{130}\alpha\norm g,
 \quad\norm{w_d}\le2^{138}\overline K\norm g,
 \quad\norm{E_{\rm a}}\le2^{141}\overline KL^2\norm g,
\]
which are smaller than the stated master bounds.

For regularity, the exact identity
\begin{equation}
 (I-WR_\sigma)^{-1}=I+Q^*\MM^{-1}QR_\sigma            \label{c:eq:t-inverse}
\end{equation}
has norm at most $1+1024\overline KL^2/\alpha$. It is checked by direct
multiplication. Translate~\eqref{c:eq:tequation}. The source derivative
is bounded using~\eqref{c:eq:modeH1}. Disjointness gives
$\norm{\nabla W}_\infty\le\Cs h^3$. The resolvent $R_\sigma$ commutes
only with period translations; conjugation by a translation $\mathcal T_a$ moves
$\varepsilon_{\rm per}$, and the resolvent identity used
for~\eqref{c:eq:modeH1} gives
$\norm{\mathcal T_aR_\sigma\mathcal T_a^{-1}-R_\sigma}\le\Cs L^4|a|$.
Subtracting the translated equation and applying~\eqref{c:eq:t-inverse}
therefore bounds its difference quotient by
$\Cs^4\overline K^2\alpha^{-1}L^4G$. The Fourier difference-quotient argument
proves $f_{\rm a}\in H^1$. Equation~\eqref{c:eq:derivativeidentity} then gives
the last bound in~\eqref{c:eq:auxnorms}. Both $f_{\rm a}$ and its weak first
derivatives are supported in the union of the well balls.
\end{proof}

\section{Finite PEC boundaries and the actual material}
\label{c:sec:boundary}

Choose an even integer-sided box containing every well ball at distance
at least $\ell_{\rm buf}+4$ from its boundary. Translate the whole layout by an even
integer vector if needed. Let $\vartheta_\Omega$ be a real $C^2$ cutoff,
equal to one at distance at least two from the boundary and zero at
distance at most one. It can be made as a product of the quintic ramp
$10s^3-15s^4+6s^5$ on $[0,1]$ and its reflected copies; its first two
derivatives are bounded by $2^{20}$. Put
\begin{equation}
 E_{\rm ref}=\vartheta_\Omega E_{\rm a}.                \label{c:eq:cutfield}
\end{equation}
It lies in $D(\Cc)$ for the finite PEC box. Indeed $E_{\rm a}\in H^1$
by~\eqref{c:eq:auxnorms},
its curl--curl belongs to $L^2$ by~\eqref{c:eq:Ridentities}, and the
commutator with a $C^2$ cutoff is first order in $E_{\rm a}$.
The resulting field has compact support inside the box.

Define the explicit majorant
\begin{equation}
 \Xi=\Cs^{32}(1+N_\Omega)^2(1+J+m)^2
                  (1+\overline K)^4(1+\alpha^{-1})^2L^{100}.   \label{c:eq:Pglobal}
\end{equation}
\begin{lemma}[A physical finite-box approximate solution]
\label{c:lem:cutresidual}
For the ideal scalar material
$\varepsilon_{\rm id}=\varepsilon_{\rm per}+z_h+\sigma+W$,
\begin{equation}
 \norm{(\Cc-\varepsilon_{\rm id})E_{\rm ref}-f_*}
 \le\Xi\big(e^{-\cs L^{1/3}}+e^{-\cs \ell_{\rm buf}/L}\big)\norm g.
                                                               \label{c:eq:cutresidual}
\end{equation}
Furthermore $\norm{E_{\rm ref}}\le\Cs \overline KL^2\norm g$, and all ideal readout
moments on the wells are unchanged by the cutoff.
\end{lemma}
\begin{proof}
Equation~\eqref{c:eq:physicalleakidentity} gives the bulk error. In
Lemma~\ref{c:lem:leakage}, $\sum_i|c_i|\le
(\alpha/\mu_h)\sqrt m\,\norm g$, so~\eqref{c:eq:auxnorms} bounds it by the
first term of~\eqref{c:eq:cutresidual}.

The cutoff commutator is
\[
 [\curl\curl,\vartheta_\Omega]E
   =\nabla\vartheta_\Omega\times\curl E
       +\curl(\nabla\vartheta_\Omega\times E).
\]
Its norm is bounded by $2^{24}$ times the $H^1$ norm of $E$ on the
boundary strip. Apply~\eqref{c:eq:cellspatial} to the compactly supported
$f_{\rm a}$ and $\partial_jf_{\rm a}$, and use~\eqref{c:eq:derivativeidentity}.
The matrix of cell-block norms between that support and the boundary
strip has norm at most its entry bound times the geometric mean of
the two cell counts, at most a fixed multiple of $N_\Omega$.
Their cell separation is at least $\ell_{\rm buf}$ up to the already reserved fixed
margin. Thus the boundary $H^1$ norm is bounded by
\[
 \Cs^8 N_\Omega \overline K^2\alpha^{-1}L^{10}
        (e^{-\cs \ell_{\rm buf}/L}+e^{-\cs L})\norm g.
\]
The constant and powers are smaller than~\eqref{c:eq:Pglobal}.
Since $L\ge1$, $e^{-\cs L}\le e^{-\cs L^{1/3}}$.
This proves the residual bound. The cutoff equals one on the wells,
so the other assertions follow immediately.
\end{proof}

The actual output material uses a finite radial polynomial $\widetilde p_{\rm r}(t)$
with dyadic coefficients, approximating $e^{-1/[2(1-t)]}$ on
$0\le t\le1$, with $\widetilde p_{\rm r}(1)=\widetilde p_{\rm r}'(1)=0$. Outside each ball it is zero.
Set
\begin{equation}
 \overline\varepsilon_0(r)=7+z_h+\varrho V_0(r)
       +\sum_i v_i \widetilde p_{\rm r}(|r-r_i|^2/L^2)^2\one_{|r-r_i|<L},
 \qquad \overline\varepsilon=\overline\varepsilon_0+\sigma.
                                                               \label{c:eq:actualmaterial}
\end{equation}
Section~\ref{c:sec:representation} constructs this polynomial with a
prescribed uniform error and proves its cost. The polynomial square is
nonnegative. The background is between six and eight; the well
increments are less than one, and all approximation budgets are less
than $1/16$. Thus $1\le\overline\varepsilon\le25$ on every draw
(in fact it stays between six and ten).

Let
\begin{equation}
 e_0=\frac{\ee}{2^{24}J_\Gamma L_b},\qquad J_\Gamma=8.
                                                               \label{c:eq:e0}
\end{equation}
Here $J_\Gamma=8$ is an upper bound for $\norm\Gamma\le7$, which
multiplies the moment errors in~\eqref{c:eq:final-logical}.
Choose the material approximation so that
\begin{equation}
 \norm{\overline\varepsilon-
          (\varepsilon_{\rm per}+z_h+\sigma+W)}_\infty
       \le\frac{e_0}{4\Cs \overline KL^2}.                    \label{c:eq:materialprecision}
\end{equation}
Its additional residual on $E_{\rm ref}$ is at most $e_0\norm g/4$.
This bounded-multiplication comparison is why the output polynomial
need not retain the ideal well's Gevrey smoothness or exact Bloch
localization. It is not justified by qualitative convergence.
The random detuning of Lemma~\ref{c:lem:detuning} is applied to the
\emph{actual} material~\eqref{c:eq:actualmaterial}, after its finite
approximation has been fixed.

\section{Terminating parameter selection}
\label{c:sec:parameters}

All choices in this section use integers and rational comparisons. In
particular, the encoder does not need a real logarithm or a decision
about an exact transcendental equality. The logical parameters
$\overline K,\ee,J$ and the occupied sites have already been computed.
Write $\ell=1+\lceil\log_2(2+m)\rceil$.

For a candidate power of eight $L\ge\max(\Ls,\ell^6)$, starting with the
least such power and multiplying by eight after each failure, choose $d$ by
starting at one and doubling until the rational tests
\eqref{c:eq:dtests} hold. The tests include $d\ge\ds$,
$d^{15}\ge\Cs^4L$, and $\alpha\le1$; the definitions are
$\zeta=3^{-d}$ and $\alpha=2^{-1026}L^2 8^{-d}$.
Let $\tau$ be the largest integral power of two no greater than
$\zeta^2/(2^{24}L^4)$. This implies
\begin{equation}
 \frac{\zeta^2}{2^{25}L^4}<\tau\le
 \frac{\zeta^2}{2^{24}L^4}.                         \label{c:eq:tau-bracket}
\end{equation}
Every center is $2LdB$ times its doubled logical coordinate, followed
by one even integer translation. The extrema of these centers are
computed directly.

For $t_{\rm b}$ a power of two, starting at one, put $\ell_{\rm buf}=Lt_{\rm b}$. Choose the smallest
even integer side lengths containing the centers with margin at least
$L+\ell_{\rm buf}+4$ on all sides. Compute $N_\Omega,L_b$ by
\eqref{c:eq:count} and~\eqref{c:eq:detuning}, and compute $e_0$ and
$\Xi$ by~\eqref{c:eq:e0} and~\eqref{c:eq:Pglobal}. Double $t_{\rm b}$ until
\begin{equation}
 \Xi\,2^{-\lfloor\cs \ell_{\rm buf}/L\rfloor}\le e_0/4.
                                                               \label{c:eq:buffer-test}
\end{equation}
The dependence of $L_b$ and $\Xi$ on the candidate box is
included at every test. Only dimensions and counts, not the ambient
array, are written during this search.

After the buffer test succeeds, require also
\begin{align}
 \Xi\,2^{-\lfloor\cs L^{1/3}\rfloor}&\le e_0/4,
                                                               \label{c:eq:leak-test}\\
 J\Cs L^5\,2^{-\lfloor\cs L\rfloor}&\le\zeta^2/4.
                                                               \label{c:eq:tail-test}
\end{align}
If either test fails, replace $L$ by $8L$ and repeat. Because $L$ is a
power of eight, its cube root is an integer. The floors here are floors
of explicit rationals; they are ordinary integer divisions. The
inequality $e^{-s}\le2^{-\lfloor s\rfloor}$ for $s\ge0$ proves that
these rational tests imply all the corresponding exponential error
budgets. Test~\eqref{c:eq:tail-test} implies
\eqref{c:eq:couplingtail}: the second spacing test gives
$\zeta^2\le c_uL^2 8^{-d}$ with ample margin.

\begin{lemma}[Termination and size]
\label{c:lem:parameters}
The preceding procedure terminates. Its output satisfies
\begin{equation}
 L=O(\ell^6),\qquad d=O(\ell),\qquad \ell_{\rm buf}=O(\ell^7).
                                                               \label{c:eq:polylogscales}
\end{equation}
Its arithmetic cost before the final geometry is written is polynomial
in $\ell$, with an effectively specified fixed multiplicative constant.
All inverse scales in the construction are polynomial in $m$.
\end{lemma}
\begin{proof}
Here are explicit bounds, including the fixed constants. Put
\begin{equation}
 Z=\Cs^{128}(1+\lceil c_\kappa\rceil+\lceil c_\eta\rceil+
                         \lceil B_{\rm in}\rceil)^8,
 \qquad Z_L=8^{Z^8}.                                \label{c:eq:ZAL}
\end{equation}
The formulas in~\eqref{c:eq:logicalparameters}, using $n\le m$ and
$N_{\rm r}\le 8+(\lceil c_\kappa\rceil+\lceil c_\eta\rceil+2)\log_2m$, give
\[
 \log_2(2+\overline K+\ee^{-1}+J)\le Z^{1/4}\ell.
\]
Here the residual factor is
$T_{\rm res}=O(m^{7/2}\ell^{5/2})$ and
$\ee^{-1}=O(m^{\lceil C\rceil+11/2}\ell^{5/2})$.
The logarithms of the fixed coefficients and the displayed exponents are
bounded by $Z^{1/4}$ with the stated value of $Z$.
The power $Z^{1/4}$ only appears in this upper bound; the encoder never
computes it.

The first spacing inequality follows once a constant multiple of
$\log_2(\ee^{-1})+\log_2L+\log_2\Cs+1$ is at most $d$.
For the second, division by $\alpha$ leaves
\[
 2^{1050}\Cs^4 L^{10}\overline K\ee^{-1}(8/9)^d\le1.
\]
The elementary inequality $(9/8)^8>2$ supplies an entirely rational
sufficient test. The constraints $d\ge\ds$, $d^{15}\ge\Cs^4L$,
and $\alpha\le1$ have the same form or the explicitly displayed
fifteenth root. Doubling therefore stops with
\begin{equation}
 d\le Z(\ell+\log_2L+L^{1/15}).                    \label{c:eq:dcoarse}
\end{equation}
For example, to avoid the root in a sufficient candidate, the least
integer whose fifteenth power exceeds $\Cs^4L$ can be found by
integer doubling and bisection. Its value is at most
$2\Cs^{4/15}L^{1/15}$, which is covered by~\eqref{c:eq:dcoarse}.

The balanced logical box and $B_{jj}<2^{53}$ bound each physical side
by
\[
 2^{70}(1+\lceil c_\kappa\rceil+\lceil c_\eta\rceil)
 Ld\sqrt{m\ell}+2\ell_{\rm buf}+16.
\]
Together with~\eqref{c:eq:tau-bracket}, this bounds the logarithms of
$N_\Omega,L_b,e_0^{-1}$ and $\Xi$ by a fixed multiple of
\[
 \log_2\Cs+\ell+d+\log_2L+\log_2(1+t_{\rm b}).
\]
Substitution of the displayed formulas, allowing coefficient $Z^{1/4}$,
is a conservative bound for that fixed multiple and all additive
constants. Since $\cs=2^{-1024}$ and $Z\ge2^{2048}$, the first power
of two $t_{\rm b}$ above $Z^3(\ell+d+\log_2L)$ satisfies
\eqref{c:eq:buffer-test}. Indeed its linear left exponent
$\lfloor\cs t_{\rm b}\rfloor$ exceeds the stated logarithmic bound, using
$\log_2t_{\rm b}\le3\log_2Z+\log_2(\ell+d+\log_2L)+1$ and
$\log_2s\le s$ for $s\ge1$. Thus
\begin{equation}
 t_{\rm b}\le Z^4(\ell+d+\log_2L).                         \label{c:eq:tcoarse}
\end{equation}
After this buffer has been chosen, all logarithmic right-hand budgets
in~\eqref{c:eq:leak-test}--\eqref{c:eq:tail-test} are at most
\begin{equation}
 Z^6(\ell+\log_2L+L^{1/15}).                        \label{c:eq:budgetcoarse}
\end{equation}
This follows by inserting~\eqref{c:eq:dcoarse}--\eqref{c:eq:tcoarse};
$\log_2t_{\rm b}$ adds only $O(\log Z+\log(\ell+d+\log_2L))$.

Since the search starts at the least power of eight at least
$\max(\Ls,\ell^6)$ and multiplies by eight, it reaches every larger
power of eight unless it has stopped before. It therefore suffices to
show that both stopping tests hold at the following candidate.
Take $L$ to be the first power of eight at least $Z_L\ell^6$.
Then $Z_L\ell^6\le L<8Z_L\ell^6$. Both
$\lfloor\cs L^{1/3}\rfloor$ and $\lfloor\cs L\rfloor$ exceed
\eqref{c:eq:budgetcoarse}. To check this without an asymptotic assertion,
use $\log_2L\le3Z^8+6\ell+3$ and
$L^{1/15}\le2Z_L^{1/15}\ell^{2/5}$. The smallest power of $Z_L$
on the left is $Z_L^{1/3}$, and
\[
 Z_L^{1/3-1/15}=2^{(4/5)Z^8},\qquad
 Z_L^{1/3}=2^{Z^8}.
\]
These exceed $2^{1028}Z^6(3Z^8+20)$ for $Z\ge2^{2048}$;
repeated squaring, or the binomial lower bound for $2^N$, verifies
this last integer inequality. Also $\ell^2\ge\ell$ and
$\ell^2\ge\ell^{2/5}$. These comparisons prove both stopping tests
at this candidate. The search must terminate no later, so
\begin{equation}
 L\le8Z_L\ell^6.                                   \label{c:eq:Lexplicit}
\end{equation}
Inserting this bound in~\eqref{c:eq:dcoarse} gives
$d\le Z^3Z_L\ell$ after increasing only the displayed conservative
constant;~\eqref{c:eq:tcoarse} then gives $t_{\rm b}\le Z^8Z_L\ell$.
This proves~\eqref{c:eq:polylogscales}.

All searched integers have $O(\ell)$ bits, with fixed constants as
above. The tested powers $2^{\lfloor\cs \ell_{\rm buf}/L\rfloor}$,
$2^{\lfloor\cs L^{1/3}\rfloor}$ and $2^{\lfloor\cs L\rfloor}$ have
polynomially many bits in $\ell$ (at most $O(\ell^6)$, with a fixed
constant of order $Z_L$). The number of doublings is $O(\ell)$, also with such a
fixed constant. Ordinary exact rational arithmetic therefore has
polynomial-in-$\ell$ cost. In particular no failed candidate requires
a large ambient matrix or continuum eigenproblem to be constructed.
Finally $\alpha^{-1}=2^{1026}L^{-2}8^d$,
$\tau^{-1}<2^{25}L^4 3^{2d}$, and
$L_b=64mN_\Omega/\tau$ are polynomial in $m$ by the bounds just proved.
The same is true of every later precision budget.
\end{proof}

The finite scalar data and templates are calibrated only after this
parameter search. The actual material approximation is fixed before
sampling $\sigma$. Consequently the counting argument in
Lemma~\ref{c:lem:detuning} is applied to one fixed self-adjoint operator,
not to an operator chosen as a function of its own random detuning.

\section{Finite calibration without a continuum-solver oracle}
\label{c:sec:calibration}

This section gives algorithms for $z_h,\mu_h,u_h$ and the three
interactions. Their error estimates refer to the infinite operators
already defined, not to an unproved convergence property of a mesh.
Only the fixed corner matrix is inverted densely. Every other operation
is a finite Fourier operation, a convergent geometric series, or a
fixed-precision elementary function calculation.

\subsection{Elementary certified arithmetic and the edge energy}
Rational enclosures for $\pi$ follow from
$\pi=16\arctan(1/5)-4\arctan(1/239)$ and the alternating arctangent
series. The next term bounds its remainder. Square roots are enclosed
by integer bisection or integer square root. Exponentials and sine and
cosine use range reduction, Taylor series, and factorial geometric-tail
bounds. On arguments of magnitude $O(N_{\rm b})$ these procedures take a fixed
polynomial number of bit operations in $N_{\rm b}$ for $N_{\rm b}$ accurate bits.
Arguments outside this range are reduced, or the exponentially small
quantity is enclosed directly. The same statements for $\chi$ were
proved in Lemma~\ref{c:lem:gevrey}. Throughout, intervals are rounded
outward to a dyadic grid. Exact Gaussian elimination on a fixed-size rational matrix has
polynomial bit cost: clearing denominators uses only a fixed number
of entries, and every intermediate determinant has fixed order.
For a real symmetric matrix its ordered eigenvalues can be enclosed
without a root-separation assumption. At a rational trial value,
compute the inertia after subtraction of that value by symmetric
elimination. A nonzero diagonal gives a one-by-one pivot. If all
remaining diagonals vanish but an off-diagonal entry is nonzero,
its two-by-two block has one positive and one negative eigenvalue
and is an invertible pivot. An entirely zero remaining matrix
contributes only zeros. Congruence preserves the inertia at each
step. Counting positive eigenvalues therefore decides on which side
of the trial value the largest eigenvalue lies, including equality.
Bisection of an initial row-sum bound gives an interval of width
$2^{-N_{\rm b}}$ in $O(N_{\rm b})$ steps, with polynomial bit cost in the entry
lengths and $N_{\rm b}$. Interval matrix errors are added by the min--max
norm bound. These are explicit finite algorithms, not assumptions
about floating-point software.

For a local fiber, retain all 24 components of the eight central
aliases. Let $D_{\rm free}(q,\mathcal E)$ be the complementary free diagonal operator.
For $|q_j|\le3q_0$ and $\mathcal E$ in the edge window or within $2h^2$ of
$E_c$, its inverse is explicitly~\eqref{c:eq:freeinverse}, with norm
at most one. The complementary material inverse is
\begin{equation}
 \big(D_{\rm free}-\varrho V_{cc}\big)^{-1}
 =\sum_{a=0}^{\infty}
       (D_{\rm free}^{-1}\varrho V_{cc})^aD_{\rm free}^{-1}.              \label{c:eq:neumanncalibration}
\end{equation}
The tail after $K$ terms has norm at most $2(7\varrho)^K\le2^{1-K}$.
Each term changes a Fourier alias by at most $K$. The 24 by 24
central Schur matrix is consequently computable with a certified
operator-norm error at most $2^{10-K}$ using only aliases within
$K+2$ steps of the central set.

At $q=0$, the largest eigenvalue of that Schur matrix is zero exactly
at $E_c$ in the interval supplied by Lemma~\ref{c:lem:edge}. In that
interval all other Schur eigenvalues are negative. Its derivative
with respect to $\mathcal E$ is
\[
 -I-A_{{\rm per},0c}(A_{{\rm per},cc}-\mathcal E)^{-2}A_{{\rm per},c0},
\]
with the central and complementary alias blocks of the fiber as subscripts,
so its quadratic-form derivative lies between $-2I$ and $-I$.
The largest Schur eigenvalue therefore decreases at a rate between
one and two. This follows also by the min--max principle applied to
the two Schur matrices, and does not assume differentiability of an
ordered multiple eigenvalue.

Choose a rational $r_c$ with
$|r_c-E_0-\varrho\lambda_*|\le2^{15}\varrho^2$, using 256-bit
enclosures of $\pi$ and $\sqrt{19}$. The initial bracket
$[a,b]=[r_c-2^{17}\varrho^2,r_c+2^{17}\varrho^2]$ has width
$2^{-182}$ and strictly contains the band enclosure.
Put $w=b-a$ and $c=(a+b)/2$. Approximate the central Schur
matrix at $c$ to operator-norm error at most $w/64$, then test the
rational matrix for negative definiteness by exact elimination.
The slope bound gives $E_c\le c+w/64$ when the test is true and
$E_c\ge c-w/64$ otherwise. Thus use the overlapping brackets
\begin{equation}
 [a,b]\longmapsto
 \begin{cases}
 [a,c+w/64],&\text{negative definite},\\
 [c-w/64,b],&\text{otherwise}.
 \end{cases}                                             \label{c:eq:edge-bracket-step}
\end{equation}
Both branches have width $33w/64$, including when $c=E_c$.
Since $(33/64)^2<1/2$, $2N_{\rm b}$ steps give width at most
$2^{-N_{\rm b}}$. If $a=a_0/d_0$ and $b=b_0/d_0$, the first branch stores
$(64a_0,31a_0+33b_0,64d_0)$ and the second stores
$(33a_0+31b_0,64b_0,64d_0)$; every step adds at most six magnitude bits.
At step $k$, matrix error $2^{-(188+k)}$ is at most $w/64$, so the
working precision and endpoint lengths grow linearly with the number
of steps. Taking $N_{\rm b}$ a fixed amount larger than
$4\log_2L$ and rounding supplies~\eqref{c:eq:z} in polynomial bit time.

\subsection{Finite approximation of the local inverse}
On the real chart at $\mathcal E=z_h$, the exact full fiber inverse has norm
at most $4L^2$. Its central Schur inverse is a compression of the
full inverse and has the same bound. Approximate the Schur matrix
to norm error at most $2^{-N_{\rm b}}/(2^{20}L^4)$, invert the rational
approximation, and enclose its inverse using the inverse-perturbation
identity. The approximate inverse exists because its perturbation
relative to $4L^2$ is less than one half. Exact elimination or a
rational inverse followed by a residual test therefore succeeds.
The complementary inverse and the off-diagonal Schur factors use
\eqref{c:eq:neumanncalibration}. The block inverse formula then
approximates the full local inverse to norm error $2^{-N_{\rm b}}$ once
\begin{equation}
 K\ge N_{\rm b}+16\lceil\log_2L\rceil+2^{20}.               \label{c:eq:NeumannN}
\end{equation}
This deliberately weak bound covers all block products. Multiplication
by an enclosure of $\chi(q)$ gives the same precision for $R_0$.

The approximation is an operator-norm statement on the \emph{whole}
fiber, including arbitrarily high longitudinal aliases. No high
frequency has been removed on the assertion that its curl energy is
large. Formula~\eqref{c:eq:freeinverse} treats those directions exactly
before the small-potential series is truncated. On a finitely supported
Fourier input, the truncated formula produces finitely supported output;
its alias radius increases by at most $2K+4$.

All computations at a rational Bloch-grid index involve $\pi$ times
rational numbers and the already enclosed $z_h$. Noncentral free
denominators are bounded away from zero. The central inverse is
bounded by $4L^2$. Thus $N_{\rm b}+O(\log L+\log(2+|\nu|))$ working bits
suffice for the free entries, and a fixed multiple of that number
suffices for the fixed-size Schur calculation. An interval straddling
the edge of the cutoff support does not require a sign decision:
the inverse is valid on the larger chart $|q_j|\le3q_0$, while the
cutoff value is enclosed using its smooth defining formula.

\subsection{Periodization, with a norm error bound}
For calibration only, take a torus of side $2M_{\rm tor}$, with $M_{\rm tor}$ a multiple
of four. Place either one well, or the two wells needed for a
particular interaction, well inside it. Use a centered placement
so that the distance between these supports and any nontrivial
periodic image is at least $M_{\rm tor}/2$. A sufficient geometric condition
is $M_{\rm tor}\ge16(L+2Ld\max B_{jj}+1)$.

The periodized kernel is the sum of all translates by $2M_{\rm tor}\Z^3$.
Lemma~\ref{c:lem:cellbounds} makes that sum converge in the relevant
cell-block norms. On one or two well supports, the norm of the image
error is at most
\begin{equation}
 \Cs^2 L^{10}
       \exp\left(-\frac{\cs}{8}\sqrt{M_{\rm tor}/L}\right).
                                                               \label{c:eq:perioderror}
\end{equation}
To verify the factor, each support has at most $2^6L^3$ period cells.
For image index $j\ne0$ the cell distance is at least
$M_{\rm tor}|j|/4$. Sum the bound~\eqref{c:eq:cellgevrey} using shells with at
most $2^{10}(s+1)^2$ indices at $\ell^\infty$ distance $s$.
The elementary substitution $u=\sqrt s$ bounds the tail by a fixed
polynomial in $\cs^{-1}$ and $M_{\rm tor}/L$, times
$e^{-(\cs/8)\sqrt{M_{\rm tor}/L}}$. Weakening the exponent once absorbs that
polynomial; the remaining constants are less than $\Cs^2$.
The row and column Schur bounds give~\eqref{c:eq:perioderror}.

Choose the least multiple of four with the geometric condition and
\begin{equation}
 M_{\rm tor}\ge\Cs^4 L(N_{\rm b}+d+\lceil\log_2L\rceil+2)^2.
                                                               \label{c:eq:Mcalibration}
\end{equation}
Then~\eqref{c:eq:perioderror} is less than $2^{-N_{\rm b}-20}$.
The side $2M_{\rm tor}$ is an auxiliary calculation domain, \emph{not} a
physical internal boundary or an additional device volume. Its
computational cost is counted below. The periodic kernel is exactly
diagonalized by the finite set of Bloch momenta $q=\pi j/M_{\rm tor}$ modulo
$\pi\Z^3$, with the antiperiodic corner included because $M_{\rm tor}$ is even.
For each such momentum its alias space is still infinite until the
certified Neumann application above is made. This is a direct
Fourier-series identity for periodization, not a quadrature assumption.

\subsection{Multiplication and the isolated eigenvector}
On this torus approximate $q_w(hx)$ by a real trigonometric polynomial
$\widetilde q_w(r)$. The tail estimate~\eqref{c:eq:gevreytail} gives a uniform
error at most $2^{-N_{\rm b}}$ with Fourier index radius
\begin{equation}
 N_Q\le\Cs^4(1+M_{\rm tor}/L)(N_{\rm b}+\lceil\log_2M_{\rm tor}\rceil+2)^2.
                                                               \label{c:eq:NQ}
\end{equation}
For clarity, the Fourier coefficient of the scaled compact bump is
$(L/(2M_{\rm tor}))^3$ times its Fourier transform at $\pi L\nu/M_{\rm tor}$ with the
fixed normalization factor. Summing the tail outside $|\nu|_\infty>N_Q$
is bounded by the rescaled integral in~\eqref{c:eq:gevreytail}, plus
one layer of cubes; increasing the fixed factor in~\eqref{c:eq:NQ}
covers that layer. The coefficients are computed by a periodic
trapezoidal rule on a grid with a fixed larger index radius. Its
alias error is bounded by the same Fourier tail. There are
$O(N_Q^3)$ nodes. Each bump value has the certified evaluation
procedure in Lemma~\ref{c:lem:gevrey}. Rounding coefficients with an
extra $O(\log N_Q)$ guard bits controls their total uniform error.
Thus multiplication is approximated in \emph{operator norm}, rather
than just on a set of sampled fields.

The exact torus compression with the compact bump acts as zero
outside its support. On that support it differs from $S_h$ by
\eqref{c:eq:perioderror}. Replacing both bump factors by $\widetilde q_w$, and
using the norm-$2^{-N_{\rm b}}$ inverse approximation, changes its norm by
at most $2^{10}L^2 2^{-N_{\rm b}}$. All these comparisons take place in the
same $L^2$ space after extension by zero; the extra zero subspace
on the torus creates no ambiguity about the isolated largest
positive eigenvalue.

Here is an explicit polynomial-in-precision eigenvector algorithm.
Scale $S_h$ to $T_h=2^{90}L^{-2}S_h$. Lemma~\ref{c:lem:effectivemass}
gives
\begin{equation}
 2^{-38}\le\lambda_{\rm top}(T_h)\le1,
 \qquad
 \max_{\lambda\ne\lambda_{\rm top}}
       \frac{|\lambda|}{\lambda_{\rm top}}
       \le1-2^{-104}.                              \label{c:eq:powergap}
\end{equation}
For the negative spectrum use the separated-band bound
$2^{112}$ on $R_0$, while the selected part is nonnegative.
Since $L\ge\Ls$, its scaled negative norm is less than
$2^{-38}(1-2^{-104})$. The positive complementary spectrum uses
the gap $2^{-192}L^2$ and the upper eigenvalue bound.

Start with the normalized vector in Lemma~\ref{c:lem:effectivemass},
whose overlap is at least $2^{-48}$. Apply an approximation to
$T_h$ successively
\begin{equation}
 t_{\rm pw}=\left\lceil2^{106}(N_{\rm b}+200)\right\rceil            \label{c:eq:powersteps}
\end{equation}
times, then normalize. The exact normalized power has error at most
$2^{-N_{\rm b}-20}$ by~\eqref{c:eq:powergap}. Its unnormalized norm is at
least $2^{-48-38t_{\rm pw}}$. Request each operator application and the
starting vector to absolute accuracy
\begin{equation}
 2^{-N_{\rm work}},\qquad
 N_{\rm work}=N_{\rm b}+40t_{\rm pw}+\lceil\log_2(t_{\rm pw}+1)\rceil+256.
                                                               \label{c:eq:workbits}
\end{equation}
The telescoping product bound, using $\|T_h\|\le1$, makes the total
unnormalized error at most
$2(t_{\rm pw}+1)2^{-N_{\rm work}}$. Dividing by the proved lower norm and
using the elementary Lipschitz bound for normalization leaves less
than $2^{-N_{\rm b}-10}$. Torus, multiplier, and Neumann errors are allocated
inside the same per-application budget. Consequently the returned
normalized vector is within $2^{-N_{\rm b}}$ of $u_h$ in $L^2$, with its
sign fixed by the starting vector. Taking the real part and
renormalizing uses another two guard bits. A Rayleigh quotient,
computed by one more certified application, then encloses $\mu_h$
to absolute error $2^{-N_{\rm b}}$ after adding $O(\log L)$ bits.
This algorithm is intentionally slow in its \emph{fixed} gap constant;
its dependence on requested precision remains polynomial.

Every vector in this procedure is a finite Fourier polynomial. A
truncated inverse application increases the ordinary torus Fourier
radius by at most $(2K+4)M_{\rm tor}$, and multiplication by $\widetilde q_w$ increases it
by at most $N_Q$. The stored initial carrier has radius at most $3M_{\rm tor}$. For its initial bump
approximation, a uniform multiplier error $e$ produces an $L^2$
error at most $2^7(1+M_{\rm tor}/L)^2e$ after the $h^{3/2}$ scaling.
The exact unnormalized starting vector has norm at least $1/4$
by the overlap calculation in Lemma~\ref{c:lem:effectivemass}.
Thus using uniform error at most
$2^{-N_{\rm work}}/[2^{12}(1+M_{\rm tor}/L)^2]$, and allocating the
same accuracy to the fixed carrier coefficients, supplies the
requested normalized starting-vector precision. These are only
$O(\log M_{\rm tor})$ extra guard bits after $M_{\rm tor}$ is chosen. Each multiplier
and inverse application is likewise assigned its stated operator
error before summation; the explicit factors $4L^2$ and $2^{90}L^{-2}$
require only $O(\log L)$ further bits.
Thus after all applications the radius is bounded by
\begin{equation}
 N_{\rm all}\le3M_{\rm tor}+(t_{\rm pw}+2)(2N_Q+(2K+4)M_{\rm tor}).
                                                               \label{c:eq:allfreq}
\end{equation}
All $M_{\rm tor}^3$ Bloch-grid entries, all Fourier coefficients, and even naive
quadratic-time convolutions therefore cost a fixed polynomial in
$L,d,N_{\rm b},\log L$. With $L=O(\ell^6)$ and $N_{\rm b},d=O(\ell)$, this is a
fixed polynomial in $\ell$. There is no dense solve of dimension
comparable to $m$ in this calculation.

\subsection{Interaction and tuning calibration}
Use the two-well torus and the same mode approximation to evaluate
\[
 h_j(d)=\langle a_h,R_0\mathcal T_{2LdBe_j}a_h\rangle.
\]
Fourier translation is multiplication by explicitly computable unit
phases, and the inner product of two finite Fourier vectors is a
finite sum. The inverse norm $4L^2$ bounds the error from either
source-mode approximation;~\eqref{c:eq:perioderror} bounds images;
and~\eqref{c:eq:NeumannN} bounds the inverse error. Requesting each
of these errors at most $\zeta^2/(2^{12}L^2)$ suffices for the
absolute interaction error $\zeta^2/16$. There is no division by a
numerically guessed coupling. Its nonzero lower bound is already
proved in Lemma~\ref{c:lem:interactions}.

The same calculation supplies $\mu_h$ to $\zeta^2/16$ and the mode
or source mode to any later required $L^2$ tolerance. The targets
\eqref{c:eq:T}, their reciprocals, and the dyadic $v_i$ are then formed
using~\eqref{c:eq:vtuning}. The three direction calibrations and the
single-well data are shared by all sites. The targets $\Theta_{ii}$ take
only finitely many values, $\alpha(\Gamma_{ii}-\deg i)$ with
half-integer $\Gamma_{ii}\in[-3,3]$ and $\deg i\le3$ for terminals
and three values $-\beta_j^2/\alpha$ for mediators, so only $O(1)$
distinct $v_i$ are computed. Every requested bit count
is $O(\ell)$ with fixed constants, by Lemma~\ref{c:lem:parameters}.
This proves both termination and the cost of every input-dependent
calibration. In particular, the effective-mass estimate, which is only
$O(h)$, is \emph{not} used as an inverse-polynomial approximation to
the shrinking couplings. It establishes isolation and overlap; the
certified calculations in this section refine the actual auxiliary
operator to the much smaller required error.

\section{Finite material, source, and readout representations}
\label{c:sec:representation}

\subsection{The representation format}
The box sides, centers, and common radius $L$ are integers. In a ball
use the local coordinates $y=(r-r_i)/L$ and $t=|y|^2$. The material
is exactly~\eqref{c:eq:actualmaterial}. Its background is the fixed
finite expression in cosines in~\eqref{c:eq:background}; its only
input-dependent background coefficient is the dyadic $z_h+\sigma$.
A ball profile is specified by dyadic coefficients in the shifted
Chebyshev basis $T_j(2t-1)$ for $\widetilde p_{\rm r}(t)$, followed by the stated
square. This factored representation avoids large coefficients from
unnecessary conversion to monomials.

A source or readout template has the form
\begin{equation}
 \widehat a(r)=\widetilde p_{\rm r}(|y|^2)\widetilde P(y)\one_{|y|<1},
                                                               \label{c:eq:templateformat}
\end{equation}
where $\widetilde P$ has three components and is a tensor-product Chebyshev
polynomial on $[-1,1]^3$, with dyadic coefficients. A source is a
sum of translates of this template with dyadic amplitudes; a readout
uses the real part of its $L^2$ pairing with $E$. Values on the sphere have
no effect on these functions as $L^2$ objects. Both the formula and
its finite coefficients, rather than a quadrature table with an
unstated error, define the output function.

The format has a direct evaluation algorithm. Integer translations,
scaling, squared radius, polynomial recurrences, and the fixed cosines
are evaluated to a requested precision by outward arithmetic. On
their real domains, $|T_j|\le1$ and their derivatives are bounded
by $j^2$. Polynomial evaluation therefore needs only logarithmically
many guard bits in the number of coefficients and their magnitude.
The indicators specify exact semialgebraic balls. Away from their
boundaries membership is decided by the rational squared-radius
comparison; arbitrary-real point evaluation at a boundary is not
used as a computation primitive. The continuum instance is defined
by these exact functions independently of any point evaluator.

\subsection{A polynomial profile with endpoint matching}
Let $p_{\rm r}(t)=e^{-1/[2(1-t)]}$ for $0\le t<1$, with $p_{\rm r}(1)=0$.
The proof of Lemma~\ref{c:lem:gevrey}, now in one variable, bounds
its derivatives by $2^{20K}(K!)^2$. Composition with
$t=(1+\cos\varphi)/2$ gives a periodic even function with bounds
$2^{30K}(K!)^2$; this follows by the product rule, or inductively
by applying $-(\sin\varphi/2)\partial_t$ and differentiating its
bounded trigonometric coefficients. Integrating its Fourier
coefficients by parts $K$ times and choosing $K$ proportional to
$\sqrt j$ gives
\begin{equation}
 |c_j|\le 2^{100}\exp(-2^{-40}\sqrt j)
                                                               \label{c:eq:chebtail}
\end{equation}
for its Chebyshev coefficients. The same bound with a fixed
polynomial factor in $j$ controls the once differentiated series,
since $\|T_j'\|_\infty\le j^2$.

Truncate at degree $D_{\rm pr}\le\Cs(N_{\rm b}+1)^2$, choosing the least larger
integer for which the explicit tail bound is below $2^{-N_{\rm b}-20}$.
The sum of $j^2e^{-2^{-40}\sqrt j}$ above $D_{\rm pr}$ is bounded by its
first term plus the corresponding integral; the substitution
$u=\sqrt j$ makes this an elementary exponential-polynomial bound.
Compute the coefficients by the discrete cosine transform on a
fixed larger grid. The aliased-coefficient error is again bounded
by~\eqref{c:eq:chebtail}, and the samples have certified elementary
function evaluations. Round coefficients with
$N_{\rm b}+O(\log(D_{\rm pr}+2))$ fractional bits. If $\widetilde p^{\rm raw}_{\rm r}$ is the result,
replace it by
\begin{equation}
 \widetilde p_{\rm r}(t)=\widetilde p^{\rm raw}_{\rm r}(t)-\widetilde p^{\rm raw}_{\rm r}(1)
                     -(\widetilde p^{\rm raw}_{\rm r})'(1)(t-1).     \label{c:eq:endpointmatch}
\end{equation}
The endpoint values are exact dyadic sums because
$T_j(1)=1$ and $T_j'(1)=j^2$; for the shifted basis the
endpoint derivative is $2j^2$. The two correction coefficients have
$O(N_{\rm b}+\log D_{\rm pr})$ bits after the allocated rounding. The exact identities
$\widetilde p_{\rm r}(1)=\widetilde p_{\rm r}'(1)=0$ hold, and the uniform function and derivative
errors are at most $2^{-N_{\rm b}}$. In particular the zero extension of
$\widetilde p_{\rm r}(|y|^2)$ is $C^1$, while its square is nonnegative. All these
operations take polynomial-in-$N_{\rm b}$ bit time.

Since the well supports are disjoint, the material approximation
error is at most $3\max_i v_i\,2^{-N_{\rm b}}$, with no factor $J$.
Choose $N_{\rm b}$ so this is within~\eqref{c:eq:materialprecision}. This is
$O(\ell)$ bits. The ball-profile degree is therefore $O(\ell^2)$.

\subsection{Finite approximation of the calibrated mode}
The calibration algorithm returns a finite real Fourier polynomial
$\widetilde u(r)$ approximating $u_h$ on the ball in $L^2$, with the accuracy
proved in Section~\ref{c:sec:calibration}. On the local cube put
$\widetilde u_L(y)=\widetilde u(Ly)$. Each coordinate frequency has magnitude at most
$\pi N_{\rm all}L/M_{\rm tor}$. Its coefficient $\ell^1$ norm is at most the
square root of its number of Fourier coefficients times its $L^2$
norm, with the explicit torus normalization. These are polynomially
bounded in $L,d,N_{\rm b}$.

A finite trigonometric polynomial of maximum coordinate frequency $k_f$
and coefficient $\ell^1$ norm $a_f$ has a tensor Chebyshev approximation
of coordinate degree
\begin{equation}
 D_f\le 16(k_f+N_{\rm b}+\log_2(2+a_f)+1)                  \label{c:eq:chebdegree}
\end{equation}
with uniform error $2^{-N_{\rm b}}$, after a fixed increase in $N_{\rm b}$.
Here is a direct bound. Substitute $y_j=(s_j+s_j^{-1})/2$ and use
$|s_j|=2$. The imaginary part of $y_j$ has magnitude at most $3/4$,
so the function on this polyannulus is bounded by
$a_f e^{(9/4)k_f}$. Cauchy's Laurent coefficient estimate gives a
factor $2^{-j}$ in each coordinate. Summing the omitted geometric
tails proves~\eqref{c:eq:chebdegree}. A tensor discrete cosine transform
with a fixed larger degree gives coefficients with a certified
alias bound by the same estimate. True coefficients are at most
eight times the real-domain supremum, so their integer bits are
bounded by $O(\log(2+a_f))$, not by $k_f$. Rounding each coefficient
with $N_{\rm b}+O(\log(D_f+2))$ fractional bits controls the total uniform
error. This gives a polynomial-cost construction, not only an
existence result for a polynomial approximation.

Use this procedure with a uniform tolerance smaller than the desired
local $L^2$ tolerance divided by $8L^{3/2}$. Multiply the resulting
$\widetilde P(y)$ by $\widetilde p_{\rm r}(|y|^2)$ as in~\eqref{c:eq:templateformat}.
The multiplier approximation is in uniform norm, the calibrated
mode approximation is in $L^2$, and the cube has volume $8L^3$.
Their triangle inequality therefore gives a certified bound
\begin{equation}
 \|\widehat a-a_h\|_{L^2(\mathbb B_L)}\le e_a          \label{c:eq:atesterror}
\end{equation}
for any prescribed inverse-polynomial $e_a$.
This does not require unproved regularity or spectral convergence
of a piecewise-polynomial Maxwell discretization.

For the chosen parameters, the index bounds are
$M_{\rm tor}=O(\ell^8)$, $N_Q=O(\ell^4)$, $K=O(\ell)$,
$t_{\rm pw}=O(\ell)$, with the conservative stored-radius bound
$N_{\rm all}=O(\ell^{11})$. For the physical frequency, use
$LN_Q/M_{\rm tor}=O(\ell^2)$ directly in~\eqref{c:eq:allfreq}:
$LN_{\rm all}/M_{\rm tor}=O(\ell^8)$, so $k_f=O(\ell^8)$.
The degree in~\eqref{c:eq:chebdegree} is therefore
$O(\ell^8)$, giving $O(\ell^{24})$ tensor coefficients and
$O(\ell^{25})$ encoded bits per template. Each coefficient, including
its integer part, has $O(\ell)$ bits. The intermediate
exact arithmetic or guard precision may have a fixed polynomial
number of bits in $\ell$, which is counted in the time bound.

\subsection{Source rounding and stable observables}
Define the rational budget
\begin{equation}
 e_f=\min\{e_0/4,\ 2^{-260}\alpha L^{-2}\},
 \qquad
 e_a\le\min\{e_f/4,e_0/4\}.                 \label{c:eq:sourcebudgets}
\end{equation}
Construct one shared template satisfying~\eqref{c:eq:atesterror} and
$\|\widehat a\|\le2$. Enclose $\alpha/\mu_h$ to absolute error
$e_f/16$. Multiply its dyadic approximation by the short rational
nonzero entries of $g_\Theta$ and round each product to a dyadic amplitude
with error at most $e_f/(16m)$. There are at most $m$ such amplitudes.
Call the resulting finite source $f$. Disjoint supports and
$\alpha/\mu_h\le1$ give
\begin{equation}
 \|f-f_*\|\le e_f \norm g\le e_0\norm g/4,
 \qquad 0<\|f\|\le2\norm g.                              \label{c:eq:finite-source-error}
\end{equation}
For example the template error contributes at most $e_f\norm g/4$, the
shared scalar error at most $e_f\norm g/8$ using $\|\widehat a\|\le2$,
and the amplitude rounding at most
$2\sqrt m\,e_f/(16m)\le e_f\norm g/8$.
The lower source bound in~\eqref{c:eq:auxnorms} and the second term
in $e_f$ ensure that the total error is at most one sixteenth of
$\|f_*\|$, so $f$ is genuinely nonzero. It is a finite sum of
bounded compactly supported functions, hence belongs to $L^2$.
No derivative of an indicator is part of its definition.

The ideal observable at well $i$ is
\begin{equation}
 z_i(E)=v_i\langle E,a_i\rangle_{L^2}.
                                                               \label{c:eq:idealreadout}
\end{equation}
It is a bounded $L^2$ functional of norm at most one. The finite
observable replaces $a_i$ by the translated template $\widehat a_i$.
Its norm is at most two. Because supports are disjoint, the norm
of the \emph{entire} template-error map is at most $e_a$:
\[
 \sum_i|\langle E,\widehat a_i-a_i\rangle|^2
 \le e_a^2\sum_i\|E\|_{L^2(\mathbb B_L+r_i)}^2
 \le e_a^2\|E\|^2.
\]
This is the stability of the prescribed volume readout. There is
no point evaluation of an $H(\curl)$ field.

Request the real part of each raw finite observable to absolute
error at most
\begin{equation}
 h_y=\frac{\ee}{2^{24}J_\Gamma\max(1,J)}.            \label{c:eq:momentaccuracy}
\end{equation}
Multiply by the computed $v_i$ and round with an additional absolute
error at most $h_y$. A dyadic grid no coarser than $h_y$ fixes the
message format. Reserve a signed integer range larger than $8L_bm$.
The decoder consumes only this fixed number of bits per message,
rejecting a malformed message or overflow. The integer and fractional
bit counts are $O(\ell)$ on every draw.

Producing the moments to this accuracy is part of the continuum
solver's output obligation, and its cost is included in the solver
bound below. No finite-time integration oracle for an arbitrary
unrepresented $L^2$ function is granted to the decoder. The theorem
specifies a finite list of stable observables, explicit kernels, and
finite accuracies; the decoder reads and processes exactly those bits.

\subsection{Geometric and profile resolution}
The only material-support interfaces are the explicit spheres and the
external box. Their radii are $L\ge1$, their mutual gaps are at least
$2Ld\min_j B_{jj}-2L\ge L$, and their boundary buffer is at least
$\ell_{\rm buf}\ge L$. The fixed microscopic background has period two. There
are no shrinking geometric supports, gaps, filaments, or layers.
The smooth within-ball variation is specified by the counted degree
and coefficients, rather than by an uncharged fine mesh.

For a quantitative resolution bound, a tensor polynomial of degree $D_f$
and coefficient sum $a_f$ has first derivative bounded by $3a_fD_f^2/L$
in physical coordinates on the local cube. The radial profile obeys
the analogous $4a_fD_f^2/L$ bound; the squared profile only changes the
fixed factor. Here $D_f$ and the coefficient sums are fixed polynomials
in $\ell$ (the radial profile is uniformly bounded with its first
derivative by the preceding construction). The source and readout
frequency and degree bounds above are also fixed polynomials in
$\ell$. Thus even an explicit representation refined at the associated
inverse-polylogarithmic resolution has only a polylogarithmic extra
cost per unit volume. The proof does not rely on such a refinement
for PDE accuracy. It relies on uniform multiplier error for material
and $L^2$ error for source and readout. Small-amplitude polynomial
oscillations or close level sets therefore hide neither a geometric
list nor an uncounted discretization requirement.

\section{Every-field recovery for the actual continuum operator}
\label{c:sec:recovery}

Fix all parameters, finite profiles, and source and test coefficients,
then sample the detuning according to Lemma~\ref{c:lem:detuning} for
$\overline\varepsilon_0$ in~\eqref{c:eq:actualmaterial}. Set
\begin{equation}
 A=\Cc-\overline\varepsilon,
 \qquad \delta=e_0=\frac{\ee}{2^{24}J_\Gamma L_b}.
                                                               \label{c:eq:finaldelta}
\end{equation}
The auxiliary estimates hold for every draw $0\le\sigma<\tau$.
Tests~\eqref{c:eq:buffer-test}--\eqref{c:eq:leak-test}, the material
budget~\eqref{c:eq:materialprecision}, and the finite source budget
prove
\begin{equation}
 \|AE_{\rm ref}-f\|\le e_0\norm g.                          \label{c:eq:comparison-residual}
\end{equation}
On the common event of Lemma~\ref{c:lem:detuning}, take any
$E\in X$ satisfying~\eqref{c:eq:physicalres}. By~\eqref{c:eq:finite-source-error},
\begin{align}
 \|E-E_{\rm ref}\|&\le L_b(\delta\|f\|+e_0\norm g)
                 \le3L_be_0\norm g,                     \label{c:eq:fieldcomparison}\\
 \|E\|&\le L_b(1+\delta)\|f\|\le3L_b\norm g.             \label{c:eq:fieldrange}
\end{align}
Here $\delta<1/2$. Both inequalities apply to every residual field,
not only the exact solution and not only a Bloch-localized field.
All polarizations and all unwanted finite-box modes are controlled
by the full inverse bound in~\eqref{c:eq:fieldcomparison}.

The ideal moments of $E_{\rm ref}$ equal those of $E_{\rm a}$ because
the cutoff is one on all well supports. Their terminal vector $z_{\rm t}$
satisfies~\eqref{c:eq:auxlogical}. At well $i$ the first-moment error is
$v_i\ip{E}{\widehat a_i}-v_i\ip{E_{\rm ref}}{a_i}
=v_i\ip{E-E_{\rm ref}}{a_i}+v_i\ip{E}{\widehat a_i-a_i}$, plus at most
$2h_y$ from rounding. The first term uses $\norm{a_i}\le1$ and
\eqref{c:eq:fieldcomparison}, the second the template-error map of
Section~\ref{c:sec:representation} with~\eqref{c:eq:atesterror}
and~\eqref{c:eq:fieldrange}; with $0<v_i<1$ and disjoint supports this
shows that the finite moments of $E$ differ from this vector by at most
\[
 3L_be_0\norm g+3L_be_a\norm g+2\sqrt J\,h_y
 \le\frac{6\ee\norm g}{2^{24}J_\Gamma}.
\]
Consequently, with $\widehat z$ the decoded moment vector,
\begin{equation}
 \|\Gamma\widehat z_{\rm t}-g\|
 \le\ee\norm g/1024+J_\Gamma\frac{6\ee\norm g}{2^{24}J_\Gamma}
 <\ee\norm g/2.                                         \label{c:eq:final-logical}
\end{equation}
Lemma~\ref{c:lem:logical} gives the desired rational answer to
$Mx=b$. Since the material, source, and test fields are real,
replacing a complex field by its real part cannot increase its
residual; requesting real parts of the moments implements exactly
that replacement. The argument also allows every set of moment
errors within the specified absolute bounds.

By~\eqref{c:eq:fieldrange} and $\|\widehat a_i\|\le2$, the raw moments
have magnitude at most $6L_b\sqrt m<8L_bm$. Thus the fixed message
range is adequate on the same event. No union over solutions or
outputs is needed. On bad draws the finite format and all costs
are unchanged; only correctness and the inverse bound are not
asserted.

\section{\texorpdfstring{Complete resource accounting and proof of Theorem~\ref{c:thm:main}}{Complete resource accounting and proof of the continuum realization theorem}}
\label{c:sec:cost}

\subsection{Physical volume and finite data}
The logical side lengths are $O(\sqrt{m\ell})$. Doubling the lattice,
inserting mediators, multiplying by $2LdB$, and adding the buffer give
\begin{equation}
 \max_j N_j=O\big(Ld\sqrt{m\ell}+\ell_{\rm buf}+L+1\big),
 \qquad
 |\Omega|=O(m^{3/2}\ell^{24}).                       \label{c:eq:volume-final}
\end{equation}
Here Lemma~\ref{c:lem:parameters} gives $L=O(\ell^6)$ and $d=O(\ell)$.
The sharper exponent from these bounds is $45/2$; 24 is a convenient
integer upper bound. The estimate counts the entire rectangular box,
including empty space between all routes and wells and the external
buffer. In vacuum wavelength units the volume is $|\Omega|/(2\pi)^3$;
using the shortest material wavelength changes this only by a fixed
factor because $1\le\varepsilon\le25$.

There are $J=O((m\ell)^{3/2})$ well records. Each records an integer
center, the common integer radius, a dyadic increment, and a pointer
to a counted polynomial profile. The background has a fixed finite
formula. Even if the common source and test templates are copied at
every site rather than shared, the degree bounds in
Section~\ref{c:sec:representation} give at most
$O(m^{3/2}\ell^{64})$ coefficient and index slots.
Only at most $m$ source amplitudes are nonzero. The readout has at
most $J$ messages, each of $O(\ell)$ bits, and the final rational
answer has $n\le m$ coordinates of $O(\ell)$ bits.

An explicit ambient format can enumerate all unit cubes of the box.
For each cube, the candidate nearby placement site is located by
integer division in the diagonal lattice $2LdB\Z^3$, followed by a
constant number of neighbor checks and a lookup in the sorted occupied
site table. The balls are separated by much more than their radii,
so at most one can meet a given cube. A partial intersection is
represented by its exact quadratic inequality, not by a polygonal or
voxel approximation to the sphere. This enumeration costs
$O(|\Omega|\operatorname{polylog}m)$ and does not hide the background
volume in a sparse list.

\subsection{Numerical scales and operation counts}
The explicit inverse quantities are
\begin{equation}
 \alpha^{-1}=2^{1026}L^{-2}8^d,
 \quad \tau^{-1}<2^{25}L^4 3^{2d},
 \quad L_b=64mN_\Omega/\tau,
 \quad \delta^{-1}=2^{24}J_\Gamma L_b\ee^{-1}.
                                                               \label{c:eq:inverse-scales}
\end{equation}
Equations~\eqref{c:eq:logicalparameters} and
\eqref{c:eq:polylogscales} show that all are polynomial in $m$.
So are the reciprocals of every material, source, mode, interaction,
and message tolerance. All output coefficients and all integer
indices therefore require $O(\ell)$ bits.

The local-calibration bounds can be checked directly from the formulas:
$M_{\rm tor}=O(\ell^8)$, $N_Q=O(\ell^4)$, $K=O(\ell)$,
$t_{\rm pw}=O(\ell)$, and $N_{\rm all}=O(\ell^{11})$.
A coefficient array therefore has $O(\ell^{33})$ entries. Using
naive convolution takes $O(\ell^{66})$ scalar operations per product;
there are polynomially many such products and Bloch-grid operations.
The tensor polynomial conversion has $O(\ell^{24})$ coefficient slots. It may
be implemented by direct discrete cosine transforms and direct Fourier
sample evaluation, whose quadratic product counts are still fixed
polynomials in $\ell$. Dyadic interval operations, elementary series,
and the 24 by 24 rational inverses use polynomial bit time in the
working precision. The bit cost also includes allocation, address
arithmetic, coefficient normalization, and serialization of the
computed arrays. With schoolbook arithmetic, the resulting bound for
the shared calibration and template construction is
$O(\ell^{4096})$. Replication over the $O(m^{3/2}\ell^{3/2})$ sites
copies the already computed coefficient lists, at
$O(m^{3/2}\ell^{27})$ bit cost; it does not rerun calibration at
every site. These costs, the parameter searches, and the ambient
enumeration fit within~\eqref{c:eq:resources}.

For fully explicit constants, use $Z,Z_L$ from~\eqref{c:eq:ZAL} and set
\begin{equation}
 C_{\rm res}=2^{Z^{16}},\qquad c_A=c_\delta=C_{\rm res}.                         \label{c:eq:theoremconstants}
\end{equation}
Here is a conservative bookkeeping check. The fixed analytic constants
are at most powers $\Cs^{1024}$. The parameter bounds are
$L\le8Z_L\ell^6$, $d\le Z^3Z_L\ell$, and
$\ell_{\rm buf}/L\le Z^8Z_L\ell$. Allowing degree 4096 in those constants,
the bit target, and the local array sizes, the logarithm of their fixed
multiplicative constant is less than
$2^{20}Z^9<Z^{16}$. The same bound covers the fixed-startup arithmetic
needed to form the constants themselves. Thus $C_{\rm res}$ bounds the
multiplicative constants in~\eqref{c:eq:resources}.
The powers of $m$ in~\eqref{c:eq:inverse-scales} are at most
$2^{20}Z^4Z_L+\lceil c_\kappa\rceil+\lceil c_\eta\rceil+100$.
Their fixed coefficients and logarithmic factors can be absorbed in
$m^{C_{\rm res}}$ for $m\ge2$: use $\ell\le3m$, and absorb a constant factor $c_{\rm f}$ by
an extra exponent $\lceil\log_2c_{\rm f}\rceil$. The resulting exponent is
strictly smaller than $C_{\rm res}$. This proves the explicit bounds
$702L_b\le m^{c_A}$ and $\delta\ge m^{-c_\delta}$, not just an unspecified
polynomial assertion.

\subsection{The uniform encoder and decoder}
The complete algorithm, in its dependency order, is as follows.
\begin{enumerate}
\item Normalize and round the input; build the finite circuit, sign
mediators, and balanced three-dimensional route of
Sections~\ref{d:app:logical}--\ref{d:app:routing}. Compute $\Gamma,g,\overline K,\ee$ and the doubled
terminal--mediator sites.
\item Use Section~\ref{c:sec:parameters} to choose $L,d,\ell_{\rm buf}$, the actual
integer box, $\tau,N_\Omega,L_b$, and all error budgets. This step
writes only counts and dimensions until its stopping tests pass.
\item Verify the fixed corner certificate. Compute $z_h$, the isolated
single-well data, and the three actual auxiliary interactions to the
prescribed tolerances by Section~\ref{c:sec:calibration}. Compute and
round every $v_i$ by~\eqref{c:eq:vtuning}.
\item Form the finite radial profile and the finite source and readout
templates, with the global budgets in
Section~\ref{c:sec:representation}. These determine the actual
$\overline\varepsilon_0$ before detuning.
\item Draw the fixed number of fair bits for the midpoint grid in
Lemma~\ref{c:lem:detuning}, form $\sigma$, and output the actual material,
source, geometry, $\delta$, test kernels, message accuracies, and
message ranges. Enumerate the ambient unit cubes as just described.
\item The decoder consumes the fixed-width moment messages, forms
$\widehat z$, and performs the coordinate selection, rational rounding,
and rescaling in Section~\ref{d:app:logical-residual}. It performs no solve
of the original system or of the assembled auxiliary system. Its side
information is $O(m)$ words passed by the encoder: the coordinate-selection
map and the scale $\nu_M/(ma_M)$ of Section~\ref{d:app:logical-residual} (and, in
the Real-RAM bridge of Section~\ref{d:app:real-transfer}, the scales
$ma_F,\nu_F$).
\end{enumerate}
The decoder can scan a complete ambient-grid message stream, including
zero entries away from the prescribed readout sites. Its description
and this stream together use $O(m^{3/2}\ell^{25})$ bits. Parsing,
addressed selection and rational arithmetic on $O(\ell)$-bit words
give the separate bound
\begin{equation}
 T_{\rm dec}=O(m^{3/2}\ell^{27}),                         \label{c:eq:decoder-cost}
\end{equation}
which is included in the joint encoding and decoding allowance.
All loops and all refinement rules above have proved stopping bounds.
The physical medium and its fixed data are given explicitly; there is
no search for an unspecified photonic crystal or an assumed numerical
spectral certificate. The fixed verification and any enormous fixed
startup cost are included in~\eqref{c:eq:theoremconstants}.

If $b=0$, return $x=0$ directly. A required physical output can be
the unit cube with $\varepsilon=1$, $f=0$, and $\delta=1$, with a
decoder that returns zero. Its curl eigenvalues are zero or at least
$2\pi^2$, so $\spec(\Cc-1)\subset\{-1\}\cup[2\pi^2-1,\infty)$, and
\eqref{c:eq:maininverse} holds with $L_b=1$. The finitely many bounded input sizes
can similarly be solved by counted exact rational elimination and
accompanied by that dummy instance. In particular the tolerance bound
also holds at $m=1$. Their lengths and costs are bounded by the
fixed input constants; the inverse-event statement of Theorem~\ref{c:thm:main}
is stated for $m\ge2$.

\begin{proof}[Proof of Theorem~\ref{c:thm:main}]
Lemma~\ref{c:lem:logical} applies to every sparse input
satisfying~\eqref{c:eq:input}, without
any geometric restriction on its graph. Lemmas~\ref{c:lem:edge}--\ref{c:lem:numerator}
prove the continuum local spectral platform, including all fiber
polarizations and aliases, a simple tunable auxiliary mode, and a
nonzero source numerator. Lemma~\ref{c:lem:interactions} supplies the
three useful couplings and the full-block upper bound with a strict
rate margin. Lemma~\ref{c:lem:assembly} includes the entire infinite
defect complement and the sum of unwanted interactions.

The auxiliary channel is not substituted for the physical equation:
Lemma~\ref{c:lem:leakage} proves its actual leakage residual, and
Lemma~\ref{c:lem:cutresidual} puts it inside the finite PEC box in the
correct operator domain. The terminating choices in
Lemma~\ref{c:lem:parameters} make both errors small relative to the
shrinking useful interactions and to the full inverse bound. Sections
\ref{c:sec:calibration}--\ref{c:sec:representation} replace every input-
dependent datum by the stated finite representation with proved
error and cost.

Apply Lemma~\ref{c:lem:detuning} to that actual material. On its one
common event, the full continuum operator has~\eqref{c:eq:maininverse},
and Section~\ref{c:sec:recovery} proves successful decoding for every
field and every finite message error in the theorem. The bounds in
this section prove all physical, computational, and numerical resource
claims, on unsuccessful draws as well. This completes the proof.
\end{proof}

\section{Explicit analytic majorants for the continuum constants}
\label{c:app:ledger}

This section supplies the numerical envelope for the analytic constants
of Part~II. It replaces ``take a sufficiently large platform constant''
by finite arithmetic bounds.

\subsection{Fixed-order derivative envelope}
Set
\begin{equation}
 M_*=2^{16384},\qquad
 M_j=(j!)^4M_*^{(j+1)^2}\quad(0\le j\le32).
                                                               \label{c:eq:Mj}
\end{equation}
All derivatives below are directional derivatives along unit vectors;
mixed derivatives can be bounded by expansion in the three coordinate
directions, adding at most $3^j$, already absorbed in~\eqref{c:eq:Mj}.
The following list gives the underlying numbers before this envelope
is used.
\begin{center}
\begin{tabular}{@{}p{0.49\textwidth}p{0.42\textwidth}@{}}
\toprule
Quantity & Explicit bound used \\
\midrule
Local analytic momentum radius & $R_q=2^{-112}$ \\
Real cutoff chart scale & $q_0=2^{-140}$ \\
Hessian and quartic constant & $2^{99}I\le H\le2^{108}I$,
 $M_4=2^{350}$ \\
Nonselected fiber inverse on the small chart & $2^{112}$ \\
Cell eigenfunction Fourier $\ell^1$ norm & $32$ \\
Scaled denominator on a complex radius-two neighborhood & $2^{256}$ \\
Scaled source numerator on that neighborhood & $2^{128}$ \\
Pole contraction radius in transverse variables & $2^{-1024}$ \\
Final pole neighborhood & $|k_\perp|\le2^{-4096}$ \\
Regular-region denominator lower factor & $M_*^{-1}(1+|k|^2)$ \\
Fixed cutoff derivative bases & at most $2^{4200}$ \\
\bottomrule
\end{tabular}
\end{center}

The first five lines follow directly from Lemma~\ref{c:lem:edge} and
its Neumann construction. For the scaled denominator, insert the
Taylor series in $h^{-2}(\lambda(hk)-z_h)$.
On a complex polydisc of radius two around the shifted origin,
$|k|<4$, its quadratic term is at most $2^{112}$ and its quartic
remainder at most $2^{358}h^2<1$. The shift $|b_{j,h}|<1$ and the
constant term are covered by $2^{256}$. The same argument on a
slightly larger polydisc bounds derivatives by Cauchy's formula;
for instance $j!2^{4j+256}$ suffices up to order 32.

For the numerator, compact support gives
\[
 h^{3/2}\int|a_h(r)|\dd r\le\sqrt{|\mathbb B_1|}<3.
\]
On the stated complex neighborhood $|e^{\pm ihk\cdot r}|\le e^8$,
and the cell Fourier $\ell^1$ norm is at most 32. Thus each scaled
source factor has magnitude less than $2^{20}$, and their product
is below $2^{40}$; the table uses a larger number. Derivatives with
respect to scaled momentum introduce bounded $hx$ factors and
$h$ times the fixed cell derivative constants. Cauchy's estimates
therefore give the claimed derivative envelope, also on the whole
real scaled chart with the allowed imaginary shifts.

\subsection{The pole chart and its remainders}
Write the shifted denominator as $D(k_\parallel,k_\perp)$, with
$D_\parallel(0)=ic$, $c\ge1$, $D_\perp(0)=0$. Its second transverse derivative
is $Q_{j,h}$, where~\eqref{c:eq:polebounds} gives
$2^{98}I<Q_{j,h}<2^{110}I$. The preceding Cauchy bound implies, for
$|k_\parallel|,|k_\perp|\le1/8$,
\[
 |D(k_\parallel,k_\perp)-icu-\tfrac12k_\perp^TQ_{j,h}k_\perp|
 \le2^{280}(|k_\parallel|^2+|k_\parallel||k_\perp|+|k_\perp|^3).
\]
Consider the map
$k_\parallel\mapsto-[D(k_\parallel,k_\perp)-icu]/(ic)$. On
$|k_\perp|\le2^{-1024}$ and $|k_\parallel|\le2^{300}|k_\perp|^2$ it maps the latter
disc strictly into itself and has derivative in $k_\parallel$ of magnitude
less than one half. Indeed the derivative remainder is at most
$2^{282}(|k_\parallel|+|k_\perp|)$, and the quadratic main term is at most
$2^{110}|k_\perp|^2$. Iteration from zero converges geometrically and
uniformly on smaller complex polydiscs. Its limit $p(k_\perp)$ is
holomorphic, is the unique zero in that disc, and obeys the same
size bound. This is a direct quantitative proof of the implicit
function used in~\eqref{c:eq:pole}.

Cauchy's estimate on the transverse radius $2^{-1024}$ bounds the
third-order remainder in $p$ by $2^{1400}|k_\perp|^3$ on the smaller
half-radius polydisc. Differentiating $D(p(k_\perp),k_\perp)=0$ at zero gives
$p(k_\perp)=iw^TQ_{j,h}k_\perp/(2c)+O(|k_\perp|^3)$. The imaginary part of the quadratic
form is at least $2^{-15}|k_\perp|^2$. On $|k_\perp|\le2^{-4096}$ the cubic
bound is less than $2^{-2696}|k_\perp|^2$, giving the weaker stated
$\operatorname{Im}p(k_\perp)\ge2^{-256}|k_\perp|^2$.
The factor
\[
 d_1(k_\parallel,k_\perp)=\int_0^1D_\parallel(p(k_\perp)+t(k_\parallel-p(k_\perp)),k_\perp)\dd t
\]
is analytic and differs from $ic$ by less than $c/2$ on this
neighborhood. Cauchy division of the numerator can consequently
be performed by subtracting its value at $k_\parallel=p(k_\perp)$ and dividing
by $k_\parallel-p(k_\perp)$; the analytic remainder is the integral of the
corresponding derivative along that segment. These formulas bound
all derivatives without division by a small, unlocated zero.

One explicit choice of the auxiliary cutoffs is
$\psi(k_\parallel/2^{-2048})\prod_{j=1}^2\psi(k_{\perp,j}/2^{-4100})$,
with smaller copies for the core where a cutoff equals one.
All poles over the transverse support have magnitude less than
$2^{-7890}$, so they lie strictly inside the longitudinal core.
The derivative bases of these cutoffs are at most $2^{4200}$.
On the regular region outside the smaller pole neighborhood,
\eqref{c:eq:tube} at its limiting value gives a denominator lower
bound $M_*^{-1}(1+|k|^2)$. For small $k$ outside that neighborhood
use $|k|\ge2^{-4104}$ and the positive quadratic real part; for
$|k|\ge1$ use the symbol's quadratic lower bound. The interval
between these regions is covered by the same inequality after
weakening to $M_*^{-1}$.

The scaled symbol derivatives obey
\[
 |\partial^jD_h(k+ib_{j,h}e_j)|
 \le j!M_*(1+|k|)^{2-j}\quad(0\le j\le32)
\]
on that regular region. For $j\le2$ this is the quadratic Taylor
estimate. For $j\ge3$, differentiation contributes $h^{j-2}$;
on the scaled chart $h|k|\le4q_0$, so the Cauchy bound
$3\varrho j!R_q^{-j}$ gives the displayed inequality. For bounded
$k$ use the already proved radius-two bounds.
The recurrence
\[
 \partial^K(D^{-1})=-D^{-1}\sum_{a=1}^K
       \binom{K}{a}(\partial^aD)(\partial^{K-a}D^{-1})
\]
then bounds its derivatives by $(K!)^4M_*^{(K+1)^2}$ times
$(1+|k|)^{-2-K}$. The same recurrence applies to an invertible
matrix or bounded operator, preserving the order of the factors;
Cauchy's estimate is an alternative for the uniformly separated
complement. Leibniz's rule and the geometric-cutoff bounds give
$M_{32}^{16}$ as a common bound for the coefficient of
$(1+|k|)^{-2}$ in the sixteenth derivative of every regular
selected-band integrand.

For the other bands a useful explicit source bound avoids assuming
that their source coefficients have rapid decay. The rescaled Bloch
transform of $a_h$ has fiber norm at most a constant: Cauchy--Schwarz
over its at most $2^6L^3$ support cells, multiplied by $h^{3/2}$,
gives this bound. Its $j$th scaled-momentum derivative inserts
$h$ times a cell coordinate, of magnitude at most a fixed constant
on the support. Thus the same bound holds for each $j\le32$ with
the factor $M_j$. The nonselected resolvent and its derivatives
are uniformly bounded, and their scaled contribution has the
extra factor $h^2$. Integrating over a chart of volume $O(h^{-3})$
gives at most $M_{32}^{16}/h$. For the selected regular part,
$\int_{|k|\le4q_0/h}(1+|k|)^{-2}\dd k\le M_*/h$ gives the
same bound. This verifies the $L d^{-16}$ term in
\eqref{c:eq:greenremainder} for the entire complement.

For the transverse Gaussian remainder, the amplitude error is
bounded by $M_3^4|k_\perp|$ and the phase error by $M_3^4T|k_\perp|^3$.
Both exact and quadratic phases have imaginary part at least
$2^{-256}|k_\perp|^2$. The identity
$e^a-e^b=(a-b)\int_0^1e^{(1-t)b+ta}\dd t$ therefore preserves
the Gaussian majorant on the segment between them. Polar
integration gives
\[
 \int_{\R^2}|k_\perp|e^{-cT|k_\perp|^2}\dd k_\perp=O(c^{-3/2}T^{-3/2}),
\]
\[
 T\int_{\R^2}|k_\perp|^3e^{-cT|k_\perp|^2}\dd k_\perp
       =O(c^{-5/2}T^{-3/2}),\qquad c\ge2^{-257}.
\]
The gamma factors at these integer and half-integer arguments are
bounded by $2^{20}$ using their elementary factorial formulas.
The omitted Gaussian tail outside the fixed cutoff has a bound
of the same form, since $\sup_{T\ge1}\sqrt T e^{-aT}\le a^{-1/2}$
and its $a$ is explicitly at least $2^{-8500}$.
The full-line longitudinal tail differentiated 16 times is bounded
by $16!$ times a power of the reciprocal longitudinal-core radius,
at most $2^{40000}$. These bounds are all below $M_{32}^{32}$.
They justify every fixed remainder coefficient in
\eqref{c:eq:greenremainder}.

\subsection{Arbitrary-order cutoff and spatial estimates}
The preceding fixed-order envelope is not used as a claim about
arbitrarily high derivatives. Those estimates have their own explicit
factorial bounds. Recall $2^{16}/q_0=2^{156}$. In
\eqref{c:eq:almostanalytic}, $2^{156}|\beta|K\le1/4$. After using
$K!\le K^K$, the defect from holomorphy is at most
\[
 2^{45}2^{156}|\beta|(K+1)^2 4^{-K}
 \le2^{60}2^{-K}.
\]
Since $K\ge2^{-159}L$ for the nonzero shifts used here, weakening
the exponent gives $\Cs e^{-\cs L}$. For a fixed additional
$q$-derivative of order $j\le32$, each term of the truncated
Taylor sum is bounded by $4^{-a}$ times a polynomial of degree
$2j$ in its index $a$, multiplied by
$2^{45}2^{156j}(j!)^2$. Summing that geometric-polynomial series is
bounded by $(2j+2)!2^{10j+100}2^{156j}$, below $M_j^4$.
This verifies the uniform fixed derivative bounds for the shifted
cutoff, independently of the truncation order or of $L$.

For the summable cell estimate, the real local inverse has an
analytic strip of width at least $2^{-256}h$. To check it, split
off the rank-one pole and use the Taylor estimate and positive
quadratic form: a complex displacement with that width changes
$h^2+q^THq$ by at most one quarter of its magnitude, by
$2ab\le a^2+b^2$ and $H\le2^{108}I$. The separated spectrum
stays separated by the complementary Neumann argument.
Cauchy's formula, one or two local inverse factors, the Gevrey
cutoff, and bounded cell gauge derivatives give
\[
 \|\partial^K(\hbox{cell integrand})\|
 \le\Cs L^4(2^{512}L)^K(K!)^2\qquad(K\ge0).
\]
Integrating by parts in its cell Fourier coefficient and choosing
$K$ proportional to $2^{-258}\sqrt{|a-b|/L}$ yields exponential
rate at least $2^{-260}$ in that square root. The chosen
$\cs=2^{-1024}$ is smaller. This proves
\eqref{c:eq:cellgevrey} with the stated constants, including the
product inverse in~\eqref{c:eq:Rderivative}. The nonsingular contour
estimate similarly has a Euclidean rate at least $2^{-256}/L$,
again larger than the stated $\cs/L$.

For the nested-cutoff leakage, the high-momentum bound $2^{192}$
follows from $H\ge2^{99}I$ and $q_0=2^{-140}$, with a factor four
for the denominator remainder. The low-input constant after the
$h^2$ cancellation is at most $2^{232}J$. With
$K\le2L^{1/3}$, the Fourier-tail exponent satisfies
\[
 \frac1{512}\sqrt{q_0L/(4K)}
 \ge2^{-82}L^{1/3}.
\]
Together with the contraction $4^{-K}$ this proves the stated
leakage rate with $\cs=2^{-1024}$. The finite geometric sums of
contraction factors are below two. Hence no factor exponential
in $K$ is hidden in the leakage constant.

\subsection{Effective comparison and regularity constants}
The Fourier-alias error in~\eqref{c:eq:umklapperror}, divided by $h$,
is bounded by $2^{100}s^8e^{-s/512}$ with $s=h^{-1/2}$.
Its maximum occurs at $s=4096$ and is below $2^{196}$.
For~\eqref{c:eq:multipliererror}, the quartic contribution is at most
\[
 h^2 M_4|p|/2^{297/2}
 \le h\,2^{350-138-297/2}<2^{64}h;
\]
the cutoff tail is at most $2^{91}h$, and the eigenvector-multiplier
error from its Cauchy derivative is below $2^{70}h$.
Thus the displayed $2^{128}h$ multiplier bound and the
$2^{513}h$ operator bound leave room for the $2^{520}h$ comparison
used in the perturbation estimates. The scalar gap, reconstruction
and real normalization give the spatial-mode bound $2^{921}h$ in
\eqref{c:eq:modeapprox}. The bounded separated part contributes only
$2^{112}h^2$ before gap division.

For translations, the material derivative is at most $7\pi\varrho$,
and the local inverse norm is at most $4L^2$. The resolvent identity
therefore bounds its translation difference by $2^{20}L^4|a|$.
The bump factors add at most $2^{15}L|a|$, so $2^{32}L^4|a|$
bounds the translated compression. The spectral gap
$2^{-192}L^2$ then gives the normalized eigenvector difference
quotient below $2^{256}L^2$, and its full $H^1$ norm below
$2^{512}L^2$. This establishes the constants used for the cutoff
field's domain, without assuming ellipticity of curl--curl on the
whole vector space.

\subsection{Closing the fixed constant ledger}
All scalar, matrix, and derivative estimates just listed use at most
32 derivatives. Their product, inverse-derivative, Gaussian, and
finite-support summations are bounded by $M_{32}^{1024}$.
For an explicit check on the combinatorics, a derivative product is
summed with the binomial theorem; the resulting factors are bounded
by $2^j(j!)^4$, already present in $M_j$. The inverse recurrence
has at most $j$ terms at its $j$th step and is bounded by induction
by $M_j^4$. Cauchy division and the two cutoff products use at most
six such families; the pole/Gaussian remainder and the regular
integral use fewer than sixteen further products of their norms.
Even assigning 32 factors $M_{32}^{16}$ to each of these stages
stays below $M_{32}^{1024}$. The large support factors are the
explicit powers of $L$, or the finite counts $J,N_\Omega$, not
part of this fixed constant.

Finally
\[
 \log_2(M_{32}^{1024})
 =1024\bigl(4\log_2(32!)+16384\cdot33^2\bigr)
 <2^{35}<2^{40}=\log_2\Cs.
\]
Thus the common analytic coefficient $\Cs$ used in the interaction,
cell, leakage, and comparison estimates exceeds this fully specified
envelope. Whenever several such estimates are composed later, the
proof uses the displayed powers $\Cs^4,\Cs^5,\Cs^{32}$ rather
than silently replacing a product by $\Cs$.
Also $L\ge\Ls\ge\Cs^{64}$ makes every fixed smallness requirement
used above strict, for example $2^{520}h<2^{-200}$ and the errors
in the positive source numerator. The value $d\ge\Cs^8$ absorbs
the fixed large-distance thresholds in the Gaussian asymptotic.
No constant in this ledger is defined by the outcome of a
nonterminating search or an inaccessible exact eigenfunction.

\sipart{III}{Yee back end}
\section{Scalar Yee operators: conventions and finite precision}\label{d:app:model}
\begin{lemma}[Finite stencil and scalar detuning]\label{d:lem:pec}
The matrix in~\eqref{d:eq:maxwell-model} is real symmetric and satisfies $\norm A_2\le16$. A uniform scalar-cell increment $\sigma/\omega^2$ changes $A$ to $A-\sigma I$, including at edges adjacent to the boundary, provided the increment respects the material interval.
\end{lemma}
\begin{proof}
Each retained edge is incident on four faces. Each face has at most four retained edges, so the diagonal of $C_{\rm Y}^{\mathsf T}C_{\rm Y}$ is four and its absolute off-diagonal row sum is at most twelve. Since $1\le(D_\eps)_{ee}\le25$, the diagonal after the mass subtraction lies strictly between zero and four. Both the row and column absolute sums are at most sixteen. Equation~\eqref{d:eq:cell-average} shows that a uniform cell increment adds exactly $\sigma I$ to $\omega^2D_\eps$.
\end{proof}

The geometry in Section~\ref{d:sec:model} fixes the finite operator completely. An oriented edge carries one electric component, and an oriented square face carries the signed sum of its boundary edges. Removing tangential boundary electric unknowns implements the relative-cochain PEC convention. Face rows containing only removed edges may be retained as zero rows. The three terms in~\eqref{d:eq:yee-dimension} count the retained orientations separately.

The four incident cells of every retained electric edge lie inside the physical domain. Thus the mass rule~\eqref{d:eq:cell-average} is the same in the bulk and next to the boundary. A translated finite defect changes the cells in its support by a common increment $v/\omega^2$. Its electric-edge effect is $vW_{\mathcal D}$, where a touched edge has weight equal to the fraction of its four incident cells that belong to the defect. After restricting to touched edges, $W_{\mathcal D}$ is diagonal with entries in $\{1/4,1/2,3/4,1\}$. In particular, $0<w_{\min}\le W_{\mathcal D}\le I$. Defect supports are placed far enough apart that their touched-edge sets are disjoint.

All material and source output is dyadic, with an explicit bit bound. The decoder's finite interface consists of approximations to the real parts of selected electric coordinates, each with a stated absolute error. On the good event, the full inverse bound supplies a polynomial magnitude bound. A fixed number of integer and fractional bits therefore suffices. Outside the good event, the encoder still writes the same bounded-size data and the decoder still reads only its prescribed finite interface; correctness is asserted only on the common good event.

The relative residual target also explains the zero-source convention. If $b=0$, the answer $x=0$ is exact. When an electromagnetic output is required in that case, use a fixed $1\times2\times2$ cell box. It has one retained electric edge, $C_{\rm Y}^{\mathsf T}C_{\rm Y}=4$, and, with $\eps_c=1$, $A=255/64$. This instance is nonsingular with condition number one. The same fixed instance can accompany a precomputed exact answer for the finitely many bounded-size input cases. Use $f=0$ and $\delta=1$, and let the decoder return the stored answer. In particular, the tolerance bound also holds at $m=1$. Those inputs have bounded dimension and bit length under the input assumptions, so their exact processing is a counted fixed cost.

The construction's fixed integer translation basis changes where defects are placed. It does not alter edge lengths, incidence coefficients, or constitutive averaging. A rectangular box containing the transformed balanced layout increases volume by a fixed factor. This property would not follow from a vertex-count bound without the balanced spatial layout established by the logical compiler.

\section{Scalar defects and controlled interactions}
\label{d:sec:physical}

The physical construction uses localized modes as variables and their mutual
interactions as matrix entries. Its main constraint is that the material
parameter is one positive scalar per cell. We therefore cannot prescribe a
general interaction matrix directly. We first construct a fixed dielectric
environment in which a scalar defect supports an isolated mode. We then show
that translated copies of that defect have quantitative interactions of the
form required by the algebraic construction. All objects in this section are
fixed before the input matrix is supplied.

The useful distinction is between an interaction that is small and one that
is both small and reliably nonzero. An upper decay estimate controls
unintended interactions. A lower estimate on selected interactions ensures
that an intended edge remains usable after the other errors are included.
The construction requires both estimates with a strict relation between
their rates.

\subsection{A periodic background with a certified band edge}

Let
\begin{equation}
 A_{\rm per}
   = C_{\rm Y}^{\mathsf T}C_{\rm Y}-\omega^2D_{\varepsilon_{\rm per}},
 \qquad \omega^2=\frac1{64},
 \label{d:eq:physical-periodic-operator}
\end{equation}
be the infinite periodic electric operator, with the cell-to-edge averaging
convention of Section~\ref{d:sec:model}. In an anchor Bloch gauge its fiber is
a finite Laurent matrix
\begin{equation}
 A_{\rm per}(k)=\sum_{\nu\in\mathcal S}A_\nu e^{i\nu\cdot k},
 \qquad A_{-\nu}=A_\nu^T\in\R^{n_{\rm f}\times n_{\rm f}},
 \qquad k\in\mathbb T^3.
 \label{d:eq:physical-laurent-fiber}
\end{equation}
The spectral input to the construction is a unique, simple, nondegenerate
minimum of a positive band. We denote its energy by \(E_c\) and its momentum
by \(k_*\). The point \(k_*\) is invariant under time reversal, so
\(2k_*\in2\pi\Z^3\), and the corresponding eigenvalue has the local expansion
\begin{equation}
 \lambda(k_*+q)
   =E_c+q^THq+O(\abs{q}^4),
 \qquad H>0.
 \label{d:eq:physical-edge-expansion}
\end{equation}
Time reversal and simplicity make the dispersion even. We also require a
complete gap below \(E_c\), a certified separation from the other bands,
and effective bounds on the analytic continuation of the selected band.
These requirements concern the whole Bloch torus.

We supply this spectral input with a fixed inversion-symmetric perturbation
of a period-20 diamond dielectric. The unperturbed scalar cells take values
1 and 12. The perturbed cells have the exact form
\begin{equation}
 \eps_{\mathrm{per}}(c)
   =\eps_{\rm dia}(c)+\frac{p_c}{2^{21}},
 \qquad 0\le p_c\le2046,
 \label{d:eq:physical-exact-material}
\end{equation}
where the integer table \(p_c\), rather than its numerical generator,
defines the material. There are 2000 cells in a primitive FCC cell and
8000 in the conventional cubic cell. The certificate verifies the lattice
translations and scalar inversion symmetry directly.

The certificate has two complementary parts. Near each of the four
inequivalent primitive \(\mathrm{L}\) points, an exact polynomial spectral test proves
\begin{equation}
 \lambda_{\mathrm{low}}(\mathrm{L}+q)-\lambda_{\mathrm{low}}(\mathrm{L})
   >\frac{\abs{q}^2}{600},
 \qquad 0<\abs{q}\le\frac1{20},
 \label{d:eq:physical-local-minimum}
\end{equation}
in Cartesian reciprocal coordinates. Exact eigenvalue enclosures order the
four center values and select \(\mathrm{L}_{++-}\). On the complement, interval
matrix certificates and an analytic treatment of the \(\mathrm{\Gamma}\) region
exclude any lower positive-band value. Section~\ref{d:app:certificate}
gives the assembly, including the distinction between symmetries of the
comparison material and those of the perturbed material.

In the conventional cell, all four primitive \(\mathrm{L}\) points fold to the
same reciprocal class. We use the representative \(k_*=(\pi,\pi,-\pi)\),
consistent with \(\mathrm{L}_{++-}\). The strict ordering of their energies preserves the
simplicity of the lowest eigenvalue under this folding. In radian Bloch
coordinates the certificate yields
\begin{equation}
 2H=\lambda''(k_*)>\frac1{12000}I,
 \qquad g_-=\frac1{128},
 \qquad h_{\mathrm{mat}}=12.
 \label{d:eq:physical-certified-constants}
\end{equation}
Here \(g_-\) separates the edge from all lower bands, and
\(\max\eps_{\mathrm{per}}+E_c/\omega^2<25-h_{\mathrm{mat}}\) leaves material
headroom for the uniform shift and the scalar defects. The finite checks
are part of a computer-assisted proof; the analytic implications below
apply to any periodic finite-range operator with the stated edge data.

\subsection{A scalar defect with a useful isolated mode}

Let \(\mathcal D\) be a finite set of scalar cells. Let \(Q_{\mathcal D}\)
restrict an electric field to affected edges, and let \(W_{\mathcal D}\) contain the
fractions of incident cells belonging to \(\mathcal D\). Its diagonal
entries lie in \(\{1/4,1/2,3/4,1\}\). Increasing every cell of
\(\mathcal D\) by \(v/\omega^2\) changes the electric operator by
\(-vK_{\mathcal D}K_{\mathcal D}^T\), where
\[
 K_{\mathcal D}=Q_{\mathcal D}^TW_{\mathcal D}^{1/2}.
\]
This representation retains the scalar-cell constraint exactly.

For \(0<\xi<g_-/2\), define the shifted background and its finite
compression by
\begin{equation}
 A_\xi=A_{\rm per}-E_\xi I,\qquad E_\xi=E_c-\xi,
 \qquad
 S_\xi=K_{\mathcal D}^TA_\xi^{-1}K_{\mathcal D}.
 \label{d:eq:physical-compression}
\end{equation}
Let \(\Phi_*\) be the real extended edge Bloch field, normalized to have
squared norm one per period cell, and put
\(\Phi_{\mathcal D}=K_{\mathcal D}^T\Phi_*\). In three dimensions the finite matrix
\(S_\xi\) has a limit as \(\xi\downarrow0\), even though the infinite
edge operator has no bounded inverse. The leading correction is
\begin{equation}
 S_\xi=S_0-
   \frac{\sqrt{\xi}}{4\pi\sqrt{\det H}}\Phi_{\mathcal D}\Phi_{\mathcal D}^T+O(\xi).
 \label{d:eq:physical-square-root}
\end{equation}
This expansion is the link between the band edge and the selected defect
mode. Its proof comes from integrating the quadratic band singularity in
three dimensions; Section~\ref{d:app:edge} gives the error estimates.

Choose \(\mathcal D\) to contain all cells of a sufficiently large fixed
cube. Coherent addition of its Bloch amplitudes gives
\begin{equation}
 \frac{\Phi_{\mathcal D}^TS_0\Phi_{\mathcal D}}{\norm{\Phi_{\mathcal D}}^2}
   \ge c_{\mathrm{cube}}L_{\rm cube}^2-\frac1{g_-},
 \label{d:eq:physical-cube-growth}
\end{equation}
where \(L_{\rm cube}\) is the number of period cells along each interior side.
Consequently \(S_0\) has a sufficiently large eigenvalue whose eigenspace
overlaps \(\Phi_{\mathcal D}\). The rank-one square-root term in
\eqref{d:eq:physical-square-root} selects a simple branch even when that
limiting eigenvalue is degenerate. For sufficiently small fixed \(\xi\),
we obtain a positive simple eigenpair \((\mu_\xi,u_\xi)\) with
\[
 \Phi_{\mathcal D}^Tu_\xi\ne0.
\]
The reciprocal \(v_*=\mu_\xi^{-1}\) is small enough to fit within the
material headroom. Birman--Schwinger algebra then gives a one-dimensional
kernel of
\(A_\xi-v_*K_{\mathcal D}K_{\mathcal D}^T\), with field
\(A_\xi^{-1}K_{\mathcal D}u_\xi\). The background gap and finite
stencil imply exponential localization. A fixed two-sided interval around
\(v_*\) remains admissible, allowing subsequent algebraic calibration.

The nonzero overlap is more than an existence property of the mode. It
prevents the leading interaction amplitude from vanishing at the complex
momenta that determine long-range transmission.

The effective algebra can already be seen before the finite-domain
estimates are introduced. For disjoint translated defects, let
\(K_{\rm all}=[K_1\ \cdots\ K_J]\) and
\(V_{\rm d}=\operatorname{diag}(v_1I,\ldots,v_JI)\).
If
\[
 (A_\xi-K_{\rm all}V_{\rm d}K_{\rm all}^T)E=f,
\]
then the defect moments \(z=V_{\rm d}K_{\rm all}^TE\) satisfy the exact
finite system
\begin{equation}
 \bigl(V_{\rm d}^{-1}
       -K_{\rm all}^TA_\xi^{-1}K_{\rm all}\bigr)z
   =K_{\rm all}^TA_\xi^{-1}f.
 \label{d:eq:physical-effective-moments}
\end{equation}
In the selected mode at each defect, the diagonal response is controlled
by \(v_i^{-1}-\mu_\xi\); the off-diagonal response is the corresponding
mode-to-mode interaction. The simple-mode gap controls the effect of
eliminating the other defect channels. This is the origin of the
variables and couplings used by the reduction.

The tunings therefore act through diagonal responses of a structured
effective system. Auxiliary mediator defects convert those responses
into the required signed logical connections. Their calibration uses the
actual directional interactions supplied below. It does not require a
separate material parameter for every matrix entry or every electric
component. The later assembly analysis also includes the corrections
generated by the eliminated channels, the finite boundary, and source
rounding. Equation~\eqref{d:eq:physical-effective-moments} identifies the
mechanism; those bounds establish that the mechanism meets the theorem's
accuracy guarantee.

\subsection{Upper and lower estimates on interactions}

For the selected fixed shift and defect, let
\begin{equation}
 G(\nu)=K_{\mathcal D}^TA_\xi^{-1}\mathcal T_\nu K_{\mathcal D},
 \qquad \nu\in\Z^3,
 \label{d:eq:physical-interaction-block}
\end{equation}
where \(\mathcal T_\nu\) translates by period cells. A complex shift of the full
Bloch integral gives, for any fixed small \(\eta_{\rm t}>0\),
\begin{equation}
 \norm{G(\nu)}
 \le C\exp\!\left[
   -(1-\eta_{\rm t})\sqrt{2\xi}\,
   \norm{(2H)^{-1/2}\nu}\right].
 \label{d:eq:physical-block-upper}
\end{equation}
This controls every interaction channel in the finite defect block.
To obtain a useful lower bound, we examine the selected scalar channel
\(h(\nu)=u_\xi^TG(\nu)u_\xi\).

For a nonzero integer direction \(\upsilon\), let \(\beta_\upsilon\) be the support
point in direction \(\upsilon\) of the small convex level surface
\(E_c-\lambda(k_*+i\beta)=\xi\), and write
\(I_\upsilon=\beta_\upsilon\cdot\upsilon\). The leading asymptotic is
\begin{equation}
 h(d\upsilon)=e^{ik_*\cdot d\upsilon-I_\upsilon d}
 \left[
 \frac{\mathcal N_\xi(k_*+i\beta_\upsilon)}
 {2\pi d\abs{\upsilon}\sqrt{\det Q_{\upsilon,\perp}}}
 +O(d^{-3/2})
 \right].
 \label{d:eq:physical-coupling-asymptotic}
\end{equation}
Here \(Q_{\upsilon,\perp}\) is the restriction of the complex-point dispersion
Hessian to the plane perpendicular to \(\upsilon\), and
\[
 \mathcal N_\xi(k)=
 (\mathcal Ff_\xi)(-k)^T\Pi(k)(\mathcal Ff_\xi)(k),
 \qquad f_\xi=K_{\mathcal D}u_\xi,
\]
uses the Bloch transform \(\mathcal Ff_\xi\) and the analytic rank-one spectral projection \(\Pi(k)\).
For the selected branch, \(\mathcal N_\xi(k_*+i\beta_\upsilon)\) is positive for all
directions once \(\xi\) is sufficiently small. The time-reversal phase in
\eqref{d:eq:physical-coupling-asymptotic} is a sign, so it does not cancel
the magnitude of the leading term.

The proof shifts the entire Fourier torus to the limiting complex
contour before introducing a smooth localization. This order ensures
that every remainder retains the exponential factor
\(e^{-I_\upsilon d}\). A longitudinal pole calculation and a transverse Gaussian
integral then produce the factor \(d^{-1}\). Section~\ref{d:app:green}
provides the contour and remainder argument. Related spectral-edge
asymptotics for periodic elliptic operators provide useful context
\cite{kha2017}; the proof here treats the finite-dimensional Bloch
matrix of the discrete operator directly.

\subsection{A placement lattice with separated interaction rates}

The anisotropy of \(H\) is handled by placing defects on a fixed skew
sublattice. Choose an invertible integer matrix \(B\) close to a large
multiple of \((2H)^{1/2}\). In these placement coordinates, the decay
metric is approximately Euclidean. The comparison
\(\norm{\nu}_2\ge\norm{\nu}_1/\sqrt3\), together with
\(2/\sqrt3>1\), leaves a strict margin between twice the available
upper decay rate and the axial lower decay rates.

\Needspace{15\baselineskip}
\begin{samepage}
\begin{proposition}[Fixed scalar interaction platform]
\label{d:prop:physical-platform}
The certified periodic background admits effectively selectable fixed
data consisting of a dyadically shifted background with spectral gap
\(g_0>0\) at zero, a finite scalar-cell defect, a normalized mode \(u\)
for a positive simple compressed eigenvalue \(\mu\) with complementary
spectral separation \(\gamma_d>0\), and a two-sided admissible scalar
tuning interval with fixed positive upper material headroom.
There are also an invertible integer placement matrix \(B\)
and rational constants \(C_{\rm d},c_{\rm d}>0\), \(0<r_{\rm d},a_{\rm d}<1\), and \(d_0\in\N\), such that
\begin{equation}
 \norm{G(B\nu)}\le C_{\rm d}r_{\rm d}^{\norm{\nu}_1},
 \qquad
 \abs{u^TG(dBe_j)u}\ge c_{\rm d}a_{\rm d}^d
 \quad (d\ge d_0,\ j=1,2,3),
 \qquad r_{\rm d}^2<a_{\rm d}.
 \label{d:eq:physical-platform-bounds}
\end{equation}
All physical cell permittivities remain in \([1,25]\), and the grid metric
and coordinate stretches remain unchanged.
\end{proposition}
\end{samepage}

The strict inequality \(r_{\rm d}^2<a_{\rm d}\) is what makes composition possible.
Intended interactions can be calibrated at a scale proportional to
\(a_{\rm d}^d\), while interactions traversing at least two placement steps are
controlled by \(r_{\rm d}^{2d}\). Their ratio decreases geometrically. The
later construction chooses \(d\) logarithmic in the input size to meet an
inverse-polynomial error budget. The matrix \(B\) specifies positions of
defects; it does not alter the Yee stencil.

\subsection{Effective selection of the fixed platform}

The edge energy \(E_c\) need not be rational. We choose a dyadic operating
shift \(E_\xi<E_c\), set \(\xi=E_c-E_\xi\), and implement the uniform scalar
increment \(E_\xi/\omega^2\). Dovetailing over such shifts and rigorous finite
eigenpair certificates selects the simple branch and its nonzero overlap.
The strict estimates above also permit certified searches for \(B,r_{\rm d},a_{\rm d}\),
the coupling constants, and the large-distance threshold. These searches
terminate and are independent of the sparse input size.

The fixed search may have very large cost. Its termination is established
analytically, and the resulting constants are included in the uniform
asymptotic bound. The finite certificates supply the periodic background
and its edge; the defect and interaction estimates supply the remaining
data needed by the search.

\begin{figure}[htbp]
  \centering
  \includegraphics[width=\linewidth]{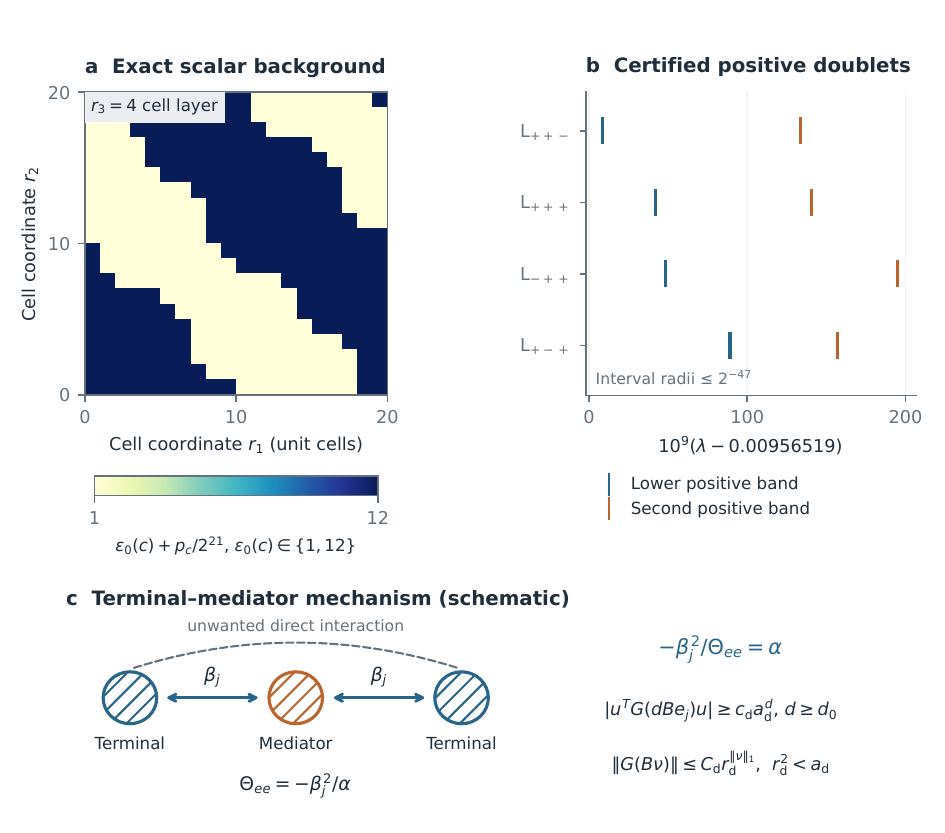}
  \caption{Certified spectral setting and analytic interaction mechanism.
  (a) The exact scalar permittivity in cell layer $r_3=4$ (cell centers at
  $r_3=9/2$) of the period-$20$ background. Each pixel is one scalar cell;
  the fixed perturbation is $p_c/2^{21}$, $0\le p_c\le2046$.
  (b) Certified intervals for the two low positive eigenvalues of
  $A_{\rm per}=C_{\rm Y}^{\mathsf T}C_{\rm Y}-D_{\varepsilon_{\rm per}}/64$ at the four
  inequivalent $\mathrm{L}$ points: the subscripts give the signs of their Cartesian
  coordinates $(\pm1/2,\pm1/2,\pm1/2)$ in $\tilde k=k/(2\pi)$.
  Primitive phase coordinates are
  $(\tilde k_x,(\tilde k_x+\tilde k_y)/2,(\tilde k_x+\tilde k_z)/2)$.
  Interval radii are at most $2^{-47}$ and are smaller than the plotted
  line width. The horizontal axis is shifted and scaled as labelled.
  (c) Schematic localized defect supports and the terminal--mediator Schur
  identity. The coupling inequalities use a fixed placement lattice $B$
  and effective constants satisfying $r_{\rm d}^2<a_{\rm d}$, with the useful lower bound
  holding for $d\ge d_0$. The drawing represents the
  analytic construction, without an instantiated defect mode or fitted
  coupling data.}
  \label{d:fig:FIG-2}
\end{figure}

\begin{figure}[htbp]
  \centering
  \includegraphics[width=\linewidth]{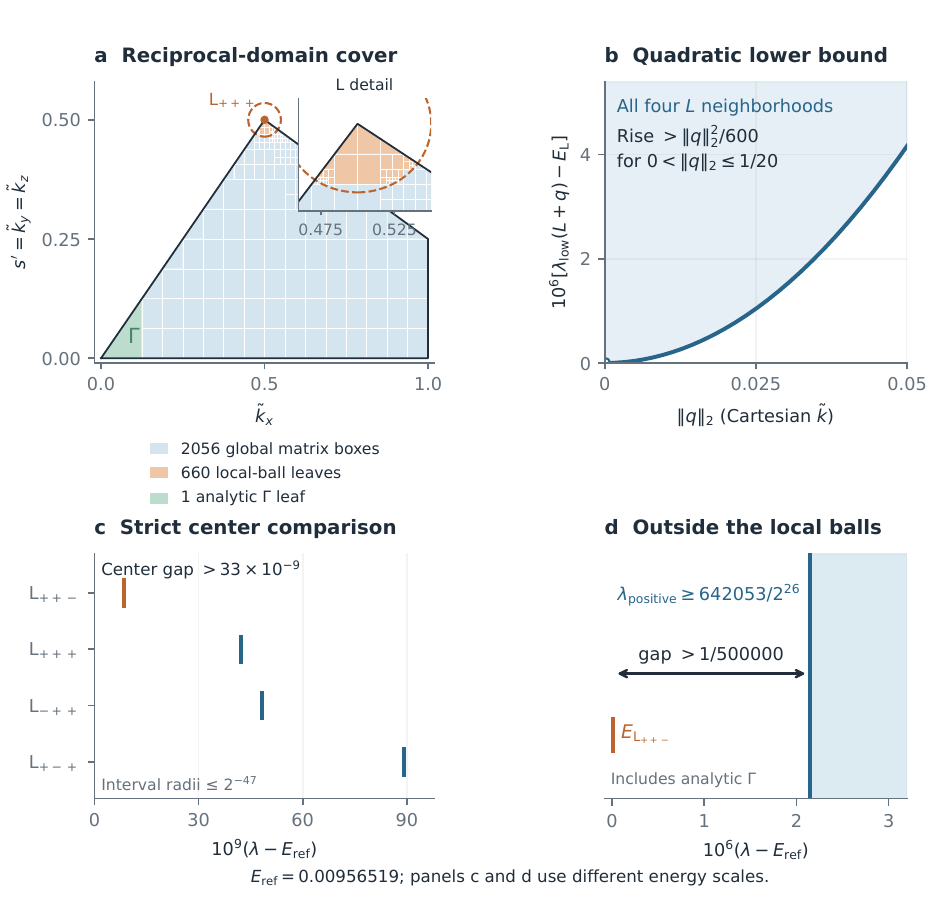}
  \caption{Exact certificates and analytic bounds select the global edge.
  (a) The canonical comparator slice
  $(\tilde k_x,\tilde k_y,\tilde k_z)=(s,s',s')$, $0\le s'\le s\le1$,
  $s+2s'\le3/2$, with exact box intersections and a local zoom.
  The legend gives full three-dimensional counts: $2056$ global matrix
  boxes, $660$ leaves wholly inside the radius-$1/20$ local ball, and one
  analytic $\mathrm{\Gamma}$ leaf. The geometric cover also classifies
  $1449$ infeasible leaves, which are not spectral regions and are omitted.
  The dashed ellipse is the ball boundary in this slice. Signed coordinate
  permutations extend the unperturbed comparator; the perturbed material
  has separate certificates for all four inequivalent $\mathrm{L}$ neighborhoods.
  (b) The lower-bound curve $\|q\|_2^2/600$ applies throughout each ball;
  it is not a sampled dispersion curve.
  (c) Exact center intervals select $\mathrm{L}_{++-}$, with every competing lower
  endpoint exceeding its upper endpoint by more than $33/10^9$.
  (d) Outside the four balls, including the analytic $\mathrm{\Gamma}$ region,
  the first positive band is at least $642053/2^{26}$, exceeding the
  selected center by more than $1/500000$. Panels (c) and (d) use the
  same reference energy and different explicitly labelled scales.
  Coordinates are Cartesian $\tilde k$; conventional radian coordinates are
  $k=2\pi\tilde k$, in which the four $\mathrm{L}$ classes fold to the class represented
  by $(\pi,\pi,-\pi)$.}
  \label{d:fig:FIG-3}
\end{figure}

Sections~\ref{d:app:edge}--\ref{d:app:certificate} give the defect,
interaction and spectral arguments. Section~\ref{d:sec:reproducibility}
specifies the finite certificate data.

\section{A scalar defect selected by a simple band edge}
\label{d:app:edge}

This section proves the defect part of
Proposition~\ref{d:prop:physical-platform}. We work with a real, periodic,
finite-range operator whose Bloch matrix has the form
\eqref{d:eq:physical-laurent-fiber}. Assume the edge data of
Section~\ref{d:sec:physical}: a unique simple edge \(E_c\) at a
time-reversal invariant momentum \(k_*\), a lower-band separation \(g_->0\),
and a positive quadratic form \(H\). The coefficients, spectral gaps,
and local analytic estimates are effectively bounded.

\subsection{The compressed resolvent expansion}

Write
\[
 d_\lambda(q)=\lambda(k_*+q)-E_c,\qquad
 d_\lambda(q)=q^THq+O(\abs{q}^4).
\]
After decreasing a fixed radius \(q_0\), choose positive rational constants
\(a_0,a_1,M_4\) such that
\[
 a_0I\le H\le a_1I,\qquad
 \abs{d_\lambda(q)-q^THq}\le M_4\abs{q}^4,\qquad
 \frac{a_0}{2}\abs{q}^2\le d_\lambda(q)\le2a_1\abs{q}^2
\]
for \(\abs{q}\le q_0\). All other spectral contributions have denominators
bounded away from zero near \(E_c\). The real normalized edge eigenvector
\(\phi_*\) extends to the non-square-summable Bloch field
\(\Phi_*(\nu)=e^{ik_*\cdot\nu}\phi_*\).

\begin{lemma}[Square-root edge expansion]
\label{d:lem:physical-square-root}
For the fixed finite defect block \(K_{\mathcal D}=Q_{\mathcal D}^TW_{\mathcal D}^{1/2}\), let
\(\Phi_{\mathcal D}=K_{\mathcal D}^T\Phi_*\) and
\[
 S_\xi=K_{\mathcal D}^TA_\xi^{-1}K_{\mathcal D}.
\]
Then \(S_0=\lim_{\xi\downarrow0}S_\xi\) exists in finite-matrix operator
norm, and
\begin{equation}
 S_\xi=S_0-c_H\sqrt{\xi}\,\Phi_{\mathcal D}\Phi_{\mathcal D}^T+R_\xi,\qquad
 c_H=\frac1{4\pi\sqrt{\det H}},\qquad
 \norm{R_\xi}\le C_{\mathcal D}\xi.
 \label{d:eq:appendix-edge-root}
\end{equation}
The constant and the permitted interval of \(\xi\) can be bounded from
the finite support and the certified edge data.
\end{lemma}

\begin{proof}
Use Bloch measure \((2\pi)^{-3}\,dq\). The selected-band contribution in
the ball \(\abs{q}<q_0\) is
\[
 \int_{\abs{q}<q_0}\frac{\mathcal B(q)}{d_\lambda(q)+\xi}
                    \frac{dq}{(2\pi)^3},
\]
where \(\mathcal B\) is the finite compressed spectral numerator,
\(\mathcal B(0)=\Phi_{\mathcal D}\Phi_{\mathcal D}^T\), and
\(\mathcal B(-q)=\mathcal B(q)^T\). For real \(q\) it is positive
semidefinite. Since \(d_\lambda\) is even, its integral is unchanged by replacing
\(\mathcal B(q)\) with its even part. Analyticity gives
\[
 \norm{\mathcal B_{\mathrm{even}}(q)-\Phi_{\mathcal D}\Phi_{\mathcal D}^T}
 \le C_{\mathcal B}\abs{q}^2.
\]
The singularity \(\abs{q}^{-2}\) is integrable in dimension three, so
the integral at \(\xi=0\) exists. The spectrally separated part also has
a finite limit and changes by \(O(\xi)\).

For the difference of the singular contributions at zero and \(\xi\),
the scalar multiplier is
\[
 \phi_\xi(s)=\frac{\xi}{s(s+\xi)}.
\]
Replacing the even numerator by \(\Phi_{\mathcal D}\Phi_{\mathcal D}^T\) introduces \(O(\xi)\), because
the resulting integrand is bounded by
\(C\xi/(c\abs{q}^2+\xi)\). Replacing \(d_\lambda(q)\) by \(q^THq\) also costs
\(O(\xi)\): the derivative estimate
\[
 \abs{\phi_\xi'(s)}
   =\frac{\xi(2s+\xi)}{s^2(s+\xi)^2}
   \le\frac{2\xi}{s^2(s+\xi)}
\]
combines with the quartic dispersion remainder to give the same integrable
majorant. Extending the remaining quadratic integral to \(\R^3\) adds
only \(O(\xi)\), since its exterior radial tail is bounded by
\(C\xi\int_{q_0}^{\infty}s^{-2}\,ds\).

The change of variables \(q=\sqrt{\xi}\,H^{-1/2}z\) gives
\[
 \int_{\R^3}\frac{\xi\,dq}
 {(q^THq)(q^THq+\xi)(2\pi)^3}
 =
 \frac{\sqrt{\xi}}{\sqrt{\det H}}
 \int_{\R^3}\frac{dz}{\abs{z}^2(\abs{z}^2+1)(2\pi)^3}
 =
 \frac{\sqrt{\xi}}{4\pi\sqrt{\det H}}.
\]
The last equality follows from
\(4\pi\int_0^\infty(1+s^2)^{-1}ds=2\pi^2\).
These estimates establish the expansion and its effective remainder.
\end{proof}

\subsection{A large scalar cube overlaps a large eigenvalue}

Choose an all-cell cube containing every cell incident to the electric
edges of \(L_{\rm cube}^3\) interior period cells, with a fixed indexing margin.
For \(L_{\rm cube}\ge8\), write \(W_\nu\) for the diagonal incident fractions in
period cell \(\nu\). Its support occupies at most \(8L_{\rm cube}^3\) period cells,
and its translation indices can be chosen to satisfy \(\abs{\nu}\le4L_{\rm cube}\).
On the interior cells \(W_\nu=I\). Hence
\begin{equation}
 a_{\rm cube}:=\norm{\Phi_{\mathcal D}}^2=\sum_\nu\phi_*^TW_\nu\phi_*,
 \qquad L_{\rm cube}^3\le a_{\rm cube}\le8L_{\rm cube}^3.
 \label{d:eq:appendix-cube-mass}
\end{equation}
The source \(f_{\mathcal D}=K_{\mathcal D}\Phi_{\mathcal D}\) has period-cell components
\(f_{\mathcal D}(\nu)=W_\nu e^{ik_*\cdot\nu}\phi_*\). Let \(\phi(q)\) be a local normalized
selected-band eigenvector with
\(\norm{\phi(q)-\phi_*}\le C_\phi\abs{q}\). Its source overlap is
\[
 \mathcal N_{\rm cube}(q)
  =\phi(q)^*\sum_\nu e^{-iq\cdot\nu}W_\nu\phi_*,
 \qquad \mathcal N_{\rm cube}(0)=a_{\rm cube}.
\]
The phase and eigenvector changes satisfy
\[
 \abs{\mathcal N_{\rm cube}(q)-a_{\rm cube}}
 \le4L_{\rm cube}\abs{q}a_{\rm cube}+8C_\phi L_{\rm cube}^3\abs{q}.
\]
Choose a fixed rational
\[
 0<q_*\le
 \min\left(q_0,\frac1{64},\frac1{64(1+C_\phi)}\right).
\]
Then \(\abs{\mathcal N_{\rm cube}(q)}\ge a_{\rm cube}/2\) for
\(\abs{q}\le q_*/L_{\rm cube}\). If \(d_{\max}\) satisfies
\(d_\lambda(q)\le d_{\max}\abs{q}^2\), the selected band contributes at least
\[
 \frac{a_{\rm cube}^2}{4}\int_{\abs{q}\le q_*/L_{\rm cube}}
    \frac{dq}{d_{\max}\abs{q}^2(2\pi)^3}
 =\frac{a_{\rm cube}^2q_*}{8\pi^2d_{\max}L_{\rm cube}}.
\]
Other upper-band contributions are nonnegative. The lower bands contribute
at least \(-\norm{f_{\mathcal D}}^2/g_-\), and \(\norm{K_{\mathcal D}}\le1\) implies
\(\norm{f_{\mathcal D}}\le\norm{\Phi_{\mathcal D}}\). Therefore
\begin{equation}
 \frac{\Phi_{\mathcal D}^TS_0\Phi_{\mathcal D}}{\norm{\Phi_{\mathcal D}}^2}
 \ge c_{\mathrm{cube}}L_{\rm cube}^2-\frac1{g_-},
 \qquad c_{\mathrm{cube}}=\frac{q_*}{80d_{\max}}>0,
 \label{d:eq:appendix-cube-rayleigh}
\end{equation}
using \(\pi^2<10\). For any fixed threshold \(\lambda_0\), choose \(L_{\rm cube}\) so that
the right side exceeds \(\lambda_0+2\). The spectral decomposition of \(S_0\)
then contains an eigenvalue \(\lambda>\lambda_0+2\) with
\(P_\lambda\Phi_{\mathcal D}\ne0\). This conclusion does not require \(\lambda\) to
be simple.

\subsection{Splitting a possibly degenerate compressed eigenvalue}

\begin{lemma}[Rank-one branch selection]
\label{d:lem:physical-branch}
Suppose real symmetric matrices obey
\[
 S_t=S_0-ct\,\Phi_{\mathcal D}\Phi_{\mathcal D}^T+R_t,\qquad
 \norm{R_t}\le Ct^2,\qquad c>0.
\]
If \(S_0\) has an eigenvalue \(\lambda\) with
\(a=P_\lambda\Phi_{\mathcal D}\ne0\), then for sufficiently small \(t>0\) a simple
eigenpair satisfies
\[
 \mu_t=\lambda-c\norm{a}^2t+O(t^2),\qquad
 u_t=\frac{a}{\norm{a}}+O(t),\qquad
 \Phi_{\mathcal D}^Tu_t=\norm{a}+O(t)>0.
\]
Its separation from the other eigenvalues is at least
\(c\norm{a}^2t/2\).
\end{lemma}

\begin{proof}
Put \(P_\lambda^\perp=I-P_\lambda\) and \(W_1=-c\,\Phi_{\mathcal D}\Phi_{\mathcal D}^T\).
If \(P_\lambda^\perp\ne0\), let
\(C_\perp=P_\lambda^\perp(S_0-\lambda I)P_\lambda^\perp\) on its range and define the
skew-symmetric off-diagonal matrix \(\mathcal X\) by
\[
 \mathcal X_{\perp\lambda}=-C_\perp^{-1}P_\lambda^\perp W_1P_\lambda,\qquad
 \mathcal X_{\lambda\perp}=P_\lambda W_1P_\lambda^\perp C_\perp^{-1},\qquad
 \mathcal X_{\lambda\lambda}=\mathcal X_{\perp\perp}=0.
\]
Use \(\mathcal X=0\) when \(P_\lambda^\perp=0\).
For \(0<t\le1\) and \(t\norm{\mathcal X}\le1\), set
\[
 \mathcal Z_t=\tfrac{t}{2}\mathcal X,\qquad
 \mathcal O_t=(I+\mathcal Z_t)(I-\mathcal Z_t)^{-1}.
\]
The Neumann series gives \(\norm{(I-\mathcal Z_t)^{-1}}\le2\).
Since \(\mathcal X^T=-\mathcal X\), the Cayley transform \(\mathcal O_t\) is orthogonal.
The exact identity
\[
 \mathcal O_t-I-2\mathcal Z_t
 =2\mathcal Z_t^2(I-\mathcal Z_t)^{-1}
\]
gives \(\norm{\mathcal O_t-I}\le2t\norm{\mathcal X}\) and
\(\norm{\mathcal O_t-I-t\mathcal X}\le t^2\norm{\mathcal X}^2\).
Conjugation yields
\[
 \mathcal O_t^TS_t\mathcal O_t
  =S_0+t(P_\lambda W_1P_\lambda+P_\lambda^\perp W_1P_\lambda^\perp)+F_t,
 \qquad \norm{F_t}\le C't^2,
\]
because the commutator \([S_0,\mathcal X]\) cancels the first-order
off-diagonal blocks. An explicit admissible constant is
\[
 C'=6(\norm{S_0}+\norm{W_1}+C)\norm{\mathcal X}^2
       +C+2(\norm{W_1}+C)\norm{\mathcal X} .
\]
Indeed, expanding \(\mathcal O_t^TS_t\mathcal O_t\) leaves a conjugation
remainder bounded by \(6\norm{S_t}t^2\norm{\mathcal X}^2\).
Replacing \(S_t\) by \(S_0\) in its first-order commutator costs
at most \(2(\norm{W_1}+C)t^2\norm{\mathcal X}\), and the original remainder
costs \(Ct^2\). These bounds use only the stated norm remainder,
without a differentiability assumption on \(t\mapsto R_t\).
On the \(P_\lambda\) block the leading matrix is
\(\lambda I-ct\,aa^T\). It has a simple eigenvalue
\(\lambda-c\norm{a}^2t\), separated by
\(c\norm{a}^2t\) from the remaining \(P_\lambda\) eigenvalues.
The fixed spectral gap to \(P_\lambda^\perp\) persists for sufficiently small
\(t\).

Choose \(t\) such that the quadratic perturbation is at most one
quarter of this linear gap. Weyl's inequalities then isolate exactly one
eigenvalue, with gap at least \(c\norm{a}^2t/2\). Projecting its
eigenvector equation onto the complement of \(a/\norm{a}\) bounds the
complementary component by \(O(t)\). Finally
\(\norm{\mathcal O_t-I}\le2t\norm{\mathcal X}\), and the sign can be chosen so that
\(\Phi_{\mathcal D}^Tu_t>0\). This proves all assertions.
\end{proof}

Apply the lemma with \(t=\sqrt{\xi}\) and \(c=c_H\).
The cube construction gives a branch \((\mu_\xi,u_\xi)\) with
\(\mu_\xi>\lambda_0+1\) and nonzero edge overlap. Choose
\[
 \lambda_0=\frac4{\omega^2 h_{\mathrm{mat}}}.
\]
Then \(v_*=\mu_\xi^{-1}<\omega^2h_{\mathrm{mat}}/4\).
Since the gap of \(A_\xi\) is at least \(\xi\),
\(\mu_\xi\le\norm{S_\xi}\le1/\xi\), hence \(v_*\ge\xi\).
For a rational \(0<\xi_{\mathrm{low}}\le\xi\), the interval
\(\abs{v-v_*}\le\xi_{\mathrm{low}}/4\) lies in
\[
 \frac34v_*\le v\le\frac54v_*<\frac{5\omega^2 h_{\mathrm{mat}}}{16}.
\]
The uniform background shift and this defect increment therefore leave
at least \(h_{\mathrm{mat}}/2\) of upper material headroom.

The kernel correspondence can be checked directly. If
\(S_\xi u=\mu_\xi u\), then
\[
 E_{\mathrm{def}}=A_\xi^{-1}K_{\mathcal D}u,\qquad
 (A_\xi-v_*K_{\mathcal D}K_{\mathcal D}^T)E_{\mathrm{def}}=0.
\]
Conversely, any field in that kernel gives a compressed eigenvector by
applying \(K_{\mathcal D}^T\); the map is injective because \(A_\xi\) is
invertible. The kernel is consequently one-dimensional. Exponential
localization follows by conjugating \(A_\xi\) with bounded
weights \(e^{a\min(\operatorname{dist}(\cdot,\operatorname{supp}K_{\mathcal D}),R)}\).
For sufficiently small fixed \(a>0\), finite range makes the conjugation
error smaller than \(\xi/2\). The inverse is uniformly bounded in \(R\),
and passage to the limit gives exponential decay.

\subsection{Nonzero analytic numerator and effective selection}

For \(f_u=K_{\mathcal D}u\), define
\[
 \mathcal N_u(k)=(\mathcal Ff_u)(-k)^T\Pi(k)(\mathcal Ff_u)(k).
\]
This is the analytic continuation of the real-band overlap and satisfies
\(\mathcal N_u(k_*)=(\Phi_{\mathcal D}^Tu)^2\). Finite source support and the analytic
Riesz projection give
\[
 \abs{\mathcal N_u(k_*+q)-\mathcal N_u(k_*)}
 \le C_{\mathcal N}\abs{q}
\]
uniformly for real unit \(u\). Since the selected branch has
\(\Phi_{\mathcal D}^Tu_\xi\to\norm{P_\lambda\Phi_{\mathcal D}}>0\), it suffices to choose \(\xi\)
so that
\[
 \abs{\Phi_{\mathcal D}^Tu_\xi}^2>2C_{\mathcal N}R\sqrt{\xi}
\]
to keep the numerator nonzero at every complex point
\(\abs{q}\le R\sqrt{\xi}\). At \(q=i\beta\) the Laurent matrix,
the isolated Riesz projection, and the paired source polynomials are real.
The numerator is therefore positive real for every real \(\beta\) with
\(\abs{\beta}\le R\sqrt{\xi}\), and nonzero throughout the stated
complex ball.

To retain dyadic material data, enumerate dyadic \(E_\xi<E_c\) and set
\(\xi=E_c-E_\xi\). The uniform increment is \(E_\xi/\omega^2\), which is dyadic at
\(\omega^2=1/64\). Certified inverse polynomials approximate the compressed
resolvent at a fixed positive distance from the spectrum. Rational
inertia or multiplicity-counting root isolation, combined with residual
and gap bounds, certifies a simple eigenvalue above \(\lambda_0+1\), its
eigenvector, and its overlap. More explicitly, a rational inverse
polynomial $p_{\ell_{\rm P}}$ gives the rational symmetric compression
$S_{\ell_{\rm P}}=Q_{\mathcal D}p_{\ell_{\rm P}}(A_{\rm per}-E_\xi I)Q_{\mathcal D}^T$. The symmetric pencil
$S_{\ell_{\rm P}}-\lambda W_{\mathcal D}^{-1}$ is congruent to
$W_{\mathcal D}^{1/2}S_{\ell_{\rm P}}W_{\mathcal D}^{1/2}-\lambda I$, so rational inertia counts its
eigenvalues without treating the square-root weights as rational entries.
Choose endpoint margins larger than twice the inverse-approximation
error; Weyl bounds then transfer the isolated eigenvalue count to
$S_\xi$. Residual and gap bounds enclose the corresponding mode.

The search acts on \(S_\xi\), never on the exact multiplicity of the
limiting computable matrix \(S_0\). The strict inequalities proved above
hold for all sufficiently small \(\xi\), so dovetailing shifts and
certificate precision terminates. This is a fixed effective construction,
and the search has not been executed.

\section{Quantitative interactions from the edge resolvent}
\label{d:app:green}

We prove the full-block upper bound and the noncancelling scalar
asymptotic used in Proposition~\ref{d:prop:physical-platform}. Recall
\(E_\xi=E_c-\xi\) and \(A_\xi=A_{\rm per}-E_\xi I\).
After translating \(k_*\) to zero in local expressions, the Laurent
coefficients remain real. Simplicity and time reversal give
\[
 \lambda(-q)=\lambda(q),\qquad
 \overline{\lambda(\bar q)}=\lambda(q),\qquad
 A_{\rm per}(q)^T=A_{\rm per}(-q).
\]
Thus \(\lambda(i\beta)\) is real, its gradient there is purely imaginary,
and its Hessian there is real symmetric.

\subsection{A nonsingular tube and the full-block estimate}

Fix \(0<\eta_{\rm t}\le1/100\). For sufficiently small \(\xi\), the matrix
\(A_{\rm per}(k+i\beta)-E_\xi I\) is invertible for every real \(k\) whenever
\begin{equation}
 \norm{(2H)^{1/2}\beta}
 \le(1-\eta_{\rm t})\sqrt{2\xi}.
 \label{d:eq:appendix-green-tube}
\end{equation}
Near the selected edge, its potentially small eigenvalue satisfies
\[
 \operatorname{Re}\bigl(\lambda(q+i\beta)-E_\xi\bigr)
 \ge\xi-\beta^TH\beta-C\abs{\beta}^4+c\abs{q}^2
 \ge c_{\eta_{\rm t}}\xi+c\abs{q}^2.
\]
The even analytic expansion absorbs the mixed remainders into the
positive quadratic term after the neighborhood is decreased.
The complementary spectral block is uniformly invertible there.
Outside that real neighborhood, the global spectral separation and a
Neumann estimate control the complex perturbation. Together these give a
uniform inverse bound \(C_{\eta_{\rm t}}/\xi\) on the tube.

For the finite source block \(K_{\mathcal D}\), its Bloch transform is a Laurent matrix.
Move the entire Fourier torus along \(s\beta\), \(0\le s\le1\).
Differentiating the integral along this path produces an integral of a
periodic total derivative, so the integral is unchanged after accounting
for the Fourier exponential. This path stays in the proved tube;
sequential coordinate shifts are unnecessary. Maximizing
\(\beta\cdot\nu\) over \eqref{d:eq:appendix-green-tube} gives
\begin{equation}
 \norm{K_{\mathcal D}^TA_\xi^{-1}\mathcal T_\nu K_{\mathcal D}}
 \le C_{\xi,\eta_{\rm t}}
 e^{-(1-\eta_{\rm t})\sqrt{2\xi}\norm{(2H)^{-1/2}\nu}}.
 \label{d:eq:appendix-green-upper}
\end{equation}
All constants are fixed after choosing \(\xi\) and \(K_{\mathcal D}\).

\subsection{The support point and the limiting contour}

Set
\[
 \Lambda(\beta)=E_c-\lambda(k_*+i\beta)
 =\beta^TH\beta+O(\abs{\beta}^4).
\]
Its Hessian is positive definite near zero. For every \(\upsilon\ne0\), the
small convex component \(\{\Lambda(\beta)<\xi\}\) has a unique support
point \(\beta_\upsilon\) satisfying
\[
 \Lambda(\beta_\upsilon)=\xi,\qquad
 \nabla\Lambda(\beta_\upsilon)=\varkappa_\upsilon\upsilon,\qquad \varkappa_\upsilon>0.
\]
Comparison with enclosing quadratic ellipsoids gives, uniformly in
direction,
\begin{equation}
 I_\upsilon:=\beta_\upsilon\cdot\upsilon
   =\sqrt{2\xi}\norm{(2H)^{-1/2}\upsilon}\,[1+O(\xi)].
 \label{d:eq:appendix-support-rate}
\end{equation}
Scaling \(\beta=\sqrt{\xi}\,\beta'\) and applying the implicit function
theorem at the ellipsoidal support point provides effective error bounds.

For every \(s<1\), the entire torus shifted by \(is\beta_\upsilon\) remains
nonsingular. Locally,
\[
 \operatorname{Re}\bigl(\lambda(q+is\beta_\upsilon)-E_\xi\bigr)
 \ge\xi-\Lambda(s\beta_\upsilon)+c\abs{q}^2>0.
\]
At \(s=1\), only \(q=0\) is singular; the inverse is bounded by
\(C(1+\abs{q}^{-2})\), uniformly as \(s\uparrow1\).
This is integrable in three dimensions. Dominated convergence therefore
gives, for the real finite source \(f_\xi=K_{\mathcal D}u_\xi\),
\begin{align}
 h(d\upsilon)
 &=\langle f_\xi,A_\xi^{-1}\mathcal T_{d\upsilon}f_\xi\rangle \nonumber\\
 &=e^{ik_*\cdot d\upsilon-I_\upsilon d}
   \frac1{(2\pi)^3}\nonumber\\
 &\quad{}\times\int_{\mathbb T^3}e^{id\upsilon\cdot q}
   (\mathcal Ff_\xi)(-k_*-q-i\beta_\upsilon)^T
   [A_{\rm per}(k_*+q+i\beta_\upsilon)-E_\xi I]^{-1}
   (\mathcal Ff_\xi)(k_*+q+i\beta_\upsilon)\,dq .
 \label{d:eq:appendix-boundary-contour}
\end{align}
All subsequent terms inherit the prefactor in this identity.

\subsection{Pole extraction and the transverse Gaussian integral}

Let \(e_\upsilon=\upsilon/\abs{\upsilon}\), and let the columns of \(S_\upsilon\) be an orthonormal
basis of \(\upsilon^\perp\). Put
\[
 Q_{\upsilon,\perp}=S_\upsilon^T\lambda''(k_*+i\beta_\upsilon)S_\upsilon>0,\qquad
 q=q_\parallel e_\upsilon+S_\upsilon q_\perp,\quad q_\perp\in\R^2.
\]
The local denominator
\[
 D(q_\parallel,q_\perp)=\lambda(k_*+i\beta_\upsilon+q_\parallel e_\upsilon+S_\upsilon q_\perp)-E_\xi
\]
has \(D(0,0)=0\), \(D_\perp(0,0)=0\), and
\(D_\parallel(0,0)=ic_\upsilon\), where \(c_\upsilon=\varkappa_\upsilon\abs{\upsilon}>0\).
Its unique nearby pole is
\begin{equation}
 q_\parallel=p(q_\perp)=\frac{i}{2c_\upsilon}q_\perp^TQ_{\upsilon,\perp}q_\perp+O(\abs{q_\perp}^3),
 \qquad \operatorname{Im}p(q_\perp)\ge c'\abs{q_\perp}^2.
 \label{d:eq:appendix-local-pole}
\end{equation}
Factor \(D(q_\parallel,q_\perp)=(q_\parallel-p(q_\perp))d_1(q_\parallel,q_\perp)\), with analytic nonzero \(d_1\).
For the analytic numerator \(\mathcal N_\xi\),
Cauchy division gives
\[
 \frac{\mathcal N_\xi(k_*+i\beta_\upsilon+q_\parallel e_\upsilon+S_\upsilon q_\perp)}{D(q_\parallel,q_\perp)}
 =\frac{\mathcal A(q_\perp)}{q_\parallel-p(q_\perp)}+\mathcal W(q_\parallel,q_\perp),\qquad
 \mathcal A(0)=\frac{\mathcal N_\xi(k_*+i\beta_\upsilon)}{ic_\upsilon}.
\]
The functions \(\mathcal A\) and \(\mathcal W\) are analytic.

Choose smooth cutoffs \(\chi_1(q_\parallel)\chi_2(q_\perp)\), equal to one near zero,
with the poles over \(\operatorname{supp}\chi_2\) lying strictly inside
the core where \(\chi_1=1\). Every remaining term in
\eqref{d:eq:appendix-boundary-contour} is smooth on the torus, and repeated
integration by parts along \(\upsilon\) bounds its Fourier coefficient by
\(O(d^{-K})\) for each fixed order \(K\). This estimate multiplies
\(e^{-I_\upsilon d}\) because the contour was shifted before localization.

For \(T=d\abs{\upsilon}>0\) and \(q_\perp\ne0\), the pole lies in the upper
half-plane, so
\[
 \int_{\R}\frac{e^{iTu}}{q_\parallel-p(q_\perp)}\,dq_\parallel
   =2\pi i e^{iTp(q_\perp)}.
\]
The full-line expression is an improper oscillatory integral.
Its tails beyond \(\abs{q_\parallel}=R\) are bounded by \(C/(TR)\), uniformly
over these poles, by integration by parts. One may therefore start with
finite truncations and pass to the limit under the compact \(q_\perp\) integral.
Replacing the full line by the cutoff \(\chi_1\) has error \(O(T^{-K})\):
the removed region stays a fixed distance from every pole, and every
positive-order derivative of \((1-\chi_1(q_\parallel))/(q_\parallel-p(q_\perp))\) is uniformly
integrable.

The singular contribution is consequently
\[
 2\pi i\int_{\R^2}\chi_2(q_\perp)\mathcal A(q_\perp)e^{iTp(q_\perp)}\,dq_\perp+O(T^{-K}).
\]
Using \eqref{d:eq:appendix-local-pole}, set \(q_\perp=T^{-1/2}y\).
The two-dimensional Gaussian integral yields
\[
 \int\chi_2(q_\perp)\mathcal A(q_\perp)e^{iTp(q_\perp)}\,dq_\perp
 =\mathcal A(0)\frac{2\pi c_\upsilon}{T\sqrt{\det Q_{\upsilon,\perp}}}
   +O(T^{-3/2}).
\]
To bound the remainder, the amplitude change is \(O(\abs{q_\perp})\),
and the exponent change is \(O(T\abs{w}^3)\). Both are integrated
against \(e^{-cT\abs{w}^2}\) after the chart is decreased, giving
\(O(T^{-3/2})\); the omitted Gaussian tails are exponentially small.
Combining the normalization factors proves
\begin{equation}
 h(d\upsilon)=e^{ik_*\cdot d\upsilon-I_\upsilon d}
 \left[
 \frac{\mathcal N_\xi(k_*+i\beta_\upsilon)}
 {2\pi d\abs{\upsilon}\sqrt{\det Q_{\upsilon,\perp}}}
 +O(d^{-3/2})
 \right].
 \label{d:eq:appendix-green-asymptotic}
\end{equation}
This holds for every sufficiently large integer \(d\), with effective
remainder for each fixed nonzero integer \(\upsilon\).

Section~\ref{d:app:edge} makes the numerator \(\mathcal N_\xi\) positive
uniformly at all support points, since
\(\abs{\beta_\upsilon}=O(\sqrt{\xi})\) uniformly in direction.
The factor \(e^{ik_*\cdot d\upsilon}\) belongs to \(\{1,-1\}\).
Thus the leading term supplies an absolute lower bound after increasing
the distance threshold. This treats the possible numerator cancellation
explicitly.

\subsection{Integer placement and a strict rate margin}

Fix \(\eta_{\rm t}=1/100\). Choose a large integer \(M_{\mathrm{lat}}\) and
an integer matrix \(B\) approximating \(M_{\mathrm{lat}}(2H)^{1/2}\)
entrywise within one. This can be achieved with certified approximations
without deciding an exact rounding tie. The error norm is at most three.
Choosing \(3\norm{(2H)^{-1/2}}/M_{\mathrm{lat}}<\eta_{\rm t}\) gives
\[
 M_{\mathrm{lat}}(1-\eta_{\rm t})\norm{y}
 \le\norm{(2H)^{-1/2}By}
 \le M_{\mathrm{lat}}(1+\eta_{\rm t})\norm{y}.
\]
In particular \(B\) is invertible. The full-block estimate and
\(\norm{\nu}_2\ge\norm{\nu}_1/\sqrt3\) yield an upper decay exponent
\[
 I_{\rm up}=\frac{(1-\eta_{\rm t})^2}{\sqrt3}\sqrt{2\xi}\,M_{\mathrm{lat}}
\]
per unit of \(\ell^1\) distance. After choosing \(\xi\) sufficiently small,
\eqref{d:eq:appendix-support-rate} bounds each axial exponent by
\[
 I_{Be_j}\le I_{\rm ax}=(1+\eta_{\rm t})^2\sqrt{2\xi}\,M_{\mathrm{lat}}.
\]
The strict inequality
\(2(1-\eta_{\rm t})^2/\sqrt3>(1+\eta_{\rm t})^2\) implies \(2I_{\rm up}>I_{\rm ax}\).
Choose rational \(r_{\rm d},a_{\rm d}\in(0,1)\) with
\[
 e^{-I_{\rm up}}<r_{\rm d}<\sqrt{a_{\rm d}},\qquad a_{\rm d}<e^{-I_{\rm ax}}.
\]
Then
\[
 \norm{G(B\nu)}\le C_{\rm d}r_{\rm d}^{\norm{\nu}_1},\qquad
 \abs{u_\xi^TG(dBe_j)u_\xi}\ge c_{\rm d}\,d^{-1}e^{-I_{Be_j}d}.
\]
The strict exponential margin \(a_{\rm d}<e^{-I_{\rm ax}}\) absorbs the factor \(d^{-1}\)
after increasing a fixed threshold, giving
\eqref{d:eq:physical-platform-bounds}. Equality of opposite-direction
couplings follows from the real symmetric resolvent.

The placement matrix is fixed and may be non-unimodular. Translating
defects on its sublattice preserves periodic compatibility and does not
change the physical grid metric. A balanced logical box with side
\(O(\sqrt S)\), followed by multiplication by \(B\), separation \(d\),
and a buffer, has an axis-aligned containing volume
\(O_B(S^{3/2}d^3)\), up to the buffer factors accounted for in the main
construction. The balance of the logical layout is used here explicitly.

\section{Certification of the periodic spectral premise}
\label{d:app:certificate}

This section connects the finite evidence to the hypotheses of
Sections~\ref{d:app:edge} and \ref{d:app:green}. The evidence uses exact
integer and rational calculations together with outward interval matrix
verification. A witness is a proposal whose properties the checker must
establish.

\subsection{Exact material and finite point certificates}

The primitive translation lattice has basis
\[
 (20,0,0),\qquad(10,10,0),\qquad(10,0,10).
\]
Cell representatives satisfy \(0\le r_1<20\), \(0\le r_2,r_3<10\).
The primitive electric fiber has dimension 6000. The exact material is
\eqref{d:eq:physical-exact-material}, with its full integer table supplied
in the reproducibility artifact. The checker verifies
\[
 p_{-c-\boldsymbol1}=p_c
\]
and all primitive translations, then applies the four-cell edge average.
The corresponding perturbation of the electric operator obeys
\begin{equation}
 0\le A_{\rm dia}-A_{\rm per}\le\delta_{\rm mat} I,
 \qquad \delta_{\rm mat}=\frac{1023}{2^{26}}<\frac1{4096}.
 \label{d:eq:appendix-material-perturbation}
\end{equation}
Here \(A_{\rm dia}\) is the unperturbed diamond operator and
\(A_{\rm per}\) is the actual perturbed operator.

At each \(\mathrm{L}\) point, primitive Bloch phases are signs. The curl incidence
and mass matrix can therefore be reconstructed with integer arithmetic.
For an electric edge in direction \(j\) anchored at \(r\), polar-vector
inversion maps the anchor to \(-r-e_j\), with its polar sign and the
primitive translation sign. The resulting signed permutation \(I_L\)
satisfies \(I_L^2=I\) and commutes with the center matrix.

Two dyadic eigenvector witnesses are projected exactly to one inversion
sector at each center. Integer residual tests give disjoint eigenvalue
intervals of radius at most \(2^{-47}\), whose total span is below
\(2^{-20}\). Independent inertia certificates isolate exactly two
positive eigenvalues in
\[
 [239/25000,\;957/100000]=[0.00956,\;0.00957],
\]
with complementary spectral separation exceeding \(11/2000\).
The residual intervals therefore contain the two simple eigenvalues
and exhaust the cluster. Both modes have the same parity at each center:
the parity is negative at \(\mathrm{L}_{+++}\) and positive at the other three.
Exact center ordering places \(\mathrm{L}_{++-}\) below every other center by more
than \(33/10^9\).

\subsection{A finite matrix test that controls every local direction}

Let \(A_{\rm cen}(q)\) denote the actual electric matrix in the centered
Yee gauge near one \(\mathrm{L}\) point, using Cartesian reciprocal coordinates
with phase \(e^{2\pi iq\cdot r/20}\). Inversion obeys
\[
 I_LA_{\rm cen}(q)I_L=A_{\rm cen}(-q)
\]
for every \(q\) in this gauge. Let \(E_{\rm L}\) be the lower center eigenvalue,
and let \(\Phi_{\rm L}\) be an exact orthonormal basis of the two-mode center
cluster. Then
\[
 (A_{\rm cen}(0)-E_{\rm L})\Phi_{\rm L}=\Phi_{\rm L}\Delta_{\rm L},\qquad
 0\le\Delta_{\rm L}\le2^{-20}I,
\]
where the eigenvalues of \(\Delta_{\rm L}\) are zero and a strictly positive
splitting. The two columns of \(\Phi_{\rm L}\) have the same inversion parity.

Write \(A^{(j)}(q)\) for the homogeneous degree-\(j\) Taylor term of
\(A_{\rm cen}(q)\). The linear term reverses parity, so the equation
\[
 (A_{\rm cen}(0)-E_{\rm L})Y(q)=A^{(1)}(q)\Phi_{\rm L}
\]
is solved entirely in the opposite parity sector. Its inverse does not
divide by the tiny internal splitting of the two-mode cluster. Define
\[
 K_2(q)=\Phi_{\rm L}^*A^{(2)}(q)\Phi_{\rm L}-\Phi_{\rm L}^*A^{(1)}(q)Y(q),\qquad
 R_2(q)=A^{(2)}(q)\Phi_{\rm L}-A^{(1)}(q)Y(q).
\]
The finite certificates establish
\begin{equation}
 \norm{Y(q)}\le3\abs{q},\qquad
 \frac{\abs{q}^2}{50}I\le K_2(q)\le\abs{q}^2I,\qquad
 \norm{R_2(q)y}^2
 \le\frac{18}{5}\abs{q}^2y^*K_2(q)y.
 \label{d:eq:appendix-local-matrix-bounds}
\end{equation}

These statements cover every real direction \(q\). The checker rebuilds
the derivative matrices from signed integer incidence operations,
encloses the exact cluster and opposite-parity solves by residual and
gap bounds, and checks rational matrix inequalities. In detail, the
two-by-two Feshbach Hessian blocks form a six-by-six Hermitian coefficient
matrix \(H_{\mathrm{coef}}\). An exact LDL test proves
\[
 \frac1{25}I_6<H_{\mathrm{coef}}<I_6.
\]
For
\[
 T_{ij}=\frac12
   (\partial_{ij}A_{\rm cen}\,\Phi_{\rm L}-\partial_iA_{\rm cen}\,\partial_jY-\partial_jA_{\rm cen}\,\partial_iY),
 \qquad \Sigma_{jk}=\sum_iT_{ij}^*T_{ik},
\]
a second rational comparison proves
\(\Sigma\le(18/5)H_{\mathrm{coef}}/2\).
Cauchy--Schwarz in the first spatial index gives the final bound in
\eqref{d:eq:appendix-local-matrix-bounds}. It controls the full quadratic
residual, including its component in the center cluster.

The dyadic witness basis is connected to the exact \(\Phi_{\rm L}\) before these
inequalities are used. Its residual is at most \(10^{-13}\), its Gram
error is at most \(10^{-12}\), and the complementary gap gives an exact
basis within \(10^{-10}\). The opposite-parity residuals and energy
enclosures supply the remaining perturbation bounds. Rational enclosures
for \(\pi\) come from Machin's identity and alternating arctangent
series. Thus no sampled directional Hessian is used as a proof premise.

\subsection{A polynomial test on the whole neighborhood}

Put \(b_{\mathrm e}=7/256\) and
\[
 \Psi(q)=\Phi_{\rm L}-Y(q),\qquad
 g_{\rm L}(\lambda)=(\lambda-E_{\rm L})(b_{\mathrm e}-\lambda).
\]
A separate uniform inertia certificate permits exactly two positive
eigenvalues below \(b_{\mathrm e}\) throughout \(\abs{q}\le1/20\).
We show that both are above the center value \(E_{\rm L}\) for \(q\ne0\).

The centered curl symbol gives explicit Taylor bounds. Since
\(\pi/20<11/70\),
\[
 \norm{A^{(j)}(q)}\le a_j\abs{q}^j,\qquad
 a_j=\frac{(4\,2^j+6)(\pi/20)^j}{j!}.
\]
The constant and degree-one residual terms are treated exactly:
if \(\mathcal R=(A_{\rm cen}(q)-E_{\rm L})\Psi\), then \(\mathcal R_1=0\).
Inversion makes the even and odd residual parts orthogonal, and gives
\begin{align*}
 \Psi^*(A_{\rm cen}-E_{\rm L})\Psi&=\Delta_{\rm L}+K_2+K_{\mathrm{rem}},\\
 \mathcal R_{\mathrm{even}}&=\Phi_{\rm L}\Delta_{\rm L}+R_2+\mathcal R_{\mathrm{even,rem}},
\end{align*}
with
\[
 \norm{K_{\mathrm{rem}}}\le\frac{13}{5}\abs{q}^4,\qquad
 \norm{\mathcal R_{\mathrm{even,rem}}}\le\frac2{25}\abs{q}^4,\qquad
 \norm{\mathcal R_{\mathrm{odd}}}\le\frac{17}{20}\abs{q}^3.
\]
For example, the leading coefficients are bounded by
\(a_4+6a_3+9a_2\), \(a_4+3a_3\), and \(a_3+3a_2\), respectively.
The ratio of each next even-degree term at \(\abs{q}=1/20\) to its predecessor
is below \(1/40000\), so an exact geometric majorant proves the displayed
tail bounds.

The identity
\[
 \Psi^*g_{\rm L}(A_{\rm cen})\Psi
 =(b_{\mathrm e}-E_{\rm L})\Psi^*(A_{\rm cen}-E_{\rm L})\Psi-\mathcal R^*\mathcal R
\]
has constant term
\(\Delta_{\rm L}[(b_{\mathrm e}-E_{\rm L})I-\Delta_{\rm L}]\ge0\).
Preserving this exact term is essential at arbitrarily small \(q\).
Set
\[
 k_{\min}=\tfrac1{50},\quad k_{\max}=1,\quad
 c_2=\tfrac{18}{5},\quad k_4=\tfrac{13}{5},\quad
 c_3=\tfrac{17}{20},\quad c_4=\tfrac2{25},\quad
 \delta_{\mathrm e}=\norm{\Delta_{\rm L}}.
\]
The preceding estimates imply, for every \(y\),
\[
 y^*\Psi^*g_{\rm L}(A_{\rm cen})\Psi y
 \ge y^*\Delta_{\rm L}[(b_{\mathrm e}-E_{\rm L})I-\Delta_{\rm L}]y
       +\mathcal L(\abs{q})\,y^*K_2(q)y,
\]
where a valid lower bound is
\begin{align}
 \mathcal L(\abs{q})={}&(b_{\mathrm e}-E_{\rm L})
 -2\delta_{\mathrm e}\frac{k_{\max}}{k_{\min}}
 -c_2\abs{q}^2-(b_{\mathrm e}-E_{\rm L})\frac{k_4}{k_{\min}}\abs{q}^2
 -\frac{2\delta_{\mathrm e}c_4}{k_{\min}}\abs{q}^2 \nonumber\\
 &-\left(
 \frac{2\sqrt{c_2}\,c_4}{\sqrt{k_{\min}}}
 +\frac{c_3^2}{k_{\min}}\right)\abs{q}^4
 -\frac{c_4^2}{k_{\min}}\abs{q}^6.
 \label{d:eq:appendix-polynomial-margin}
\end{align}
The \(\Delta_{\rm L},K_2\) anticommutator is bounded in norm; commutation is
not assumed. The other terms enumerate the full even residual square,
its cross terms, and the odd residual square.

Substitution of \(\abs{q}=1/20\), \(\delta_{\mathrm e}=2^{-20}\), the exact
interval for \(E_{\rm L}\), and \(\sqrt{180}<27/2\) gives
\(\mathcal L(\abs{q})>1/400\) by rational arithmetic. Therefore
\[
 \Psi^*g_{\rm L}(A_{\rm cen})\Psi\ge\frac{\abs{q}^2}{20000}I.
\]
Parity gives \(\Phi_{\rm L}^*Y=0\), so
\[
 I\le\Psi^*\Psi=I+Y^*Y\le\frac{409}{400}I.
\]
The min--max principle gives two eigenvalues of
\(g_{\rm L}(A_{\rm cen}(q))\) at least \(\abs{q}^2/20450\).
The scalar polynomial \(g_{\rm L}\) is positive exactly on
\((E_{\rm L},b_{\mathrm e})\). Uniform inertia allows at most two eigenvalues
there, so both low positive bands obey
\[
 (\lambda-E_{\rm L})(b_{\mathrm e}-\lambda)\ge\frac{\abs{q}^2}{20450}.
\]
Since \(b_{\mathrm e}-\lambda<b_{\mathrm e}\), this proves
\[
 \lambda(q)-E_{\rm L}>\frac{\abs{q}^2}{600}
 \qquad(0<\abs{q}\le1/20).
\]
Center simplicity and inversion then imply a Hessian at least
\(I/300\) in these Cartesian reciprocal coordinates.

\subsection{Global exclusion and coordinate folding}

The original diamond certificate establishes a complete gap of at least
\(1/4096\) around zero. For the global comparator, put
\[
 s_0=\frac{157}{16384},\qquad
 \sigma_0=2^{-24},\qquad
 \delta_{\rm mat}=\frac{1023}{2^{26}},\qquad
 D'=D_{\varepsilon_{\rm dia}}+64s_0I.
\]
The shifted material satisfies \(I\le D'<13I\).
The magnetic interval certificate excludes generalized curl eigenvalues
from \([\omega^2-\sigma_0,\omega^2+\sigma_0]\) and counts two negative transverse
directions. Nonzero electric and magnetic spectra agree with
multiplicity after mass normalization. Including the electric gradient
sector gives 2002 negative electric directions. Congruence yields
\[
 A_{\rm dia}-s_0I
 =D'^{1/2}
  \bigl(D'^{-1/2}C_{\rm Y}^{\mathsf T}C_{\rm Y}D'^{-1/2}-\omega^2I\bigr)D'^{1/2}.
\]
Its inverse norm is at most \(1/\sigma_0\), and its inertia is unchanged.
The rank-2003 electric eigenvalue therefore satisfies
\(\lambda_{2003}(A_{\rm dia})\ge s_0+\sigma_0\).
An analytic \(\mathrm{\Gamma}\)-cube argument gives the same conclusion at the
singular magnetic-projector center and throughout its neighborhood.

The material perturbation in \eqref{d:eq:appendix-material-perturbation}
is smaller than the original gap, so the actual matrix has the same
2002 negative eigenvalues everywhere. Its first positive band is the
same ordered eigenvalue. Weyl monotonicity gives
\begin{equation}
 \lambda_{\mathrm{positive}}(A_{\rm per})
 \ge s_0+\sigma_0-\delta_{\rm mat}
 =\frac{642053}{2^{26}}
 >\frac{23913}{2500000}+\frac1{500000}
 >E_c+\frac1{500000}
 \label{d:eq:appendix-global-margin}
\end{equation}
on the certified complement.

The primitive reciprocal Voronoi zone is
\(\abs{\tilde k_j}\le1\), \(\sum_j\abs{\tilde k_j}\le3/2\).
The original comparator admits 48 signed-coordinate permutations.
An exact polytope cover accounts for the closed irreducible wedge,
using 2056 matrix boxes, 660 local-neighborhood leaves, and one analytic
\(\mathrm{\Gamma}\) leaf, with no uncovered region. Each excluded local box lies
wholly within a radius-\(1/20\) ball. The eight signed \(\mathrm{L}\) points form
four inequivalent classes modulo the primitive reciprocal lattice.
Separate actual-material local certificates apply at those four classes.
The comparison \eqref{d:eq:appendix-global-margin} is pointwise and does
not assign the original material's broken symmetries to the perturbation.

The whole-ball estimates, the strict center ordering, and the global
complement bound prove a unique positive-band minimum at \(\mathrm{L}_{++-}\).
Passing to the conventional period-20 cell decomposes the fiber over the
four characters of the finite lattice quotient. It is a unitary direct
sum. We choose \(k_*=(\pi,\pi,-\pi)\) as the conventional-cell
representative. The four \(\mathrm{L}\) classes fold to this same point modulo
\(2\pi\mathbb Z^3\), and their
strict energy ordering leaves the lowest eigenvalue simple. With
\(k=2\pi\tilde k\), its Hessian satisfies
\[
 2H=\lambda''(k_*)=\frac{\lambda''_{\rm Cart}}{(2\pi)^2}
 \ge\frac{I}{1200\pi^2}>\frac{I}{12000},
\]
where \(\lambda''_{\rm Cart}\ge I/300\) is the Hessian in the Cartesian coordinates \(\tilde k\).
All lower bands are negative and \(E_c>0.00956>1/128\), so
\(g_-=1/128\) is valid. Finally
\[
 \max\eps_{\mathrm{per}}+64E_c
 \le12+\frac{1023}{2^{20}}+64E_c<13=25-12,
\]
which supplies \(h_{\mathrm{mat}}=12\).

\subsection{Effective analytic data}

The simple finite-dimensional edge eigenvalue supplies a contour for a
Riesz projection. A Neumann estimate for the explicit Laurent fiber
provides a complex analytic neighborhood, and Cauchy estimates bound its
derivatives. Time reversal gives the quartic dispersion remainder.
These operations are effective once the finite spectral separations have
strict rational bounds.

The compact real separation can also be made explicit. In the centered
coordinates,
\(\norm{A_{\rm per}(\mathrm{L}+q)-A_{\rm per}(\mathrm{L})}<3\abs{q}\).
If \(\delta_{\mathrm{cert}}>0\) bounds the center doublet splitting
from below, its second band remains at least
\(E_c+\delta_{\mathrm{cert}}/2\) for
\(\abs{q}\le\delta_{\mathrm{cert}}/6\).
For any smaller radius \(q_{\mathrm{inner}}>0\), the local lower-band
bound contributes \(q_{\mathrm{inner}}^2/600\) outside that inner ball.
The other \(\mathrm{L}\) neighborhoods contribute \(33/10^9\), and the global
complement contributes \(1/500000\). Their minimum gives a positive
effective separation on the required compact complement.

These matrix inequalities, local estimates and covering identities supply
the spectral premise for Sections~\ref{d:app:edge} and \ref{d:app:green}.
Section~\ref{d:sec:reproducibility} specifies the certificate inputs and
arithmetic used to verify the finite inequalities.

Figure~\ref{d:fig:FIG-3} displays the reciprocal-space cover and the
rigorous enclosure data used in this section.

\section{Finite assembly and stable recovery}\label{d:sec:composition}

The physical platform and logical compiler have compatible interfaces, but a finite Maxwell instance contains many more degrees of freedom than the intended logical variables. We now control those additional variables and every approximation used by the encoder and decoder. The decisive estimates concern the full operator and the total interaction error.

\subsection{Separation and mediator elimination}

Let $\Gamma$ be the logical matrix from Section~\ref{d:sec:logical}, with positive unit off-diagonal edges, maximum degree three, $\norm\Gamma\le7$, and a known bound $\overline K\ge\max(1,\norm{\Gamma^{-1}})$. Let $g$ be its source and $\eta_*$ the physical accuracy budget selected in~\eqref{d:eq:selected-accuracy}. The compiler only appends zeros and places coordinates of the normalized input source, so $\norm g_\infty=1$ and $1\le\norm g_2\le\sqrt m$. In this section $A_\xi$ is the selected fixed operating background, and its interactions are reindexed by $G(\nu):=G_{\rm previous}(B\nu)$, where $B$ is the placement matrix in Proposition~\ref{d:prop:physical-platform}. The physical platform supplies fixed constants $C_{\rm d},c_{\rm d}>0$, $0<r_{\rm d},a_{\rm d}<1$, and a fixed translation lattice such that
\begin{equation}\label{d:eq:composition-couplings}
 \norm{G(\nu)}\le C_{\rm d}r_{\rm d}^{\norm \nu_1},\qquad
 \abs{u^{\mathsf T}G(de_j)u}\ge c_{\rm d}a_{\rm d}^d,\qquad r_{\rm d}^2<a_{\rm d}.
\end{equation}
Here $u$ is the normalized isolated defect mode and $G(\nu)$ is the full interaction block between two translated defects in the placement coordinates. We use $J_{\rm Y}=16$ for the admissible Yee row-sum bound. Set
\begin{equation}\label{d:eq:coupling-scales}
 \zeta=r_{\rm d}^d,\qquad \alpha=\frac {c_{\rm d}}4a_{\rm d}^d.
\end{equation}
Double the logical lattice, place terminal defects at its vertices, and place one mediator defect at the midpoint of each actual logical edge. Scale the resulting positions by $d$ in the fixed translation lattice. For sufficiently large $d$, all defect supports are disjoint.

For each axial direction, approximate the signed scalar interaction $-u^{\mathsf T}G(de_j)u$ by $\beta_j$ with absolute error at most $\zeta^2$. Choosing $\zeta^2\le\alpha$ ensures $\abs{\beta_j}\ge3\alpha$. Define the target matrix $\Theta$ on terminal and mediator amplitudes by
\begin{equation}\label{d:eq:target-mediator}
 \Theta_{ii}=\alpha\bigl(\Gamma_{ii}-\deg(i)\bigr),\qquad
 \Theta_{ee}=-\frac{\beta_j^2}{\alpha},
\end{equation}
with both incidences of a $j$-directed mediator equal to $\beta_j$. Eliminating the mediator gives $+\alpha$ at each endpoint pair and a compensating diagonal contribution. Consequently the exact terminal Schur complement is $\alpha\Gamma$, independent of the sign of $\beta_j$.

The auxiliary algebra is stable. If $\Theta_{\rm tm},\Theta_{\rm mm}$ are the terminal--mediator and mediator blocks, then
\begin{equation}\label{d:eq:mediator-norms}
 \norm{\Theta_{\rm mm}^{-1}}\le\alpha^{-1},\qquad
 \norm{\Theta_{\rm tm}\Theta_{\rm mm}^{-1}}\le\sqrt6,\qquad
 \norm{\Theta^{-1}}\le\frac{16\overline K}{\alpha}.
\end{equation}
The largest required diagonal shift is bounded by
\begin{equation}\label{d:eq:tuning-size}
 \max\left\{7\alpha,(C_{\rm d}+1)^2\frac{\zeta^2}{\alpha}\right\}.
\end{equation}
Both terms tend to zero. Thus the scalar tuning interval supplied by the fixed platform suffices once the separation is large enough.

\subsection{The sum of unwanted interactions}

The occupied sites form a subset of an integer lattice. For $\zeta\le1/2$, geometric series give
\begin{equation}\label{d:eq:lattice-tail}
 \sum_{\nu\ne0}\zeta^{\norm \nu_1}\le64\zeta,
 \qquad
 \sum_{\norm \nu_1\ge2}\zeta^{\norm \nu_1}\le256\zeta^2.
\end{equation}
A matrix of block norms bounds the complete interaction operator through its row and column sums. Hence the aggregate tail is controlled independently of the number of defects. This estimate is the reason a local interaction picture remains valid as the input grows.

Let $\gamma_d>0$ be the separation of the chosen single-defect mode from the remaining compressed spectrum. Eliminating those nonresonant directions produces a bright block
\begin{equation}\label{d:eq:bright-error}
 S_b=\Theta+\Delta,\qquad
 \norm\Delta\le C_\Delta\zeta^2,
 \quad C_\Delta=256C_{\rm d}+5+\frac2{\gamma_d}(64C_{\rm d}+1)^2.
\end{equation}
The complementary block may be indefinite; its invertibility follows from its spectral separation. We choose $d$ to satisfy
\begin{equation}\label{d:eq:spacing-budget}
 C_\Delta\zeta^2\le\frac{\alpha\eta_*}{4096\overline K}.
\end{equation}
Since $r_{\rm d}^2/a_{\rm d}<1$, this requires only
\begin{equation}\label{d:eq:log-spacing}
 d=O\bigl(1+\log\overline K+\log(1/\eta_*)\bigr).
\end{equation}
The encoder selects an integer using certified rational geometric inequalities; exact real logarithms are unnecessary.

\subsection{Finite boundaries and a common good event}

A bulk gap does not by itself establish a gap for a truncated PEC box. We therefore control its actual finite matrix. Surround the full defect and source support by a physical buffer of width $\ell_{\rm buf}$, and let $K_0$ be the finite background of dimension $N$. Choose a dyadic $\tau$ within fixed factors of $\zeta^2$, sufficiently small for the bulk gap and the reserved material headroom. Independently sample a uniform midpoint-grid shift $\sigma\in(0,\tau)$ from a power-of-two grid with at least $16(m+N)^2$ points. Lemma~\ref{d:lem:pec} realizes
\begin{equation}\label{d:eq:finite-background}
 K=K_0-\sigma I,\qquad L_b=\frac{16(m+N)^2}{\tau}
\end{equation}
by a uniform scalar-cell increment. Counting grid points near the real eigenvalues of $K_0$ gives
\begin{equation}\label{d:eq:good-event}
 \Pr\{\norm{K^{-1}}>L_b\}
 \le\frac{3N}{16(m+N)^2}\le\frac3{64m}.
\end{equation}
Singular draws are included in this exceptional event.

Let $Q$ restrict to the entire union of defect coordinates. A set-to-set exponential resolvent estimate and the finite/infinite restriction identity imply, on the complementary good event,
\begin{equation}\label{d:eq:boundary-error}
 \norm{QK^{-1}Q^{\mathsf T}-Q_\infty A_\xi^{-1}Q_\infty^{\mathsf T}}
 \le\frac{2\tau}{g_0^2}+\frac{4J_{\rm Y}L_b}{g_0}e^{-\mu_b\ell_{\rm buf}},
\end{equation}
where $g_0>0$ and $\mu_b>0$ depend only on the fixed platform. The same bound applies to the whole defect union, without a multiplicative defect count. The encoder determines $N$ and $\ell_{\rm buf}$ by a terminating doubling calculation before writing the box. A width $\ell_{\rm buf}=O(d+\log m)$ makes~\eqref{d:eq:boundary-error} at most $\zeta^2$.

\subsection{Conditioning of the full Maxwell matrix}

Let $W_{\mathcal D}$ be the diagonal edge-weight matrix induced by one scalar-cell defect, $W_{\rm all}$ its block sum, and $v_i/\omega^2$ the scalar increment on defect $i$. Choose each $v_i$ so that
\begin{equation}\label{d:eq:scalar-tuning}
 \abs{v_i^{-1}-(\mu+\Theta_{ii})}\le\zeta^2,
\end{equation}
where $\mu$ is the chosen compressed-resolvent eigenvalue. The values are rounded at the cell level. With $V_{\rm d}=\operatorname{diag}(v_iW_{\mathcal D})$, the actual operator is
\begin{equation}\label{d:eq:assembled-A}
 A=K-Q^{\mathsf T}V_{\rm d}Q.
\end{equation}
The weighted Birman--Schwinger matrix is
\begin{equation}\label{d:eq:bs-matrix}
 \mathcal M=W_{\rm all}^{1/2}
       (V_{\rm d}^{-1}-QK^{-1}Q^{\mathsf T})W_{\rm all}^{1/2}.
\end{equation}
Its bright Schur block is~\eqref{d:eq:bright-error}. Triangular factorization and the inverse perturbation bound give
\begin{equation}\label{d:eq:full-inverse-budget}
 Q_M=4\left(\frac{32\overline K}{\alpha}+\frac2{\gamma_d}\right),\qquad
 \norm{A^{-1}}\le Q_A:=L_b+L_b^2Q_M,
 \qquad \kappa_2(A)\le J_{\rm Y}Q_A.
\end{equation}
This bound includes the background, all auxiliary modes, and all nonresonant defect coordinates. Section~\ref{d:app:composition} supplies the block calculation and the exact inverse identity.

\subsection{A local source and an every-solution decoder}

Let $U$ insert the normalized mode $u$ into every defect block and set $g_\Theta=(g,0)$ on terminal and mediator sites. Define
\begin{equation}\label{d:eq:ideal-maps}
 f_*=\alpha KQ^{\mathsf T}W_{\rm all}^{-1/2}Ug_\Theta,
 \qquad z=U^{\mathsf T}W_{\rm all}^{-1/2}V_{\rm d}QE.
\end{equation}
The source uses one finite stencil application to a defect-supported vector, and the readout uses only defect coordinates. Write $[U\ U^\perp]^{\mathsf T}\mathcal M[U\ U^\perp]=\left[\begin{smallmatrix}\MM_{bb}&\MM_{bc}\\\MM_{bc}^{\mathsf T}&\MM_{cc}\end{smallmatrix}\right]$. Direct multiplication yields the exact residual identity
\begin{equation}\label{d:eq:exact-residual}
 S_bz=\alpha g_\Theta+
 (U^{\mathsf T}-\MM_{bc}\MM_{cc}^{-1}(U^\perp)^{\mathsf T})W_{\rm all}^{1/2}QK^{-1}(AE-f_*).
\end{equation}
The residual multiplier has norm at most $2L_b$. This identity transfers a solver's residual to the logical equations without assumptions on the solver's method.

Put $c_W=1/\sqrt{w_{\min}}$, where $w_{\min}$ is the smallest nonzero defect weight. It is sufficient to choose
\begin{equation}\label{d:eq:source-budget}
 \delta=\frac{\eta_*}{8192J_{\rm Y}c_WL_b},\qquad
 \norm{f-f_*}\le\frac{\alpha\eta_*\norm g}{8192L_b}.
\end{equation}
Every solution of the output residual condition then has ideal logical
terminal residual below $\eta_*\norm g/16$. The decoder reads only
the terminal coordinates corresponding to the original input variables.
Mode calibration and finite coordinate access contribute at most
$11\eta_*\norm g/1024$ to their Euclidean readout error.
The accuracy choice below pays for both residual transport and this
absolute coordinate error. Complex fields are reduced to their real parts,
which cannot increase the residual of a real matrix and real source.

The actual output source must satisfy the global budget in~\eqref{d:eq:source-budget}. If $p_f\le N$ is its total support size, a dyadic coordinate spacing at most
\begin{equation}\label{d:eq:source-grid}
 \frac{\alpha\eta_*}{32768L_b\max(1,p_f)}
\end{equation}
is sufficient together with the mode calibration. Counting only one defect's source support would not justify this estimate.

\subsection{\texorpdfstring{Precision and proof of Theorem~\ref{d:thm:main}}{Precision and proof of the Yee realization theorem}}

The needed spectral data can be refined locally. Since the spectrum of $A_\xi$ avoids $(-g_0,g_0)$ and $\norm{A_\xi}\le J_{\rm Y}$,
\begin{equation}\label{d:eq:inverse-polynomial}
 p_{\ell_{\rm P}}(A_\xi)=\frac {A_\xi}{J_{\rm Y}^2}\sum_{j=0}^{\ell_{\rm P}}(I-A_\xi^2/J_{\rm Y}^2)^j
\end{equation}
approximates $A_\xi^{-1}$ with error at most
$g_0^{-1}(1-g_0^2/J_{\rm Y}^2)^{\ell_{\rm P}+1}$. Its finite propagation radius is $O(\ell_{\rm P})$. Thus inverse-polynomial calibration requires $O(\log m)$ stencil iterations in an $O(\log m)$-radius neighborhood. The isolated finite-dimensional eigendata are refined using certified isolation. All platform dimensions are fixed before $m$ varies.

\begin{proof}[Proof of Theorem~\ref{d:thm:main}]
Apply the logical compiler with $\eta_0=m^{-\lceil c_\eta\rceil}$. It supplies
a matrix $\Gamma$ in a balanced box of volume $O(S^{3/2})$, where
$S=O(m\log m)$, a polynomial bound $\overline K$, and the residual
amplification factor $T_{\rm res}$ in~\eqref{d:eq:logical-tolerance}. Choose
\begin{equation}\label{d:eq:selected-accuracy}
 \eta_*=\min\left\{\frac12,\,
 \frac{\eta_0}{\max\{1,T_{\rm res}\}\max\{1,m^2\}}\right\}.
\end{equation}
Then $T_{\rm res}\eta_*\le\eta_0$ and
$m^2\eta_*\le\eta_0$. The first inequality turns the ideal terminal
residual into the compiler tolerance $\eta_{\rm res}\norm g$.
Since $\norm g\le\sqrt m\le m$, the second implies
$\eta_*\norm g\le\eta_0/m$. Thus the readout error is less than
$\eta_0/(32m)$ per selected coordinate, as required before the final
rounding in Section~\ref{d:app:logical-residual}.
The physical platform of Section~\ref{d:sec:physical} supplies all fixed
constants in~\eqref{d:eq:composition-couplings} and the admissible tuning interval.

Choose $d$ by~\eqref{d:eq:spacing-budget} and the finitely many ancillary platform inequalities. Then $d=O(\log m)$. Determine the physical buffer, select the random shift, calibrate the local mode and the three axial interactions, round scalar cells and the total source, and emit every cell and source coordinate. On the common good event~\eqref{d:eq:good-event}, equations~\eqref{d:eq:bright-error}--\eqref{d:eq:source-budget} control the full inverse and every residual solution. Finite readout gives the logical residual required by the compiler, whose final coordinate selection, rounding, and rescaling yield~\eqref{c:eq:target}.

Balanced routing, logarithmic separation, and logarithmic buffering give
\begin{equation}\label{d:eq:volume-complete}
 N=O\!\left(S^{3/2}(20d+\ell_{\rm buf}+O(1))^3\right)=\Ot(m^{3/2}),
\end{equation}
with $20$ the conventional period side in Yee cells and a fixed factor for the translation lattice. Each inverse scale in the construction is polynomial in $m$. The explicit budget in Section~\ref{d:app:precision} therefore gives effectively computable $c_A,c_\delta$ and logarithmic output bit lengths. Calibration and readout require polynomial-in-$\log m$ work per fixed-support local block. Their total cost, together with writing the full box and source, is $\Ot(m^{3/2})$. All these operations have the same bound on every random draw. The fixed startup search is a terminating effective procedure independent of $m$; its cost is included in the constant. The zero-source and bounded-size cases are treated as in Section~\ref{d:app:model}. This proves all claims.
\end{proof}

\section{Finite assembly and operator bounds}\label{d:app:composition}

This section gives the quantitative composition argument under the fixed physical platform established in Section~\ref{d:sec:physical}. Let $A_\xi$ be the real periodic infinite Yee operator at the operating energy, with $\norm{A_\xi}\le J_{\rm Y}$, stencil radius $r_{\rm st}$, and $\operatorname{dist}(0,\operatorname{spec}A_\xi)\ge g_0>0$. Let $Q_{\mathcal D}$ select the touched edges of a fixed scalar defect, let $W_{\mathcal D}$ be its positive diagonal weight matrix, and let $(\mu,u)$ be the chosen normalized simple eigenpair of its compressed resolvent. Its separation from the remaining compressed spectrum is $\gamma_d>0$. The tuning $v=1/\mu$ lies in the interior of the admissible scalar interval, with separate headroom for uniform positive detuning. All these data are fixed independently of $m$.

\subsection{Exact mediator algebra}

Order terminals first and mediators second. For one edge $e=\{i,i'\}$ in direction $j$, its mediator has diagonal $-\beta_j^2/\alpha$ and incidences $\beta_j$ at $i,i'$. Its elimination contributes
\[
 -\begin{bmatrix}\beta_j\\\beta_j\end{bmatrix}
   \left(-\frac{\beta_j^2}{\alpha}\right)^{-1}
   \begin{bmatrix}\beta_j&\beta_j\end{bmatrix}
 =\alpha\begin{bmatrix}1&1\\1&1\end{bmatrix}.
\]
Summing the diagonal contributions over incident edges cancels the $-\alpha\deg(i)$ term in~\eqref{d:eq:target-mediator}. The terminal Schur complement is therefore exactly $\alpha\Gamma$.

The mediator inverse has norm at most $1/\alpha$ because $\abs{\beta_j}\ge3\alpha$. Every entry of $\Theta_{\rm tm}\Theta_{\rm mm}^{-1}$ has magnitude $\alpha/\abs{\beta_j}\le1/3$. Degree three at terminals and degree two at mediators give the conservative bound $\norm{\Theta_{\rm tm}\Theta_{\rm mm}^{-1}}\le\sqrt6$. Triangular factorization with diagonal blocks $\alpha\Gamma,\Theta_{\rm mm}$ then yields $\norm{\Theta^{-1}}\le16\overline K/\alpha$. The same elimination applied to an arbitrary residual $\rho_\Theta=\Theta z-\alpha g_\Theta$ gives
\begin{equation}\label{d:eq:app-mediator-residual}
 \norm{\Gamma z_{\rm t}-g}
 \le\frac{1+\sqrt6}{\alpha}\norm{\rho_\Theta}
 <\frac4\alpha\norm{\rho_\Theta}.
\end{equation}
This identity is used after all physical perturbations have been bounded; it does not assume that the raw electromagnetic interaction is exactly $\Theta$.

\subsection{Summed interactions and the complementary space}

For $0<\zeta\le1/2$,
\[
 \sum_{\nu\in\Z^3}\zeta^{\norm \nu_1}
 =\left(\frac{1+\zeta}{1-\zeta}\right)^3.
\]
Subtracting the constant term gives numerator $6\zeta+2\zeta^3$ over $(1-\zeta)^3$. Subtracting the six nearest-neighbor terms as well gives numerator $18\zeta^2-16\zeta^3+6\zeta^4$. Bounding these on $[0,1/2]$ gives~\eqref{d:eq:lattice-tail}. Restricting to an occupied subset only decreases the absolute row sums. Symmetry provides the same column sums, so the induced Euclidean operator norm is bounded by the row-sum estimate for the matrix of block norms.

Let $[U\ U^\perp]$ be a blockwise orthogonal decomposition into the chosen mode and its complement. In the weighted matrix~\eqref{d:eq:bs-matrix}, write
\[
 [U\ U^\perp]^{\mathsf T}\mathcal M[U\ U^\perp]
 =\begin{bmatrix}\MM_{bb}&\MM_{bc}\\\MM_{bc}^{\mathsf T}&\MM_{cc}\end{bmatrix}.
\]
At infinite separation, the complementary block is the block sum of $\mu I-S_{\rm def}$ restricted to $u^\perp$, where $S_{\rm def}=W_{\mathcal D}^{1/2}Q_{\mathcal D}A_\xi^{-1}Q_{\mathcal D}^{\mathsf T}W_{\mathcal D}^{1/2}$ is the single-defect compressed resolvent at the selected operating point. This notation distinguishes it from the edge-limit matrix $S_0$ in Section~\ref{d:app:edge}. Its smallest singular value on the complement is at least $\gamma_d$. Scalar on-site tuning contributes a multiple of the identity in each weighted defect block, so it introduces no mixed mode/complement term.

The interaction row sums, scalar rounding, and finite-background approximation give, after a fixed increase of the minimum separation,
\begin{equation}\label{d:eq:app-complement}
 \norm{\MM_{cc}^{-1}}\le D_*:=2/\gamma_d,\qquad
 \norm{\MM_{bc}}\le64C_{\rm d}\zeta+\zeta^2,\qquad
 1+\norm{\MM_{bc}} D_*\le2.
\end{equation}
The inverse perturbation lemma applies to an indefinite $\MM_{cc}$: if $\MM_{cc}^{(0)}$ is invertible and $\norm{(\MM_{cc}^{(0)})^{-1}(\MM_{cc}-\MM_{cc}^{(0)})}\le1/2$, then the Neumann series gives the stated inverse bound.

The bright-block error before complementary elimination has one unit of on-site rounding, three incidence units, one finite-background unit, and the non-neighbor tail. The correction $\MM_{bc}\MM_{cc}^{-1}\MM_{bc}^{\mathsf T}$ is bounded by $D_*(64C_{\rm d}+1)^2\zeta^2$. This proves~\eqref{d:eq:bright-error}. Since
\[
 \norm{\Theta^{-1}\Delta}\le\frac{16\overline K}{\alpha}
 \frac{\alpha\eta_*}{4096\overline K}<\frac12,
\]
we have $\norm{S_b^{-1}}\le32\overline K/\alpha$.

For completeness, a second elimination of the physical mediator amplitudes gives terminal response $\alpha\Gamma+\Delta_{\rm eff}$ with
\begin{equation}\label{d:eq:app-effective-error}
 \norm{\Delta_{\rm eff}}
 \le12\norm\Delta+288\norm\Delta^2/\alpha
 \le13\norm\Delta.
\end{equation}
The unperturbed mediator triangular factor has squared norm below twelve; its lower-right perturbation is still the mediator block of $\Delta$. The perturbed mediator inverse has norm at most $2/\alpha$. Multiplying the two mixed transformed errors gives the quadratic term. The final inequality follows from $\norm\Delta/\alpha\le1/8192$. The universal residual argument below can use~\eqref{d:eq:app-mediator-residual} directly and does not require replacing the terminal response by an exact one.

\subsection{Finite spectral detuning}

Let $J_{\rm grid}$ be the least power of two at least $16(m+N)^2$. Choose a dyadic $\tau>0$ with
\[
 \tau\le g_0/4,\qquad \tau\le g_0^2\zeta^2/16,
\]
and with $\tau/\omega^2$ inside the reserved uniform material headroom. Sample
\[
 \sigma=(j+1/2)\tau/J_{\rm grid},\qquad
 j\in\{0,\ldots,J_{\rm grid}-1\}.
\]
Set $L_b=16(m+N)^2/\tau$. For any eigenvalue of the real symmetric $K_0$, the interval of radius $1/L_b$ contains at most $2J_{\rm grid}/(L_b\tau)+1$ grid points. The union bound over the $N$ eigenvalues gives
\[
 \Pr\{\norm{(K_0-\sigma I)^{-1}}>L_b\}
 \le\frac{2N}{L_b\tau}+\frac{N}{J_{\rm grid}}
 \le\frac{3N}{16(m+N)^2}.
\]
The final estimate in~\eqref{d:eq:good-event} follows from $(m+N)^2\ge4mN$. This event controls every finite boundary mode because the count uses the full finite spectrum. The argument uses self-adjointness of the unstretched operator.

\subsection{The finite and infinite restriction identity}

For $0\le\sigma\le g_0/4$, set $R_\sigma=(A_\xi-\sigma I)^{-1}$. Conjugate by bounded clipped exponential distance weights. With
\[
 \mu_b=r_{\rm st}^{-1}\log\left(1+\frac{g_0}{4J_{\rm Y}}\right),
\]
the change in operator row and column sums is at most $g_0/4$. A Neumann-series estimate gives weighted inverse norm at most $2/g_0$. Removing the weights and taking the clipping parameter to infinity yields the set-to-set bound
\begin{equation}\label{d:eq:app-set-resolvent}
 \norm{P_{S_1}R_\sigma P_{S_2}^{\mathsf T}}
 \le\frac2{g_0}\exp[-\mu_b\operatorname{dist}_1(S_1,S_2)].
\end{equation}

Let $P_{\rm box}$ restrict the infinite edge space to the retained finite edges, and put $T_\partial=K_0P_{\rm box}-P_{\rm box}A_\xi$. Its input support lies in a fixed-width boundary strip, and $\norm{T_\partial}\le2J_{\rm Y}$. Taking the buffer from that strip absorbs stencil-width constants. Direct multiplication gives
\[
 K^{-1}Q^{\mathsf T}-P_{\rm box}R_\sigma Q_\infty^{\mathsf T}
 =-K^{-1}T_\partial R_\sigma Q_\infty^{\mathsf T}.
\]
Together with $R_\sigma-A_\xi^{-1}=\sigma R_\sigma A_\xi^{-1}$, this gives~\eqref{d:eq:boundary-error}. Because $Q_\infty$ selects the whole defect union, the set distance in~\eqref{d:eq:app-set-resolvent} is sufficient and no defect-count factor appears.

There is no circular choice of buffer and dimension. For a proposed $\ell_{\rm buf}$, calculate the box dimensions and $N$ arithmetically. They are bounded by a fixed polynomial in $m,d,\ell_{\rm buf}$. The exponential decrease in~\eqref{d:eq:boundary-error} eventually dominates the resulting polynomial factor in $L_b$. Doubling the buffer therefore terminates at $\ell_{\rm buf}=O(d+\log m)$. A rational geometric upper bound for the exponential permits a certified stopping test. Only the final successful box is materialized.

\subsection{The full inverse identity}

Triangular factorization in the bright/complement basis, using~\eqref{d:eq:app-complement}, gives
\[
 \norm{\mathcal M^{-1}}
 \le4\left(\frac{32\overline K}{\alpha}+D_*\right)=Q_M.
\]
Direct multiplication verifies
\begin{equation}\label{d:eq:app-woodbury}
 A^{-1}=K^{-1}+K^{-1}Q^{\mathsf T}W_{\rm all}^{1/2}
 \mathcal M^{-1}W_{\rm all}^{1/2}QK^{-1}.
\end{equation}
The factors $Q,Q^{\mathsf T},W_{\rm all}^{1/2}$ have norm at most one, since the defect edge supports are disjoint and $W_{\mathcal D}\le I$. Hence $\norm{A^{-1}}\le L_b+L_b^2Q_M$. The admissible scalar material and the finite-stencil estimate give $\norm A\le J_{\rm Y}=16$, establishing the condition bound for every Maxwell unknown.

\section{Source precision and complete resource accounting}\label{d:app:precision}

\subsection{An exact identity for arbitrary residual solutions}

Let $g_\Theta=(g,0)$, and define $f_*,z$ by~\eqref{d:eq:ideal-maps}. For $\rho_E=AE-f_*$, put $w_{\rm d}=W_{\rm all}^{-1/2}V_{\rm d}QE$. Multiplying the definition of $\mathcal M$ gives
\[
 \mathcal Mw_{\rm d}=W_{\rm all}^{1/2}QK^{-1}(f_*+\rho_E)
 =\alpha Ug_\Theta+W_{\rm all}^{1/2}QK^{-1}\rho_E.
\]
Projection onto $U,U^\perp$ and elimination of the second block establish~\eqref{d:eq:exact-residual}. Its multiplier norm is bounded by $(1+\norm{\MM_{bc}} D_*)L_b\le2L_b$.

Suppose the actual source satisfies~\eqref{d:eq:source-budget}. Since $\norm{f_*}\le\alpha J_{\rm Y}c_W\norm g$, we have $\norm{f}\le2\alpha J_{\rm Y}c_W\norm g$. Every vector with residual at most $\delta\norm{f}$ obeys
\begin{align*}
 \norm{AE-f_*}&\le\frac{3\alpha\eta_*\norm g}{8192L_b},\\
 \norm{S_bz-\alpha g_\Theta}&\le\frac{3\alpha\eta_*\norm g}{4096},\\
 \norm z&<33\overline K \norm g.
\end{align*}
Adding $\Delta z$ and using~\eqref{d:eq:spacing-budget} gives
\[
 \norm{\Theta z-\alpha g_\Theta}\le\frac{36\alpha\eta_*\norm g}{4096}
 <\frac{\alpha\eta_*\norm g}{64}.
\]
Equation~\eqref{d:eq:app-mediator-residual} implies logical terminal residual below $\eta_*\norm g/16$. These estimates use only the residual inequality and operator bounds, so they apply to every such $E$. If $E$ is complex, $A$ and $f$ are real and $\norm{A\operatorname{Re}E-f}\le\norm{AE-f}$. It is therefore sufficient to read real coordinates.

\subsection{Calibration and the output source}

Write $u_W=W_{\mathcal D}^{1/2}u$ and let $v_{\max}$ be a fixed upper bound on $\abs{v_i}$. A sufficient calibration tolerance is
\begin{equation}\label{d:eq:app-q-calibration}
 \norm{\widehat u_W-u_W}\le
 \min\left\{
 \frac{w_{\min}\eta_*}{32768J_{\rm Y}L_b},
 \frac{\eta_*}{1024v_{\max}\alpha J_{\rm Y}c_WQ_A}
 \right\}.
\end{equation}
The injection in~\eqref{d:eq:ideal-maps} uses $W_{\mathcal D}^{-1}u_W$. The first term makes its coefficient error at most $\alpha\eta_*\norm g/(32768L_b)$. Round every output source coordinate to a dyadic grid of spacing at most~\eqref{d:eq:source-grid}, using the total support count $p_f$ across all defects. The Euclidean rounding error is no greater than another $\alpha\eta_*\norm g/(32768L_b)$ because $\norm g\ge1$. Their sum is strictly inside~\eqref{d:eq:source-budget}.

For a real scalar spectral value $\lambda$ with $g_0\le\abs\lambda\le J_{\rm Y}$,
\[
 \lambda^{-1}-\frac\lambda{J_{\rm Y}^2}\sum_{j=0}^{\ell_{\rm P}}(1-\lambda^2/J_{\rm Y}^2)^j
 =\lambda^{-1}(1-\lambda^2/J_{\rm Y}^2)^{\ell_{\rm P}+1}.
\]
The spectral theorem yields the inverse-polynomial error bound following~\eqref{d:eq:inverse-polynomial}. Its propagation radius is at most $(2\ell_{\rm P}+1)r_{\rm st}$. A compressed resolvent or interaction block can consequently be calibrated to inverse-polynomial accuracy using $O(\log m)$ iterations on $O(\log^3m)$ coordinates, with polynomial-in-$\log m$ bit cost for the fixed platform. Couplings separated by $O(\log m)$ are covered by enlarging the same radius by a fixed factor.

The finite-dimensional isolated eigendata are refined using certified inertia and root isolation, or a squared spectral filter centered in the isolating interval. The isolation and fixed complementary gap are premises of this step. They justify its accuracy and cost even when the compressed operator is indefinite.

\subsection{Finite readout}

The readout on defect $i$ simplifies to
\begin{equation}\label{d:eq:app-implemented-readout}
 z_i=v_iu_W^{\mathsf T}Q_i E.
\end{equation}
On the good event, $\norm E\le3\alpha J_{\rm Y}c_WQ_A\norm g$. Replacing $u_W$ by
$\widehat u_W$ contributes at most $3\eta_*\norm g/1024$ by the second
bound in~\eqref{d:eq:app-q-calibration}. Allow an additional
$\eta_*\norm g/128$ for accessed-coordinate and readout arithmetic errors.
The combined readout error is at most $11\eta_*\norm g/1024$.
With~\eqref{d:eq:selected-accuracy} and $\norm g\le m$, this is less than
$\eta_0/(32m)$. Every selected original coordinate is therefore read to the
accuracy required in Section~\ref{d:app:logical-residual}.
Separately, the ideal terminal residual is at most
$\eta_*\norm g/16\le\eta_{\rm res}\norm g$. Apply the logical decoder to those ideal
coordinates and their finite approximations. This pays for both errors
without requiring a residual test on a rounded vector of all auxiliary
coordinates.

For $p_{\rm acc}\ge1$ accessed real coordinates, use absolute coordinate error at most
\begin{equation}\label{d:eq:app-coordinate-grid}
 h_y=\frac{\eta_*}{8192\max(1,v_{\max})p_{\rm acc}}.
\end{equation}
The first calibration bound ensures $\norm{\widehat u_W}\le2$. Thus the coordinate-access contribution is at most $2v_{\max}\sqrt{p_{\rm acc}}\,h_y\le\eta_*/4096$, inside the arithmetic allowance. The bound on $\norm E$ controls the integer range, and~\eqref{d:eq:app-coordinate-grid} controls the fractional bits. Both are $O(\log m)$. The final compiler decoder selects original coordinates, rounds, and rescales; it performs no original-system solve on unbounded inputs.

\subsection{Explicit polynomial exponents}

The logical proof supplies safe fixed exponents
\[
 p_K=\lceil c_\kappa\rceil+8,\qquad
 p_\eta=\lceil c_\eta\rceil+7,
\]
such that $\overline K=O(m^{p_K})$ and
$\eta_*^{-1}=O(m^{p_\eta})$. In fact
$T_{\rm res}=O(m^{7/2}\log^{5/2}(2+m))$, so
\eqref{d:eq:selected-accuracy} gives
$\eta_*^{-1}=O(m^{\lceil c_\eta\rceil+11/2}\log^{5/2}(2+m))$;
the displayed $p_\eta$ absorbs the logarithm. Recall $r_{\rm d}^2/a_{\rm d}<1$.
Apart from fixed ancillary bounds, the spacing condition follows from
\[
 d\ge\frac{\log[(16384C_\Delta/c_{\rm d})\overline K/\eta_*]}{\log(a_{\rm d}/r_{\rm d}^2)}.
\]
Choose a fixed $c_{d,1}>(p_K+p_\eta)/\log(a_{\rm d}/r_{\rm d}^2)$, enlarge the additive constant as needed, and obtain $d\le c_{d,0}+c_{d,1}\log m$. For fixed
\[
 p_\zeta>c_{d,1}\log(1/r_{\rm d}),\qquad
 p_\alpha>c_{d,1}\log(1/a_{\rm d}),
\]
we have $\zeta^{-1}=O(m^{p_\zeta})$ and $\alpha^{-1}=O(m^{p_\alpha})$. Counting the final box gives
\begin{align}\label{d:eq:app-exponents}
 N&=\Ot(m^{3/2}), & L_b&=\Ot(m^{3+2p_\zeta}),\notag\\
 \kappa_2(A)&=\Ot(m^{6+4p_\zeta+p_\alpha+p_K}), &
 \delta^{-1}&=\Ot(m^{3+2p_\zeta+p_\eta}).
\end{align}
All scalar precision requirements are inverse-polynomial, so their dyadic representations have logarithmic length. Intermediate exact local arithmetic can use a fixed power of $\log m$ bits without changing the soft asymptotic encoder bound.

Explicit bounds of the form $\kappa_2(A)\le m^{c_A}$ and $\delta\ge m^{-c_\delta}$ require absorbing fixed coefficients as well as logarithms. If a quantity is bounded by $K_*m^{a_1}(\log(2+m))^{a_2}$ for $m\ge2$, choose integers $a_3\ge\max(0,a_2)$ and $a_4$ with $2^{a_4}\ge\max(1,K_*)$. Since $\log(2+m)\le m$, the exponent $\lceil a_1+a_3\rceil+a_4$ suffices. Rational doubling selects $a_4$ effectively. Apply this separately to conditioning and inverse tolerance. Fixed base cases are covered by Section~\ref{d:app:model}.

\subsection{Uniform time and probability}

The encoder selects its fixed platform by an effective terminating enumeration justified by strict certified inequalities. The startup cost may be very large but is independent of $m$. For each input it constructs the balanced logical network, chooses separation and buffer, calibrates fixed-dimensional data, samples a bounded number of random bits, and writes every scalar cell and source coordinate. The decoder performs only prescribed finite-prefix reads and short local arithmetic. These operations cost $\Ot(m^{3/2})$ on every draw.

The single event that $\norm{K^{-1}}\le L_b$ controls the finite-background approximation, complementary elimination, full Maxwell inverse, and every-solution residual transfer. No union over returned solutions is needed. The probability statement is consequently exactly~\eqref{d:eq:good-event}, with resources still bounded on the exceptional draws.

\sipart{IV}{Algorithm transfer and conditional hardness}
\section{Solver calls, verification and amplification}\label{si:sec:transfer-proof}

This section and the next prove Theorems~\ref{d:thm:algorithm-transfer}
and~\ref{c:thm:transfer}. An algorithm for every Maxwell instance in
the constructed class can be applied to the encoding of a sparse linear system, after
which the decoder recovers the algebraic solution. The resource bounds of
the realization theorems hold on every random draw, while conditioning
and correctness hold on a common good event; the arguments below turn
this into a randomized algorithm with verified output.

\subsection{Rational inputs, capped calls, and verification}
For rational inputs, encode, cap the solver at its known good-instance
bound, and decode. On the common event, $N=\tO(m^{3/2})$, $B_{\rm out}=O(\ell)$,
and both $\kappa_X$ and $\delta^{-1}$ are polynomial in $m$.
Substitution in~\eqref{c:eq:solverbound} gives the second term of
\eqref{c:eq:transfer}; preparation and decoding give the first.
A bad draw is not allowed an unbounded solver call: a timeout or an
invalid output format is rejected.

Verify a candidate using the original sparse residual with outward
short-dyadic arithmetic. This costs $O(m\operatorname{polylog}m)$,
without constructing a common denominator for all rational entries.
For an explicit acceptance rule, put $\nu_M=\|b\|_\infty$ and enclose
$\rho_M=\|Mx-b\|^2/\nu_M^2$ and $b_M=\|b\|^2/\nu_M^2$. Their intermediate
magnitudes are bounded by an explicit polynomial from the input and
fixed output ranges. If that bound is $P_{\rm mag}\ge1$, rounding every one
of the $O(m)$ additions and products on a grid at most
$\eta_0^2/(2^{20}m^2P_{\rm mag}^4)$ gives interval widths at most
$\eta_0^2/64$ for $\rho_M$ and $1/64$ for $b_M$; the elementary product
error bound proves this by induction over the operations. Accept
only when the upper bound for $\rho_M$ is at most $\eta_0^2$ times the
lower bound for $b_M$. Acceptance is sound. The successful decoder
has residual less than $\eta_0\|b\|/4$ by
Remark~\ref{si:rem:continuum-compiler}, so it passes with this margin and
$b_M\ge1$.

Fix a positive rational lower bound for the solver's uniform success
probability, as part of its specification. A good draw and successful
call then have a fixed positive joint probability. Independent capped
trials and the sound acceptance test give inverse-polynomial failure
with $O(\log m)$ repetitions, preserving~\eqref{c:eq:transfer}. The
finite-event quantifier is essential here: correctness does not depend
on which admissible residual field the solver chose.

\subsection{Randomization and verifiable acceptance}

The Maxwell reduction has resource bounds on every random draw, while
its conditioning and correctness guarantees hold on a common good
event. A bad draw may be singular or poorly conditioned, so a runtime
bound for well-conditioned instances cannot control an unrestricted
call on such a draw.

We run each Maxwell solve with a cap equal to a proved upper bound for
a good input of the relevant size and input class. A timeout is rejected.
For a returned field, we decode the saddle coordinates and evaluate
the residual of the original, unrounded real saddle system. This costs
$O(m)$ Real-RAM operations. An accepted result satisfies the desired
projection-relative error regardless of whether its physical draw was
good. Good draws and successful solver calls pass with a fixed positive
probability, after allocating the rounding margins in
Section~\ref{d:app:real-transfer}. Independent capped repetitions amplify
success, and $O(\log m)$ attempts yield any prescribed inverse-polynomial
failure bound. Their cost remains~\eqref{d:eq:sparse-composition-time}.

\section{A transfer from rectangular real inputs}
\label{d:app:real-transfer}

Theorem~\ref{d:thm:main} starts with a short rational square system.
Assumption~\ref{d:ass:slth} permits rectangular matrices and
unrestricted real entries. We bridge these models explicitly.

After removing empty rows, the active row count is at most $m$. Normalize
$F=F_0/(ma_F)$ with $a_F=\max|F_{0,ij}|$ and compute $b_F=F^\top b$ with exact Real-RAM
operations before rounding either the matrix or the source. If $b_F=0$,
then $\Pi_{F_0}b=0$ and $x=0$ is an exact answer. Otherwise let
$\nu_F=\norm{b_F}_\infty$, $g_F=b_F/\nu_F$, and construct the sparse saddle system
\begin{equation}
 K_s\begin{pmatrix}y_{\rm s}\\x_{\rm s}\end{pmatrix}=c_{\rm s},
 \qquad
 K_s=\begin{pmatrix}I&F\\F^\top&0\end{pmatrix},
 \qquad c_{\rm s}=\begin{pmatrix}0\\-g_F\end{pmatrix}.
 \label{d:eq:complexity-saddle}
\end{equation}
Its entries are bounded by one, its dimension is at most $2m$, and it has
at most $3m$ nonzeros. The normal matrix $F^\top F$ is used to analyze
the system, but need not be formed by the algorithm.

Let $P=m^{\lceil c_\kappa\rceil+1}$. The inverse of $K_s$ has norm
at most $4P^2$. A relative saddle residual $\eta_s$ implies a
projection-relative error at most $2P\eta_s$ after rescaling the $x_{\rm s}$
coordinates. Taking $\eta_s=m^{-(6\lceil c_\kappa\rceil+3)}$ therefore suffices for the
squared-error target~\eqref{d:eq:slth-target}.

We round the bounded saddle coefficients and normalized source to
$O_{c_\kappa}(\log m)$ bits, with an explicit perturbation budget that preserves
both inverse bounds and the original saddle residual. An additional
$m$-dimensional identity block ensures that the rounded input's own
nonzero count lies between $m$ and $4m$. This counted padding makes its
condition and accuracy bounds fixed polynomials in its own input
size. The rounded square system is then a valid input to
Theorem~\ref{d:thm:main}. The subsections below supply the
normalization, residual identity, rounding inequalities, and parameter
choices.

Computing $b_F$ before rounding matters when it is small because of
cancellation. Rounding $b$ relative to $\norm{b}_2$ would not in general
control the much smaller projected source. Here $g_F$ is normalized only
after the exact projection-relevant quantity has been formed. The scale
factors retained for final output are Real-RAM words; they do not become
large-bit physical coefficients. This is a Real-RAM reduction for
unrestricted real inputs and a finite-bit reduction after the stated
rounding step.

We prove Theorem~\ref{d:thm:algorithm-transfer}, including the error
criterion and randomized runtime model. The source problem and its
projection-relative target are those in
Assumption~\ref{d:ass:slth}. The same bridge, with
Theorem~\ref{c:thm:main} in place of Theorem~\ref{d:thm:main} and the
moment-message decoder of Section~\ref{c:sec:recovery}, proves the
Real-RAM part of Theorem~\ref{c:thm:transfer}.

\subsection{Input access and exact source normalization}

Let $F_0\in\R^{n_0\times n}$ have full column rank, $m\ge2$
nonzeros, and condition number at most $m^{c_\kappa}$. A sparse nonzero list
identifies all nonempty rows. Delete empty rows and relabel the active
indices in $O(m\log m)$ operations. The remaining row count $n_1$
satisfies $n\le n_1\le m$. Empty rows contribute zero to the column
space and the projected right-hand side, and deleting them does not
change the nonzero singular values. We access $b$ only at active row
indices. This uses the random-access input model of the assumption;
a sequential representation containing additional empty rows would
require its reading cost to be charged separately.

Set
\begin{equation}
 a_F=\max_{i,j}|F_{0,ij}|,\quad F=F_0/(ma_F),\quad b_F=F^\top b.
\end{equation}
Compute $b_F$ with exact Real-RAM operations before any rounding. Let
$P=m^{\lceil c_\kappa\rceil+1}$. Then
$\norm{F}\le1$ and $\sigma_{\min}(F)\ge P^{-1}$.
If $b_F=0$, full column rank gives $\Pi_{F_0}b=0$, so $x=0$ is exact.
Otherwise let $\nu_F=\norm{b_F}_\infty>0$ and $g_F=b_F/\nu_F$.
Construct $K_s,c_{\rm s}$ as in~\eqref{d:eq:complexity-saddle}.
Every entry has absolute value at most one,
$1\le\norm{c_{\rm s}}=\norm{g_F}\le\sqrt n$, the dimension is at most $2m$,
and the nonzero count is at most $3m$.

The exact saddle solution is
$(-Fx_{\rm s*},x_{\rm s*})$, where $F^\top Fx_{\rm s*}=g_F$, so
\begin{equation}
 x_{\rm s*}=F^\dagger b/\nu_F,
 \qquad x_*=(\nu_F/(ma_F))x_{\rm s*}=F_0^\dagger b.
 \label{d:eq:app-saddle-exact}
\end{equation}
The algorithm never forms $F^\top F$. It retains $ma_F,\nu_F$ as Real-RAM
words for final scaling; they are not short-rational physical input
coefficients.

\subsection{Saddle conditioning and the target error}

With $\Pi_{F_0}=F(F^\top F)^{-1}F^\top$, direct multiplication verifies
\begin{equation}
 K_s^{-1}=
 \begin{pmatrix}
 I-\Pi_{F_0}&F(F^\top F)^{-1}\\
 (F^\top F)^{-1}F^\top&-(F^\top F)^{-1}
 \end{pmatrix}.
 \label{d:eq:app-saddle-inverse}
\end{equation}
The four blocks have norm at most $1,P,P,P^2$, respectively. Hence
\begin{equation}
 \norm{K_s^{-1}}\le1+2P+P^2\le4P^2=:Q_s,
 \qquad \norm{K_s}\le2.
 \label{d:eq:app-saddle-bounds}
\end{equation}
For an approximate solution $(y_{\rm s},x_{\rm s})$, write
$\rho_1=y_{\rm s}+Fx_{\rm s}$ and $\rho_2=F^\top y_{\rm s}+g_F$. Eliminating $y_{\rm s}$ gives the exact
identity
\begin{equation}
 F(x_{\rm s}-x_{\rm s*})=\Pi_{F_0}\rho_1-F(F^\top F)^{-1}\rho_2.
 \label{d:eq:app-saddle-residual-identity}
\end{equation}
Therefore $\norm{F(x_{\rm s}-x_{\rm s*})}\le2P\norm{(\rho_1,\rho_2)}$.
Since $g_F=F^\top Fx_{\rm s*}$ and $\norm{F}\le1$,
$\norm{g_F}\le\norm{Fx_{\rm s*}}$. If the relative saddle residual is at
most $\eta_s$, the decoded vector $x=(\nu_F/(ma_F))x_{\rm s}$ satisfies
\begin{equation}
 \frac{\norm{F_0x-\Pi_{F_0}b}}{\norm{\Pi_{F_0}b}}
 =\frac{\norm{F(x_{\rm s}-x_{\rm s*})}}{\norm{Fx_{\rm s*}}}
 \le2P\eta_s.
 \label{d:eq:app-projection-relative}
\end{equation}
No factor $\norm{b}/\norm{\Pi_{F_0}b}$ occurs.

Choose $\eta_s=m^{-(6\lceil c_\kappa\rceil+3)}$. For $m\ge2$,
\begin{equation}
 2P\eta_s=2m^{-5\lceil c_\kappa\rceil-2}\le m^{-5\lceil c_\kappa\rceil}
                 \le n^{-5\lceil c_\kappa\rceil}\le n^{-5c_\kappa}.
\end{equation}
Squaring~\eqref{d:eq:app-projection-relative} proves exactly
\eqref{d:eq:slth-target}. Integer exponents are used to make the
parameters computable with basic arithmetic operations.

\subsection{Rounding to the square theorem's input class}

Choose a dyadic spacing
\begin{equation}
 u_s\le\frac{\eta_s}{1024mQ_s}.
\end{equation}
Round each existing nonzero of $K_s$ to that grid, using identical
roundings for the two occurrences of each $F$ entry, and round $g_F$ on
the same grid. Existing zeros remain zero. Write the resulting matrix
and source as $\widehat K_s,\widehat c_{\rm s}$. A conservative count of
the changed entries gives
\begin{align}
 \norm{\widehat K_s-K_s}
   &\le2mu_s\le\frac{\eta_s}{512Q_s},\\
 \norm{\widehat c_{\rm s}-c_{\rm s}}
   &\le mu_s\le\frac{\eta_s}{1024Q_s}
         \le\frac{\eta_s}{1024}\norm{c_{\rm s}}.
 \label{d:eq:app-saddle-rounding}
\end{align}
By the Neumann series and~\eqref{d:eq:app-saddle-bounds},
$\norm{\widehat K_s^{-1}}\le2Q_s$ and
$\norm{\widehat K_s}\le3$.

Suppose a returned vector $(y_{\rm s},x_{\rm s})$ satisfies
\begin{equation}
 \norm{\widehat K_s(y_{\rm s},x_{\rm s})-\widehat c_{\rm s}}
 \le\frac{\eta_s}{64}\norm{\widehat c_{\rm s}}.
\end{equation}
Then
\begin{equation}
 \norm{(y_{\rm s},x_{\rm s})}
 \le2Q_s(1+\eta_s/64)(1+\eta_s/1024)\norm{c_{\rm s}}
 <3Q_s\norm{c_{\rm s}}.
\end{equation}
Restoring the unrounded coefficients therefore gives
\begin{align}
 \norm{K_s(y_{\rm s},x_{\rm s})-c_{\rm s}}
 &\le\norm{\widehat K_s(y_{\rm s},x_{\rm s})-\widehat c_{\rm s}}
       +\norm{\widehat K_s-K_s}\norm{(y_{\rm s},x_{\rm s})}
       +\norm{\widehat c_{\rm s}-c_{\rm s}}\\
 &\le\eta_s\left(
       \frac{1+1/1024}{64}+\frac{3}{512}+\frac{1}{1024}
       \right)\norm{c_{\rm s}}
 <\frac{\eta_s}{16}\norm{c_{\rm s}}.
 \label{d:eq:app-unrounded-margin}
\end{align}
This margin permits any additional finite-precision decoding budget
to be allocated explicitly before testing the original residual.

Append an $m$ by $m$ identity block with zero source. The padded
matrix has nonzero count $m'$ between $m$ and $4m$, dimension at most
$3m$, and
\begin{equation}
 \kappa_2(\widehat K_s\oplus I_m)
 \le6Q_s=24m^{2\lceil c_\kappa\rceil+2}\le(m')^{2\lceil c_\kappa\rceil+7}.
\end{equation}
Its requested relative accuracy $\eta_s/64$ is at least
$(m')^{-(6\lceil c_\kappa\rceil+9)}$. Thus its condition and accuracy bounds have the
fixed-polynomial form required by the square realization theorem,
expressed in terms of the rounded input's own sparsity. All entries
have $O_{c_\kappa}(\log m)$ bits, and the padding costs $O(m)$.

The dyadic approximations can be computed in the basic Real RAM by
bisection and comparisons on $[-1,1]$, using $O(\log m)$ operations
per entry. No floor operation on unrestricted reals is needed.
Repeated doubling chooses the spacing using the explicit integer
exponent bounds. After this step, the Maxwell encoder operates on
short rational data with its usual bit-operation bounds.

\subsection{Verification, time caps, and amplification}

The decoded square-system answer has a prescribed finite rational
representation. Discard identity-padding coordinates and test the
original residual $\norm{K_s(y_{\rm s},x_{\rm s})-c_{\rm s}}\le\eta_s\norm{c_{\rm s}}$ using exact
Real-RAM operations. Comparing squared norms avoids a square-root
primitive. This test costs $O(m)$. Passing the test implies
\eqref{d:eq:slth-target}; the good-output bound
\eqref{d:eq:app-unrounded-margin} supplies substantial acceptance margin.
Final output uses $n$ multiplications by $\nu_F/(ma_F)$.

For each physical draw, apply a time cap equal to the solver's known
runtime bound for a good Maxwell instance at the relevant parameters. Reject
a timeout or an output that fails the prescribed interface or residual
check. Finite-bit subprograms use a private bounded address region.
Check each proposed index before storing or dereferencing it, and abort
the trial if it exceeds the prescribed polynomial address bound.
Retain the original input and verifier data outside that region; clear
the touched private cells on exit. The tests and cleanup are charged
to the capped trial. On executions within the input class the index
bounds make every guard inactive. On other executions the guards prevent a time cap from
being mistaken for a bound on the bit length of an index.
A good draw has probability at least $1-1/m$, and a successful
solver call on it has a fixed positive probability. Independent
repetition therefore gives a fixed positive acceptance probability
per trial for all sufficiently large inputs. A constant multiple of
$\log m$ capped trials makes the probability of no accepted answer
smaller than any chosen inverse polynomial. An accepted answer is
always valid, because acceptance uses the original real saddle system.
Finitely many small sizes can be handled directly.

The encoder and decoder cost $\Ot((m')^{3/2})=\Ot(m^{3/2})$ on every
draw. On a good draw,
$N=\Ot(m^{3/2})$, $B_{\rm out}=O(\log m)$, and both the condition number and
inverse residual tolerance are polynomial in $m$. Substituting these
quantities into~\eqref{d:eq:maxwell-solver-model} gives
$\Ot(m^{3\theta/2})$ for the capped solve. Verification costs
$O(m)$, and logarithmically many trials do not change the soft-order
bound. We obtain~\eqref{d:eq:sparse-composition-time}, completing the
transfer theorem.

\section{Conditional exponent restrictions}\label{si:sec:hardness}

\begin{corollary}[A conditional exponent restriction]
\label{d:cor:complexity}
If Assumption~\ref{d:ass:slth} holds for some $c_\kappa>0$ and
$\gamma>3/2$, a solver satisfying~\eqref{d:eq:maxwell-solver-model} cannot
exist with $\theta<2\gamma/3$ on the complete constructed Maxwell family.
\end{corollary}

The first term in~\eqref{d:eq:sparse-composition-time} is the encoding and
decoding cost. Thus the implication requires $\gamma>3/2$ even if the
Maxwell solve itself were arbitrarily fast. Under this condition and
$\theta<2\gamma/3$, both polynomial exponents on the right side of
\eqref{d:eq:sparse-composition-time} are strictly below $\gamma$;
logarithmic factors can be absorbed into a smaller positive exponent
gap. This proves the corollary from the transfer theorem.

The hardness exponent must be compatible with upper bounds for the same
input and accuracy model. Exact-arithmetic conjugate gradients on the
normal operator costs $O(mn)\le O(m^2)$, and its conditioning-dependent
convergence bound further restricts the admissible exponent to
$\gamma\le\min\{2,1+c_\kappa\}$; the derivation follows
Corollary~\ref{c:cor:hardness} below. Any stronger applicable upper bound further narrows this
necessary range. Figure~\ref{fig:consequence} of the main text displays the exponent
comparison; values outside an admissible hardness range have no
corresponding lower-bound consequence.

\subsection{The continuum corollary and known upper bounds}
Use exactly the randomized exponent-gap hypothesis of
Assumption~\ref{d:ass:slth}: for fixed $c_\kappa$ and $\gamma$, every
$\widetilde O(m^{\gamma'})$ randomized Real-RAM algorithm,
$\gamma'<\gamma$, fails to solve all instances of
\eqref{c:eq:projectiontarget} with success probability at least $2/3$.
It is a hypothesis, independent of Theorem~\ref{c:thm:main}. Hardness
statements in the unit-cost Real-RAM model depend on the set of unit-cost
operations; here it consists of $+,-,\times,\div$ and comparisons, and
excludes the floor function.

\begin{corollary}
\label{c:cor:hardness}
If that matching hypothesis holds for an admissible exponent
$\gamma>3/2$, a solver satisfying~\eqref{c:eq:solverbound} cannot have
$\theta<2\gamma/3$ on the complete constructed continuum family.
\end{corollary}
\begin{proof}
Under both strict inequalities, $3/2<\gamma$ and
$3\theta/2<\gamma$. The two exponents in~\eqref{c:eq:transfer} are
strictly smaller than $\gamma$; logarithmic factors can be absorbed
in a smaller positive exponent gap. The amplified algorithm has
success above $2/3$, contradicting the stated hypothesis.
\end{proof}

The admissibility qualification cannot be omitted. In the exact-arithmetic
Real-RAM model of Assumption~\ref{d:ass:slth}, conjugate gradients applied to the positive
 definite normal operator $F_0^TF_0$ solves the projected problem
exactly in at most $n$ iterations. Each normal-operator product is
two sparse products and costs $O(m)$; the remaining vector work is
$O(n)\le O(m)$. There is no need to form the normal matrix.
For completeness the recurrence uses only $+,-,\times,\div$:
\[
 a^{\rm cg}_j=\frac{\rho_j^T\rho_j}{p_j^T(F_0^TF_0)p_j},\quad
 x_{j+1}=x_j+a^{\rm cg}_jp_j,\quad
 \rho_{j+1}=\rho_j-a^{\rm cg}_j(F_0^TF_0)p_j,
\]
\[
 b^{\rm cg}_j=\frac{\rho_{j+1}^T\rho_{j+1}}{\rho_j^T\rho_j},\qquad
 p_{j+1}=\rho_{j+1}+b^{\rm cg}_jp_j,
 \quad \rho_0=p_0=F_0^Tb.
\]
Stop when $\rho_j=0$. The usual induction, substituting these identities,
makes the nonzero directions mutually $F_0^TF_0$-orthogonal and
makes the residual orthogonal to their span. Thus each nonterminal
step adds an independent direction, so after at most $n$ steps the
residual is zero. Positivity makes all used denominators nonzero.
This gives $O(mn)\le O(m^2)$ exact Real-RAM operations, including
zero-projection sources. The same exact-arithmetic upper bound is
explicitly recorded in~\citep[p.~3:2]{andoni2019}.

The conditioning assumption gives a further restriction when $c_\kappa<1$.
Let $x_*=F_0^\dagger b$.
The same orthogonality induction shows that the $t$th CG iterate
minimizes the energy error over its Krylov space. Therefore, for
every polynomial $p$ of degree at most $t$ with $p(0)=1$,
\[
 \frac{\|F_0(x_t-x_*)\|}{\|\Pi_{F_0}b\|}
 \le \max_{\lambda\in\spec F_0^TF_0}|p(\lambda)|.
\]
For completeness, the error candidates are $p(F_0^TF_0)x_*$: writing
$p(\lambda)=1-\lambda p_1(\lambda)$ makes the corresponding iterate
$p_1(F_0^TF_0)F_0^Tb$ belong to the Krylov space. Orthogonality to that
space proves the energy-minimization statement by Pythagoras.
On the interval $[\lambda_-,\lambda_+]$ containing the spectrum, use
\[
 p(\lambda)=
 \frac{T_t((\lambda_-+\lambda_+-2\lambda)/(\lambda_+-\lambda_-))}{T_t((\lambda_-+\lambda_+)/(\lambda_+-\lambda_-))}.
\]
The recurrence for Chebyshev polynomials gives $|T_t|\le1$ on
$[-1,1]$ and $T_t(s)=(w^t+w^{-t})/2$ for
$w=s+\sqrt{s^2-1}>1$. Hence, with
$\kappa=\kappa_2(F_0)$,
\[
 \frac{\|F_0(x_t-x_*)\|}{\|\Pi_{F_0}b\|}
 \le2\left(\frac{\kappa-1}{\kappa+1}\right)^t
 \le2e^{-2t/\kappa}.
\]
The case $\kappa=1$ terminates in one step. Thus
$O_{c_\kappa}(m^{c_\kappa}\log(2+m))$ iterations suffice for the target
$n^{-5c_\kappa}$, and finite termination also caps the count at $n$.
No square root is used in the CG algorithm; the displayed polynomials
only analyze it. For an arbitrary fixed real exponent $c_\kappa$,
one may choose any fixed rational $c_\kappa'>c_\kappa$ in a counted iteration
cap. The integer ceiling of $m^{c_\kappa'}$ is computable by integer
bisection on a fixed-power inequality, so this uses only the stated
Real-RAM operations. Taking $c_\kappa'$ arbitrarily close to $c_\kappa$ is enough
for the exponent-gap comparison.

Accordingly, the matching hypothesis is contradicted whenever
$\gamma>\min\{2,1+c_\kappa\}$. Its necessary upper-bound-compatible range
in this corollary is
\begin{equation}
 3/2<\gamma\le\min\{2,1+c_\kappa\},\qquad \theta\ge2\gamma/3
 \quad\hbox{conditionally on the matching hypothesis}.          \label{c:eq:gamma-range}
\end{equation}
For $c_\kappa\le1/2$ that range is empty; for $c_\kappa\ge1$ its displayed
upper endpoint is two. Admissibility also requires consistency
with any stronger algorithmic upper bound in the same model.
No hardness assertion for any member of this range is made here.
The finite-bit algorithms for sparse systems with short encoded
inputs~\cite{pengvempala2021,nie2022}, including the recent
$\widetilde O(mn\log(\kappa/\eta))$ result~\cite{kelner2026}, are
consistent with this restriction; it is not substituted for the unrestricted-real,
projection-relative arithmetic model above. The continuum construction
and its quantitative realization are independent of accepting any
sparse-system hardness hypothesis.

\subsection{Interpretation of the consequence}

The corollary concerns a uniform worst-case solver over the constructed
family. It identifies the computational consequences of supporting every
instance in that family with the stated dependence on conditioning and
accuracy. An algorithm whose cost is $N$ times a polynomial in
$\kappa_2(A)$ does not satisfy~\eqref{d:eq:maxwell-solver-model}: on these
instances the condition number is allowed to grow polynomially with the
input size. The distinction between polynomially bounded instances and
polylogarithmic algorithmic dependence on their condition number must be
retained when stating the result.

The theorem leaves room for faster algorithms on additional restricted
families, for instance when the admissible geometry or the conditioning
assumption is stronger. It also does not measure the time required by an
analog device to prepare a field, settle, or expose the prescribed output
precision. Those questions require a physical resource model in addition
to the linear-system realization proved here.

\sipart{V}{Computer-assisted certificates and reproducibility}
\section{Computer-assisted certificates and reproducibility}\label{d:sec:reproducibility}

Two finite computer-assisted proofs enter the construction: the
spectral certificate of the period-20 Yee background (Part~III) and the
corner certificate of the continuum background (Part~II). Each checks
finitely many exact inequalities; the analytic arguments of the
respective part extend them to the infinite-dimensional operators. The
logical compiler of Part~I uses no certificate. Its worked example in
Figure~\ref{d:fig:overview} is an exact finite computation.

\subsection{The Yee spectral certificate}\label{si:sec:yee-certificate}

The spectral construction uses exact finite inequalities for the specified
periodic Maxwell operator. Its certificate data comprise 604 bulk-gap
matrices, 2056 global-comparator matrices, local point and neighborhood
data, and a reciprocal-space cover. The matrix and local estimates enter
the spectral assembly in Section~\ref{d:app:certificate}; the resulting
band edge is the premise for the analytic defect and interaction proofs.

The checker reconstructs the incidence and material operators, symmetry
maps, interval inequalities and geometric coverage from the certificate
inputs. Integer and rational operations are exact. Interval computations
use outward bounds with their rounding errors included. Proposed
enclosures are supplied as data arrays, and the required matrix
inequalities are tested against the reconstructed operator.

In summary, the Yee certificate checks:
\begin{itemize}
\item the exact integer material table~\eqref{d:eq:physical-exact-material},
its primitive translations and its scalar inversion symmetry;
\item the eigenvalue enclosures and inertia counts at the four
inequivalent $\mathrm{L}$ points, the center ordering that selects $\mathrm{L}_{++-}$,
and the two-mode cluster data;
\item the local matrix inequalities~\eqref{d:eq:appendix-local-matrix-bounds}
and the uniform inertia count on each radius-$1/20$ neighborhood;
\item the bulk-gap and global-comparator interval matrices and the exact
polytope cover of the irreducible wedge, which give the global
margin~\eqref{d:eq:appendix-global-margin}.
\end{itemize}
All groups are needed for the global spectral conclusion.

\paragraph{Status.} The checker has been run on the complete certificate data,
and all finite inequalities listed above pass. The final step of the argument,
i.e., a complete and independently verified binding of the checked point,
center, derivative and global matrix data to the exact Yee operator, including
the argument in the punctured neighborhoods of the selected band minimum, is
being completed as part of an ongoing formal verification. Until this step is
complete, Theorem~\ref{d:thm:main} and the Yee transfer
Theorem~\ref{d:thm:algorithm-transfer} are conditional on it. The continuum
results of Part~II do not depend on the Yee certificate.

\subsection{The continuum corner certificate}\label{c:app:certificate}

The continuum checker is a standard-library program in exact rational
arithmetic. It represents $s+t\sqrt{19}$ by the exact rational pair
$(s,t)$. Multiplication uses $\sqrt{19}^2=19$, and inversion uses
$(s-t\sqrt{19})/(s^2-19t^2)$ after checking a nonzero denominator.
It reconstructs the 24-component corner projection $P_\varsigma$ and the
actual neighboring-wavevector coupling $-2^{j-2}P_\varsigma P_{\varsigma'}$.
For each of the eight reflection parities it verifies its
intertwining with~\eqref{c:eq:parityblock}, projection invariance,
the Gram normalization, pairwise parity orthogonality, and the
trace and trace-square identities
for the two transverse roots. The velocity Gram identities are
checked on the reconstructed corner vector, not assumed from a
list of eigenvalues. It then checks the three exact Hessian
coefficients~\eqref{c:eq:explicitA0} by inversion in
$\Q(\sqrt{19})$.

For the order and placement comparisons, rational interval
arithmetic encloses all sixteen roots in~\eqref{c:eq:clusterroots},
$\pi$, and the necessary square roots. Square-root endpoints are
obtained by integer square root after scaling by $2^{440}$;
rounding the lower endpoint down and the upper endpoint up
rigorously encloses the true value. Machin's formula uses 128 terms
of each alternating arctangent series and the next-term error.
Logarithms use
\[
 \log s=2\sum_{j=0}^{127}\frac{w^{2j+1}}{2j+1}+R,
 \qquad w=\frac{s-1}{s+1},\qquad
 |R|\le\frac{2|w|^{257}}{257(1-w^2)}.
\]
This formula follows by integrating the geometric series and is
used only for $s=2,3$. Every interval operation is exact rational
endpoint arithmetic. Induction on the operations proves enclosure;
no machine floating-point value enters the calculation.

The accepted inequalities verify the cluster gap $>1/20$, the
finite-contour perturbation inequalities in Lemma~\ref{c:lem:edge},
the Hessian and quartic bounds used there, the fixed integer
placement~\eqref{c:eq:Bmetric} including the $2^{-60}$ Hessian error,
and the strict rate comparisons
$2(.99)^2/(7/4)>\log3$, which with $\sqrt3<7/4$ gives
$2(.99)^2/\sqrt3>\log3$, and
$\log8-2(1.01)^2>1/50$. The verifier contains all its finite input
data. It also checks the small-integer logarithmic inequality for the
fixed majorant ledger of Section~\ref{c:app:ledger}. It uses no
downloaded data, mesh eigensolver or special-function package.

These finite algebraic inequalities enter the continuum estimates at the
places specified above. The resolvent, auxiliary-mode, interaction,
leakage and boundary arguments control their infinite-dimensional
extensions. Sections~\ref{c:sec:parameters} and~\ref{c:sec:cost} give the
terminating parameter choices and account for the complete encoding cost.

\subsection{Formal proofs and verification records}

The formal verification of the continuum construction in Lean is
substantially complete. The companion repository includes Lean proofs for
the continuum operator and its PEC domain, the finite-box spectral count and
detuning, the local Bloch band, the isolated defect mode and interactions,
leakage and boundary estimates, finite function representations, and
residual-to-solution recovery. Shared finite algebra, routing and
arithmetic routines also support the Yee construction. The
\href{https://github.com/yaugenst/maxwell-universality/blob/main/lean/proof-map.md}{theorem map}
links these statements to their source declarations and records the
hypotheses at each interface.

The verification records identify the checked declarations, their
dependency closure, source hashes and pinned Lean and library versions.
Kernel replay checks the proof terms and restricts the accepted axioms
to propositional extensionality, classical choice and quotient soundness.
The records distinguish a component check from a replay of its complete
dependency closure, making the scope of each verification reproducible.

\subsection{Rerunning the checks}

The accompanying research materials contain both checkers, their
complete input data, and the data behind the figures. The Yee checker
has a single entry point that runs the matrix, local and coverage
checks; an input manifest identifies the coefficient and enclosure data
used by each check. The continuum checker requires only a standard
Python interpreter and its standard library. It must be run with
assertions enabled, and it refuses to run otherwise. On success it
prints a final confirmation line stating that all exact rational checks
passed and writes a receipt recording the rational intervals and exact
algebraic identities. The receipt can be regenerated and is not a
substitute for running the checks.

The exact logical example in Figure~\ref{d:fig:overview} includes its sparse
matrices, paths, residuals and decoded coordinates.
Figures~\ref{d:fig:FIG-2} and~\ref{d:fig:FIG-3} use the material and spectral
certificate data.
The geometric and defect drawings are schematic.

\sipart{VI}{Numerical illustrations}
\section{Numerical realizations and controls}\label{n:sec:numerical}

We consider two finite dielectric realizations of prescribed symmetric systems
$\Gamma x=b$. The first uses whole-cavity calibration in a three-dimensional
periodic medium; the second uses one isolated-component calibration to assemble
a family of chains in a rectangular waveguide. Both solve the lossless,
full-vector Maxwell equation~\eqref{eq:maxwell}, with scalar positive
permittivity and PEC outer walls. Frequency and permeability are one, and all
lengths below are in the corresponding dimensionless units.

\subsection{Sources, readout and error}

Let $B$ collect compact source profiles at the terminal sites, let $V$ be a
diagonal matrix of fixed terminal gains and let $\alpha>0$ set the coupling
scale. The input and readout are
\begin{equation}
 f=\alpha BVb,\qquad x=\alpha Vb+VB^{\mathsf T}E.
 \label{n:eq:source-readout}
\end{equation}
In the discretization, $E$ and the columns of $B$ include the square root of
cell volume, so $B^{\mathsf T}E$ approximates the corresponding volume
integrals. These signed overlaps preserve the sign of the solution. Once the
material and gains are chosen, changing $b$ changes only the coefficients of
the source profiles. Source programming and decoding use only local operations;
the target solution shown in the figures is computed afterward for comparison.

Solving for all basis sources gives a decoded response matrix $X$. We report
\begin{equation}
 e_{\rm eq}=\norm{\Gamma X-I}_2,
 \qquad
 \frac{\norm{\Gamma Xb-b}_2}{\norm b_2}\le e_{\rm eq}
 \quad (b\ne0).
 \label{n:eq:error}
\end{equation}
Quantitative readout uses the original staggered fields and probes. The
interpolation to a central section and the field normalization in
\cref{fig:numerical} serve only to display the computed phasors.

\subsection{A calibrated periodic cavity}

The first cavity has dimensions $8\times6\times8$ and background
\begin{equation}
 \varepsilon_{\rm bg}(r)=7+E_c-0.3+
 \varrho\bigl[\cos(\pi r_1)+2\cos(\pi r_2)+4\cos(\pi r_3)\bigr],
 \qquad \varrho=\tfrac12,
 \label{n:eq:periodic-background}
\end{equation}
where $E_c=1.5725344295489645$ is the eigenvalue at $k_*$ of the selected
corner branch of ${\nabla\times\nabla\times-\varepsilon_{\rm per}}$ for this
modulation, computed in a truncated plane-wave basis. The selected branch
of ${\nabla\times\nabla\times-\varepsilon_{\rm bg}}$ therefore has eigenvalue
$0.3$ at $k_*$. Six radius-one bumps add
$v_i w(\lVert r-c_i\rVert)$, where
$w(s)=\exp[-1/(1-s^2)]$ for $|s|<1$ and zero otherwise. The terminal centers
are $(-2,0,-2)$, $(2,0,-2)$ and $(-2,0,2)$; the auxiliary centers are
$(0,0,-2)$, $(-2,0,0)$ and $(0,0,0)$. The target is
\begin{equation}
 \Gamma=\begin{pmatrix}3&-0.6&0.4\\-0.6&2.5&0.8\\0.4&0.8&3.5\end{pmatrix},
 \qquad \kappa_2(\Gamma)=2.367.
 \label{n:eq:periodic-target}
\end{equation}
The full-cavity calibration changes only the six bump strengths, keeping the
source profiles, $\alpha=0.01804944288849926$ and the terminal gains fixed.
The final strengths are approximately
$(14.87272,23.66068,12.48289,3.17042,25.71297,5.06464)$, with continuous
material enclosure $[4.7725344,21.2318084]$.

We use compatible staggered derivatives of order eight on a uniform grid of
spacing $h$. Material calibration
at $h=1/6$ gives $e_{\rm eq}=3.68\times10^{-12}$. Holding the material,
source functions and decoder fixed gives $e_{\rm eq}=3.9657\times10^{-3}$
at $h=2/13$ and $7.8782\times10^{-4}$ at $h=1/7$, the latter on
379\,610 electric unknowns. For the
displayed input $b=(1,0,0)$, the decoded answer is
$(0.3638851,0.1083593,-0.0663881)$ and its relative equation error is
$5.6211\times10^{-4}$. The latter is one input's error, while
\eqref{n:eq:error} measures the complete response. Algebraic Maxwell residuals
are below $2\times10^{-13}$. The change between the calibration grid and the
finer grid measures the sensitivity to discretization at this fixed design.

\subsection{Local assembly in a waveguide}

The second platform has a $1\times0.5$ rectangular PEC cross-section, with
\begin{equation}
 \varepsilon(z)=1+\sum_i v_i w(z-z_i).
 \label{n:eq:waveguide-material}
\end{equation}
Each well fills the transverse cross-section and varies smoothly along $z$.
The field sector
$E(r)=\hat y\sin[\pi(x+1/2)]e(z)$ is invariant, and its equation is
\begin{equation}
 \left[-\frac{d^2}{dz^2}+\pi^2-\varepsilon(z)\right]e(z)=f(z).
 \label{n:eq:channel}
\end{equation}
The background is below cutoff, with decay rate
$\kappa=\sqrt{\pi^2-1}=2.978188\ldots$. We calibrate one well in a
length-12 interval at spacing $0.005$, using the largest eigenvalue of the
compressed background resolvent
$w^{1/2}(-d^2/dz^2+\kappa^2)^{-1}w^{1/2}$. This gives
$\mu=0.02909782699697072$, with resonant strength $1/\mu=34.36682746\ldots$.
The isolated field has $99.0009\%$ of its unweighted electric-field norm
squared inside the well (\cref{n:fig:controls}a). A sine-series interpolation
of this field, multiplied by the compact bump and divided by $\mu$, supplies
one smooth source profile $q$ that is translated to every site.

The infinite-background Green function is
$G(z,z')=e^{-\kappa|z-z'|}/(2\kappa)$. For well separation $d>2$, it gives
the nearest-neighbor coupling
\begin{equation}
 t(d)=\frac{e^{-\kappa d}}{2\kappa}
 \left(\int q(s)e^{\kappa s}\,ds\right)
 \left(\int q(s)e^{-\kappa s}\,ds\right).
 \label{n:eq:pair-coupling}
\end{equation}
At $d=4$, $t=1.2284541758091573\times10^{-6}$; a separate two-well background
solve agrees to relative error below $7\times10^{-13}$. Retaining the
calibrated local modes gives diagonal entries $v_i^{-1}-\mu$ and neighboring
entries $-t$, which yields the rule
\begin{equation}
 v_i=(\mu+t\Gamma_{ii})^{-1},\qquad \alpha=t.
 \label{n:eq:strength}
\end{equation}
The transverse factor in each column of $B$ is normalized in the cross-section,
and $V$ contains the corresponding terminal strengths. The full Maxwell solve
then measures the errors due to the omitted local modes and non-neighboring
interactions.

We prescribe three- and four-site chains with diagonal 3, a three-site chain
with diagonal $(2.5,3,4)$, and an indefinite three-site chain with diagonal 1;
each has neighboring entries $-1$. The remaining target uses three wells with
reduced diagonal $(2,-1,2)$ and neighboring entries $-1$. Eliminating the
unforced middle well gives the terminal Schur complement
\begin{equation}
 \Gamma_{\rm term}=\begin{pmatrix}3&1\\1&3\end{pmatrix}.
 \label{n:eq:mediator}
\end{equation}
For $b=(1,0)$, its measured terminal answer is approximately
$(0.37550,-0.12507)$, compared with $(0.375,-0.125)$. The negative reduced
diagonal is a detuning of a positive dielectric: the five designs all have
$1\le\varepsilon<13.644$. In particular, the coefficient $v_i\simeq34.36$
multiplies a bump of peak $1/e$.

\subsection{Refinement and component transfer}

All five designs use the same isolated calibration and coupling rule, with
no fitting against their assembled responses. Wells are centered at
$z_i=4[i-(n-1)/2]$, $i=0,\ldots,n-1$, in a waveguide of length $12+4(n-1)$.
For verification, we assemble the complete three-component matrix
$C^{\mathsf T}C-\operatorname{diag}(\varepsilon)$ with compatible derivatives
of order 12, and solve for every basis input. A design meets the accuracy
criterion if its equation error is below $1\%$ on the two finer of three
grids. The invariant channel follows from
the material and source symmetry; the numerical operator has no modal
projection. The largest system has 380\,100 electric unknowns, and all
algebraic residuals are below $10^{-8}$.

\begin{table}[tbp]
\centering
\caption{\textbf{Fixed-design waveguide refinement.} Entries are the percentage
equation error $100\norm{\Gamma X-I}_2$. Each row holds its material,
continuous source functions and decoder fixed across the three full-vector
grids. The final column uses the same design and source in the independent
one-dimensional channel calculation at longitudinal spacing $0.005$.
The coarsest indefinite-chain error exceeds the $1\%$ criterion; all five
targets meet it on both finer full-vector grids.}
\label{n:tab:waveguide-errors}
\small
\setlength{\tabcolsep}{4pt}
\begin{tabular}{@{}lccccc@{}}
\toprule
Target & $(h_1^{-1},h_2^{-1},h_3^{-1})$ & $h_1$ & $h_2$ & $h_3$ & Fine channel \\
\midrule
Three-site chain & $(16,20,22)$ & 0.6679 & 0.3219 & 0.1443 & 0.0158 \\
Unequal diagonals & $(16,20,22)$ & 0.6853 & 0.3301 & 0.1478 & 0.0160 \\
Four-site chain & $(16,20,22)$ & 0.7684 & 0.3707 & 0.1667 & 0.0189 \\
Two terminals + mediator & $(16,20,22)$ & 0.7906 & 0.3822 & 0.1727 & 0.0189 \\
Indefinite chain & $(20,22,24)$ & 1.2132 & 0.5482 & 0.1889 & 0.0587 \\
\bottomrule
\end{tabular}

\end{table}

The full-vector errors decrease on refinement (\cref{n:fig:controls}b,
\cref{n:tab:waveguide-errors}). The independent finer channel calculation
separates the remaining model error from the observed grid dependence.
Halving the isolated calibration spacing to $0.0025$ changes $\mu$ by
$7.66\times10^{-14}$, or $6.23\times10^{-8}$ of the useful coupling. Keeping
the sources and material fixed while comparing channel spacings $0.01$ and
$0.005$ changes the equation errors in the 70 spacing-and-rounding cases
below by at most $6.6\times10^{-9}$. These are numerical refinement checks;
a small algebraic solve residual alone does not certify continuum accuracy.

Extending the three-site waveguide from length 20 to 24, with unchanged
centers, materials, source functions and decoder, changes $X$ by relative
spectral norm $2.00\times10^{-11}$ at $h=1/20$ and $1.20\times10^{-11}$ at
$h=1/22$. This insensitivity to the outer walls supports the use of isolated
components in this platform.

For comparison, compact periodic components at the modulation
$\varrho=0.5$ of~\eqref{n:eq:periodic-background} did not transfer reliably.
We prescribed $\Gamma=\bigl(\begin{smallmatrix}a&1\\1&a\end{smallmatrix}\bigr)$,
$a=3,4$, and calibrated the material against single- and two-component
cavities at defect radii $R=3,4,5$, with calibration-box side $2R+4$ and
detuning $0.3/R^2$. Five local models fit their targets to
equation error below $3\times10^{-13}$, but their assembled finest-grid
errors were $59.0\%$, $82.8\%$, $180.6\%$, $98.4\%$ and $97.7\%$
(\cref{n:fig:controls}c). The remaining case, $R=3$ and $a=3$, stopped with
local-model error $35.2\%$, so no assembled verification was performed.
The archived single- and pair-component tests also show wall dependence and
direct interactions that bypass the intended mediator. These compact
parameters are outside the sufficient scales of the continuum construction.

\subsection{Spacing and material precision}

Increasing the well separation suppresses distant interactions and weakens
the intended coupling at the same time. For the four-site chain, changing
$d=3$ to $d=4$ lowers the nominal fine-channel error from $0.3714\%$ to
$0.0189\%$, while the useful coupling decreases by a factor of 19.65. A
full-vector solve at $d=3$ and $h=1/22$ gives error $0.3788\%$ without
retuning.

\begin{figure}[!htbp]
\centering
\includegraphics[width=\linewidth]{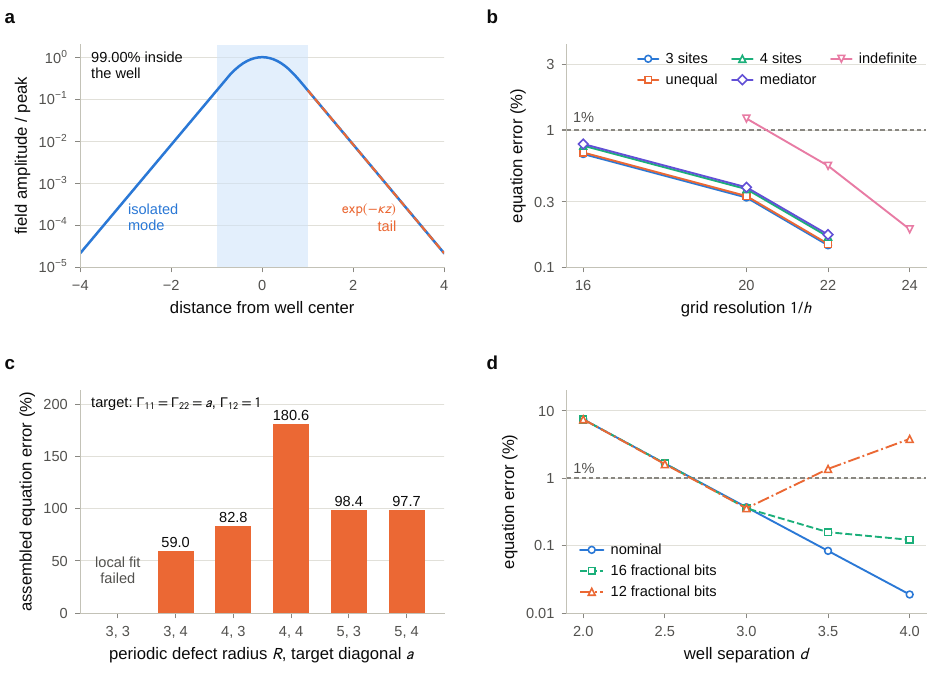}
\caption{\textbf{Confinement, assembly and material precision in the numerical examples.}
\textbf{a}, Isolated-waveguide field amplitude, normalized to its peak;
shading marks the well support. The $99.00\%$ fraction refers to the unweighted
electric-field norm squared inside that support. The dashed tail has the
background cutoff exponent $\kappa=\sqrt{\pi^2-1}$.
\textbf{b}, All-input equation errors of the five prescribed waveguide targets
under full-vector refinement, including the coarsest indefinite case above
$1\%$.
\textbf{c}, Assembled periodic-device errors after calibration against local
single- and pair-component responses. The labels give defect radius $R$ and
target diagonal $a$; the first case did not meet its local-fit criterion.
\textbf{d}, Four-site waveguide errors versus separation in the fine channel
calculation. Nominal strengths and strengths rounded to 12 or 16 fractional
binary bits use identical source functions and decoder gains. Markers denote
computed cases, with connecting lines as guides.}
\label{n:fig:controls}
\end{figure}

We round only the material strengths to a grid of spacing $2^{-p}$, keeping
the source profiles, source gains and decoder at their nominal values.
For $p=12$, the four-site errors are $0.3585\%$ at $d=3$ and $3.8246\%$ at
$d=4$; at $d=4$, using $p=16$ gives $0.1220\%$
(\cref{n:fig:controls}d). Thus the more isolated assembly requires finer
material settings to preserve the accuracy of its weaker coupling. The
rounding study contains 70 cases: five separations from 2 to 4 in steps of
$0.5$, the four-site
chain and the two-terminal mediator, and nominal strengths or
$p\in\{8,12,16,20,24,28\}$. These particular roundings can cancel other
errors and do not give worst-case tolerance bounds. Geometric errors,
source and detector hardware, dielectric loss and finite wall conductivity
remain outside the numerical model.

The accompanying numerical archive contains every prescribed target,
component calibration, material design, verification result and failed
transfer test used here. The figure script reconstructs the readouts from
the saved full-grid fields and probes, checks them against the reported
responses, and stores compact plotting data with hashes of the original
files. Regenerating the figures from these bundled data requires no Maxwell
solve.

Local calibration and figure generation run on the CPU. The full-vector
calculations use a GPU direct solver in 64-bit arithmetic, with two host
numerical-library threads and one concurrent job per GPU. For the
379\,610-unknown periodic case, the factorization plan estimated 112.6~GB of
peak device memory. This estimate concerns the chosen matrix and backend;
other solvers can have different memory requirements.

\end{document}